\documentclass{article}
\usepackage{amssymb}

\input{tcilatex}

\begin{document}

\begin{center}
\bigskip

{\LARGE QUANTUM\ ENTANGLEMENT\ FOR\bigskip }

{\LARGE SYSTEMS\ OF\ IDENTICAL\ BOSONS\bigskip }

{\LARGE I. GENERAL\ FEATURES\bigskip }

{\LARGE \ \ }

\bigskip

B. J. Dalton$^{1,2}$, J. Goold$^{1,3}$, B. M. Garraway$^{4}$ and M. D. Reid$%
^{2}${\LARGE \bigskip }

$^{1}$\textit{Physics Department, University College Cork, Cork City, Ireland%
}

$^{2}$\textit{Centre for Quantum and Optical Science}$^{\ast }$\textit{,
Swinburne University of Technology, Melbourne, Victoria 3122, Australia}

$^{3}$\textit{The Abdus Salam International Centre for Theoretical Physics
(ICTP), Trieste 34151, Italy}

$^{4}$\textit{Department of Physics and Astronomy, University of Sussex,
Falmer, Brighton BN19QH, United Kingdom}

\textit{\bigskip \bigskip \bigskip }
\end{center}

Email: bdalton@swin.edu.au

$\ast $ Formerly \textit{Centre for Atom Optics and Ultrafast Spectroscopy}

\pagebreak

$\mathbf{Abstract}$

These two accompanying papers are concerned with two mode entanglement for
systems of identical massive bosons and the relationship to spin squeezing
and other quantum correlation effects. Entanglement is a key quantum feature
of composite systems in which the probabilities for joint measurements on
the composite sub-systems are no longer determined from measurement
probabilities on the separate sub-systems. There are many aspects of
entanglement that can be studied. This two-part review focuses on the
meaning of entanglement, the quantum paradoxes associated with entangled
states, and the important tests that allow an experimentalist to determine
whether a quantum state - in particular, one for massive bosons is
entangled. An overall outcome of the review is to distinguish criteria (and
hence experiments) for entanglement that fully utilise the symmetrisation
principle and the super-selection rules that can be applied to bosonic
massive particles.

In the first paper (\textbf{I}), the background is given for the meaning of
entanglement in the context of systems of identical particles. For such
systems, the requirement is that the relevant quantum density operators must
satisfy the symmetrisation principle and that global and local
super-selection rules prohibit states in which there are coherences between
differing particle numbers. The justification for these requirements is
fully discussed. In the second quantisation approach that is used, both the
system and the sub-systems are modes (or sets of modes) rather than
particles, particles being associated with different occupancies of the
modes. The definition of entangled states is based on first defining the
non-entangled states - after specifying which modes constitute the
sub-systems. This work mainly focuses on the two mode entanglement for
massive bosons, but is put in the context of tests of local hidden variable
theories, where one may not be able to make the above restrictions. The
review provides the detailed arguments necessary for the conclusions of a
recent paper, where the question of how to rigorously demonstrate the
entanglement of a two-mode Bose-Einstein condensate (BEC) has been examined.

In the accompanying review paper (\textbf{II}), we consider spin squeezing
and other tests for entanglement that have been proposed for two-mode
bosonic systems. We apply the approach of review (I) to determine which
tests, and which modifications of the tests, are useful for detecting
entanglement in massive bosonic (BEC), as opposed to photonic, systems.
Several new inequalities are derived, a theory for the required two-mode
interferometry is presented, and key experiments to date are analysed.

\bigskip

\bigskip


\begin{center}
\pagebreak
\end{center}

{\Large Contents\bigskip }

{\Large 1. Introduction}

\qquad\ \textbf{1.1 Entanglement: Definition and Historical Context}

\textbf{\qquad\ 1.2 Measures and Tests for Entanglement}

\textbf{\qquad\ 1.3 Particle versus Mode Entanglement}

\textbf{\qquad\ 1.4 Symmetrization and Super-Selection Rules}

\textbf{\qquad\ 1.5 Entanglement Tests and Experiments - Paper II}

\textbf{\qquad\ 1.6 Outline of Papers I and II}{\Large \medskip }

{\Large 2. Entanglement - General Features}

\qquad\ \textbf{2.1 Quantum States}

\textbf{\qquad 2.2 Entangled and Non-Entangled States}

\qquad \qquad \textit{2.2.1 General Considerations}

\qquad \qquad \textit{2.2.2 Local Systems and Operations}

\qquad \qquad \textit{2.2.3 Constraints on Sub-System Density Operators}

\qquad \qquad 2.2.4 \textit{Classical Entanglement}

\qquad \textbf{2.3 Separate, Joint Measurements. Reduced Density Operator}

\qquad \qquad \textit{2.3.1 Joint Measurements on Sub-Systems}

\qquad \qquad \textit{2.3.2 Single Sub-System Measurements. Reduced Density
Operator}

\qquad \qquad \textit{2.3.3 Mean Value and Variance}

\qquad \qquad \textit{2.3.4 Conditional Probabilities}

\qquad \qquad \textit{2.3.5 Conditional Mean and Variance}

\qquad \textbf{2.4 Non-Entangled States}

\qquad \qquad \textit{2.4.1 Non-Entangled States - Joint Measurements on
Sub-Systems}

\qquad \qquad \textit{2.4.2 Non-Entangled States - Single Sub-System
Measurements}

\qquad \qquad \textit{2.4.3 Non-Entangled States - Conditional Probability}

\qquad \qquad \textit{2.4.4 Non-Entangled States - Mean Values and
Correlations}

\qquad \textbf{2.5 Local Hidden Variable Theory (LHV)}

\qquad \qquad \textit{2.5.1 LHV - Mean Values and Correlations}

\qquad \qquad \textit{2.5.2 LHV - GHZ State}

\qquad \textbf{2.6 Paradoxes}

\qquad \qquad \textit{2.6.1 EPR Paradox}

\qquad \qquad \textit{2.6.2 Schrodinger Cat Paradox}

\qquad \textbf{2.7 Bell Inequalities}

\qquad \qquad \textit{2.7.1 LHV Result}

\qquad \qquad \textit{2.7.2 Non-Entangled State Result}

\qquad \qquad \textit{2.7.3 Bell Inequality Violation and Entanglement}

\qquad \textbf{2.8 Non-Local Correlations}

\qquad \qquad \textit{2.8.1 LHV Result}

\qquad \qquad \textit{2.8.2 Non-Entangled State Result}

\qquad \qquad \textit{2.8.3 Correlation Violation and Entanglement}{\Large %
\medskip }

{\Large 3. Identical Particles and Entanglement}

\qquad \textbf{3.1 Symmetrisation Principle}

\qquad \qquad \textit{3.1.1 Sub-Systems -\ Particles or Modes ?}

\qquad \qquad \textit{3.1.2 Multi-Mode Sub-Systems}

\qquad \textbf{3.2 Super-Selection Rules - General}

\qquad \qquad \textit{3.2.1 Global Particle Number SSR}

\qquad \qquad \textit{3.2.2 Examples of Global SSR Compliant States}

\qquad \qquad \textit{3.2.3 SSR and Conservation Laws}

\qquad \qquad \textit{3.2.4 Global SSR Compliant States and Quantum
Correlation Functions}

\qquad \qquad \textit{3.2.5 Testing the SSR}

\qquad \qquad \textit{3.2.6 SSR Justification - No Suitable Phase Reference}

\qquad \qquad \textit{3.2.7 SSR Justification - Physics Considerations}

\qquad \qquad \textit{3.2.8 SSR Justification - Galilean Frames ?}

\qquad \qquad \textit{3.2.9 SSR and Photons}

\qquad \textbf{3.3 Reference Frames and SSR Violation}

\qquad \qquad \textit{3.3.1 Linking SSR and Reference Frames}

\qquad \qquad \textit{3.3.2 \textbf{Can C}oherent Superpositions of an Atom
and a Molecule Occur ?}

\qquad \qquad \textit{3.3.3 Detection of SSR Violating States}

\qquad \textbf{3.4 Super-Selection Rules - Separable States}

\qquad \qquad \textit{3.4.1 Local Particle Number SSR}

\qquad \qquad \textit{3.4.2 Criteria for Local and Global SSR in Separable
States}

\qquad \qquad \textit{3.4.3 States that are Global but not Local SSR
Compliant}

\qquad \qquad \textit{3.4.4 Particle Entanglement Measure}

\qquad \qquad \textit{3.4.5 General Form of Non-Entangled States}

\qquad \textbf{3.5 Bipartite Systems}

\qquad \qquad \textit{3.5.1 Two Single Modes}\textbf{\ - }\textit{Coherence
Terms}

\qquad \qquad \textit{3.5.2 Two Pairs of Modes - Coherence Terms\medskip }

{\Large 4. Discussion and Summary of Key Results\bigskip }

\textbf{References}

\textbf{Acknowledgements\bigskip }

{\Large Appendices\bigskip }

{\Large A. Projective Measurements and Conditional Probabilities}

\qquad \textbf{A.1 Projective Measurements}

\qquad \textbf{A.2 Conditional Probabilities}

\qquad \textbf{A.3 Conditional Mean and Variance\medskip }

{\Large B. Inequalities}

\qquad \textbf{B.1 Integral Inequality}

\qquad \textbf{B.2 Sum Inequality\medskip }

{\Large C. EPR Spin Paradox}

\qquad \textbf{C.1 Local Spin Operators}

\qquad \textbf{C.2 Conditioned Variances\medskip }

{\Large D. Extracting Entanglement due to Symmetrisation}

\qquad \textbf{D.1 Three Particle Case - Bosons}

\qquad \textbf{D.2 Two Particle Case - Bosons}

\qquad \textbf{D.3 Two Particle Case - Fermions \medskip }

{\Large E. Reference Frames and Super-Selection Rules}

\qquad \textbf{E.1 Two Observers with Different Reference Frames}

\qquad \textbf{E.2 Symmetry Groups}

\qquad \textbf{E.3 Relationships - Situation A}

\qquad \textbf{E.4 Relationships - Situation B}

\qquad \textbf{E.5 Dynamical and Measurement Considerations}

\qquad \textbf{E.6 Nature of Reference Frames}

\qquad \textbf{E.7 Relational Description of Phase References}

\qquad \textbf{E.8 Irreducible Matrix Representations and SSR}

\qquad \textbf{E.9 Non-Entangled States\medskip }

{\Large F. Super-Selection Rule Violations ?}

\qquad \textbf{F.1 Preparation of a Coherent Superposition of an Atom and a
Molecule ?}

\qquad \qquad \textit{F.1.1 Hamiltonian}

\qquad \qquad \textit{F.1.2 Initial State}

\qquad \qquad \textit{F.1.3 Implicated Reference Frame}

\qquad \qquad \textit{F.1.4 Process - Alice and Charlie Descriptions}

\qquad \qquad \textit{F.1.5 Interference Effects Without SSR Violation}

\qquad \qquad \textit{F.1.6 Conclusion}

\qquad \textbf{F.2 Detection of Coherent Superposition of a Vacuum and
One-Boson State ?}{\Large \medskip }

\pagebreak

\section{Introduction}

\label{Section- - Intrduction}

Since the paradoxes of Einstein-Podolski-Rosen (EPR) \cite{Einstein35a},
Schrodinger \cite{Schrodinger35a} and Bell \cite{Bell65a}, \emph{entanglement%
} has been recognised as a key feature that distinguishes quantum physics
from classical physics. Entangled quantum states underlie the EPR and Bell
paradoxes, which reveal the conflict between quantum mechanics and local
realism, and the famous Schrodinger cat paradox, where a cat is apparently
prepared in a state simultaneously both dead and alive. Entanglement not
only provides a way to rigorously test quantum principles, but is the basis
for the many quantum information tasks like quantum cryptography. Despite
the fundamental interest, there have been only a few experimental tests of
entanglement for systems of massive particles. Yet, the substantial recent
progress in cooling atomic systems, in particular to form Bose-Einstein
condensates, makes such entanglement tests much more feasible.

In this review (\textbf{I}), we explain the meaning of entanglement, and
examine how to verify entanglement, for systems of identical boson
particles. This leads us to focus on symmetrisation and superselection
rules, and to consider their implication for entanglement criteria when
applied to massive bosonic particles. This paper provides the theoretical
background for a recent paper \cite{Dalton14a} and a subsequent review (%
\textbf{II}) that analyses the suitability of specific criteria, new and
old, to detect entanglement in bosonic systems, and applies the criteria to
interpret experimental findings.

As well as reviewing the topic and presenting some new results in review II,
these two articles are intended as comprehensive papers for post docs and
postgraduates who are changing field or starting work in a new one and need
to gain a thorough understanding of the present state of knowledge. With
this aim in mind we have not followed the conventional approach in review
articles of merely quoting formulae and giving references, but instead have
presented full proofs of the key results. To really understand a field, we
believe it is necessary to work through the derivations. However, in order
to shorten the main body of the articles, we have included many of the
details in Appendices.

\subsection{Entanglement: Definitions and Historical Context}

Entanglement arises in the context of \emph{composite} quantum systems
composed of distinct components or \emph{sub-systems} and is distinct from
other features of quantum physics such as \emph{quantization} for measured
values\textbf{\ }of physical quantities, \emph{probabilistic} \emph{outcomes}
for such measurements, \emph{uncertainty principles} involving pairs of
physical quantities and so on. Such sub-systems are usually associated with
sub-sets of the physical quantities applying to the overall system, and in
general more than one choice of sub-systems can be made. The formalism of
quantum theory treats \emph{pure states} for systems made up of two or more
distinct sub-systems via tensor products of sub-system states, and since
these product states exist in a Hilbert space, it follows that linear
combinations of such products could also represent possible pure quantum
states for the system. Such \emph{quantum superpositions} which cannot be
expressed as a \emph{single} product of sub-system states are known as \emph{%
entangled} (or \emph{non-separable}) states.

The concept of entanglement can then be extended to \emph{mixed states},
where quantum states for the system and the sub-systems are specified by
density operators. The detailed definition of entangled states is set out in
Section \ref{Section - Entanglement}. This definition is based on first
carefully defining the \emph{non-entangled} (or \emph{separable}) states.
The set of non-entangled states must allow\emph{\ all} possible quantum
states for the given sub-system, but in addition these states must be \emph{%
preparable} via processes involving \emph{separate operations} on each
sub-system after which correlated sub-system quantum states are combined in
accordance with \emph{classical probabilities}. Thus, although the
sub-system states retain their quantum natures the combination resulting in
the overall system state is formed classically rather than quantum
mechanically. This overall process then involves \emph{local operations} on
the distinct sub-systems and \emph{classical communication} (\emph{LOCC})\
to prepare a general non-entangled state. The entangled states are then just
the quantum states which are \emph{not} non-entangled states. The general
idea that in all composite systems the non-entangled states all involve LOCC
preparation processes was first suggested by Werner \cite{Werner89a}. The
notion of \emph{quantum states}, the nature of the systems and \emph{%
sub-systems} involved and the specific features required in the definition
of non-entangled states when \emph{identical particles} are involved is
discussed in detail in Section \ref{Section - Identical Particles and
Entanglement}. Entangled states underlie a number of effects that cannot be
interpreted in terms of classical physics, including \emph{spin squeezing},
non-local measurement\emph{\ correlations} - such as for the
Einstein-Podolski-Rosen \emph{(EPR) paradox} and violations of \emph{Bell
Inequalities }(\cite{Einstein35a}, \cite{Schrodinger35a}, \cite{Bell65a}, 
\cite{Brunner14a}, \cite{Hensen15a}). The quantum theory of measurement \cite%
{vonNeumann32a}, \cite{Wheeler83a}, \cite{Zurek81a}, \cite{Peres93a} invokes
entangled states of the system and measuring apparatus as key concepts in
the theory.\textbf{\ }More recently, entangled states have been recognised
as a resource that can be used in various \emph{quantum technologies} for
applications such as teleportation, quantum cryptography, quantum computing
and so on. Recent expositions on the effects of entanglement and its role in 
\emph{quantum information science} include \cite{Peres93a}, \cite{Nielsen00a}%
, \cite{Vedral07a}, \cite{Barnett09a}, \cite{Reid09a}, \cite{Reid12a}.

It would be pointless to characterise states as entangled unless such states
have some important properties. The key requirement is that entangled states
exhibit a novel \emph{quantum feature} that is only found in \emph{composite}
systems. As will be seen in SubSection \ref{SubSection - Joint Measurements}%
, \emph{separable} states are such that the \emph{joint probability} for
measurements of all physical quantities associated with the sub-systems can
be found from separate measurement probabilities obtained from the
sub-system density operators and the overall classical probability for
creating particular products of sub-system states. In general, entangled
states do not exhibit this feature of separable probabilities, and it is
this key non-separability feature that led Schrodinger to call these states
"entangled". Where the sub-systems are spatially separated, one can define
spacelike separated local measurement events on each. This was historically
the reason why the sub-systems and their measurements are often referred to
as\emph{\ local}. The EPR paper \cite{Einstein35a} suggested the possibility
that although the predictions of quantum theory were correct, the theory was 
\emph{incomplete} and there was an underlying reality in the form of \emph{%
classical hidden variables}. Averaging over the unknown values of the hidden
variables would be required to produce the same measurement probabilities as
quantum theory. \emph{Local hidden variable} theories (LHV) are discussed in
SubSection \ref{SubSection - Hidden Variable Theory}, and it will be seen
that the joint probabilities for measurements of sub-system physical
quantities are of the same form as for\emph{\ separable} states. As will be
seen in SubSections \ref{SubSection - Paradoxes} and \ref{SubSection - Bell
Inequalities}, EPR or Bell inequality violations do \emph{not} occur for
states described either by LHV theories or as quantum separable states.
Hence there is a direct link between EPR and Bell violations and both the 
\emph{failure} of LHV theories and the presence of \emph{entanglement}. The
fact that certain entangled states do not exhibit the feature of separable
probabilities shown in classical LHV theories highlights entanglement being
a \emph{non-classical }feature found in composite systems.

Note that although an EPR or Bell inequality violation requires the quantum
state to be entangled, there are examples of mixed entangled states that do 
\emph{not} violate a Bell inequality. For \emph{pure} states of qubits Gisin 
\cite{Gisin91a} showed that entangled states always violated Bell
inequality, but for\ \emph{mixed} states Werner \cite{Werner89a} and others 
\cite{Barrett02a}, \cite{Toth06a}, \cite{Almeida07a} have shown there are
entangled states (\emph{Werner states} \cite{Vedral07a}) for which a hidden
variable theory can be constructed that gives the same joint probability
function for measurement outcomes as quantum theory. Of course the quantum
measurement outcomes must be given before the hidden variable model is
constructed - there is no known way to determine the LHV theory distribution
functions independently. These specific entangled states will therefore
satisfy Bell inequalities.

The mixed entangled states considered by Werner \cite{Werner89a} for which a
hidden variable theory could be constructed were of a special form. Two
distinguishable sub-systems each with $d$ basis states $\left\vert
u_{r}\right\rangle $ were considered, for which the states could be
transformed by unitary operators $\widehat{U}$, and the combined density
operator $\widehat{\rho }$ was required to be invariant under all unitary
transformations of the form $\widehat{U}\otimes \widehat{U}$, so that $%
\widehat{\rho }=(\widehat{U}\otimes \widehat{U})\widehat{\rho }(\widehat{U}%
^{\dag }\otimes \widehat{U}^{\dag })$. Werner \cite{Werner89a} considered
the following unitary operators: (a) $\widehat{U}_{-r}$ such that $\widehat{U%
}_{-r}\left\vert u_{r}\right\rangle =-\left\vert u_{r}\right\rangle $, $%
\widehat{U}_{-r}\left\vert u_{s}\right\rangle =+\left\vert
u_{s}\right\rangle $ for $s\neq r$ (b) $\widehat{P}(r\rightarrow \mu r)$,
which permute basis states $\widehat{P}(r\rightarrow \mu r)\left\vert
u_{r}\right\rangle =\left\vert u_{\mu r}\right\rangle \ $(c) $\widehat{U}%
_{rot}(n,m)$, which transform basis states $\left\vert u_{n}\right\rangle $, 
$\left\vert u_{m}\right\rangle $ into linear combinations of each other $%
\widehat{U}_{rot}(n,m)\left\vert u_{n}\right\rangle =U_{nn}\left\vert
u_{n}\right\rangle +U_{nm}\left\vert u_{m}\right\rangle $, $\widehat{U}%
_{rot}(n,m)\left\vert u_{m}\right\rangle =U_{mn}\left\vert
u_{n}\right\rangle +U_{mm}\left\vert u_{m}\right\rangle $. As a consequence
of these invariances Werner \cite{Werner89a} showed that the density
operator could be expressed in terms of a single parameter $\Phi $ in the
form $\widehat{\rho }=(d^{3}-d)^{-1}\left\{ (d-\Phi )\widehat{1}+(d\Phi -1)%
\widehat{V}\right\} $, where $\widehat{1}$ is the unit operator and $%
\widehat{V}$ is the flip operator. These have matrix elements $\left( 
\widehat{1}\right) _{rs,\,nm}=\delta _{rn}\,\delta _{sm}$\ and $\left( 
\widehat{V}\right) _{rs,\,nm}=\delta _{rm}\,\delta _{sn}$\ . From this form
of the density operator Werner \cite{Werner89a} showed that the probability
function for joint measurement outcomes on the two sub-systems could be
expressed in the same form as applied in hidden variable theory. So although
the mixed entangled state Werner considered were of a restricted type, the
work demonstrated that entanglement did not preclude all hidden variable
theory interpretations of the joint measurements. The fact that some
entangled states do not violate a Bell inequality is another consequence of
Werner's result.

The issue of how best to treat the quantum aspects of \emph{correlations} in
measurement outcomes in composite quantum systems is still an active area of
research and is beyond the scope of these two papers. Quantum entanglement
is clearly relevant to the discussion, but concepts such as \emph{quantum
discord} \cite{Ollivier02a}, \cite{Modi12a} and \emph{EPR\ steering} \cite%
{Wiseman07a}, \cite{Jones07a}, \cite{Cavalcanti09a} are now being used to
describe quantum correlations. The link between these concepts is discussed
in \cite{He15a}. In these recent discussions of quantum correlation, it
turns out that some separable states are regarded as exhibiting quantum
correlations.

It is now generally recognised that entanglement is a \emph{relative}
concept (\cite{Simon02a}, \cite{Hines03a}, \cite{TerraCunha07a}), \cite%
{Vedral07a}, \cite{Horodecki09a}, \cite{Guhne09a} and not only depends on
the quantum state under discussion but also on which \emph{sub-systems} are
being considered as entangled (or non-entangled). A quantum state may be
entangled for one choice of the sub-systems but may be non-entangled if
another choice of sub-systems is made. A simple example often cited is that
for the hydrogen atom \cite{TerraCunha07a}, a system made up of two
distinguishable particles, a proton and an electron. Here the energy
eigenstates are non-entangled if the sub-systems refer to the centre of mass
of the entire atom and the relative position of the electron and the proton,
but which would be entangled if the sub-systems were the positions of the
electron and proton. It could be argued that the centre of mass and the
relative position are not really independent sub-systems - one always
accompanies the other - but as unrelated centre of mass and relative
position quantum states can be prepared, they can be regarded as distinct
sub-systems. The individual positions of the electron and the proton are
also distinct sub-systems, and the ground state of the hydrogen atom is
indeed entangled - the electron position is tightly correlated with the
proton position. Another example involves two different choices of single
particle states in a two mode Bose-Einstein condensate (BEC) - a system with
a large number of identical particles. The issue of defining sub-systems
will be dealt with below, but taking the original two sub-systems to be
bosonic modes (or single particle states) denoted\textbf{\ }$\left\vert \phi
_{A}\right\rangle $\textbf{\ }and\textbf{\ }$\left\vert \phi
_{B}\right\rangle $, a well known $N$\ boson entangled state of these two
modes $A$\ and $B$\ (with mode annihilation operators\textbf{\ }$\widehat{a}$
and\textbf{\ }$\widehat{b}$) is the\emph{\ binomial state} given by $%
\left\vert \Phi \right\rangle =((\cos \theta \,\exp (i\chi /2)\,\widehat{a}%
+\sin \theta \,\exp (-i\chi /2)\,\widehat{b})^{\dag })^{N}/\sqrt{N!}%
\;\left\vert 0\right\rangle $\ (see \cite{Dalton12a} and Paper II,\textbf{\
Section 3.7}) which is a quantum superposition of Fock states $(\,\widehat{a}%
^{\dag })^{k}/\sqrt{k!}\;(\,\widehat{b}^{\dag })^{N-k}/\sqrt{(N-k)!}%
\;\left\vert 0\right\rangle $ with $k=0,\,...\,,N$. Introducing new modes
via $\widehat{c}=(\cos \theta \,\exp (i\chi /2)\,\widehat{a}+\sin \theta
\,\exp (-i\chi /2)\,\widehat{b})$ and $\widehat{d}=(-\sin \theta \,\exp
(i\chi /2)\,\widehat{a}+\cos \theta \,\exp (-i\chi /2)\,\widehat{b})$ we see
that we can also write $\left\vert \Phi \right\rangle =(\,\widehat{c}^{\dag
})^{N}/\sqrt{N!}\;\left\vert 0\right\rangle $, so that the same quantum
state is a separable state if the sub-systems are chosen to be the new modes 
$C$\ and $D$. Another example is the ground state of the single mode
non-interacting BEC trapped in a harmonic oscillator (HO) potential. This is
a separable state, with all bosons in the lowest energy mode if the
sub-systems are chosen as the HO modes. However, if \textbf{single} particle
position states spatially localised in two different regions are chosen as
two sub-systems, then the same ground state for the identical particle
system is spatially entangled, as pointed out by Goold et al \cite{Goold09a}.

\subsection{Measures and Tests for Entanglement}

Various \emph{measures} of entanglement have been defined for certain types
of quantum state - see \cite{Vedral07a}, \cite{Barnett09a}, \cite%
{Horodecki09a}, \cite{Guhne09a}, \cite{Tichy11a}, \cite{Modi12a}, \cite%
{Amico08a}, for details of these, and are aimed at quantifying entanglement
to determine which states are more entangled than others. This is important
since entanglement is considered as a resource needed in various quantum
technologies. Calculations based on such measures of entanglement confirm
that for some choices of sub-systems the quantum state is entangled, for
others it is non-entangled. For two mode pure states the \emph{entanglement
entropy} - being the difference between the entropy for the pure state
(zero) and that associated with the reduced density operator for either of
the two sub-systems - is a useful entanglement measure. As entropy and
information changes are directly linked \cite{Vedral07a}, \cite{Barnett09a},
this measure is of importance to \emph{quantum information science}.
Measurements of entanglement based on \emph{Renyi entropy} and \emph{purity}
are discussed in \cite{Islam15a}, \cite{Cramer13a} and \cite{Daley12a}.
Another entanglement measure is \emph{particle entanglement}, defined by
Wiseman et al \cite{Wiseman03a}, \cite{Dowling06b}, \cite{Tichy11a} for
identical particle systems and based on projecting the quantum state onto
states with definite particle numbers. One of the problems with entanglement
measures is that there is often no simple way to measure the quantities
required.

In the case of \emph{bipartite} entanglement in qubit systems \cite{Peres96a}
and \cite{Horodecki96a} obtained\ a sufficient condition for a quantum state
to be entangled (\emph{PPT condition}) (see \cite{Guhne09a}, \cite%
{Horodecki09a} for details). Suppose the density operator $\widehat{\rho }$
is changed into $\widehat{\rho }^{T}$ by mapping the matrix elements
associated with one of the sub-systems into their \emph{transpose}. Then
provided the new operator $\widehat{\rho }^{T}$ is a valid density operator
(with real, non-negative eigenvalues that add to unity), the original
density operator represents a separable state. Thus, if it is shown that
some of the eigenvalues of $\widehat{\rho }^{T}$\ are negative, then the
state $\widehat{\rho }$\ is entangled.\textbf{\ }However, it is often not
practical to use this as an entanglement test for systems with large numbers
of basis states, as it requires being able to measure all the density matrix
elements. It was also later realised \cite{Horodecki97a} that in general,
the PPT condition was not a necessary condition for entanglement, apart from
cases of $2\times 2$\ and $2\times 3$\ subsystems - that is, showing that $%
\widehat{\rho }^{T}$\ has only positive eigenvalues will not guarantee that $%
\widehat{\rho }$\ is separable, as counter-examples for $2\times 4$\ and $%
3\times 3$\ subsystems showed.

Although not directly related to the various quantitative measures of
entanglement, the results for certain measurements can play the role of
being \emph{signatures} or \emph{witnesses} or \emph{tests} of entanglement 
\cite{Horodecki09a}, \cite{Guhne09a}, \cite{Tichy11a}. These are in the form
of \emph{inequalities} for \emph{variances} and \emph{mean values} for
certain physical quantities, which are dependent on the inequalities
applying for non-entangled quantum states. If such inequalities are \emph{%
violated} then it can be concluded that the state is \emph{entangled} for
the relevant sub-systems. In the case of entanglement witnesses, the idea is
to find a hermitian operator $\widehat{W}$\ such that for separable states $%
Tr(\widehat{W}\widehat{\rho })\geq 0$, so that if $Tr(\widehat{W}\widehat{%
\rho })<0$\ the state must be entangled. Here we note that the density
operator occurs linearly when evaluating the quantities involved. Some of the%
\emph{\ correlation}\textbf{\ }tests discussed in paper II are cases
involving entanglement witnesses. However, in more general tests for
entanglement the density operator appears non-linearly. For example, a%
\textbf{\ }\emph{spin squeezing}\textbf{\ }test for entanglement may require
showing that the variance for a spin operator is less than \ a multiple of
the magnitude of the mean value of another spin operator - thus for example $%
\left\langle \Delta \widehat{S}_{x}^{2}\right\rangle <|\left\langle \widehat{%
S}_{z}\right\rangle |/2$. This could be written as $Tr((\widehat{S}%
_{x}^{2}\pm \widehat{S}_{z}/2)\widehat{\rho })-(Tr(\widehat{S}_{x})\widehat{%
\rho }))^{2}<0$, which is of a more general form than for an entanglement
witness. Non-linear tests are discussed in Ref. \cite{Guhne09a}. One of the
advantages of entanglement tests is that the quantities involved can be
measured.\textbf{\ }It cannot be emphasised enough that these tests provide 
\emph{sufficiency conditions} for establishing that a state is entangled. So
if the test is satisfied we can conclude that the state is not separable.%
\textbf{\ }The failure of a test does \emph{not} mean that the state is not
entangled - sufficiency does not imply \emph{necessity}. The violation of a 
\emph{Bell inequality} is an example of such a signature of entanglement,
and the demonstration of \emph{spin squeezing} is regarded as another.
However, the absence of spin squeezing (for example) does not guarantee
non-entanglement, as the case of the \emph{NOON} state in SubSection \textbf{%
3.6 }of the accompanying paper II shows. A significant number of such
inequalities have now been proposed and such signatures of entanglement are
the primary focus of the accompanying paper, which is aimed at identifying
which of these inequalities really do identify entangled states, especially
in the context of \emph{two mode} systems of \emph{identical bosons}.

At present there is\emph{\ no clear linkage} between quantitative measures
of entanglement (such as entanglement entropy) and the quantities used in
conjunction with the various entanglement tests (such as the relative spin
fluctuation in spin squeezing experiments). Results for experiments
demonstrating such non-classical effects cannot yet be used to say much more
than the state \emph{is} entangled, whereas ideally these experiments should
determine \emph{how} entangled the state is. Again we emphasise that neither
the entanglement tests nor the entanglement measures are being used to \emph{%
define} entanglement. Entanglement is defined first as being the quantum
states that are non-separable, the tests for and measures of entanglement
are \emph{consequential} on this definition.

\subsection{Particle versus Mode Entanglement}

These two papers deal with\emph{\ identical} particles - bosons or fermions.
In the \emph{second quantisation} approach used here the system is regarded
as a set of \emph{quantum fields}, each of which may be considered as a
collection of single particle states or \emph{modes}. We now take into
account the situation where systems of \emph{identical particles} are
involved. This requires us to give special consideration to the requirement
that quantum states in such cases must conform to the \emph{symmetrisation
principle}. What sub-systems are possible must take into account that
entanglement requires the specification of sub-systems that are\textbf{\ }%
\emph{distinguishable} from each other and on which\textbf{\ }\emph{%
measurements}\textbf{\ }can be made. In addition, the sub-systems must be
able to exist as\textbf{\emph{\ }}\emph{separate}\textbf{\ }systems which
can be prepared in quantum states for that sub-system alone. These key
requirements that the sub-systems must be distinguishable, susceptible to
measurements and can exist in separate quantum states are necessary for the
concept of entanglement to make\textbf{\ }\emph{physical} sense, and have
important consequences for the choice of sub-systems when identical
particles are involved. These three key logical requirement for sub-systems
rule out considering\textbf{\emph{\ }}\emph{labelled identical particles} as
sub-systems and lead to the conclusion that sub-systems must be\textbf{\ }%
\emph{modes}. Thus both the system and sub-systems will be specified via the
modes that are involved, so here the sub-systems in terms of which
non-entangled (and hence entangled) states are defined are \emph{modes} or 
\emph{sets} of modes, not particles \cite{Simon02a}, \cite{Hines03a}, \cite%
{TerraCunha07a}), \cite{Vedral07a}, \cite{Benatti10a}, \cite{Benatti11a}. In
this approach, \emph{particles} are associated with the \emph{occupancies}
of the various modes, so that situations with differing numbers of particles
will be treated as differing quantum \emph{states} of the same system, not
as different systems - as in the \emph{first quantisation} approach. Note
that the choice of modes is \emph{not unique} - original sets of orthogonal
one particle states (modes) may be replaced by other orthogonal sets. An
example is given in Section \textbf{2 }of accompanying paper II. Modes can
often be categorised as \emph{localised} modes, where the corresponding
single particle wavefunction is confined to a restricted spatial region, or
may be categorised as \emph{delocalised} modes, where the opposite applies.
Single particle harmonic oscillator states are an example of localised
modes, momentum states are an example of delocalised modes. This distinction
is significant when phenomena such as EPR violations and teleportation are
considered.

However, even if the system consists entirely of\textbf{\emph{\ }}\emph{%
distinguishable} particles\textbf{\ }we can still regard the sub-systems as
collections of modes. Each distinguishable particle is still associated with
a set of single particle states or modes (momentum eigenstates, harmonic
oscillator eigenstates, \textbf{etc.}) that can be occupied. More general
states associated with a single particle may be quantum superposition states
of those with a single particle occupancy of the modes. If the overall
system consists of a number of distinguishable particles each of which is
considered as a sub-system, then each such sub-system can equally be
regarded as the set of modes associated with the particular distinguishable
particle. Overall system states involving just one particle of each type
would be simultaneous eigenstates of the number operators for each of the
distinguishable particles, with an eigenvalue of unity corresponding to
there being only one particle of each type. The second quantisation approach
can still be used, but is somewhat superfluous when the modes for each
particle are only occupied once.

Although \emph{multi-mode} systems are also considered, in this paper we
mainly focus on \emph{two mode} systems of identical \emph{bosonic} atoms,
where the atoms at most occupy only two single particle states or modes. For
bosonic atoms this situation applies in two mode interferometry, where if a
single hyperfine component is involved the modes concerned may be two
distinct spatial modes, such as in a double well magnetic or optical trap,
or if two hyperfine components are involved in a single well trap each
component has its own spatial mode. Large numbers of bosons may be involved
since there is no restriction on the number of bosons that can occupy a
bosonic mode. For fermionic atoms each hyperfine component again has its own
spatial mode. However, if large numbers of \emph{fermionic} atoms are
involved then as the Pauli exclusion principle only allows each mode to
accommodate one fermion, it follows that a large number of modes must
considered and two mode systems would be restricted to at most two fermions.
Consideration of multi-mode entanglement for large numbers of fermions is 
\emph{outside} the scope of the present paper (see \cite{Lunkes05a} for a
treatment of this), and unless otherwise indicated the focus will be on 
\emph{bosonic modes}. The paper focuses on identical bosonic \emph{atoms} -
whether the paper also applies to \emph{photons} is less clear and will be
discussed below.

\subsection{Symmetrization and Super-Selection Rules}

The work presented here begins with the \emph{fundamental issue} of how an
entangled state should be \emph{defined} in the context of systems involving 
\emph{identical particles}. To reiterate - in the commonly used \emph{%
mathematical approach} for defining entangled states, this requires \emph{%
first} defining a general \emph{non-entangled} state, all \emph{other}
states therefore being entangled. We adhere to the original definition of
Werner \cite{Werner89a} in which the separable states are those that can be
prepared by \emph{local operations} and \emph{classical communication} (%
\emph{LOCC}). This approach is adopted by other authors, see for example 
\cite{Bartlett06b}, \cite{Jones07a}, \cite{Masanes08a}. However, in other
papers - see for example \cite{Verstraete03a}, \cite{Schuch04a} so-called 
\emph{separable non-local} states are introduced in which LOCC is \emph{not}
required (see SubSection.\ref{SubSection - Two Mode Coherent State Mixture}
for an example). By contrast (and consistent with Werner's approach), in the
present paper it is contended that the \emph{density operators} both for the 
\emph{quantum states}\textbf{\ }of the overall system and those for the
non-entangled (local) sub-systems in the context of \emph{non-entangled
states} must be compatible with certain principles and rules that have been
found to be both necessary and sufficient for understanding\textbf{\ }\emph{%
physical experiments}\textbf{\ }in non-relativistic many-body systems. In
some other work (discussed below) this has not been the case. A key feature
required of all quantum states for systems involving \emph{identical
particles}, entangled or not is that they satisfy the \emph{symmetrization
principle }\cite{Dirac30a}. This places restrictions both on the form of the
overall density operator and also on what can be validly considered to be a
sub-system. In particular this rules out \emph{individual} identical
particles being treated as sub-systems, as is done in some papers (see
below). If the system consists entirely of\ \emph{distinguishable particles}%
\textbf{\ }then the symmetrization principle is not relevant. In addition, 
\emph{super-selection rules }(SSR)\emph{\ }\cite{Wick52a} only allow density
operators which have \emph{zero coherences} between states with \emph{%
differing total numbers} of particles to represent valid \emph{quantum states%
}, and this will be taken into account for \emph{all} quantum states of the
overall system, entangled or not. This is referred to as the \emph{global
particle number super-selection rule} \ In \emph{non-entangled} or \emph{%
separable} states the density operator is a sum over products of sub-system
density operators, each product being weighted by its probability of
ocurring (see below for details). For the non-entangled or separable quantum
states, a so-called\emph{\ local particle number} \emph{super-selection rule}
will \emph{also} be applied to the density operators describing each of the 
\emph{sub-systems}. These sub-system density operators must then have have
zero coherences between states with differing numbers of \emph{sub-system
particles}. This additional restriction excludes density operators as
defining non-entangled states when the sub-system density operators do not
conform to the local particle number super-selection rule. Consequently,
density operators where the local particle number SSR does not apply would
be regarded as entangled states. This viewpoint is discussed in papers by
Bartlett et al \cite{Bartlett06b}, \cite{Bartlett07a} as one of several
approaches for defining entangled states. However, other authors such as 
\cite{Verstraete03a}, \cite{Schuch04a} state on the contrary that states
where the sub-system density operators do \emph{not} conform to the local
particle number super-selection rule \emph{are} still separable, others such
as \cite{Hillery06a}, \cite{Hillery09a} do so by implication - the latter
papers applied to atomic as well as photon modes.\textbf{\ }So in these two
papers we are advocating a \emph{different definition }to some \emph{other} 
\emph{definitions} of entanglement in identical particle systems, the
consequence being that the set of entangled states is now much \emph{larger}%
. This is a \emph{key idea} in this paper - not only should super-selection
rules on particle numbers be applied to the the \emph{overall} quantum
state, entangled or not, but it \emph{also} should be applied to the density
operators that describe states of the modal \emph{sub-systems} involved in
the general definition of \emph{non-entangled} states.

Note that for systems entirely consisting of\ $N$\ \emph{distinguishable}
particles the super-selection rules are still true, but are now \emph{%
superfluous}. Each sub-system is the set of modes or \emph{one particle\
states} of the \emph{specific} distinguishable particle and the overall
state is an $N$\ particle state in which the sub-systems only contain \emph{%
one} particle. Consequently there are no sub-system or system coherences
between states with differing particle number.

The detailed reasons for adopting the viewpoint that the entanglement
criteria be compliant with the requirement of the local particle number
super-selecrtion rule (SSR) for the sub-system are set out below. As will be
seen, the local particle number super-selection rule restriction \emph{%
firstly} depends on the \emph{fundamental requirement} that for \emph{all}
composite systems - whether identical particles are involved or not -
non-entangled states are only those that can be prepared via processes that
involve only \emph{LOCC}. The requirement that the sub-system density
operators in identical particle cases satisfy the local particle number SSR
is \emph{consequential} on the sub-system states being possible sub-system
quantum states. As mentioned before, the general definition of non-entangled
states based on LOCC preparation processes was first suggested by Werner 
\cite{Werner89a}. Apart from the papers by Bartlett et al \cite{Bartlett06b}%
, \cite{Bartlett07a} we are not aware that this LOCC/SSR based criteria for
non-entangled states has been invoked previously for identical particle
systems, indeed the opposite approach has been proposed \cite{Verstraete03a}%
, \cite{Schuch04a}. However, the idea of considering whether sub-system
states should satisfy the local particle number SSR has been presented in
several papers -\emph{\ }\cite{Verstraete03a}, \cite{Schuch04a}, \cite%
{Bartlett06b}, \cite{Bartlett07a}, \cite{Vaccaro08a}, \cite{White09a}, \cite%
{Paterek11a}, mainly in the context of pure states for bosonic systems,
though in these papers the focus is on issues other than the definition of
entanglement - such as quantum communication protocols \cite{Verstraete03a},
multicopy distillation \cite{Bartlett06b}, mechanical work and accessible
entanglement \cite{Vaccaro08a}, \cite{White09a} and Bell inequality
violation \cite{Paterek11a}. The consequences for entanglement of applying
this super-selection rule requirement to the sub-system density operators
are quite \emph{significant}, and in the accompanying paper II important 
\emph{new} \emph{entanglement tests} are determined. Not only can it
immediately be established that spin squeezing \emph{requires} entangled
states, but though several of the other inequalities (accompanying paper II)
that have been used as signatures of entanglement are still valid, \emph{%
additional tests} can be obtained which only apply to entangled states that
are defined to conform to the symmetrisation principle and the
super-selection rules.

It is worth emphasising that requiring the sub-system density operators
satisfy the local particle number SSR means that there are less states than
otherwise would be the case which are classed as non-entangled, and \emph{%
more states} will be regarded as \emph{entangled}. It is therefore not
surprising that additional tests for entanglement will result. If \emph{%
further restrictions} are placed on the sub-system density operator - such
as requiring them to correspond to a fixed number of bosons again there will
be more states regarded as entangled, and even more entanglement tests will
apply. A particular example is given in SubSection \textbf{4.3 }of paper II,
where the sub-systems are restricted to one boson states.

The \emph{symmetrisation} requirement for systems involving identical
particles is well established since the work of Dirac. There are two types
of justification for applying the \emph{super-selection rules} for systems
of identical particles\ (both massive and otherwise). The \emph{first}
approach for invoking the superselection rule to exclude quantum
superposition states with differing numbers of identical particles is based
on simple considerations and may be summarised as:

1. No way is known for creating such SSR non-compliant states.

2. No way is known for measuring the properties of such states.

3. Coherence and interference effects can be understood\ in terms of SSR
compliant states.

The \emph{second} approach is more sophisticated and involves linking the
absence or presence of SSR to whether or not there is a suitable \emph{%
reference frame} in terms of which the quantum state is described \cite%
{Aharonov67a}, \cite{Bartlett03a}, \cite{Sanders03a}, \cite{Verstraete03a}, 
\cite{Schuch04a}, \cite{Kitaev04a}, \cite{van Enk05a}, \cite{Bartlett06a}, 
\cite{Bartlett07a}, \cite{Vaccaro08a}, \cite{White09a}, \cite{Tichy11a}.
This approach will be described in SubSection \ref{SubSection -
Super-Selection Rule} and Appendix \ref{Appendix - Reference Frames and SSR}%
, the key idea being that SSR are a consequence of considering the
description of a quantum state by a real observer (Charlie) whose phase
reference frame has an \emph{unknown phase difference} from that of a
hypothetical observer ((Alice),\textbf{\ }both studying the same system.
Alice is assumed to possess a phase reference frame such that her
description of the quantum state of the system violates the SSR. Charlie, on
the other hand is an actual observer with no such phase reference frame.
Thus, whilst Alice's description of the system involves a quantum state may
violate the SSR, the description of the same system by Charlie will involve
a quantum state that is SSR compliant. In the main part of this paper the
density operator $\widehat{\rho }$ used to describe the various quantum
states will be that of the external observer (Charlie). Note that if
well-defined phase\textbf{\ }references do\textbf{\ }\emph{exist}\textbf{\ }%
and the relationship\textbf{\ }between them is\textbf{\ }\emph{known}, then
the SSR can be challenged (see SubSection \ref{SubSection - Challenges to
SSR} and Appendix \ref{Appendix - Reference Frames and SSR}), but this
situation does not apply in the case of massive bosons (or fermions).

It should be noted that both of these justifications for applying the SSR
are dependent on what is practical in terms of measurements in \emph{%
non-relativistic} quantum physics. Here the situation is clearer for systems
of massive particles such as atoms than for massless particles such as
photons. Applying SSR for photons is discussed in SubSubSection \ref%
{SubSubSection - SSR and Photons}.

However, to allow for quantum states that as far as we know cannot be made
or measured, and for which there are no known physical effects that require
their presence\ is an unnecessary feature to add to the non-relativistic
quantum physics of many body systems or to quantum optics. Considerations
based on the general principle of simplicity (Occam's razor) would suggest
not doing so until a clear physical justification for including them is
found. The quantum state is intended to specify what is known about a
quantum system and how it was prepared. It is used to determine the
probabilities for possible measurements on the system. Clearly there is no
point in including non-SSR compliant terms in the density operator for the
quantum state. Such terms would neither allow for possible preparation
processes, or contribute to measurement probabilities associated with
physical efects. Furthermore, experiments can be carried out on each of the
mode sub-systems considered as a \emph{separate} system, and essentially the 
\emph{same reasons} that justify applying the super-selection rule to the
overall system also apply to the separate mode sub-systems in the context of
defining \emph{non-entangled states}. Hence, unless it can be justified to
ignore the super-selection rule for the overall system it would be \emph{%
inconsistent} not to apply it to the sub-system as well. As we will see, for
separable states the requirement that the overall state is SSR compliant
generally implies that the sub-system states are SSR compliant - though in
some special cases this is not the case (see SubSection 4.3.3\textbf{\ }of
paper II). The \emph{onus} is on those who wish to ignore the
super-selection rule for the separate sub-systems to justify why it is being
applied to the overall system. In addition, \emph{joint} \emph{measurements}
on \emph{all} the sub-systems can be carried out, and the interpretation of
the measurement probabilities requires the density operators for the
sub-system \ states to be physically based. The general application of
super-selection rules has however been challenged (see SubSection \ref%
{SubSection - Super-Selection Rule}) on the basis that super-selection rules
are not a fundamental requirement of quantum theory, but are restrictions
that could be lifted if there is a suitable system that acts as a \emph{%
reference} for the coherences involved. In Section \ref{Section - Identical
Particles and Entanglement} and in Appendix \ref{Appendix - Super-Selection
Rule Violations ?} an analysis of these objections to the super-selection
rule is presented, and in Appendix \ref{Appendix - Reference Frames and SSR}
we see that the approach based on phase reference frames does indeed justify
the application of the SSR both to the general quantum states for multi-mode
systems of identical particles and to the sub-system states for
non-entangled states of these systems.

The sceptic who wishes to ignore the super-selection rules in the definition
of entanglement - and consequently only consider as valid tests for
entanglement where SSR compliance is \emph{not} used in their derivation -
needs to carry out a research program analogous to that which resulted in 
\emph{parity non-conservation} becoming a basic feature of \emph{weak
interaction theory}. The successful incorporation of parity non-conservation
involved first proposing (on symmetry grounds) possible interactions in
which parity was not conserved, second working out possible experiments that
could confirm parity non conservation and third carrying out key experiments
that did confirm this. At this stage no such work in regard to SSR violation
in non-relativistic many body physics has been carried out or is likely to
be in the near future (except possibly for photons). As we will see in paper
II, none of the experimental methods for entanglement tests that we examine
can detect SSR non-compliance - none involve a suitable\emph{\ phase
reference}. To ignore SSR in non-relativistic entanglement theory and
experiment on the grounds of scepticism would be analogous to ignoring
parity conservation in quantum chemistry or atomic physics - areas which are
well-understood in terms of parity being conserved (apart from the
well-known parity violating effects of external electric fields). When and
if SSR violation in non-relativistic many body physics is found would then
be the time to revise the definition of quantum entanglement. In these two
papers we will utilise the definition of entanglement and derive tests based
on SSR compliance, though of course recognising that there are also tests
that do not require SSR compliance which are \emph{also} valid for SSR
compliant states. Although other definitions of entanglement will be
considered for comparison, to avoid confusion the SSR compliant definition
will be the one which we mean when we refer to entanglement.

A further sound scientific argument can be presented in favour of studying
SSR compliant entanglement tests (as is our aim in these papers (I) and
(II)). This involves a consideration of what can be concluded from such
tests by suppporters or sceptics of SSR. For example, one such test (see
paper II) involves spin squeezing in two mode systems. If the state is
separable \emph{and} the sub-system states comply with local SSR then there
is \emph{no} spin squeezing. However, if experimental tests \emph{do}
demonstrate spin squeezing, then what can we conclude? The supporters of SSR
compliance being required for the sub-systems would conclude that the state
was not separable and hence \emph{entanglement} is present between the
subsystems. On the other hand, the sceptic who does not believe local SSR
compliance is required would have no option but to conclude \emph{either}
that entanglement is present between the subsystems \emph{or} (if they argue
there is no entanglement) the state is separable but one or both of the
quantum subsystems violates the SSR. The sceptic may favour the second
conclusion, but that would then imply an \emph{actual} experimental
circumstance where superselection rules did \emph{not} apply to the
sub-system states. In that case, the issues raised in the last four
paragraphs regarding \emph{lack} of \emph{phase references} or \emph{SSR
violating preparation processes} etc. must be addressed directly. Either
way, the study of such SSR based experiments is clearly important. Put
another way, suppose the sceptic were to derive a \emph{different} test
using the separability requirement \emph{alone,} for which an experimental
outcome shows that the two subsystems were\emph{\ }indeed\emph{\ not }%
entangled. This would seem to require a test for entanglement which is \emph{%
necessary} as well as being sufficient - the latter alone being usually the
case for entanglement tests.\textbf{\ }Such criteria and measurements are a
challenge, but not impossible even though we have not met this challenge in
these two papers. If the conclusion from the earlier \emph{SSR\ based}
experiment was \emph{either} entanglement \emph{or} separability with non
SSR compliance, then if the result from the different test based \emph{only}
on separability ruled entanglement \emph{out}, it follows that the system
must be in a separable state in which the sub-system states violate the SSR.
Conversely, the latter test may confirm the entanglement possibility found
in the earlier test. Thus, \emph{in principle} there could be a\emph{\ pair}
of experiments that give evidence of entanglement, \emph{or} failure of the
Super Selection Rule. For such investigations to be possible, the use of
entanglement criteria that \emph{do} invoke the local super-selection rules
is \emph{also} required.

\subsection{Entanglement Tests and Experiments - Paper II}

The main focus of the accompanying paper II is to derive the SSR compliant
criteria and to consider the experimental implementation. This leads to
important links between\textbf{\ }\emph{spin squeezing }and entanglement.
The link with \emph{quantum correlation functions} (as proposed in Refs.\cite%
{Hillery06a}, \cite{Hillery09a}) is also treated. Heisenberg Uncertainty
Principle inequalities involving spin operators \cite{Wodkiewicz85a} and the
consequent property of spin squeezing have been well-known in quantum optics
for many years. The importance of spin squeezing in quantum metrology is
discussed in the paper by Kitagawa et al \cite{Kitagawa93a} for general spin
systems. It was suggested in this paper that correlations between the
individual spins was needed to produce spin squeezing, though no
quantitative proof was presented and the more precise concept of
entanglement was not mentioned. For the case of two mode systems the
earliest paper linking spin squeezing to entanglement is that of Sorensen et
al \cite{Sorensen01a}, which considers a system of identical bosonic atoms,
each of which can occupy one of two internal states. This paper states that
spin squeezing requires the quantum state to be entangled, with a proof
given in the Appendix. A consideration of how such spin squeezing may be
generated via collisional interactions is also presented. The paper by
Sorensen et al is often referred to as establishing the link between spin
squeezing and entanglement - see for example Micheli et al \cite{Micheli03a}%
, Toth et al \cite{Toth07a}, Hyllus et al \cite{Hyllus12a}. However, the
paper by Sorensen et al \cite{Sorensen01a} is based on a definition of
non-entangled states in which the sub-systems are the identical particles,
and this is inconsistent with the symmetrization principle. However, the
accompanying paper II establishes the link between spin squeezing and
entanglement based on a definition of entanglement consistent with the
system and sub-system density operators representing quantum states.

It is also important to consider which \emph{components} of the spin
operator vector are squeezed, and this issue is also considered in the
accompanying paper. In the context of the present second quantisation
approach to identical particle systems the three spin operator components
for two mode systems are expressed in terms of the annihilation, creation
operators for the two chosen modes. Spin squeezing can be defined (see
Section \textbf{2} in the accompanying paper) in terms of the variances of
these spin operators, however the \emph{covariance matrix} for the three
spin operators will in general have off-diagonal elements, and spin
squeezing is also defined in terms of rotated spin operators referred to as 
\emph{principal spin operators} for which the covariance matrix is \emph{%
diagonal}. The principal spin operators are related to new mode
annihilation, creation operators in the same form as for the original spin
operators, where the \emph{new modes} are two orthogonal linear combinations
of the originally chosen modes. In discussing the relationship between spin
squeezing and entanglement, the modes which may be entangled are generally
those associated with the definition of the spin operators.

A further focus of the accompanying paper is on the relationship between
entanglement and certain \emph{correlation properties} of sub-system
operators. Tests for entanglement based on such correlations have also been
published - see for example \cite{Hillery06a}, \cite{Hillery09a}. These
tests were based on ignoring the super-selection rules, so in the
accompanying paper we present revised correlation tests for entanglement
when the super-selection rules are definitely complied with. We also show
the link between correlation tests and tests involving spin operators.

The accompanying paper also deals with the important question of what\emph{\
measurement systems}\textbf{\ }are suitable for making spin and correlation
tests for entanglement. We first\textbf{\ }consider a \emph{simple two mode}%
\textbf{\emph{\ }}\emph{interferometer}\textbf{\ }which involves coupling
the two modes employing a\textbf{\ }\emph{resonant classical field pulse}%
\textbf{\ }which is associated with a variable\emph{\ pulse area}\textbf{\ }%
for its amplitude and has an adjustable\textbf{\ }\emph{phase}. It is shown
that measurement of the mean value and variance of the\textbf{\ }\emph{%
population difference}\textbf{\ }between the two modes\textbf{\ }\emph{after}%
\textbf{\ }the interferometer pulse enables measurements of the\textbf{\ }%
\emph{mean value}\textbf{\ }and\textbf{\ }\emph{covariance matrix}\textbf{\ }%
elements of the\textbf{\ }\emph{spin operators}\textbf{\ }for the quantum
state that existed\textbf{\ }\emph{before}\textbf{\ }the pulse was applied.
The mean values and variances of certain spin operators are relevant for
correlation and spin squeezing entanglement tests.

Paper II is focused on\emph{\ two mode} bosonic systems. These are of
particular interest because cold atomic gases cooled well below the
Bose-Einstein condensation (BEC) transition temperature can be prepared
where essentially only two modes are occupied (\cite{Leggett01a}, \cite%
{Dalton12a}). This can be achieved for cases involving a single hyperfine
components using a double well trap potential or for two hyperfine
components using a single well. At higher temperatures more than two modes
may be occupied, so\emph{\ multi-mode} systems are also of importance and
thus are considered in paper II.

\subsection{Outlines of Papers I and II}

The plan of the present\textbf{\ }paper is as follows. In Section \ref%
{Section - Entanglement} the key definitions of entangled states are
covered, and a detailed discussion on why the symmetrisation principle and
the super-selection rule is invoked in discussed in Section \ref{Section -
Identical Particles and Entanglement}. Challenges to the necessity of the
super-selection rule are outlined, with arguements against such challenges
dealt with in Appendices \ref{Appendix - Reference Frames and SSR} and \ref%
{Appendix - Super-Selection Rule Violations ?}. Two key mathematical
inequalities are derived in Appendix \ref{Appendix - Inequalities}. Details
for the spin EPR paradox are given in Appendix \ref{Appendix - Spin EPR
Paradox}. The final Section \ref{Section - Discussion & Summary of Key
Results} summarises and discusses the key features about entanglement
treated in this paper.

In the accompanying paper II, Section \textbf{2} sets out the definitions of
spin squeezing and in the following Section \textbf{3} it is shown that spin
squeezing is a signature of entanglement, both for the original spin
operators with entanglement of the original modes and the principle spin
operators with entanglement of the two new modes, and also for multi-mode
cases. Details are in Appendices\textbf{\ A }and\textbf{\ B.} A number of
other tests for entanglement proposed by other authors are considered in
Sections \textbf{4, 5 }and\textbf{\ 6}, with details of these treatments set
out in Appendices \textbf{B, C }and\textbf{\ D}. In Section\textbf{\ 7 }it
is shown that a simple two mode interferometer can be used to measure the
mean values and covariance matrix for the spin operators involved in
entanglement tests. The treatment is then generalised to situations
involving measurements on multi-mode systems. Details are covered in
Appendices \textbf{G }and\textbf{\ H}. \textbf{\ }Actual experiments aimed
at detecting entanglement via spin squeezing tests are examined in Section 
\textbf{8. }The final Section \textbf{9} summarises and discusses the key
results regarding entanglement tests. Appendices \textbf{E, F }and\textbf{\
I }provide details regarding certain important states whose features are
discussed in the paper - the "separable but non-local " states and the
relative phase eigenstate.

\pagebreak

\section{Entanglement - General Features}

\label{Section - Entanglement}

\subsection{Quantum States}

The standard Copenhagen quantum theory notions of \emph{physical} \emph{%
systems} that can exist in various \emph{states} and have associated \emph{%
properties} on which \emph{measurements} can be made are presumed in this
paper. The measuring system may be also treated via quantum theory, but
there is always some component that behaves \emph{classically,} so that
quantum fluctuations in the quantity recorded by the \emph{observer} are
small. The term \emph{quantum state} (or "physical quantum state" or just
"state" for short) refers to a state that can either be prepared via a
process consistent with the laws of quantum physics and on which
measurements can be then performed and the probabilistic results predicted
from this state (\emph{prediction}), or a state whose existence can be
inferred from later quantum measurements (\emph{retrodiction}). We may also
refer to such states as\textbf{\ }\emph{allowed}\textbf{\ }quantum states,
and our approach is intended to be \emph{physically}\textbf{\ }based. In
quantum theory, quantum states are \emph{represented} mathematically by 
\emph{density operators} for mixed states or \emph{state vectors} for pure
states. For identical particle systems these representations must satisfy
symmetrisation and other basic requirements in accordance with the laws of
quantum theory. The probabilities of measurement outcomes and the
probabilities associated with retrodiction can be interpreted as\textbf{\ }%
\emph{Bayesian probabilities}\textbf{\ }\cite{Pegg02a}\textbf{, }\cite%
{Caves02a}, and the quantum state is\textbf{\ }\emph{observer dependent.}%
\textbf{\ }The quantum state, the system it is associated with and the
quantities that can be measured are considered here as entities that are
viewed as being \emph{both} ontological and epistimological. Different
observers may have different information about how the quantum state was
prepared, hence the quantum state is in part \emph{epistimological}, and
would be described differently by different observers. Hence the observer is
important, but as there is actually something out there to be studied,
quantum states also have an \emph{ontological}\textbf{\ }aspect. We will
avoid the unqualified term "physical state" because this term is generally
invoked in discussions about the pre-Copenhagen notion of \emph{reality}%
\textbf{\ }and refers to some as yet unknown but more fundamental
description of the system which underlies the quantum state \cite{Pusey12a}.
Hidden variable theories attempt to describe this more fundamental physical
state that is assumed to exist - attempts that so far have been unsuccessful
if locality is also invoked (see below). In addition to those associated
with physical quantum states, other density operators and state vectors may
be introduced for mathematical convenience. For physical quantum states, the
density operator is determined from either the preparation process or
inferred from the measurement process \textbf{- }\emph{quantum tomography}%
\textbf{\ -\ }and in general it is a statistical mixture of density
operators for possible preparation processes. Measurement itself constitutes
a possible preparation process. Following preparation, further experimental
processes may change the quantum state and dynamical equations give the time
evolution of the density operator between preparation and measurement, the
simplest situation being where measurement takes place immediately after
preparation. A full discussion of the predictive and retrodictive aspects of
the density operator is given in papers by Pegg et al \cite{Pegg02a}, \cite%
{Pegg05a}. Whilst there are often different mathematical forms for the
density operator that lead to the same predictive results for subsequent
measurements, the results of the measurements can also be used to
retrodictively determine the\emph{\ preferred form} of the density operator
that is consistent with the available preparation and measurement operators.
An example is given in \cite{Pegg05a}.

\subsection{Entangled and Non-Entangled States}

\label{SubSection - Entangled and Non-Entangled States}

\subsubsection{General Considerations}

Here the commonly applied \emph{physically-based approach} to mathematically
defining entangled states will be described \cite{Barnett09a}. The
definition involves vectors and density operators that represent states than
can be prepared in real experiments, so the mathematical approach is to be 
\emph{physically based}. The concept of quantum entanglement involves \emph{%
composite systems} made up of component \emph{sub-systems} each of which are 
\emph{distinguishable} from the other sub-systems, and where each could
constitute a stand-alone quantum system. This means the each sub-system will
have its own set of physically realisable quantum states - mixed or pure -
which could be prepared independently of the quantum states of the other
sub-systems. As will be seen, the requirement that sub-systems be
distinguishable and their states be physically preparable will have
important consequences, especially in the context of identical particle
systems.\textbf{\ }The formal definition of what is meant by an entangled
state starts with the pure states, described via a vector in a Hilbert
space. The formalism of quantum theory allows for \emph{pure states} for
composite systems made up of two or more distinct sub-systems via tensor
products of sub-system states 
\begin{equation}
\left\vert \Phi \right\rangle =\left\vert \Phi _{A}\right\rangle \otimes
\left\vert \Phi _{B}\right\rangle \otimes \left\vert \Phi _{C}\right\rangle
...  \label{Eq.NonEntangledPureState}
\end{equation}%
Such products are called \emph{non-entangled} or \emph{separable} states.
However, since these product states exist in a Hilbert space, it follows
that linear combinations of such products of the form%
\begin{equation}
\left\vert \Phi \right\rangle =\tsum\limits_{\alpha \beta \gamma
..}C_{\alpha \beta \gamma ..}\left\vert \Phi _{A}^{\alpha }\right\rangle
\otimes \left\vert \Phi _{B}^{\beta }\right\rangle \otimes \left\vert \Phi
_{C}^{\gamma }\right\rangle .  \label{Eq.EntangledPureState}
\end{equation}%
could also represent possible pure quantum states for the system. Such \emph{%
quantum superpositions} which cannot be expressed as a \emph{single} product
of sub-system states are known as \emph{entangled} (or \emph{non-separable})
states.

The concept of entanglement can be extended to \emph{mixed states}, which
are described via density operators in the Hilbert space. If $A$, $B$, ...
are the sub-systems with $\widehat{\rho }_{R}^{A}$, $\widehat{\rho }_{R}^{B}$%
, being density operators the sub-systems $A$, $B$, .then a \emph{general
non-entangled }or \emph{separable} state is one where the overall density
operator $\widehat{\rho }$ can be written as the weighted sum of tensor
products of these sub-system density operators in the form \cite{Werner89a} 
\begin{equation}
\widehat{\rho }=\sum_{R}P_{R}\,\widehat{\rho }_{R}^{A}\otimes \widehat{\rho }%
_{R}^{B}\otimes \widehat{\rho }_{R}^{C}\otimes ...
\label{Eq.NonEntangledState}
\end{equation}%
with $\sum_{R}P_{R}=1$ and $P_{R}\geq 0$ giving the probability that the
specific product state $\widehat{\rho }_{R}=\widehat{\rho }_{R}^{A}\otimes 
\widehat{\rho }_{R}^{B}\otimes \widehat{\rho }_{R}^{C}\otimes ...$ occurs.
It is assumed that at least in principle such separable states can be
prepared \cite{Werner89a}. This implies the possibility of turning off the
interactions between the different sub-systems, a task that may be difficult
in practice except for well-separated sub-systems.\textbf{\ }\emph{Entangled 
}states (or \emph{non-separable} states) are those that cannot be written in
this form, so in this approach knowing what the term entangled state refers
to is based on \emph{first} knowing what the general form is for a
non-entangled state. The density operator $\widehat{\rho }=\left\vert \Phi
\right\rangle \left\langle \Phi \right\vert $ for the pure state in (\ref%
{Eq.EntangledPureState}) is not of the form (\ref{Eq.NonEntangledState}), as
there are cross terms of the form $C_{\alpha \beta \gamma ..}C_{\theta
\lambda \eta ..}^{\ast }(\left\vert \Phi _{A}^{\alpha }\right\rangle
\left\langle \Phi _{A}^{\theta }\right\vert )\otimes (\left\vert \Phi
_{B}^{\beta }\right\rangle \left\langle \Phi _{B}^{\lambda }\right\vert
)\otimes ...$ involved.

The concepts of separability and entanglement based on the Eqs. (\ref%
{Eq.NonEntangledPureState}) and (\ref{Eq.NonEntangledState}) for
non-entangled states do not however just rest on the mathematical forms
alone. Implicitly there is the \emph{assumption} that separable quantum
states described by the two expressions can actually be created in \emph{%
physical processes}. The sub-systems involved must therefore be \emph{%
distinguishable quantum systems} in their own right, and the sub-system
states $\left\vert \Phi _{A}\right\rangle ,\left\vert \Phi _{B}\right\rangle
,\,...$ or $\widehat{\rho }_{R}^{A},\widehat{\rho }_{R}^{B},\,...$ must also
be \emph{possible quantum states} for the sub-systems. We will return to
these requirements later. The issue of the physical preparation of
non-entangled (separable) states starting from some uncorrelated fiducial
state for the separate sub-systems was introduced by Werner \cite{Werner89a}%
, and discussed further by Bartlett et al (see \cite{Bartlett06b}, Section
IIB). This involves the ideas of \emph{local operations} and \emph{classical
communication (LOCC) }dealt with in the next SubSection.

The key requirement is that entangled states exhibit a novel \emph{quantum
feature} that is only found in \emph{composite} systems. Separable states
are such that the \emph{joint probability} for measurements of all physical
quantities associated with the sub-systems can be found from separate
measurement probabilities obtained from the sub-system density operators $%
\widehat{\rho }_{R}^{A}$, $\widehat{\rho }_{R}^{B}$, etc and the overall
classical probability $P_{R}$ (see SubSection\ref{SubSection - Joint
Measurements}). This feature of separable probabilities is absent in \emph{%
certain} entangled states, and because of this key \emph{non-separability
feature} Schrodinger called these states "entangled". The separability
feature for the joint probabilities is essentially a classical feature and
applies in hidden variable theories (HVT) (see SubSection\ref{SubSection -
Hidden Variable Theory}) applied to quantum systems - as well as to quantum
separable states. The fact that entangled states are quantum states that can
exhibit the failure of this separability feature for classical LHV theories
highlights entanglement being a non-classical feature for composite systems.

An alternative \emph{operational approach} to defining entangled states
focuses on whether or not they exhibit certain non-classical features such
as Bell Inequality violation or whether they satisfy certain mathematical
tests such as having a non-negative partial transpose \cite{Peres96a}, \cite%
{Horodecki09a}, and a \emph{utilititarian approach} focuses or whether
entangled states have technological applications such as in various quantum
information protocols. As will be seen in SubSection \ref{SubSection - SSR
Separate Modes}, the particular definition of entangled states based on
their non-creatability via LOCC essentially coincides with the approach used
in the present paper. It has been realised for some time that different
types of entangled states occur, for example states in which a Bell
inequality is violated or states demonstrating an EPR paradox \cite{Reid89a}%
. Wiseman et al \cite{Wiseman07a}, \cite{Jones07a}, \cite{Cavalcanti09a} and
Reid et al \cite{Cavalcanti11a}, \cite{Reid09a}, \cite{Reid12a}, \cite{He15a}
discuss the concept of a \emph{heirarchy} of \emph{entangled states}, with
states exhibiting \emph{Bell nonlocality} being a subset of states for which
there is \emph{EPR steering}, which in turn is a subset of all the \emph{%
entangled states}, the latter being defined as states whose density
operators cannot be written as in Eq. (\ref{Eq.NonEntangledState}) though
without further consideration if additional properties are required for the
sub-system density operators. The operational approach could lead into a
quagmire of differing interpretations of entanglement dependng on which
non-classical feature is highlighted, and the utilitarian approach implies
that all entangled states have a technological use - which is by no means
the case. For these reasons, the present physical approach based on the
quantities involved representing allowed sub-system states is generally
favoured \cite{Barnett09a}. It is also compatible with later classifying
entangled states in a heirarchy.

\subsubsection{Local Systems and Operations}

As pointed out by Vedral \cite{Vedral07a}, one reason for calling states
such as in Eqs.(\ref{Eq.NonEntangledPureState}) and (\ref%
{Eq.NonEntangledState}) separable is associated with the idea of performing
operations on the separate sub-systems that do not affect the other
sub-systems. Such operations on such \emph{local systems} are referred to as 
\emph{local operations} and include unitary operations $\widehat{U}_{A}$, $%
\widehat{U}_{B}$, that change the states via $\widehat{\rho }%
_{R}^{A}\rightarrow \widehat{U}_{A}\widehat{\rho }_{R}^{A}\widehat{U}%
_{A}^{-1}$, $\widehat{\rho }_{R}^{B}\rightarrow \widehat{U}_{B}\widehat{\rho 
}_{R}^{B}\widehat{U}_{B}^{-1}$, etc as in a time evolution, and could
include processes by which the states $\widehat{\rho }_{R}^{A}$, $\widehat{%
\rho }_{R}^{B}$, are separately prepared from suitable initial states.

We note that performing local operations on a separable state only produces
another separable state, not an entangled state. Such local operations are
obviously faciltated in experiments if the sub-systems are essentially \emph{%
non-interacting} - such as when they are spatially \emph{well-separated},
though this does not have to be the case. The local systems and operations
could involve sub-systems whose quantum states and operators are just in
different parts of Hilbert space, such as for cold atoms in different
hyperfine states even when located in the same spatial region. Note the
distinction between \emph{local} and \emph{localised}. As described by
Werner \cite{Werner89a}, if one observer (Alice) is associated with
preparing separate sub-system $A$ in an allowed quantum state $\widehat{\rho 
}_{R}^{A}$ via local operations with a probability $P_{R}$, a second
observer (Bob) could be then advised via a \emph{classical communication}
channel to prepare sub-system $B$ in state $\widehat{\rho }_{R}^{B}$ via
local operations. After repeating this process for different choices $R$ of
the correlated pairs of sub-system states, the overall quantum state
prepared by both observers via this local operation and classical
communication protocol \emph{(}$\emph{LOCC)}$ would then be the bipartite
non-entangled state $\widehat{\rho }=\sum_{R}P_{R}\,\widehat{\rho }%
_{R}^{A}\otimes \widehat{\rho }_{R}^{B}$. Multipartite non-entangled states
of the form (\ref{Eq.NonEntangledState}) can also be prepared via LOCC
protocols involving further observers. As will be seen, the separable or
non-entangled states are just those that can be prepared by LOCC protocols.

\subsubsection{Constraints on Sub-System Density Operators}

A key issue however is whether density operators $\widehat{\rho }$ and $%
\widehat{\rho }_{R}^{A}$, $\widehat{\rho }_{R}^{B}$, in Eq. (\ref%
{Eq.NonEntangledState}) always represent possible \emph{quantum states},
even if the operators $\widehat{\rho }$ and $\widehat{\rho }_{R}^{A}$, $%
\widehat{\rho }_{R}^{B}$, etc satisfy all the standard mathematical
requirements for density operators - Hermitiancy, positiveness, trace equal
to unity, trace of density operator squared being not greater than unity. In
this paper it will be argued that for systems of identical massive particles%
\textbf{\ }there are further requirements not only on the overall density
operator, but also (for separable states)\textbf{\ }on those for the
individual sub-systems that are imposed by \emph{symmetrisation} and \emph{%
super-selection} rules.

\subsubsection{Classical Entanglement}

\label{SubSubSection - Classical Entanglement}

In addition to quantum entanglement there is a body of work (see \cite%
{Spreeuw98a}, \cite{Borges10a}), \cite{Aiello14a}\ dealing with so-called%
\textbf{\ }\emph{classical entanglement}\textbf{. }Here the\textbf{\ }\emph{%
states}\textbf{\ }of\textbf{\ }\emph{classical}\textbf{\ }systems - such as
a classical EM field - are represented via a formalism involving\textbf{\ }%
\emph{linear vector spaces}\textbf{\ }and classical entanglement is defined 
\emph{mathematically}. \ A discussion of classical entanglement is beyond
the scope of this paper. Although there are some formal similarities with
quantum entanglement - and even Bell type inequalites which can be violated,
there are key features that is not analogous to that for composite quantum
systems - quantum non-locality being one \cite{Spreeuw98a}. In the end,
rather than just focusing on similarities in the mathematical formalisms,
classical and quantum entanglement are seen as fundamentally different when
the physics of the two different types of system - one classical and
deterministic, the other quantum and probabilistic are taken into account.
In particular, the key feature of quantum entanglement relating to joint
measurement probailities is quite different to the corresponding one for
classical entanglement.

\subsection{Separate and Joint Measurements, Reduced Density Operator}

\label{SubSection - Joint Measurements}

In this SubSection we consider separate and joint measurements on systems
involving several sub-systems and introduce results for probabilities, mean
values for measurements on one of the sub-systems which are conditional on
the results for measurements on another of the sub-systems. This will
require consideration of quantum theoretical \emph{conditional probabilities}%
. The measurements involved will be assumed for simplicity to be von Neumann 
\emph{projective measurements} for physical quantities represented by
Hermitian operators $\widehat{\Omega }$, which project the quantum state
into subspaces for the eigenvalue $\lambda _{i}$ that is measured, the
subspaces being associated with Hermitian, idempotent \emph{projectors} $%
\widehat{\Pi }_{i}$ whose sum over all eigenvalues is unity. These concepts
are treated in several quantum theory textbooks, for example \cite{Peres93a}%
, \cite{Isham95a}. For completeness, an account setting out the key results
is presented in Appendix \ref{Appendix - Projective Measurements}.

\subsubsection{Joint Measurements on Sub-Systems}

For situations involving distinct sub-systems measurements can be carried
out on all the sub-systems and the results expressed in terms of the \emph{%
joint probability} for various outcomes. If $\widehat{\Omega }_{A}$ is a
physical quantity associated with sub-system $A$, with eigenvalues $\lambda
_{i}^{A}$ and with $\widehat{\Pi }_{i}^{A}$ the projector onto the subspace
with eigenvalue $\lambda _{i}^{A}$, $\widehat{\Omega }_{B}$ is a physical
quantity associated with sub-system $B$, with eigenvalues $\lambda _{j}^{B}$
and with $\widehat{\Pi }_{j}^{B}$ the projector onto the subspace with
eigenvalue $\lambda _{j}^{B}$ etc., then the \emph{joint probability} $%
P_{AB..}(i,j,\,...)$ that measurement of $\widehat{\Omega }_{A}$ leads to
result $\lambda _{i}^{A}$ , measurement of $\widehat{\Omega }_{B}$ leads to
result $\lambda _{j}^{B}$ ,etc is given by%
\begin{equation}
P_{AB..}(i,j,\,...)=Tr(\widehat{\Pi }_{i}^{A}\,\widehat{\Pi }_{j}^{B}\,...%
\widehat{\rho })  \label{Eq.JointProb}
\end{equation}%
This joint probability depends on the full density operator $\widehat{\rho }$
representing the allowed quantum state as well as on the quantities being
measured. Here the projectors (strictly $\widehat{\Pi }_{i}^{A}\otimes 
\widehat{1}^{B}\otimes \,...\,$, $\widehat{1}^{A}\otimes \widehat{\Pi }%
_{j}^{B}\otimes ...\,$, etc) commute, so the order of measurements is
immaterial. An alternative notation in which the physical quantities are
also specified is $P_{AB..}(\widehat{\Omega }_{A},i;\widehat{\Omega }%
_{B},j;\,...)$.

\subsubsection{Single Measurements on Sub-Systems and Reduced Density
Operator}

The \emph{reduced density operator} $\widehat{\rho }_{A}$ for sub-system $A$
given by 
\begin{equation}
\widehat{\rho }_{A}=Tr_{B,C,...}(\widehat{\rho })
\label{Eq.ReducedDensityOpr}
\end{equation}%
and enables the results for measurements on sub-system $A$ to be determined
for the situation where the results for all joint measurements involving the
other sub-systems are \emph{discarded}. The probability $P_{A}(i)$ that
measurement of $\widehat{\Omega }_{A}$ leads to result $\lambda _{i}^{A}$
irrespective of the results for meaurements on the other sub-systems is
given by%
\begin{eqnarray}
P_{A}(i) &=&\tsum\limits_{j,k,...}P_{AB..}(i,j,\,...)  \nonumber \\
&=&Tr(\widehat{\Pi }_{i}^{A}\,\widehat{\rho })  \label{Eq.SeparateProbModeA0}
\\
&=&Tr_{A}(\widehat{\Pi }_{i}^{A}\,\widehat{\rho }_{A})
\label{Eq.SeparateProbModeA}
\end{eqnarray}%
using $\tsum\limits_{j}\widehat{\Pi }_{j}^{B}=\widehat{1}$, etc. Hence the
reduced density operator $\widehat{\rho }_{A}$ plays the role of specifying
the quantum state for mode $A$ considered as a separate sub-system, even if
the original state $\widehat{\rho }$ is entangled. An alternative notation
in which the physical quantity is also specified is $P_{A}(\widehat{\Omega }%
_{A},i)$.

\subsubsection{Mean Value and Variance}

The \emph{mean value} for measuring a physical quantity $\widehat{\Omega }%
_{A}$ will be given by 
\begin{eqnarray}
\left\langle \widehat{\Omega }_{A}\right\rangle &=&\sum_{\lambda
_{i}^{A}}\lambda _{i}^{A}P_{A}(i)  \nonumber \\
&=&Tr_{A}(\widehat{\Omega }^{A}\,\widehat{\rho }_{A})  \label{Eq.Mean}
\end{eqnarray}%
where we have used $\widehat{\Omega }^{A}=\sum_{\lambda _{i}^{A}}\lambda
_{i}^{A}\widehat{\Pi }_{i}^{A}$.

The \emph{variance} of measurements of the physical quantity $\widehat{%
\Omega }_{A}$ will be given by 
\begin{eqnarray}
\left\langle (\Delta \widehat{\Omega }^{A})^{2}\right\rangle
&=&\sum_{\lambda _{i}^{A}}(\lambda _{i}^{A}-\left\langle \widehat{\Omega }%
_{A}\right\rangle )^{2}P_{A}(i)  \nonumber \\
&=&Tr_{A}(\left( \widehat{\Omega }^{A}-\left\langle \widehat{\Omega }%
_{A}\right\rangle \right) ^{2}\,\widehat{\rho }_{A})  \label{Eq.Variance}
\end{eqnarray}%
so both the mean and variance only depend on the reduced density operator $%
\widehat{\rho }_{A}$.

On the other hand the \emph{mean value} of a \emph{product} of sub-system
operators $\widehat{\Omega }_{A}\otimes \widehat{\Omega }_{B}\otimes 
\widehat{\Omega }_{C}\otimes ...$, where $\widehat{\Omega }_{A}$, $\widehat{%
\Omega }_{B}$, $\widehat{\Omega }_{C}$, .. are Hermitian operators
representing physical quantities for the separate sub-systems, is given by%
\begin{eqnarray}
\left\langle \widehat{\Omega }_{A}\otimes \widehat{\Omega }_{B}\otimes 
\widehat{\Omega }_{C}\otimes .\right\rangle &=&\sum_{\lambda
_{i}^{A}}\sum_{\lambda _{j}^{B}}...\lambda _{i}^{A}\lambda
_{j}^{B}...P_{AB..}(i,j,\,...)  \nonumber \\
&=&Tr\left( \widehat{\Omega }_{A}\otimes \widehat{\Omega }_{B}\otimes 
\widehat{\Omega }_{C}\otimes .\right) \widehat{\rho }
\label{Eq.MeanProductGeneral}
\end{eqnarray}%
which involves the overall system density operator, as expected.

\subsubsection{Conditional Probabilities}

Treating the case of two sub-systems for simplicity we can use Bayes theorem
(see Appendix \ref{Appendix - Projective Measurements}, Eq.(\ref{Eq.BayesThm}%
)) to obtain expressions for \emph{conditional probabilities} \cite%
{Barnett09a}. The conditional probability that if measurement of $\widehat{%
\Omega }_{B}$ associated with sub-system $B$ leads to eigenvalue $\lambda
_{j}^{B}$ then measurement of $\widehat{\Omega }_{A}$ associated with
sub-system $A$ leads to eigenvalue $\lambda _{i}^{A}$ is given by 
\begin{equation}
P_{AB}(i|j)=Tr(\widehat{\Pi }_{i}^{A}\,\widehat{\Pi }_{j}^{B}\,\widehat{\rho 
})/Tr(\widehat{\Pi }_{j}^{B}\widehat{\rho })  \label{Eq.CondProbGeneralCase}
\end{equation}%
In general, the overall density operator is required to determine the
conditional probability. An alternative notation in which the physical
quantities are also specified is $P_{AB}(\widehat{\Omega }_{A},i|\widehat{%
\Omega }_{B},j)$.

As shown in Appendix \ref{Appendix - Projective Measurements} the
conditional probability is given by 
\begin{equation}
P_{AB}(i|j)=Tr(\widehat{\Pi }_{i}^{A}\widehat{\rho }_{cond}(\widehat{\Omega }%
_{B},\lambda _{j}^{B}))  \label{Eq.ConProbGC2}
\end{equation}%
where 
\begin{equation}
\widehat{\rho }_{cond}(\widehat{\Omega }_{B},\lambda _{j}^{B})=\widehat{\Pi }%
_{j}^{B}\,\widehat{\rho }\,\widehat{\Pi }_{j}^{B}/Tr(\widehat{\Pi }_{j}^{B}%
\widehat{\rho })  \label{Eq.CondDensityOprMeastB}
\end{equation}%
is the so-called \emph{conditioned density operator}, corresponding the
quantum state produced following the measurement of $\widehat{\Omega }_{B}$
that obtained the result $\lambda _{j}^{B}$. The conditional probability
result is the same as%
\begin{equation}
P_{AB}(i|j)=Tr(\widehat{\Pi }_{i}^{A}\widehat{\rho }_{cond}(\widehat{\Omega }%
_{B},\lambda _{j}^{B}))  \label{Eq.CondProbGeneralCase2}
\end{equation}%
which is the same as the expression (\ref{Eq.SeparateProbModeA0}) with $%
\widehat{\rho }$ replaced by $\widehat{\rho }_{cond}(\widehat{\Omega }%
_{B},\lambda _{j}^{B})$. This is what would be expected for a conditioned
measurement probability.

Also, if the measurement results for $\widehat{\Omega }_{B}$ are not
recorded the conditioned density operator now becomes 
\begin{eqnarray}
\widehat{\rho }_{cond}(\widehat{\Omega }_{B}) &=&\sum_{\lambda
_{j}^{B}}P_{B}(j)\widehat{\rho }_{cond}(\widehat{\Omega }_{B},\lambda
_{j}^{B})  \nonumber \\
&=&\sum_{\lambda _{j}^{B}}\widehat{\Pi }_{j}^{B}\,\widehat{\rho }\,\widehat{%
\Pi }_{j}^{B}  \label{Eq.UnrecordedCondDensOpr}
\end{eqnarray}%
This is still different to the original density operator $\widehat{\rho }$
because a measurement of $\widehat{\Omega }_{B}$ has occured, even if we
dont know the outcome. However, the measurement probability for $\widehat{%
\Omega }_{A}$ is now%
\begin{eqnarray}
P_{AB}(i|Any\,j) &=&Tr(\widehat{\Pi }_{i}^{A}\widehat{\rho }_{cond}(\widehat{%
\Omega }_{B}))  \nonumber \\
&=&Tr(\widehat{\Pi }_{i}^{A}\widehat{\rho })  \label{Eq.NoSignalling0} \\
&=&P_{A}(i)  \label{Eq.NoSignalling}
\end{eqnarray}%
where we have used the cyclic properties of the trace, $\left( \widehat{\Pi }%
_{j}^{B}\right) ^{2}=\widehat{\Pi }_{j}^{B}$ and $\sum_{\lambda _{j}^{B}}%
\widehat{\Pi }_{j}^{B}=\widehat{1}$. The results in Eqs. (\ref%
{Eq.NoSignalling0}) and (\ref{Eq.NoSignalling}) are the same as the
measurement probability for $\widehat{\Omega }_{A}$ if no measurement for $%
\widehat{\Omega }_{B}$ had taken place at all. This is perhaps not
surprising, since the record of the latter measurements was discarded.
Another way of showing this result is that Bayes Theorem tells us that $%
\sum_{j}P_{AB}(i|j)P_{B}(j)=\sum_{j}P_{AB}(i,j)=P_{A}(i)$, since $%
\sum_{j}P_{AB}(i,j)$ is the probability that measurement of $\widehat{\Omega 
}_{A}$ will lead to $\lambda _{i}^{A}$ and measurement of $\widehat{\Omega }%
_{B}$ will lead to any of the $\lambda _{j}^{B}$. This result is called the
no-signalling theorem \cite{Barnett09a}.

Also, as $P_{AB}(i|Any\,j)=Tr(\widehat{\Pi }_{i}^{A}\widehat{\rho }_{cond}(%
\widehat{\Omega }_{B}))$\ we see from (\ref{Eq.SeparateProbModeA}) that 
\begin{equation}
\widehat{\rho }_{A}=Tr_{B}(\widehat{\rho }_{cond}(\widehat{\Omega }_{B}))
\label{Eq.RelnReducedDensOprCondDensOpr}
\end{equation}%
showing that the trace over $B$\ of the conditioned density operator for the
state obtained by measuring \emph{any} observable $\widehat{\Omega }_{A}$\
and then discarding the results just gives the \emph{reduced density operator%
} for sub-system $A$.

\subsubsection{Conditional Mean and Variance}

As explained in Appendix \ref{Appendix - Projective Measurements}, to
determine the \emph{conditioned mean value} of $\widehat{\Lambda }$ after
measurement of $\widehat{\Omega }$ has led to the eigenvalue $\lambda _{i}$
we use $\widehat{\rho }_{cond}(\widehat{\Omega },i)$ rather than $\,\widehat{%
\rho }$ in the mean formula $\left\langle \widehat{\Lambda }\right\rangle
=Tr(\widehat{\Lambda }\widehat{\rho })$ and the result is given in terms of
the conditional probability $P(\widehat{\Lambda }j|\widehat{\Omega }i)$.
Here we refer to two commuting observables and include the operators in the
notation to avoid any misinterpretation. Hence%
\begin{eqnarray}
\left\langle \widehat{\Lambda }\right\rangle _{\widehat{\Omega },i} &=&Tr(%
\widehat{\Lambda }\widehat{\rho }_{cond}(\widehat{\Omega },i))  \nonumber \\
&=&\dsum\limits_{j}\mu _{j}\,P(\widehat{\Lambda },j|\widehat{\Omega },i)
\label{Eq.CondMean0}
\end{eqnarray}

For the \emph{conditioned variance} of $\widehat{\Lambda }$ after
measurement of $\widehat{\Omega }$ has led to the eigenvalue $\lambda _{i}$
we use $\widehat{\rho }_{cond}(\widehat{\Omega },i)$ rather than $\,\widehat{%
\rho }$ and the conditioned mean $\left\langle \widehat{\Lambda }%
\right\rangle _{\widehat{\Omega },i}$rather than $\left\langle \widehat{%
\Lambda }\right\rangle $ in the variance formula $\left\langle \Delta 
\widehat{\Lambda }^{2}\right\rangle =Tr((\widehat{\Lambda }-\left\langle 
\widehat{\Lambda }\right\rangle )^{2}\widehat{\rho })$. Hence%
\begin{eqnarray}
\left\langle \Delta \widehat{\Lambda }^{2}\right\rangle _{\widehat{\Omega }%
,i} &=&Tr((\widehat{\Lambda }-\left\langle \widehat{\Lambda }\right\rangle _{%
\widehat{\Omega },i})^{2}\widehat{\rho }_{cond}(\widehat{\Omega },i)) 
\nonumber \\
&=&\dsum\limits_{j}(\mu _{j}-\left\langle \widehat{\Lambda }\right\rangle _{%
\widehat{\Omega },i})^{2}\,P(\widehat{\Lambda },j|\widehat{\Omega },i)
\label{Eq.CondVariance0}
\end{eqnarray}

If we weighted the conditioned mean by the probability $P(\widehat{\Omega }%
,i)$ that measuring $\widehat{\Omega }$ has led to the eigenvalue $\lambda
_{i}$ and summed over the possible outcomes $\lambda _{i}$ for the $\widehat{%
\Omega }$ measurement, then we obtain the mean for measurements of $\widehat{%
\Lambda }$ after un-recorded measurements of $\widehat{\Omega }$ have
occured. From Bayes theorem $\tsum\limits_{i}P(\widehat{\Lambda },j|\widehat{%
\Omega },i)P(\widehat{\Omega },i)=P(\widehat{\Lambda },j)$ so this gives the 
\emph{unrecorded mean} $\left\langle \widehat{\Lambda }\right\rangle _{%
\widehat{\Omega }}$ as 
\begin{eqnarray}
\left\langle \widehat{\Lambda }\right\rangle _{\widehat{\Omega }}
&=&\tsum\limits_{i}\left\langle \widehat{\Lambda }\right\rangle _{\widehat{%
\Omega },i}P(\widehat{\Omega },i)  \nonumber \\
&=&\dsum\limits_{j}\mu _{j}\,P(\widehat{\Lambda },j)  \nonumber \\
&=&\left\langle \widehat{\Lambda }\right\rangle  \label{Eq.UnrecordedMean}
\end{eqnarray}%
which is the usual mean value for measurements of $\widehat{\Lambda }$ when
no measurements of $\widehat{\Omega }$ have occured. Note that no such
similar result occurs for the \emph{unrecorded} \emph{variance }$%
\left\langle \Delta \widehat{\Lambda }^{2}\right\rangle _{\widehat{\Omega }}$%
\begin{eqnarray}
\left\langle \Delta \widehat{\Lambda }^{2}\right\rangle _{\widehat{\Omega }}
&=&\tsum\limits_{i}\left\langle \Delta \widehat{\Lambda }^{2}\right\rangle _{%
\widehat{\Omega },i}P(\widehat{\Omega },i)  \nonumber \\
&\neq &\left\langle \Delta \widehat{\Lambda }^{2}\right\rangle
\label{Eq.UnrecordedVariance}
\end{eqnarray}

\pagebreak

\subsection{Non-Entangled States}

\label{SubSection - Non-Entangled States}

In this SubSection we will set out the key results for measurements on
non-entangled states.

\subsubsection{Non-Entangled States - Joint Measurements on Sub-Systems}

In the case of the general \emph{non-entangled state} we find \ that the
joint probability is 
\begin{equation}
P_{AB..}(i,j,\,...)=\sum_{R}P_{R}\,P_{A}^{R}(i)P_{B}^{R}(j)\,...
\label{Eq.JointProbNonEntState}
\end{equation}%
where%
\begin{equation}
P_{A}^{R}(i)=Tr(\widehat{\Pi }_{i}^{A}\,\widehat{\rho }_{R}^{A})\qquad
P_{B}^{R}(j)=Tr(\widehat{\Pi }_{j}^{B}\,\widehat{\rho }_{R}^{B})\qquad ..
\label{Eq.SepProb}
\end{equation}%
are the probabilities for measurement results for $\widehat{\Omega }_{A}$, $%
\widehat{\Omega }_{B}$, ... on the separate sub-systems with density
operators $\,\widehat{\rho }_{R}^{A}$, $\widehat{\rho }_{R}^{B}$, etc and
the overall joint probability is given by the products of the probabilities $%
P_{A}^{R}(i)$, $P_{B}^{R}(j)$, .. for the measurement results $\lambda
_{i}^{A}$, $\lambda _{j}^{B}$, ... for physical quantities $\widehat{\Omega }%
_{A}$, $\widehat{\Omega }_{B}$, ... if the sub-systems are in the states $%
\widehat{\rho }_{R}^{A}$, $\widehat{\rho }_{R}^{B}$, etc. Note that here $%
P_{A}^{R}(i)$, $P_{B}^{R}(j)$, are given by quantum theory formulae for the
sub-systenm states. For simplicity only quantized measured values will be
considered - the extension to continuous values is straightforward. Thus the
results for the probabilities of joint measurements when the system is in a
separable quantum state are determined by the measurement probabilities in 
\emph{possible} quantum states for the sub-systems, combined with a\ \emph{%
classical}\textbf{\ }probability for creating the particular pair of
sub-system quantum states. Note the emphasis on "possible" - some of the
separable states described in \cite{Verstraete03a} are not possible.

Furthermore, if we consider measurements of the physical quantity $\widehat{%
\Omega }_{A}\otimes \widehat{\Omega }_{B}$ then for a separable state the%
\emph{\ mean value} for measurement of this quantity is given by%
\begin{equation}
\left\langle \widehat{\Omega }_{A}\otimes \widehat{\Omega }_{B}\right\rangle
=Tr\left( \widehat{\Omega }_{A}\otimes \widehat{\Omega }_{B}\,\widehat{\rho }%
\right) =\dsum\limits_{R}P_{R}\,\left\langle \widehat{\Omega }%
_{A}\right\rangle _{R}^{A}\,\left\langle \widehat{\Omega }_{B}\right\rangle
_{R}^{B}\,  \label{Eq.MeanJointMeastSepState}
\end{equation}%
where $\left\langle \widehat{\Omega }_{A}\right\rangle _{R}^{A}=Tr_{a}\left( 
\widehat{\Omega }_{A}\,\widehat{\rho }_{R}^{A}\right) $ and $\left\langle 
\widehat{\Omega }_{B}\right\rangle _{R}^{B}=Tr_{b}\left( \widehat{\Omega }%
_{B}\,\widehat{\rho }_{R}^{B}\right) $ are the mean values of $\widehat{%
\Omega }_{A}$ and $\widehat{\Omega }_{B}$ for the sub-system states $%
\widehat{\rho }_{R}^{A}$ and $\widehat{\rho }_{R}^{B}$ respectively. If $%
\left\langle \widehat{\Omega }_{A}\otimes \widehat{\Omega }_{B}\right\rangle
=\left\langle \widehat{\Omega }_{A}\right\rangle \left\langle \widehat{%
\Omega }_{B}\right\rangle $ then the state is said to be \emph{uncorrelated}%
. Separable states are \emph{correlated} except for the case where $\widehat{%
\rho }_{sep}=\widehat{\rho }^{A}\otimes \widehat{\rho }^{B}$, but the
correlation is essentially \emph{non-quantum} and attributable to the
classical probabilities $P_{R}$. However, for separable states the
inequality $|\left\langle \widehat{\Omega }_{A}\otimes \widehat{\Omega }%
_{B}^{\dag }\right\rangle |^{2}\leq \left\langle \widehat{\Omega }_{A}^{\dag
}\widehat{\Omega }_{A}\otimes \widehat{\Omega }_{B}^{\dag }\widehat{\Omega }%
_{B}\right\rangle $ applies, so that if $|\left\langle \widehat{\Omega }%
_{A}\otimes \widehat{\Omega }_{B}^{\dag }\right\rangle |^{2}>\left\langle 
\widehat{\Omega }_{A}^{\dag }\widehat{\Omega }_{A}\otimes \widehat{\Omega }%
_{A}^{\dag }\widehat{\Omega }_{A}\right\rangle $ then the state is entangled.

In the simple non-entangled\textbf{\ }\emph{pure state}\textbf{\ }situation
in Eq.(\ref{Eq.NonEntangledPureState}) the joint probabilty only involves a
single product of sub-system probabilities\textbf{\ }%
\begin{equation}
P_{AB..}(i,j,\text{\thinspace }...)=P_{A}(i)P_{B}(j)...
\label{Eq.JointProbPureNonEntState}
\end{equation}%
where\textbf{\ }%
\begin{equation}
P_{A}(i)=\left\langle \Phi _{A}\right\vert \widehat{\Pi }_{i}^{A}\,\left%
\vert \Phi _{A}\right\rangle \qquad P_{B}(j)=\left\langle \Phi
_{B}\right\vert \widehat{\Pi }_{j}^{B}\,\left\vert \Phi _{B}\right\rangle
\qquad ..  \label{Eq.SubSystProb}
\end{equation}%
just give the probabilities for measurements in the separate sub-systems.

This \emph{key result} (\ref{Eq.JointProbNonEntState}) showing that the
joint measurement probability for a separable state only depends on \emph{%
separate measurement probabilities} for the sub-systems, together with the
classical probability for preparing correlated product states of the
sub-systems, does \emph{not} necessarily apply for entangled states \cite%
{Werner89a}. However the \emph{key quantum feature} for \emph{composite
systems} of non-separability for joint measurement probabilites applies only
to entangled states. This strange quantum feature of entangled states has
been regarded as particularly unusual when the sub-systems are\textbf{\ }%
\emph{spatially well-separated} (or non-local) because then measurement
events can become space-like separated. This is relevant to quantum
paradoxes such as Einstein-Poldolsky-Rosen (EPR) and Bell's theorem which
aim to show there could be no causal classical theory explaining quantum
mechanics \cite{Einstein35a}, \cite{Schrodinger35a}\textbf{.} Measurements
on sub-system $A$ of physical quantity $\widehat{\Omega }_{A}$ affect the
results of measurements of $\widehat{\Omega }_{B}$ at the same time on a
distant sub-system $B$, even if the choice of measured quantity $\widehat{%
\Omega }_{B}$ is unknown to the experimenter measuring $\widehat{\Omega }%
_{A} $. As will be shown below, a similar result to (\ref%
{Eq.JointProbNonEntState}) also occurs in \emph{hidden variable theory} - a
classical theory - so non-separability for joint measurements resulting from
entanglement is a truly \emph{non-classical feature} of composite systems.

\subsubsection{Non-Entangled States - Single Sub-System Measurements}

For the general non-entangled state, the reduced density operator for
sub-system $A$ is given by 
\begin{equation}
\widehat{\rho }_{A}=\sum_{R}P_{R}\,\widehat{\rho }_{R}^{A}
\label{Eq.ReducedDensityOprNonEntState}
\end{equation}%
A key feature of a non-entangled state is that the results of a measurement
on any \emph{one} of the sub-systems is \emph{independent} of the states for
the \emph{other} subsystems. From Eqs.(\ref{Eq.SeparateProbModeA}) and (\ref%
{Eq.ReducedDensityOprNonEntState}) the probability $P_{A}(i)$ that
measurement of $\widehat{\Omega }_{A}$ leads to result $\lambda _{i}^{A}$ is
given by 
\begin{equation}
P_{A}(i)=\sum_{R}P_{R}\,P_{A}^{R}(i)  \label{Eq.MeasProbNonEntState}
\end{equation}%
where the reduced density operator is given by Eq. (\ref%
{Eq.ReducedDensityOprNonEntState}) for the non-entangled state in Eq. (\ref%
{Eq.NonEntangledState}). This result only depends on the reduced density
operator $\widehat{\rho }_{A}$, which represents a state for sub-system $A$
and which is a statistical mixture of the sub-system states $\,\widehat{\rho 
}_{R}^{A}$, with a probability $P_{R}$ that is the \emph{same} for all
sub-systems. The result for the measurement probability $P_{A}(i)$ is just
the statistical average of the results that would apply if sub-system $A$
were in possible states $\,\widehat{\rho }_{R}^{A}$. For all quantum states
the final expression for the measurement probability $P_{A}(i)$ only
involves a trace of quantities $\widehat{\Pi }_{i}^{A}$, $\widehat{\rho }%
_{A} $ that apply to sub-system $A$, but for a non-entangled state the
reduced density operator $\widehat{\rho }_{A}$ is given by an expression (%
\ref{Eq.ReducedDensityOprNonEntState}) that does \emph{not} involve density
operators for the other sub-systems. Thus for a non-entangled state, the
probability $P_{A}(i)$ is \emph{independent} of the states $\widehat{\rho }%
_{R}^{B}$, $\widehat{\rho }_{R}^{C}$, associated with the other sub-systems.
Analogous results apply for measurements on the other sub-systems.

\subsubsection{Non-Entangled States - Conditional Probability}

For a general non-entangled bipartite mixed state the conditional
probability is given by%
\begin{equation}
P_{AB}(i|j)=\sum_{R}P_{R}\,P_{A}^{R}(i)P_{B}^{R}(j)/\sum_{R}P_{R}%
\,P_{B}^{R}(j)  \label{Eq.CondProbNonEntangledState}
\end{equation}%
which in general depends on $\widehat{\Omega }_{B}$ associated with
sub-system $B$ and the eigenvalue $\lambda _{j}^{B}$. This may seem
surprising for the case where $A$ and $B$ are localised sub-systems which
are well separated. Even for separable states a measurement result for
sub-system $B$ will give \emph{immediate} information about a totally
separated measurement on sub-system $A$ - which is \emph{space-like}
separated. However it should be remembered that the general separable system
can still be a \emph{correlated} state, since each sub-system density
operator $\widehat{\rho }_{R}^{B}$ for sub-system $B$ is matched with a
corresponding density operator $\widehat{\rho }_{R}^{A}$ for sub-system $A$.
Results at $A$ can be correlated with those at $B$, so the observer at $A$
can potentially infer from a local measurement on the sub-system $A$ the
result of a local measurement on sub-system $B$. It is therefore not
necessarily the case that measurement results for $A$ are independent of
those for $B$. However, as we will see below, such correlations (usually)
hve a classical interpretation. Result (\ref{Eq.CondProbNonEntangledState})
is \emph{not} a case of the "spooky action at a distance" that Einstein \cite%
{Einstein35a} referred to.

However, for a non-entangled pure state where $\widehat{\rho }=\widehat{\rho 
}^{A}\otimes \widehat{\rho }^{B}$ we do find that 
\begin{equation}
P_{AB}(i|j)=P_{A}(i)  \label{Eq.CondProbNonEntPureState}
\end{equation}%
where $P_{A}(i)=Tr(\widehat{\Pi }_{i}^{A}\widehat{\rho }^{A})$. For
separable pure states the conditional probability is independent of $%
\widehat{\Omega }_{B}$ associated with sub-system $B$ and the eigenvalue $%
\lambda _{j}^{B}$.

Also of course $\sum_{j}P_{AB}(i|j)P_{B}(j)=P_{A}(i)$ is true for separable
states since it applies to general bipartite states. Hence if the
measurement results for $\widehat{\Omega }_{B}$ are discarded then the
probability distribution for measurements on $\widehat{\Omega }_{A}$ will be
determined from the conditioned density operator $\widehat{\rho }_{cond}(%
\widehat{\Omega }_{B})$ and just result in $P_{A}(i)$ - as in shown in Eq.(%
\ref{Eq.NoSignalling}) for any quantum state.

\subsubsection{Non-Entangled States - Mean Values and Correlations}

For non-entangled states as in Eq. (\ref{Eq.NonEntangledState}) the mean
value for measuring a physical quantity $\widehat{\Omega }_{A}\otimes 
\widehat{\Omega }_{B}\otimes \widehat{\Omega }_{C}\otimes ...$, where $%
\widehat{\Omega }_{A}$, $\widehat{\Omega }_{B}$, $\widehat{\Omega }_{C}$, ..
are Hermitian operators representing physical quantities for the separate
sub-systems can be obtained from Eqs.(\ref{Eq.NonEntangledState}) and (\ref%
{Eq.MeanProductGeneral}) and is given by 
\begin{equation}
\left\langle \widehat{\Omega }_{A}\otimes \widehat{\Omega }_{B}\otimes 
\widehat{\Omega }_{C}\otimes .\right\rangle =\sum_{R}P_{R}\,\left\langle 
\widehat{\Omega }_{A}\right\rangle _{R}^{A}\,\left\langle \widehat{\Omega }%
_{B}\right\rangle _{R}^{B}\,\left\langle \widehat{\Omega }_{C}\right\rangle
_{R}^{C}...  \label{Eq.MeanValueNonEntState}
\end{equation}%
where 
\begin{equation}
\left\langle \widehat{\Omega }_{K}\right\rangle _{R}^{K}=Tr(\widehat{\Omega }%
_{K}\,\widehat{\rho }_{R}^{K}),\qquad (K=A,B,\,...)
\label{Eq.SubSysMeanValueSepState}
\end{equation}%
is the mean value for measuring $\widehat{\Omega }_{K}$ in the $K$
sub-system when its density operator is $\widehat{\rho }_{R}^{K}$. Since the
overall mean value is not equal to the product of the separate mean values,
the measurements on the sub-systems are said to be \emph{correlated}.
However, for the general non-entangled state as the mean value is just the
products of mean values weighted by the probability of preparing the
particular product state - which involves a LOCC\ protocal, as we have seen
- the correlation is \emph{classical} rather than \emph{quantum} \cite%
{Barnett09a}. In the case of a single product state where $\widehat{\rho }=%
\widehat{\rho }^{A}\otimes \widehat{\rho }^{B}\otimes \widehat{\rho }%
^{C}\otimes ...$ we have $\left\langle \widehat{\Omega }_{A}\otimes \widehat{%
\Omega }_{B}\otimes \widehat{\Omega }_{C}\otimes .\right\rangle
=\left\langle \widehat{\Omega }_{A}\right\rangle ^{A}\,\left\langle \widehat{%
\Omega }_{B}\right\rangle ^{B}\,\left\langle \widehat{\Omega }%
_{C}\right\rangle ^{C}...$ which is just the product of mean values for the
separate sub-systems, and in this case the measurements on the sub-systems
are said to be \emph{uncorrelated}. For entangled states however the last
result for $\left\langle \widehat{\Omega }_{A}\otimes \widehat{\Omega }%
_{B}\otimes \widehat{\Omega }_{C}\otimes .\right\rangle $ does not apply,
and the correlation is strictly quantum.

\subsection{Local Hidden Variable Theories}

\label{SubSection - Hidden Variable Theory}

In a general \emph{local hidden variable theory} as envisaged by Einstein et
al \cite{Einstein35a} and Bell \cite{Bell65a}, physical quantities
associated with the sub-systems are denoted $\Omega _{A}$, $\Omega _{B}$
etc, which are real numbers not operators. Their values are assumed to be $%
\lambda _{i}^{A}$, $\lambda _{j}^{B}$ etc - having the same ranges as in
quantum theory, since HVT does not challenge the quantization feature. In
the \emph{realist} viewpoint of HVT all the physical quantities have \emph{%
definite values} at \emph{any} time, the probabilities for measuring these
values being determined from a set of hidden variables $\xi $, which are
themselves given by a probability function $P(\xi )$\ for each state
preparation process. Measurement is \emph{not} required for the values for
physical quantities to be created, as in quantum theory, nor do the hidden
variables change as a result of the act of measurement itself (though they
may change as a result of local interactions of the system with the
measurement apparatus \cite{Clauser69a}, \cite{Clauser78a}. As in classical
physics, ideal measurement is assumed\textbf{\ }\emph{not}\textbf{\ }to
change the state of the system - the hidden variables would only change in
accord with the (as yet unknown) dynamical equations that govern their
evolution. The hidden variables are regarded as the\textbf{\ }\emph{elements
of reality}\textbf{\ }that constitute the fundamental way of describing the
system \cite{Einstein35a}. There may be just a single hidden variable or a
set, and the hidden variables could be discrete or continuous - these
details do not matter in a general HVT. In the original treatment of Bell 
\cite{Bell65a} the hidden variables uniquely determine the \emph{actual}
values that physical quantities would have when measured. However, in a
so-called "\emph{fuzzy}" hidden variable theory \cite{Clauser69a}, \cite%
{Clauser78a}, \cite{Reid00a}, \cite{Reid03a}, \cite{Reid09a}, \cite%
{Brunner14a} (see also Section 7.1 of \cite{Vedral07a}) the values for $%
\Omega _{A}$, $\Omega _{B}$ etc are determined \emph{probabilistically} from
the hidden variables, the probability functions being\emph{\ classical}%
\textbf{\ }and allow for the hidden variables not being known - just as in
classical statistical mechanics, where the unknown (but real) positions and
momenta of the classical particles are described via probabilities. The
probabilistic treatment of the hidden variables attempts to replicate the
probabilistic nature of quantum theory. For our purposes we will consider
only local hidden variable theories (LHV) - this is sufficient to
demonstrate key results such as the Bell inequalities. For local hidden
variable theories although the hidden variables\ $\xi $\ are \emph{global},
they act locally even for spatially separable sub-systems. For particular
hidden variables $\xi $\ the probability that $\Omega _{A}$\ has value $%
\lambda _{i}^{A}$\ will be given by $P_{A}(i,\xi )$\ and the probability
that $\Omega _{B}$\ has value $\lambda _{j}^{B}$\ will be given by $%
P_{B}(j,\xi )$, etc. The LHVT joint probability for measurement outcome for $%
\Omega _{A}$, $\Omega _{B},$ etc will be given by\textbf{\ }%
\begin{equation}
P_{AB..}(i,j,..)=\tint d\xi \,P(\xi )\,P_{A}(i,\xi )P_{B}(j,\xi )...
\label{Eq.JointProbLHVT}
\end{equation}%
Here $P(\xi )d\xi $\ is the probability that the hidden variables are in the
range $d\xi $\ around $\xi $, the HV being assumed continuous - which is not
a requirement \cite{Bell65a}. The probabilities satisfy the usual sum rules
for all outcomes giving unity, thus $\sum_{i}P_{A}(i,\xi )=1$, etc., $\int
d\xi \,P(\xi )=1$. The sub-system probabilities $P_{A}(i,\xi )$, $%
P_{B}(j,\xi )$\ etc only depend on the hidden variables $\xi $. Bell
inequalities are constraints derived on the basis of the assumption (\ref%
{Eq.JointProbLHVT}), and if violated therefore falsify all LHV\ theories.

The \emph{formal similarity} between the hidden variable theory expression
for the joint probability (\ref{Eq.JointProbLHVT}) and the quantum
expression (\ref{Eq.JointProbNonEntState}) for a separable state is
noticable. We could map $\xi \rightarrow R$, $P(\xi )\rightarrow P_{R}$, $%
\tint d\xi \Rightarrow \tsum\limits_{R}$, $P_{A}(i,\xi )\rightarrow
P_{A}^{R}(i)$\ and $P_{B}(j,\xi )\rightarrow P_{B}^{R}(j)$. The Werner
preparation process \cite{Werner89a} would then determine the setting for
the hidden variables $\xi $\textbf{.} If a hidden variable theory \emph{%
underpinned} quantum theory, it follows that the quantum probabilities $%
P_{A}^{R}(i)$ and $P_{B}^{R}(j)$ would always be equivalent to hidden
variable probabilites $P_{A}(i,\xi )$ or $P_{B}(j,\xi )$ for each of the
sub-systems (it would not be consistent to only have this apply to one of
the sub-systems and not the other). From the expression (\ref%
{Eq.JointProbLHVT}) for the joint probability general hidden variable theory
expressions for the mean value $\left\langle \Omega _{A}\times \Omega
_{B}\right\rangle _{HVT}$ for the product of the measurement results for
observables $\Omega _{A}$ and $\Omega _{B}$ for subsystems $A$, $B$
respectively (see (\ref{Eq.MeanValueHVT}) below)\ can be obtained that are
analogous to the quantum expression (\ref{Eq.MeanValueNonEntState}) for a
separable state. There is of course no independent fully developed classical
HVT that can actually predict the $P_{A}(i,\xi )$, $P_{B}(j,\xi ).$etc.

However, as we will see both the HVT (see \cite{Barnett09a} for a proof) and
the quantum separable state predictions are consistent with Bell
Inequalities, and it therefore requires a quantum entangled state to violate
Bell inequalities and to demonstrate failure of the LHV theory model (\ref%
{Eq.JointProbLHVT}). Naturally it follows that such quantum entangled states
cannot be described via a LHV theory. Hence the \emph{experimental violation}
of Bell inequalities would \emph{also} show that the particular quantum
state must be\emph{\ entangled}.

\subsubsection{LHV - Mean Values and Correlation}

The actual values that would be assigned to the physical quantities $\Omega
_{A}$, $\Omega _{B}$ etc will depend on the hidden variables but can be
taken as the \emph{mean values} of the possible values $\lambda _{i}^{A}$,$%
\lambda _{i}^{A}$ etc. We denote these mean values as $\left\langle \Omega
_{A}(\xi )\right\rangle $, $\left\langle \Omega _{B}(\xi )\right\rangle $
etc where%
\begin{equation}
\left\langle \Omega _{K}(\xi )\right\rangle =\tsum\limits_{\lambda
_{k}^{K}}\lambda _{k}^{K}\,P_{K}(k,\xi )\qquad (K=A,B,\,...)
\label{Eq.SubSysMeanValueHVT}
\end{equation}%
These expressions my be compared to Eq.(\ref{Eq.SubSysMeanValueSepState})
for the mean values of physical quantities $\widehat{\Omega }_{A}$, $%
\widehat{\Omega }_{B}$ etc in quantum separable states.

We can then obtain an expression for the mean value in HVT of the physical
quantity $\Omega _{A}\times \Omega _{B}\times \Omega _{C}\times ...$, where $%
\Omega _{A}$, $\Omega _{B}$, etc. are physical quantities for the separate
sub-systems. This is obtained from Eqs.(\ref{Eq.JointProbLHVT}) and (\ref%
{Eq.SubSysMeanValueHVT}) and is given by 
\begin{equation}
\left\langle \Omega _{A}\times \Omega _{B}\times \Omega _{C}\times
.\right\rangle _{LHV}=\tint d\xi \,P(\xi )\,\left\langle \Omega _{A}(\xi
)\right\rangle \,\left\langle \Omega _{B}(\xi )\right\rangle \,\left\langle
\Omega _{C}(\xi )\right\rangle ..  \label{Eq.MeanValueHVT}
\end{equation}%
This may be compared to Eq.(\ref{Eq.MeanValueNonEntState}) for the mean
value of the physical quantity $\widehat{\Omega }_{A}\otimes \widehat{\Omega 
}_{B}\otimes \widehat{\Omega }_{C}\otimes ..$ in quantum separable
states.\smallskip

\subsubsection{LHV - GHZ State}

The GHZ state \cite{Greenberger03a}, \cite{Greenberger90a} is an entangled
state of three sub-systems $A$, $B$ and $C$, each of which is associated
with two quantum states $\left\vert +1\right\rangle $ and $\left\vert
-1\right\rangle $. Each sub-system has three physical quantities, which
correspond to Pauli spin operators $\widehat{\sigma }_{x}$, $\widehat{\sigma 
}_{y}$ and $\widehat{\sigma }_{z}$. The quantum states $\left\vert
+1\right\rangle $ and $\left\vert -1\right\rangle $ are eigenstates of $%
\widehat{\sigma }_{z}$ with eigenvalues $+1$ and $-1$ respectively. Note
that the eigenvalues of the other two Pauli spin operators are also $+1$ and 
$-1$. The GHZ\ state is defined by%
\begin{equation}
\left\vert \Psi \right\rangle _{GHZ}=(\left\vert +1\right\rangle
_{A}\left\vert +1\right\rangle _{B}\left\vert +1\right\rangle
_{C}+\left\vert -1\right\rangle _{A}\left\vert -1\right\rangle
_{B}\left\vert -1\right\rangle _{C})/\sqrt{2}  \label{Eq.GHZState}
\end{equation}

The GHZ state provides a clear example of an entangled quantum state which
cannot be described via local hidden variable theory \cite{Greenberger90a}, 
\cite{Mermin90a}. In a \emph{non-fuzzy} version of the LHV model each of the
nine physical quantities $\sigma _{x}^{A}$, $\sigma _{y}^{A}$, $\sigma
_{z}^{A}$, $\sigma _{x}^{B}$, $\sigma _{y}^{B}$, $\sigma _{z}^{B}$, $\sigma
_{x}^{C}$, $\sigma _{y}^{C}$, $\sigma _{z}^{C}$ will be associated with
hidden variables that \emph{directly} specify the values $+1$ and $-1$ that
each one of these physical quantities may have. We denote these hidden
variables as $M_{\alpha }^{K}$, where $K=A,B,C$ and $\alpha =x,y,z$ and we
have $M_{\alpha }^{K}=+1$ or $-1$. With this direct specification of the
physical values Eq.(\ref{Eq.SubSysMeanValueHVT}) just becomes $\left\langle
\sigma _{\alpha }^{K}(M^{K})\right\rangle =M_{\alpha }^{K}$ and Eq.(\ref%
{Eq.MeanValueHVT}) becomes $\left\langle \sigma _{\alpha }^{A}\times \sigma
_{\beta }^{B}\times \sigma _{\gamma }^{C}.\right\rangle _{LHV}=M_{\alpha
}^{A}$ $M_{\beta }^{B}$ $M_{\gamma }^{C}$ $.$We can then derive a
contradiction with quantum theory regarding the LHV description of the GHZ
state.

Firstly, using the Pauli spin matrices for the $\left\vert +1\right\rangle $
and $\left\vert -1\right\rangle $ basis states 
\begin{equation}
\left[ \widehat{\sigma }_{x}\right] =\left[ 
\begin{tabular}{ll}
$0$ & $1$ \\ 
$1$ & $0$%
\end{tabular}%
\right] \quad \left[ \widehat{\sigma }_{y}\right] =\left[ 
\begin{tabular}{ll}
$0$ & $-i$ \\ 
$i$ & $0$%
\end{tabular}%
\right] \quad \left[ \widehat{\sigma }_{x}\right] =\left[ 
\begin{tabular}{ll}
$1$ & $0$ \\ 
$0$ & $-1$%
\end{tabular}%
\right]  \label{Eq.PauliMatrices}
\end{equation}%
it is straightforward to show that the GHZ state satisfies three eigenvalue
equations 
\begin{eqnarray}
\widehat{\sigma }_{x}^{A}\widehat{\sigma }_{y}^{B}\widehat{\sigma }%
_{y}^{C}\,\left\vert \Psi \right\rangle _{GHZ} &=&(-1)\left\vert \Psi
\right\rangle _{GHZ}  \nonumber \\
\widehat{\sigma }_{y}^{A}\widehat{\sigma }_{x}^{B}\widehat{\sigma }%
_{y}^{C}\,\left\vert \Psi \right\rangle _{GHZ} &=&(-1)\left\vert \Psi
\right\rangle _{GHZ}  \nonumber \\
\widehat{\sigma }_{y}^{A}\widehat{\sigma }_{y}^{B}\widehat{\sigma }%
_{x}^{C}\,\left\vert \Psi \right\rangle _{GHZ} &=&(-1)\left\vert \Psi
\right\rangle _{GHZ}  \label{Eq.GHZEigenV}
\end{eqnarray}%
Hence in LHV the three quantities $\sigma _{x}^{A}\sigma _{y}^{B}\sigma
_{y}^{C}$, $\sigma _{y}^{A}\sigma _{x}^{B}\sigma _{y}^{C}$ and $\sigma
_{y}^{A}\sigma _{y}^{B}\sigma _{x}^{C}$ must all have value $-1$ in the GHZ
state, so that as the values for these quantities are just the products of
the values for each of the factors we get three equations 
\begin{equation}
M_{x}^{A}\,M_{y}^{B}M_{y}^{C}=-1\quad M_{y}^{A}\,M_{x}^{B}M_{y}^{C}=-1\quad
M_{y}^{A}\,M_{y}^{B}M_{x}^{C}=-1  \label{Eq.HVTResultsGHZ}
\end{equation}%
Secondly, if we apply all three operators $\widehat{\sigma }_{x}^{A}\widehat{%
\sigma }_{x}^{B}\widehat{\sigma }_{x}^{C}$to the GHZ state we find another
eigenvalue equation 
\begin{equation}
\widehat{\sigma }_{x}^{A}\widehat{\sigma }_{x}^{B}\widehat{\sigma }%
_{x}^{C}\,\left\vert \Psi \right\rangle _{GHZ}=(+1)\left\vert \Psi
\right\rangle _{GHZ}  \label{Eq.GHZEigenV2}
\end{equation}%
which leads to 
\begin{equation}
M_{x}^{A}\,M_{x}^{B}M_{x}^{C}=+1  \label{Eq.HVTResultsGHZ2}
\end{equation}%
However, if we multiply the three equations in Eq.(\ref{Eq.HVTResultsGHZ})
together and use $(M_{y}^{K})^{2}=+1$ we find that $M_{x}^{A}%
\,M_{x}^{B}M_{x}^{C}=-1$, in direct contradiction to the last equation. Thus
the assignment of hidden variables for all the physical quantities $\sigma
_{\alpha }^{K}$ fails to describe the GHZ state. As we will see in the next
SubSection, there are tests involving the violation of Bell Inequalities
that are satisfied by some entangled states which allow a demonstration of
the failure of more general local LHV theories, even allowing for
correlations that are less than ideal.

The assumption of non-fuzzy LHV\ theories is not essential for the GHZ
arguement in the case of the ideal GHZ state (\ref{Eq.GHZState}). This is
because one may use the correlations of (\ref{Eq.GHZEigenV}) to establish a 
\emph{precise} prediction of one of the spins at $A$, by measuring the spins
at the other two locations. The assumption of local realism (on which the
LHV theory is based) then establishes a precise value for the hidden
variable \cite{Greenberger03a}, \cite{Greenberger90a}, \cite{Mermin90a}%
\textbf{. }In a more realistic scenario where the GHZ correlations are not
perfect, the "elements of reality" established this way becme fuzzy, and in
that case Mermin's Bell inequality \cite{Mermin90a}\textbf{\ }can be used to
establish a contradiction with LHV models.

\subsection{Paradoxes}

\label{SubSection - Paradoxes}

The EPR and Schrodinger Cat paradoxes figured prominently in early
discussions about entanglement. Both paradoxes involve \emph{composite
systems }and the consideration of quantum states which are entangled \ Both
these paradoxes reflect the conflict between \emph{quantum theory}, in which
the values for physical quantities only take on definite values when
measurement occurs and \emph{classical theory}, in which the values for
physical quantities always exist even when measurement is not involved. The
latter viewpoint is referred to as \emph{realism}. Quantum theory is also
probabalistic, so although the possible outcomes for measuring a physical
quantity can be determined prior to measurement, the actual outcome in a
given quantum state for any measurement is only known in terms of a \emph{%
probability}. However, from the realist viewpoint, quantum theory is \emph{%
incomplete} and a \emph{future theory} based around \emph{hidden variables}
would determine the actual values of the physical quantities, as well as the
quantum probabilities that particular values will be found via measurement.

Whilst the EPR and Schrodinger Cat paradoxes are of historical interest and
have provoked much debate, it was the formulation of the \emph{Bell
inequalities} (which are described in the next SubSection \ref{SubSection -
Bell Inequalities}) and the conditions under which they could be violated
that provided the first clear case of where the predictions of quantum
theory could differ from those of hidden variable theories. It then became
possible to carry out actual experiments to distinguish these two
fundamentally different theories. The actual experimental evidence is
consistent with quantum theory and (apart from a small number of remaining
loopholes) rules out local hidden variable theories.\smallskip

\subsubsection{EPR Paradox}

In the original version of the EPR paradox, Einstein et al \cite{Einstein35a}
considered a two-particle system $A$, $B$ in which the particles were
associated with \emph{positions} $\widehat{x}_{A}$, $\widehat{x}_{B}$ and 
\emph{momenta} $\widehat{p}_{A}$, $\widehat{p}_{B}$. They envisaged a
quantum state in which the pairs of physical quantities $\widehat{x}_{A}$, $%
\widehat{x}_{B}$ or $\widehat{p}_{A}$, $\widehat{p}_{B}$ had highly \emph{%
correlated }values - measured or otherwise. To be specific, one may consider
a simultaneous eigenstate $\left\vert \Phi \right\rangle $ of the two
commuting operators $\widehat{x}_{A}-\widehat{x}_{B}$ and $\widehat{p}_{A}+%
\widehat{p}_{B}$, where $(\widehat{x}_{A}-\widehat{x}_{B})\left\vert \Phi
\right\rangle =2x\left\vert \Phi \right\rangle $ and $(\widehat{p}_{A}+%
\widehat{p}_{B})\left\vert \Phi \right\rangle =0\left\vert \Phi
\right\rangle $. This state is an example of an entangled state, as may be
seen if it is expanded in terms of position eigenstates $\left\vert
x_{A}\,x_{B}\right\rangle $. If the system is in state $\left\vert \Phi
\right\rangle $ then from standard quantum theory if $A$ had a mean momentum 
$p$ then $B$ would have a mean momentum $-p$. Alternatively, if $A$ had a
mean position $x$ then $B$ would have a mean position $-x$. Then if the
eigenvalue $2x$ is very large so that\textbf{\ }the two particles will be
well-separated (in quantum theory their spatial wave functions would be
localised in separate spatial regions) it follows that if the position of $B$
was measured then the position of $A$ would be immediately known, even if
the particles were light years apart. On the other hand, if the momentum of $%
B$ was measured instead, then the momentum of $A$ would immediately be
known. From the realist point of view both $A$ and $B$ always have definite
positions and momenta, even if these are not known, so all these
measurements do is reveal these (hidden) values. It would seem then that
measurements of position and momentum on particle $B$ \emph{could }lead to a
knowledge of the position and momentum at a far distant particle $A$,
perhaps with an accuracy that would violate the Heisenberg Uncertainty
Principle (HUP). As we will see, this is not the case when quantum theory is
applied correctly. However, what Einstein et al pointed out as being
particularly strange was that the choice of whether the momentum or position
of $B$ was measured (and found to have a definite value) would instantly
determine which of the position or momentum of $A$ would then have a
definite value - even if $A$ and $B$ were separated by such a large distance
that no signal could have been passed from $B$ to $A$ regarding which
quantity was measured. Einstein referred to this as "spooky action at a
distance" to highlight the strangeness of what came to be referred to as
entangled states.\textbf{\ }Thus a somewhat paradoxical situation would seem
to arise. Einstein stated that this did not demonstrate that quantum theory
was wrong, only that it was incomplete.

The EPR argument assumes\emph{\ local realism}, to justify that the
posibility of an exact \emph{prediction} of the postion of the far-away
particle $A$ (based on the measurement of\textbf{\ }the position for the
particle $B$) implies the realist viewpoint that the position of particle $A$
was predetermined. The same argument applies to the momentum of particle $A$%
, and hence EPR conclude that \emph{both} the position and momentum of
particle $A$ are precisely predetermined \textbf{- }in conflict with the
Heisenberg Uncertainty Principle derived from quantum mechanics. Since the
argument is based on the assumption of local realism, the modern
interpretation of the EPR analysis is that it reveals (for the appropriate
entangled state) the inconsistency of local realism with the completeness of
quantum mechanics.

Discussions of the \emph{EPR paradox} \cite{Einstein35a} in terms of hidden
variable theories has been given by numerous authors (see \cite{Reid03a}, 
\cite{Vedral07a}, \cite{Barnett09a}, \cite{Reid09a} for example). The papers
and reviews by Reid et al \cite{Reid89a}, \cite{Reid03a}, \cite{Reid09a},
give a full account taking into consideration the "fuzzy" version of \emph{%
local} HVT (LHV) and determining the predictions for the conditional
variances for $x_{A}$ and $p_{A}$ based both on separable quantum states and
states described via local HVT. This treatment successfully quantifies the
somewhat qualitative considerations described in the previous paragraph. If
the position for particle $B$ is measured and the result is $x$, then the
original density operator $\widehat{\rho }$ for the two particle system is
changed into the conditional density operator $\widehat{\rho }_{cond}(%
\widehat{x}_{B},x)=\widehat{\Pi }_{x}^{B}\,\widehat{\rho }\,\widehat{\Pi }%
_{x}^{B}/Tr(\widehat{\Pi }_{x}^{B}\widehat{\rho })$, where $\widehat{\Pi }%
_{x}^{B}=(\left\vert x\right\rangle \left\langle x\right\vert )_{B}$ is the
projector onto the eigenvector $\left\vert x\right\rangle _{B}$ (the
eigenvalues $x$ are assumed for simplicity to form a quasi-continuum).
Similarly, if the momentum for particle $B$ is measured and the result is $p$%
, then the original density operator $\widehat{\rho }$ for the two particle
system is changed into the conditional density operator $\widehat{\rho }%
_{cond}(\widehat{p}_{B},p)=\widehat{\Pi }_{p}^{B}\,\widehat{\rho }\,\widehat{%
\Pi }_{p}^{B}/Tr(\widehat{\Pi }_{p}^{B}\widehat{\rho })$, where $\widehat{%
\Pi }_{p}^{B}=(\left\vert p\right\rangle \left\langle p\right\vert )_{B}$ is
the projector onto the eigenvector $\left\vert p\right\rangle _{B}$ (the
eigenvalues $p$ are assumed for simplicity to form a quasi-continuum). Here
we outline the discussion based on quantum separable states. Conditional
variances for position and momentum for sub-system $A$ are considered based
on measurements for sub-system $B$ of position. It is shown that for these
conditional variances the Heisenberg uncertainty principle still applies.
The same conclusion is obtained if the measurements on sub-system $B$ had
been the momentum. As the experimenter on sub-system $A$ could not know
whether the measurement on sub-system $B$ was on position or momentum, the
action at a distance feature of quantum entanglement is confirmed.

The question is whether the conditional variances $\left\langle \Delta 
\widehat{x}_{A}^{2}\right\rangle _{\widehat{x}_{B}}$\ for measuring $%
\widehat{x}_{A}$\ for sub-system $A$\ having measured $\widehat{x}_{B}$\ for
sub-system $B$, and $\left\langle \Delta \widehat{p}_{A}^{2}\right\rangle _{%
\widehat{p}_{B}}$\ for measuring $\widehat{p}_{A}$\ for sub-system $A$\
having measured $\widehat{p}_{B}$\ for sub-system $B$\ violate the
Heisenberg Uncertainty Principle \cite{Reid89a} 
\begin{equation}
\left\langle \Delta \widehat{x}_{A}^{2}\right\rangle _{\widehat{x}%
_{B}}\left\langle \Delta \widehat{p}_{A}^{2}\right\rangle _{\widehat{p}_{B}}<%
\frac{1}{4}\hbar ^{2}  \label{Eq.EPRViolation}
\end{equation}%
where the measurements on sub-system $B$\ are left unrecorded. If this
inequality holds we have an \emph{EPR\ violation}.

For\emph{\ separable states} the \emph{conditional probability} that
measurement of $\widehat{x}_{A}$ on sub-system $A$ leads to eigenvalue $%
x_{A} $ given that measurement of $\widehat{x}_{B}$ on sub-system $B$ leads
to eigenvalue $x_{B}$ is obtained from Eq.(\ref{Eq.CondProbNonEntangledState}%
) as 
\begin{equation}
P(\widehat{x}_{A},x_{A}|\widehat{x}_{B},x_{B})=\sum_{R}P_{R}\,P_{A}^{R}(%
\widehat{x}_{A},x_{A})P_{B}^{R}(\widehat{x}_{B},x_{B})/\sum_{R}P_{R}%
\,P_{B}^{R}(\widehat{x}_{B},x_{B})
\end{equation}%
where 
\begin{equation}
P_{A}^{R}(\widehat{x}_{A},x_{A})=Tr_{A}(\widehat{\Pi }_{x_{A}}^{A}\widehat{%
\rho }_{R}^{A})\qquad P_{B}^{R}(\widehat{x}_{B},x_{B})=Tr_{B}(\widehat{\Pi }%
_{x_{B}}^{B}\widehat{\rho }_{R}^{B})
\end{equation}%
are the probabilities for position measurements in the separate sub-systems.
The probability that measurement of $\widehat{x}_{B}$ on sub-system $B$
leads to eigenvalue $x_{B}$ is 
\begin{equation}
P(\widehat{x}_{B},x_{B})=\sum_{R}P_{R}\,P_{B}^{R}(\widehat{x}_{B},x_{B})
\end{equation}

The \emph{mean} result for measurement of $\widehat{x}_{A}$ for this \emph{%
conditional} measurement is from Eq.(\ref{Eq.CondMean0}) 
\begin{eqnarray}
\left\langle \widehat{x}_{A}\right\rangle _{\widehat{x}_{B},x_{B}}
&=&\dsum\limits_{x_{A}}x_{A}\,P(\widehat{x}_{A},x_{A}|\widehat{x}_{B},x_{B})
\nonumber \\
&=&\sum_{R}P_{R}\,\left\langle \widehat{x}_{A}\right\rangle _{R}P_{B}^{R}(%
\widehat{x}_{B},x_{B})/P(\widehat{x}_{B},x_{B})
\end{eqnarray}%
where 
\begin{equation}
\left\langle \widehat{x}_{A}\right\rangle
_{R}=\dsum\limits_{x_{A}}x_{A}P_{A}^{R}(\widehat{x}_{A},x_{A})
\end{equation}%
is the \emph{mean} result for measurement of $\widehat{x}_{A}$ when the
sub-system is in state $\widehat{\rho }_{R}^{A}$.

The \emph{conditional variance} for measurement of $\widehat{x}_{A}$ for the
conditional measurement of $\widehat{x}_{B}$ on sub-system $B$ which led to
eigenvalue $x_{B}$ is from Eq.(\ref{Eq.CondVariance0}) 
\begin{eqnarray}
\left\langle \Delta \widehat{x}_{A}^{2}\right\rangle _{\widehat{x}%
_{B},x_{B}} &=&\dsum\limits_{x_{A}}(x_{A}-\left\langle \widehat{x}%
_{A}\right\rangle _{\widehat{x}_{B},x_{B}})^{2}\,P(\widehat{x}_{A},x_{A}|%
\widehat{x}_{B},x_{B})  \nonumber \\
&=&\sum_{R}P_{R}\,\left\langle \Delta \widehat{x}_{A}^{2}\right\rangle _{%
\widehat{x}_{B},x_{B}}^{R}P_{B}^{R}(\widehat{x}_{B},x_{B})/P(\widehat{x}%
_{B},x_{B})
\end{eqnarray}%
where 
\[
\left\langle \Delta \widehat{x}_{A}^{2}\right\rangle _{\widehat{x}%
_{B},x_{B}}^{R}=\dsum\limits_{x_{A}}(x_{A}-\left\langle \widehat{x}%
_{A}\right\rangle _{\widehat{x}_{B},x_{B}})^{2}\,P_{A}^{R}(\widehat{x}%
_{A},x_{A}) 
\]%
is a variance for measurement of $\widehat{x}_{A}$ for when the sub-system
is in state $\widehat{\rho }_{R}^{A}$ but now with the fluctuation about the
mean $\left\langle \widehat{x}_{A}\right\rangle _{\widehat{x}_{B},x_{B}}$
for measurements conditional on measuring $\widehat{x}_{B}$.

However, for each sub-system state $R$ the quantity $\left\langle \Delta 
\widehat{x}_{A}^{2}\right\rangle _{\widehat{x}_{B},x_{B}}^{R}$ is \emph{%
minimised} if $\left\langle \widehat{x}_{A}\right\rangle _{\widehat{x}%
_{B},x_{B}}$ is replaced by the unconditioned mean $\left\langle \widehat{x}%
_{A}\right\rangle _{R}$ just determined from $\widehat{\rho }_{R}^{A}$. Thus
we have an inequality%
\begin{equation}
\left\langle \Delta \widehat{x}_{A}^{2}\right\rangle _{\widehat{x}%
_{B},x_{B}}^{R}\geq \left\langle \Delta \widehat{x}_{A}^{2}\right\rangle ^{R}
\label{Eq.VarianceIneqEPR}
\end{equation}%
where 
\begin{equation}
\left\langle \Delta \widehat{x}_{A}^{2}\right\rangle
^{R}=\dsum\limits_{x_{A}}(x_{A}-\left\langle \widehat{x}_{A}\right\rangle
)^{2}\,P_{A}^{R}(\widehat{x}_{A},x_{A})
\end{equation}%
is the \emph{normal variance} for measurement of $\widehat{x}_{A}$ for when
the sub-system is in state $\widehat{\rho }_{R}^{A}$.

Now if the measurements of $\widehat{x}_{B}$ are \emph{unrecorded} \textbf{-}
as would be the case from the point of view of the experimenter on spatially
well-separated sub-system $A$ when measurements on this sub-system take
place at the same time - then the \emph{conditioned variance} is 
\begin{eqnarray}
\left\langle \Delta \widehat{x}_{A}^{2}\right\rangle _{\widehat{x}_{B}}
&=&\tsum\limits_{x_{B}}\left\langle \Delta \widehat{x}_{A}^{2}\right\rangle
_{\widehat{x}_{B},x_{B}}P(\widehat{x}_{B},x_{B})  \nonumber \\
&=&\tsum\limits_{x_{B}}\sum_{R}P_{R}\,\left\langle \Delta \widehat{x}%
_{A}^{2}\right\rangle _{\widehat{x}_{B},x_{B}}^{R}P_{B}^{R}(\widehat{x}%
_{B},x_{B})
\end{eqnarray}%
which in view of inequality (\ref{Eq.VarianceIneqEPR}) satisfies 
\begin{eqnarray}
\left\langle \Delta \widehat{x}_{A}^{2}\right\rangle _{\widehat{x}_{B}}
&\geq &\tsum\limits_{x_{B}}\sum_{R}P_{R}\,\left\langle \Delta \widehat{x}%
_{A}^{2}\right\rangle ^{R}\,P_{B}^{R}(\widehat{x}_{B},x_{B})  \nonumber \\
&=&\sum_{R}P_{R}\,\left\langle \Delta \widehat{x}_{A}^{2}\right\rangle ^{R}
\end{eqnarray}%
using $\tsum\limits_{x_{B}}P_{B}^{R}(\widehat{x}_{B},x_{B})=1$. Thus the
variance for measurement of position $\widehat{x}_{A}$ conditioned on
unrecorded measurements for position $\widehat{x}_{B}$ satisfies an
inequality that only depends on the variances for measurements of $\widehat{x%
}_{A}$ in the possible sub-system $A$ states $\widehat{\rho }_{R}^{A}$.

Now exactly the same treatment can be carried out for the variance of \emph{%
momentum} $\widehat{p}_{A}$ also\textbf{\ }conditioned on unrecorded
measurements of measurements for momentum $\widehat{x}_{B}$. Details are
given in Appendix \ref{Appendix - Projective Measurements}. We have with%
\begin{eqnarray*}
\left\langle \Delta \widehat{p}_{A}^{2}\right\rangle _{\widehat{p}_{B}}
&=&\tsum\limits_{p_{B}}\left\langle \Delta \widehat{p}_{A}^{2}\right\rangle
_{\widehat{p}_{B},p_{B}}P(\widehat{p}_{B},p_{B}) \\
\left\langle \Delta \widehat{p}_{A}^{2}\right\rangle _{\widehat{p}%
_{B},p_{B}} &=&\dsum\limits_{p_{A}}(p_{A}-\left\langle \widehat{p}%
_{A}\right\rangle _{\widehat{p}_{B},p_{B}})^{2}\,P(\widehat{p}_{A},p_{A}|%
\widehat{p}_{B},p_{B}) \\
\left\langle \widehat{p}_{A}\right\rangle _{\widehat{p}_{B},p_{B}}
&=&\dsum\limits_{p_{A}}p_{A}\,P(\widehat{p}_{A},p_{A}|\widehat{p}_{B},p_{B})
\end{eqnarray*}%
the inequality 
\begin{equation}
\left\langle \Delta \widehat{p}_{A}^{2}\right\rangle _{\widehat{p}_{B}}\geq
\sum_{R}P_{R}\,\left\langle \Delta \widehat{p}_{A}^{2}\right\rangle ^{R}
\label{Eq.InequaltiyEPR2}
\end{equation}%
with 
\begin{equation}
\left\langle \Delta \widehat{p}_{A}^{2}\right\rangle
^{R}=\dsum\limits_{p_{A}}(p_{A}-\left\langle \widehat{p}_{A}\right\rangle
)^{2}\,P_{A}^{R}(\widehat{p}_{A},p_{A})
\end{equation}%
is the normal variance for measurement of $\widehat{p}_{A}$ for when the
sub-system is in state $\widehat{\rho }_{R}^{A}$.

We now multiply the two conditional variances, which it is important to note
were associated with two \emph{different conditioned states} based on two 
\emph{different} measurements - position and momentum - carried out on
sub-system $B$. 
\begin{equation}
\left\langle \Delta \widehat{x}_{A}^{2}\right\rangle _{\widehat{x}%
_{B}}\left\langle \Delta \widehat{p}_{A}^{2}\right\rangle _{\widehat{p}%
_{B}}\geq \sum_{R}P_{R}\,\left\langle \Delta \widehat{x}_{A}^{2}\right%
\rangle ^{R}\sum_{S}P_{S}\,\left\langle \Delta \widehat{p}%
_{A}^{2}\right\rangle ^{S}
\end{equation}%
However, from the general inequality in Eq.(\ref{Eq.SumInequality0})%
\begin{equation}
\tsum\limits_{R}\,P_{R}\,C_{R}\,\tsum\limits_{R}\,P_{R}\,D_{R}\geq \left(
\tsum\limits_{R}\,P_{R}\,\sqrt{C_{R}D_{R}}\right) ^{2}
\end{equation}%
we then have%
\begin{eqnarray}
\left\langle \Delta \widehat{x}_{A}^{2}\right\rangle _{\widehat{x}%
_{B}}\left\langle \Delta \widehat{p}_{A}^{2}\right\rangle _{\widehat{p}_{B}}
&\geq &\left( \tsum\limits_{R}\,P_{R}\,\sqrt{\left\langle \Delta \widehat{x}%
_{A}^{2}\right\rangle ^{R}\left\langle \Delta \widehat{p}_{A}^{2}\right%
\rangle ^{R}}\right) ^{2}  \nonumber \\
&=&\left( \tsum\limits_{R}\,P_{R}\,\sqrt{\left\langle \Delta \widehat{x}%
_{A}^{2}\right\rangle ^{R}}\times \sqrt{\left\langle \Delta \widehat{p}%
_{A}^{2}\right\rangle ^{R}}\right) ^{2}
\end{eqnarray}%
But we know from the HUP that for \emph{any} given state $\widehat{\rho }%
_{R}^{A}$ that $\left\langle \Delta \widehat{x}_{A}^{2}\right\rangle
^{R}\left\langle \Delta \widehat{p}_{A}^{2}\right\rangle ^{R}\geq \frac{1}{4}%
\hbar ^{2}$, so for the conditioned variances associated with a \emph{%
separable} state 
\begin{equation}
\left\langle \Delta \widehat{x}_{A}^{2}\right\rangle _{\widehat{x}%
_{B}}\left\langle \Delta \widehat{p}_{A}^{2}\right\rangle _{\widehat{p}%
_{B}}\geq \frac{1}{4}\hbar ^{2}  \label{Eq.HUPResultSepStates}
\end{equation}%
showing that for a separable state the conditioned variances $\left\langle
\Delta \widehat{x}_{A}^{2}\right\rangle _{\widehat{x}_{B}}$ and $%
\left\langle \Delta \widehat{p}_{A}^{2}\right\rangle _{\widehat{p}_{B}}$%
still satisfy the HUP. It is important to note that these variances were
associated with two different conditioned states based on two different
measurements - position and momentum - carried out on sub-system $B$, the
results of which the observer for sub-system $A$ would be unaware of. Thus
if the EPR violations as defined in Eq.(\ref{Eq.EPRViolation}) are to occur
then the state must be entangled. Progress towards experimental confirmation
of EPR violations is reviewed in Refs. \cite{Reid09a}, \cite{Brunner14a}.

In \cite{Reid03a} an analogous treatment based on \emph{local hidden
variable theory} (LHV)\ also shows that the HUP is satisfied for the
conditioned variances. The details of this treatment will not be given here,
but the formal similarity of expressions for conditional probabilities in
LHV theories and for separable states indicates the steps involved.

The EPR paradox is not confined to position and momentum measurements on two
sub-systems. A related paradox \cite{Bohm51a} occurs in the case of
measurements on \emph{spin components }$\widehat{S}_{\alpha 1}$ and $%
\widehat{S}_{\alpha 2}$ - with $\alpha =x,y,z$ - associated with two
sub-systems $1$ and $2$. The spin operators also satisfy non-zero
commutation rules (see paper II for details) 
\begin{equation}
\lbrack \widehat{S}_{\alpha 1},\widehat{S}_{\beta 1}]=i\widehat{S}_{\gamma
1}\qquad \lbrack \widehat{S}_{\alpha 2},\widehat{S}_{\beta 2}]=i\widehat{S}%
_{\gamma 2}  \label{Eq.SpinCommRules}
\end{equation}%
where $\alpha ,\beta ,\gamma $ are $x,y,z$ in cyclic order. Different spin
components for each sub-system do not have simultaneous precise measurements
leading to Heisenberg Uncertainty principle relations involving the \emph{%
variances} and \emph{mean} values%
\begin{equation}
\left\langle \Delta \widehat{S}_{\alpha 1}^{2}\right\rangle \left\langle
\Delta \widehat{S}_{\beta 1}^{2}\right\rangle \geq \frac{1}{4}|\left\langle 
\widehat{S}_{\gamma 1}\right\rangle |^{2}\qquad \left\langle \Delta \widehat{%
S}_{\alpha 2}^{2}\right\rangle \left\langle \Delta \widehat{S}_{\beta
2}^{2}\right\rangle \geq \frac{1}{4}|\left\langle \widehat{S}_{\gamma
2}\right\rangle |^{2}  \label{Eq.SpinHUP}
\end{equation}%
As in the case of position and momentum a special state of the combined
system has interesting features. For the case where the spin quantum number
of each sub-system is $1/2$ the measured values for \emph{any} spin
component of either system is either $+1/2$ or $-1/2$. In terms of
eigenstates for $\widehat{S}_{x1}$ and $\widehat{S}_{x2}$ we consider the
state 
\begin{equation}
\left\vert \Psi ^{-}\right\rangle =\frac{1}{\sqrt{2}}\left( \left\vert
x,+\right\rangle _{1}\otimes \left\vert x,-\right\rangle _{2}-\left\vert
x,-\right\rangle _{1}\otimes \left\vert x,+\right\rangle _{2}\right)
\label{Eq.SingletState}
\end{equation}%
This is actually one of the Bell states. In this form it shows that
measurements of the $x$ components of the spins are perfectly correlated, so
that for example if the measurement of $\widehat{S}_{x2}$ for sub-system $2$
results in the value $-1/2$, then a subsequent measurement of $\widehat{S}%
_{x1}$ for sub-system $1$ must result in the value $+1/2$. However, the same
state can be expressed in terms of eigenstates for $\widehat{S}_{y1}$ and $%
\widehat{S}_{y2}$ as 
\begin{equation}
\left\vert \Psi ^{-}\right\rangle =\frac{1}{\sqrt{2}}\left( \left\vert
y,+\right\rangle _{1}\otimes \left\vert y,-\right\rangle _{2}-\left\vert
y,-\right\rangle _{1}\otimes \left\vert y,+\right\rangle _{2}\right)
\end{equation}%
and analogous statements regarding measurement correlations apply if the
measurements were for $\widehat{S}_{y2}$ on sub-system $2$ with a subsequent
measurement of $\widehat{S}_{y1}$ on sub-system $1$. If the two sub-systems
were well-separated it might be expected that first measuring $\widehat{S}%
_{x2}$ for sub-system $2$ would determine the result of measuring $\widehat{S%
}_{x1}$ for sub-system $1$, and then measuring $\widehat{S}_{y2}$ for
sub-system $2$ would determine the result of measuring $\widehat{S}_{y1}$
for sub-system $1$ - and as the second ($\widehat{S}_{y2}$) measurement on
far distant sub-system $2$ should not affect the former measurement on
sub-system $1$ this would appear to result in precise measured values for $%
\widehat{S}_{x1}$ and $\widehat{S}_{y1}$ on sub-system $1$, which conflicts
with the Heisenberg Uncertainty principle requirement that $\left\langle
\Delta \widehat{S}_{x1}^{2}\right\rangle \left\langle \Delta \widehat{S}%
_{y1}^{2}\right\rangle \geq \frac{1}{4}|\left\langle \widehat{S}%
_{z1}\right\rangle |^{2}$.

However we can consider the variances for $\widehat{S}_{x1}$ and $\widehat{S}%
_{y1}$ which are conditional on measurements for $\widehat{S}_{x2}$ and $%
\widehat{S}_{y2}$ for sub-system $2$ and show that for \emph{separable}
states of the two sub-systems we have%
\begin{equation}
\left\langle \Delta \widehat{S}_{x1}^{2}\right\rangle _{\widehat{S}%
_{x2}}\left\langle \Delta \widehat{S}_{y1}^{2}\right\rangle _{\widehat{S}%
_{y2}}\geq \frac{1}{4}|\left\langle \widehat{S}_{z1}\right\rangle |^{2}
\label{Eq.EPRSepSpin}
\end{equation}%
Thus if we find that 
\begin{equation}
\left\langle \Delta \widehat{S}_{x1}^{2}\right\rangle _{\widehat{S}%
_{x2}}\left\langle \Delta \widehat{S}_{y1}^{2}\right\rangle _{\widehat{S}%
_{y2}}<\frac{1}{4}|\left\langle \widehat{S}_{z1}\right\rangle |^{2}
\label{Eq.EPRSpinViol}
\end{equation}%
then we have an example of a \emph{spin EPR violation}. Such a violation
requires that the quantum state is \emph{entangled}. The derivation \ of the
result (\ref{Eq.EPRSepSpin}) for separable states is set out in Appendix \ref%
{Appendix - Spin EPR Paradox}.

An effect related to the EPR paradox is \emph{EPR Steering}. As we have
seen, the measurement of the position for particle $B$ changes the density
operator and consequently the probability distributions for measurements on
particle $A$ will now be determined from the conditional probabilities, such
as $P_{AB}(\widehat{x}_{A},x_{A}|\widehat{x}_{B},x_{B})$ or $P_{AB}(\widehat{%
p}_{A},p_{A}|\widehat{x}_{B},x_{B}).$Thus measurements on $B$ are said to
steer the results for measurements on $A$. Steering will of course only
apply if the measurement results for $\widehat{x}_{B}$ are \emph{recorded},
and not discarded. A discussion of EPR Steering (see \cite{Reid09a}, \cite%
{He15a}) is beyond the scope of this article.

\subsubsection{Schrodinger Cat Paradox}

The Schrodinger Cat Paradox \cite{Schrodinger35a}, \cite{Rinner08a} relates
to composite systems where one sub-system (the cat) is macroscopic and the
other sub-system is microscopic (the radioactive atom). The paradox is a
clear consequence of quantum theory allowing the existence of entangled
states. Schrodinger envisaged a state in which an alive cat and an undecayed
atom existed at an initial time, and because the decayed atom would be
associated with a dead cat, the system after a time corresponding to the
half-life for radioactive decay would be described in quantum theory via the
entangled state 
\begin{equation}
\left\vert \Psi \right\rangle =\frac{1}{\sqrt{2}}(\left\vert e\right\rangle
_{Atom}\left\vert Alive\right\rangle _{Cat}+\left\vert g\right\rangle
_{Atom}\left\vert Dead\right\rangle _{Cat})  \label{Eq.SchrodingerCat}
\end{equation}%
in an obvious notation. The quantum state defined by (\ref{Eq.SchrodingerCat}%
) represents the knowledge that an observer outside the box would have about
the combined atom-cat system one hour after the live cat was placed in the
box along with an undecayed atom. The combined system is in an enclosed box,
and opening the box and observing what is inside constitutes a measurement
on the system. According to quantum theory if the box was opened at this
time there would be a probability of $1/2$ of finding the atom undecayed and
the cat alive, with the same probability for finding a decayed atom and a
dead cat. From the realist viewpoint the cat should be \emph{either} dead 
\emph{or} it should be alive \emph{irrespective} of whether the box is
opened or not, and it was regarded as a paradox that in the quantum theory
description of the state prior to measurement the cat is in some sense \emph{%
both} dead \emph{and} alive. This paradox is made worse because the cat is a
macroscopic system - how could a cat be either dead or alive at the same
time, it must be one or the other? From the quantum point of view in which
the actual values of physical quantities \emph{only} appear when measurement
occurs, the Schrodinger cat presents no paradox. The two possible values
signifying the health of the cat are "alive" and "dead", and these values
are found with a probability of $1/2$ when measurement takes place on
opening the box, and this would entirely explain the results if such an
experiment were to be performed. There is of course no paradox if the
quantum state is only considered to represent the observer's information
about what is inside the box. If the box is closed then at one half life
after the cat was put into the box, the state vector (\ref{Eq.SchrodingerCat}%
) enables the outside observer to correctly assess the probability that the
cat will be alive is $1/2$. If the box is then opened and the cat is found
to be dead, then the observer's information changes and the state vector for
the cat-atom system is now 
\begin{equation}
\left\vert \Psi ^{^{\prime }}\right\rangle =\left\vert g\right\rangle
_{Atom}\left\vert Dead\right\rangle _{Cat}  \label{Eq.DeadCat}
\end{equation}%
In this interpretation of quantum states, the notion of there being some
sort of underlying reality that exists\emph{\ prior}\textbf{\ }to
measurement is rejected. It is only this notion that such a reality must
exist - perhaps described via hidden variables - that leads to the paradox.
EPR paradoxes can also be constructed from the entangled state (\ref%
{Eq.SchrodingerCat}), as outlined in Refs.\textbf{\ }\cite{Reid05a}, \cite%
{Rosales15a}.

In recent times, experiments based on a \emph{Rydberg atom} in a \emph{%
microwave cavity} \cite{Haroche} involving states such as (\ref%
{Eq.SchrodingerCat}) have been performed showing that entanglement can occur
between macroscopic and microscopic systems, and it is even possible to
prepare states analogous to $\frac{1}{\sqrt{2}}(\left\vert
Alive\right\rangle _{Cat}+\left\vert Dead\right\rangle _{Cat})$ in the
macroscopic system itself. In such experiments the different macroscopic
states are large amplitude coherent states of the cavity mode. Coherent
states are possible for \emph{microwave photons} as they are created from 
\emph{classical currents} wiith \emph{well-defined phases}. A coherent
superposition of an alive and dead cat within the cat sub-system itself can
be \emph{created} by measurement. The entangled state in (\ref%
{Eq.SchrodingerCat}) can also be written as%
\begin{eqnarray}
\left\vert \Psi \right\rangle &=&\frac{1}{\sqrt{2}}\{\frac{1}{\sqrt{2}}%
\left( \left\vert e\right\rangle _{Atom}+\left\vert g\right\rangle
_{Atom}\right) \frac{1}{\sqrt{2}}(\left\vert Alive\right\rangle
_{Cat}+\left\vert Dead\right\rangle _{Cat})  \nonumber \\
&&+\frac{1}{\sqrt{2}}\left( \left\vert e\right\rangle _{Atom}-\left\vert
g\right\rangle _{Atom}\right) \frac{1}{\sqrt{2}}(\left\vert
Alive\right\rangle _{Cat}-\left\vert Dead\right\rangle _{Cat})\}
\label{Eq.CatNewBasis}
\end{eqnarray}%
so that measurement on the atom for an observable in which the superposition
states $\frac{1}{\sqrt{2}}\left( \left\vert e\right\rangle _{Atom}\pm
\left\vert g\right\rangle _{Atom}\right) $ are the eigenstates for this
observable would result in the cat \emph{then} being in the
corresponding.macroscopic superposition states $\frac{1}{\sqrt{2}}%
(\left\vert Alive\right\rangle _{Cat}\pm \left\vert Dead\right\rangle
_{Cat}) $ of an alive and dead cat.

\subsection{Bell Inequalities}

\label{SubSection - Bell Inequalities}

Violations of Bell's Inequalities represent situations where neither hidden
variable theory nor quantum theory based on separable states can account for
the result, and therefore provide a clear case where an entangled quantum
state is involved.

\subsubsection{Local Hidden Variable Theory Result}

A key feature of entangled states is that they are associated with \emph{%
violations of Bell inequalities} \cite{Bell65a} and hence can exhibit this
particular \emph{non-classical} feature. The Bell inequalities arise in
attempts to restore a \emph{classical} interpretation of quantum thory via
hidden variable treatments, where actual values are assigned to all
measureable quantities - including those which in quantum theory are
associated with non-commuting Hermitian operators. In this case we consider
two different physical quantities $\Omega _{A}$ for sub-system $A$, which
are listed $A_{1}$, $A_{2}$, etc , and two $\Omega _{B}$ for sub-system $B$,
which are listed $B_{1}$, $B_{2}$, etc$.$ The corresponding quantum
Hermitian operators $\widehat{\Omega }_{A}$, $\widehat{\Omega }_{B}$, etc
are $\widehat{A}_{1}$, $\widehat{A}_{2}$ and$,\widehat{B}_{1}$, $\widehat{B}%
_{2}$. The Bell inequalities involve the \emph{mean value} $\left\langle
A_{i}\times B_{j}\right\rangle _{HVT}$ of the product of observables $A_{i}$
and $B_{j}$ for subsystems $A$, $B$ respectively, for which there are two
possible measured values, $+1$ and $-1$. For simplicity we consider a local
HVT. In a local hidden variable theory.(LHV) we see using (\ref%
{Eq.JointProbLHVT}) that the mean values $\left\langle A_{i}\times
B_{j}\right\rangle _{LHV}$ are given by 
\begin{equation}
\left\langle A_{i}\times B_{j}\right\rangle _{LHV}=\dint d\xi \,P(\xi
)\,\left\langle A_{i}(\xi )\right\rangle \,\left\langle B_{j}(\xi
)\right\rangle  \label{Eq.HiddenVarMean}
\end{equation}%
where $\left\langle A_{i}(\xi )\right\rangle $ and $\left\langle B_{j}(\xi
)\right\rangle $\textbf{\ }(as in Eq.(\ref{Eq.SubSysMeanValueHVT})) are the
values are assigned to $A_{i}$ and $B_{j}$ when the hidden variables are $%
\xi $, and $P(\xi )$ is the hidden variable probability distribution
function. If the corresponding quantum Hermitian operators are such that
their eigenvalues are $+1$ and $-1$ - as in the case of Pauli spin operators
- then the only possible values for $\left\langle A_{i}(\xi )\right\rangle $
and $\left\langle B_{j}(\xi )\right\rangle $ are betweem $+1$ and $-1$,
since HVT does not conflict with quantum theory regarding allowed values for
physical quantities. However, local hidden variable theory predicts certain
inequalities for the mean values of products of physical quantities for the
two sub-systems.

The form given by Clauser et al \cite{Clauser69a} for \emph{Bell's inequality%
} is 
\begin{equation}
|S|\leq 2  \label{Eq.BellInequality}
\end{equation}%
where 
\begin{equation}
S=\left\langle A_{1}\times B_{1}\right\rangle _{LHV}+\left\langle
A_{1}\times B_{2}\right\rangle _{LHV}+\left\langle A_{2}\times
B_{1}\right\rangle _{LHV}-\left\langle A_{2}\times B_{2}\right\rangle _{LHV}
\label{Eq.ClassicalBellQuantity}
\end{equation}%
The minus sign can actually be attached to any one of the four terms.

Following the proof of the Bell inequalities in \cite{Barnett09a} we have%
\begin{eqnarray}
\left\langle A_{2}\times B_{1}\right\rangle _{LHV}-\left\langle A_{2}\times
B_{2}\right\rangle _{LHV} &=&\dint d\xi \,P(\xi )\,(\left\langle A_{2}(\xi
)\right\rangle \,\left\langle B_{1}(\xi )\right\rangle -\left\langle
A_{2}(\xi )\right\rangle \,\left\langle B_{2}(\xi )\right\rangle )  \nonumber
\\
&=&\dint d\xi \,P(\xi )\,(\left\langle A_{2}(\xi )\right\rangle
\,\left\langle B_{1}(\xi )\right\rangle (1\pm \left\langle A_{1}(\xi
)\right\rangle \,\left\langle B_{2}(\xi )\right\rangle )  \nonumber \\
&&-\dint d\xi \,P(\xi )\,(\left\langle A_{2}(\xi )\right\rangle
\,\left\langle B_{2}(\xi )\right\rangle (1\pm \left\langle A_{1}(\xi
)\right\rangle \,\left\langle B_{1}(\xi )\right\rangle )  \nonumber \\
&&
\end{eqnarray}%
Now all the quantities $\left\langle A_{i}(\xi )\right\rangle $, $%
\left\langle B_{j}(\xi )\right\rangle $ are bounded by $+1$ or $-1$, so the
expressions $(1\pm \left\langle A_{1}(\xi )\right\rangle \,\left\langle
B_{2}(\xi )\right\rangle )$ and $(1\pm \left\langle A_{1}(\xi )\right\rangle
\,\left\langle B_{1}(\xi )\right\rangle )$ are never negative. Taking the
modulus of the left side leads to an equality%
\begin{eqnarray}
&&\left\vert \left\langle A_{2}\times B_{1}\right\rangle _{LHV}-\left\langle
A_{2}\times B_{2}\right\rangle _{LHV}\right\vert  \nonumber \\
&\leq &\dint d\xi \,P(\xi )\,(\left\vert \left\langle A_{2}(\xi
)\right\rangle \right\vert \,\left\vert \left\langle B_{1}(\xi
)\right\rangle \right\vert (1\pm \left\langle A_{1}(\xi )\right\rangle
\,\left\langle B_{2}(\xi )\right\rangle )  \nonumber \\
&&+\dint d\xi \,P(\xi )\,(\left\vert \left\langle A_{2}(\xi )\right\rangle
\right\vert \,\left\vert \left\langle B_{2}(\xi )\right\rangle \right\vert
(1\pm \left\langle A_{1}(\xi )\right\rangle \,\left\langle B_{1}(\xi
)\right\rangle )  \nonumber \\
&\leq &\dint d\xi \,P(\xi )\,(1\pm \left\langle A_{1}(\xi )\right\rangle
\,\left\langle B_{2}(\xi )\right\rangle )+\dint d\xi \,P(\xi )\,(1\pm
\left\langle A_{1}(\xi )\right\rangle \,\left\langle B_{1}(\xi
)\right\rangle )  \nonumber \\
&=&2\pm (\dint d\xi \,P(\xi )\,\left\langle A_{1}(\xi )\right\rangle
\,\left\langle B_{2}(\xi )\right\rangle +\dint d\xi \,P(\xi )\,\left\langle
A_{1}(\xi )\right\rangle \,\left\langle B_{1}(\xi )\right\rangle )  \nonumber
\\
&=&2\pm (\left\langle A_{1}\times B_{2}\right\rangle _{LHV}+\left\langle
A_{1}\times B_{1}\right\rangle _{LHV})
\end{eqnarray}%
where we have used the results that $\left\vert \left\langle A_{2}(\xi
)\right\rangle \right\vert \,,\left\vert \left\langle B_{1}(\xi
)\right\rangle \right\vert $ and $\left\vert \left\langle B_{2}(\xi
)\right\rangle \right\vert $ are all less than unity and that\textbf{\ }$%
\dint d\xi \,P(\xi )=1$\textbf{. }Hence since $\left\vert \left\langle
A_{1}\times B_{2}\right\rangle _{LHV}+\left\langle A_{1}\times
B_{1}\right\rangle _{LHV}\right\vert =+(\left\langle A_{1}\times
B_{2}\right\rangle _{LHV}+\left\langle A_{1}\times B_{1}\right\rangle
_{LHV}) $ or $-(\left\langle A_{1}\times B_{2}\right\rangle
_{LHV}+\left\langle A_{1}\times B_{1}\right\rangle _{LHV})$ we have%
\begin{equation}
\left\vert \left\langle A_{2}\times B_{1}\right\rangle _{LHV}-\left\langle
A_{2}\times B_{2}\right\rangle _{LHV}\right\vert \pm \left\vert \left\langle
A_{1}\times B_{2}\right\rangle _{LHV}+\left\langle A_{1}\times
B_{1}\right\rangle _{LHV}\right\vert \leq 2  \label{Eq.BellIneq0}
\end{equation}%
But since $\left\vert X-Y\right\vert \leq \left\vert X\right\vert
+\left\vert Y\right\vert $ we see that from the $+$ version of the last
inequality that 
\begin{equation}
\left\vert \left\langle A_{2}\times B_{1}\right\rangle _{LHV}-\left\langle
A_{2}\times B_{2}\right\rangle _{LHV}+\left\langle A_{1}\times
B_{2}\right\rangle _{HVT}+\left\langle A_{1}\times B_{1}\right\rangle
_{HVT}\right\vert \leq 2  \label{Eq.BellInequality1}
\end{equation}%
This is a Bell inequality. Interchanging $A_{2}\leftrightarrow A_{1}$ and
repeating the derivation gives $\left\vert \left\langle A_{1}\times
B_{1}\right\rangle _{HVT}-\left\langle A_{1}\times B_{2}\right\rangle
_{LHV}+\left\langle A_{2}\times B_{2}\right\rangle _{LHV}+\left\langle
A_{2}\times B_{1}\right\rangle _{LHV}\right\vert \leq 2$, which is another
Bell inequality. Interchanging $B_{1}\leftrightarrow B_{2}$ and repeating
the derivation gives $\left\vert \left\langle A_{2}\times B_{2}\right\rangle
_{LHV}-\left\langle A_{2}\times B_{1}\right\rangle _{LHV}+\left\langle
A_{1}\times B_{1}\right\rangle _{LHV}+\left\langle A_{1}\times
B_{2}\right\rangle _{LHV}\right\vert \leq 2$, and interchanging $%
A_{2}\leftrightarrow A_{1}$ and $B_{1}\leftrightarrow B_{2}$ and repeating
the derivation gives $\left\vert \left\langle A_{1}\times B_{2}\right\rangle
_{LHV}-\left\langle A_{1}\times B_{1}\right\rangle _{LHV}+\left\langle
A_{2}\times B_{1}\right\rangle _{LHV}+\left\langle A_{2}\times
B_{2}\right\rangle _{LHV}\right\vert \leq 2$. Thus the minus sign can be
attached to any one of the four terms. An example of an entangled state that
violates the Bell inequality is given in SubSection \ref{SubSubSection -
Bell Violation}.

\subsubsection{Non-Entangled State Result}

It can be shown that the Bell inequalities \emph{also} \emph{always} occur
for \emph{non-entangled} states (see Section 7.3 of the book by Vedral \cite%
{Vedral07a}). For Bell's inequalities we consider Hermitian operators $%
\widehat{A}_{i}$ and $\widehat{B}_{j}$ for subsystems $A$, $B$ respectively,
for which there are two eigenvalues $+1$ and $-1$, where examples of the
operators are given by the components $\widehat{A}_{i}=a_{i}\cdot \widehat{%
\sigma }_{A}$ and $\widehat{B}_{j}=b_{j}\cdot \widehat{\sigma }_{B}$ of
Pauli spin operators $\widehat{\sigma }_{A}$ and $\widehat{\sigma }_{B}$
along directions with unit vectors $a_{i}$ and $b_{j}$. The corresponding
quantum theory quantity for the Bell inequality is%
\begin{equation}
S=E(\widehat{A}_{1}\otimes \widehat{B}_{1})+E(\widehat{A}_{1}\otimes 
\widehat{B}_{2})+E(\widehat{A}_{2}\otimes \widehat{B}_{1})-E(\widehat{A}%
_{2}\otimes \widehat{B}_{2})  \label{Eq.BellQuantity}
\end{equation}%
where in quantum theory the mean value is given by $E(\widehat{A}_{i}\otimes 
\widehat{B}_{j})=\left\langle \widehat{A}_{i}\otimes \widehat{B}%
_{j}\right\rangle =Tr(\widehat{\rho }\,\widehat{A}_{i}\otimes \widehat{B}%
_{j})$. For the general bipartite non-entangled state given by \ref%
{Eq.NonEntangledState} it is easy to show that%
\begin{equation}
S=\sum_{R}P_{R}\,\left( \left\langle \widehat{A}_{1}\right\rangle
_{R}^{A}\left\langle \widehat{B}_{1}+\widehat{B}_{2}\right\rangle
_{R}^{B}+\left\langle \widehat{A}_{2}\right\rangle _{R}^{A}\left\langle 
\widehat{B}_{1}-\widehat{B}_{2}\right\rangle _{R}^{B}\right)
\label{Eq.BellQNonEntState}
\end{equation}%
where $\left\langle \widehat{A}_{i}\right\rangle _{R}^{A}=Tr(\widehat{A}%
_{i}\,\widehat{\rho }_{R}^{A})$ and $\left\langle \widehat{B}%
_{j}\right\rangle _{R}^{B}=Tr(\widehat{B}_{j}\,\widehat{\rho }_{R}^{B})$ are
the expectation values of $\widehat{A}_{i}$ and $\widehat{B}_{j}$ for the
sub-systems $A$, $B$ in states $\widehat{\rho }_{R}^{A}$ and $\widehat{\rho }%
_{R}^{B}$ respectively. Now $\left\langle \widehat{A}_{i}\right\rangle
_{R}^{A}$ and $\left\langle \widehat{B}_{j}\right\rangle _{R}^{B}$ must lie
in the range $-1$ to $+1$, so that $\left\langle \widehat{B}_{1}\pm \widehat{%
B}_{2}\right\rangle _{R}^{B}$ must each lie in the range $-2$ to $+2$. Hence%
\begin{eqnarray}
|S| &\leq &\sum_{R}P_{R}\,\left( |\left\langle \widehat{A}_{1}\right\rangle
_{R}^{A}|\,|\left\langle \widehat{B}_{1}+\widehat{B}_{2}\right\rangle
_{R}^{B}|+|\left\langle \widehat{A}_{2}\right\rangle
_{R}^{A}|\,|\left\langle \widehat{B}_{1}-\widehat{B}_{2}\right\rangle
_{R}^{B}|\right)  \nonumber \\
&\leq &\sum_{R}P_{R}\,\left( |\left\langle \widehat{B}_{1}+\widehat{B}%
_{2}\right\rangle _{R}^{B}|\,+\,|\left\langle \widehat{B}_{1}-\widehat{B}%
_{2}\right\rangle _{R}^{B}|\right)  \nonumber \\
&\leq &2  \label{Eq.BellInequalityNonEntState}
\end{eqnarray}%
since to obtain $|\left\langle \widehat{B}_{1}+\widehat{B}_{2}\right\rangle
_{R}^{B}|=2$ requires $\left\langle \widehat{B}_{1}\right\rangle
_{R}^{B}=\left\langle \widehat{B}_{2}\right\rangle _{R}^{B}=\pm 1$ and then $%
|\left\langle \widehat{B}_{1}-\widehat{B}_{2}\right\rangle
_{R}^{B}|=|\left\langle \widehat{B}_{1}\right\rangle _{R}^{B}-\left\langle 
\widehat{B}_{2}\right\rangle _{R}^{B}|=0$, or to obtain $|\left\langle 
\widehat{B}_{1}-\widehat{B}_{2}\right\rangle _{R}^{B}|=2$ requires $%
\left\langle \widehat{B}_{1}\right\rangle _{R}^{B}=-\left\langle \widehat{B}%
_{2}\right\rangle _{R}^{B}=\pm 1$ and then $|\left\langle \widehat{B}_{1}+%
\widehat{B}_{2}\right\rangle _{R}^{B}|=|\left\langle \widehat{B}%
_{1}\right\rangle _{R}^{B}+\left\langle \widehat{B}_{2}\right\rangle
_{R}^{B}|=0$.

\subsubsection{Bell Inequality Violation and Entanglement}

\label{SubSubSection - Bell Violation}

It follows that for a general two mode non-entangled state $|S|$ cannot
violate the Bell inequality upper bound of $2$. Thus, the violation of Bell
inequalities proves that the quantum state must be entangled for the
sub-systems involved, so Bell inequality violations are a test of
entanglement.\textbf{\ }For entangled states such as the Bell state $%
\left\vert \Psi _{-}\right\rangle $\ (see \cite{Barnett09a}, Section 2.5)
written in terms of eigenstates of $\widehat{\sigma }_{z}^{A}$\ and $%
\widehat{\sigma }_{z}^{B}$\textbf{\ }%
\begin{equation}
\left\vert \Psi _{-}\right\rangle =\frac{1}{\sqrt{2}}(\left\vert
+1\right\rangle _{A}\otimes \left\vert -1\right\rangle _{B}-\left\vert
-1\right\rangle _{A}\otimes \left\vert +1\right\rangle _{B})
\end{equation}%
we find that\textbf{\ }%
\begin{equation}
E(\underrightarrow{a}\cdot \underrightarrow{\widehat{\sigma }}^{A}\otimes 
\underrightarrow{b}\cdot \underrightarrow{\widehat{\sigma }}^{B})=-%
\underrightarrow{a}\cdot \underrightarrow{b}  \label{Eq.BellSingletResult}
\end{equation}%
The Bell inequality (\ref{Eq.BellInequality1}) can be violated for the
choice where $b_{1}$and $b_{2}$\ are orthogonal and $a_{1}$, $a_{2}$\ are
parallel to $b_{1}+b_{2}$, $b_{1}-b_{2}$\ respectively (see \cite{Barnett09a}%
, Section 5.1). Furthermore, such a quantum state cannot be described via a
hidden variable theory, since Bell inequalities are always satisfied using a
hidden variable theory. Experiments have been carried out in optical systems
providing strong evidence for the existence of quantum states that violate
Bell inequalities with only a few loopholes remaining(see \cite{Horodecki09a}%
, \cite{Reid09a} and \cite{Brunner14a} for references to experiments). Such
violation of Bell inequalities is clearly a \emph{non-classical} feature,
since the experiments rule out all local hidden variable theories. As Bell
inequalities do not occur for separable states, the experimental observation
of a Bell inequality indicates the presence of an entangled state. These
violations are not without applications, since such Bell entangled states
can be useful in device-independent quantum key distribution\ \cite%
{Horodecki09a}, \cite{Barnett09a}, \cite{Vedral07a}.

\subsection{Non-local Correlations}

\label{SubSection - Non-local Correlations}

Another feature of entangled states is that they are associated with \emph{%
strong correlations} for \emph{observables} associated with \emph{localised
sub-systems} that are \emph{well-separated}, a particular example being 
\emph{EPR} \emph{correlations} between non-commuting observables. Entangled
states can exhibit this particular \emph{non-classical} feature, which again
cannot be accounted for via a hidden variable theory.

\subsubsection{Local Hidden Variable Theory}

Consider two operators $\widehat{\Omega }_{A}$ and $\widehat{\Omega }_{B}$
associated with sub-systems $A$ and $B$. These would be Hermitian if
observables are involved, but for generality this is not required. In a
local hidden variable theory these would be associated with functions $%
\Omega _{C}(\xi )$ $(C=A,B)$ of the local hidden variables $\xi $, with the
Hermitean adjoints $\widehat{\Omega }_{C}^{\dag }$ being associated with the
complex conjugates $\Omega _{C}^{\ast }(\xi )$. In local hidden variable
theory \emph{correlation functions} are given by the following mean values 
\begin{eqnarray}
\left\langle \Omega _{A}^{\ast }\times \Omega _{B}\right\rangle _{LHV}
&=&\dint d\xi \,P(\xi )\,\Omega _{A}^{\ast }(\xi )\Omega _{B}(\xi ) 
\nonumber \\
\left\langle \Omega _{A}^{\ast }\Omega _{A}\times \Omega _{B}^{\ast }\Omega
_{B}\right\rangle _{LHV} &=&\dint d\xi \,P(\xi )\,\Omega _{A}^{\ast }(\xi
)\Omega _{A}(\xi )\,\,\Omega _{B}^{\ast }(\xi )\Omega _{B}(\xi )
\label{Eq.CorrelnFunctions}
\end{eqnarray}%
which then can be shown to satisfy the following \emph{correlation inequality%
}%
\begin{equation}
|\,\left\langle \Omega _{A}^{\ast }\times \Omega _{B}\right\rangle
_{LHV}\,|^{2}\leq \left\langle \Omega _{A}^{\ast }\Omega _{A}\times \Omega
_{B}^{\ast }\Omega _{B}\right\rangle _{LHV}  \label{Eq.CorrelnInequal}
\end{equation}

This result is based on the inequality%
\begin{equation}
\tint d\xi \,P(\xi )C(\xi )\geq \left( \tint d\xi \,P(\xi )\sqrt{C(\xi )}%
\right) ^{2}  \label{Eq.IntegralInequalityB}
\end{equation}%
for real, positive functions $C(\xi ),P(\xi )$ and where $\dint d\xi \,P(\xi
)=1$, and which is proved in Appendix \ref{Appendix - Inequalities}. In the
present case we have $C(\xi )=\Omega _{A}^{\ast }(\xi )\Omega _{A}(\xi
)\,\,\Omega _{B}^{\ast }(\xi )\Omega _{B}(\xi )$, which is real, positive. A
violation of the inequality in Eq. (\ref{Eq.CorrelnInequal}) is an
indication of strong correlation between sub-systems $A$ and $B$.

\subsubsection{Non-Entangled State Result}

It can be shown that the correlation inequalities are \emph{always}
satisfied for non-entangled states. In quantum theory the correlation
functions are given by $\left\langle \widehat{\Omega }_{A}^{\dag }\otimes 
\widehat{\Omega }_{B}\right\rangle =Tr(\widehat{\rho }\,\widehat{\Omega }%
_{A}^{\dag }\otimes \widehat{\Omega }_{B})$ and $\left\langle \widehat{%
\Omega }_{A}^{\dag }\widehat{\Omega }_{A}\otimes \widehat{\Omega }_{B}^{\dag
}\widehat{\Omega }_{B}\right\rangle =Tr(\widehat{\rho }\,\widehat{\Omega }%
_{A}^{\dag }\widehat{\Omega }_{B}\otimes \widehat{\Omega }_{B}^{\dag }%
\widehat{\Omega }_{B})$. For a non-entangled state of sub-systems $A$ and $B$
we have%
\begin{eqnarray}
\left\langle \widehat{\Omega }_{A}^{\dag }\otimes \widehat{\Omega }%
_{B}\right\rangle &=&\sum_{R}P_{R}\,\left\langle \widehat{\Omega }_{A}^{\dag
}\right\rangle _{R}^{A}\left\langle \widehat{\Omega }_{B}\right\rangle
_{R}^{B}  \nonumber \\
\left\langle \widehat{\Omega }_{A}^{\dag }\widehat{\Omega }_{A}\otimes 
\widehat{\Omega }_{B}^{\dag }\widehat{\Omega }_{B}\right\rangle
&=&\sum_{R}P_{R}\,\left\langle \widehat{\Omega }_{A}^{\dag }\widehat{\Omega }%
_{A}\right\rangle _{R}^{A}\left\langle \widehat{\Omega }_{B}^{\dag }\widehat{%
\Omega }_{B}\right\rangle _{R}^{B}  \label{Eq.QuantumCorrelFns}
\end{eqnarray}

Now 
\begin{equation}
|\left\langle \widehat{\Omega }_{A}^{\dag }\otimes \widehat{\Omega }%
_{B}\right\rangle |\leq \sum_{R}P_{R}\,|\left\langle \widehat{\Omega }%
_{A}^{\dag }\right\rangle _{R}^{A}|\;|\left\langle \widehat{\Omega }%
_{B}\right\rangle _{R}^{B}|  \label{Eq.ResultE1}
\end{equation}%
since the modulus of a sum is always less than the sum of the moduli. Using $%
\left\langle \left( \widehat{\Omega }_{C}^{\dag }-\left\langle \widehat{%
\Omega }_{C}^{\dag }\right\rangle \right) \left( \widehat{\Omega }%
_{C}-\left\langle \widehat{\Omega }_{C}\right\rangle \right) \right\rangle
\geq 0$ with $(C=A,B)$, we obtain the Schwarz inequality - which is true for
all states - $\left\langle \widehat{\Omega }_{C}^{\dag }\widehat{\Omega }%
_{C}\right\rangle \geq \left\langle \widehat{\Omega }_{C}^{\dag
}\right\rangle \left\langle \widehat{\Omega }_{C}\right\rangle
=|\left\langle \widehat{\Omega }_{C}\right\rangle |^{2}=|\left\langle 
\widehat{\Omega }_{C}^{\dag }\right\rangle |^{2}$, and hence 
\begin{equation}
|\left\langle \widehat{\Omega }_{A}^{\dag }\otimes \widehat{\Omega }%
_{B}\right\rangle |\leq \sum_{R}P_{R}\,\sqrt{\left\langle \widehat{\Omega }%
_{A}^{\dag }\widehat{\Omega }_{A}\right\rangle _{R}^{A}}\;\sqrt{\left\langle 
\widehat{\Omega }_{B}^{\dag }\widehat{\Omega }_{B}\right\rangle _{R}^{B}}
\label{Eq.ResultE2}
\end{equation}

Next we use the inequality 
\begin{equation}
\tsum\limits_{R}\,P_{R}\,C_{R}\geq \left( \tsum\limits_{R}\,P_{R}\,\sqrt{%
C_{R}}\right) ^{2}  \label{Eq.SumInequalityB}
\end{equation}%
for real, positive functions $C_{R},P_{R}$ and where $\sum_{R}P_{R}=1$. This
inequality, which was used in the paper by Hillery et al \cite{Hillery06a},
is proved in Appendix \ref{Appendix - Inequalities}. In the present case we
have $C_{R}=\left\langle \widehat{\Omega }_{A}^{\dag }\widehat{\Omega }%
_{A}\right\rangle _{R}^{A}\left\langle \widehat{\Omega }_{B}^{\dag }\widehat{%
\Omega }_{B}\right\rangle _{R}^{B}$ so that 
\begin{equation}
|\left\langle \widehat{\Omega }_{A}^{\dag }\otimes \widehat{\Omega }%
_{B}\right\rangle |^{2}\leq \sum_{R}P_{R}\,\left\langle \widehat{\Omega }%
_{A}^{\dag }\widehat{\Omega }_{A}\right\rangle _{R}^{A}\left\langle \widehat{%
\Omega }_{B}^{\dag }\widehat{\Omega }_{B}\right\rangle _{R}^{B}=\left\langle 
\widehat{\Omega }_{A}^{\dag }\widehat{\Omega }_{A}\otimes \widehat{\Omega }%
_{B}^{\dag }\widehat{\Omega }_{B}\right\rangle  \label{Eq.ResultE3}
\end{equation}

Thus for a \emph{non-entangled state} we obtain the \emph{correlation
inequality}%
\begin{equation}
|\left\langle \widehat{\Omega }_{A}^{\dag }\otimes \widehat{\Omega }%
_{B}\right\rangle |^{2}=|\left\langle \widehat{\Omega }_{A}\otimes \widehat{%
\Omega }_{B}^{\dag }\right\rangle |^{2}\leq \left\langle \widehat{\Omega }%
_{A}^{\dag }\widehat{\Omega }_{A}\otimes \widehat{\Omega }_{B}^{\dag }%
\widehat{\Omega }_{B}\right\rangle  \label{Eq.QuantumCorrenInequality}
\end{equation}%
where the general result $\left\langle \widehat{\Omega }_{A}^{\dag }\otimes 
\widehat{\Omega }_{B}\right\rangle =\left\langle \widehat{\Omega }%
_{A}\otimes \widehat{\Omega }_{B}^{\dag }\right\rangle ^{\ast }$ has been
used. Thus non-entangled states have correlation functions that are
consistent with hidden variable theory.

\subsubsection{Correlation Violation and Entanglement}

Hence if it is found that the correlation inequality is violated $%
|\left\langle \widehat{\Omega }_{A}^{\dag }\otimes \widehat{\Omega }%
_{B}\right\rangle |^{2}=|\left\langle \widehat{\Omega }_{A}\otimes \widehat{%
\Omega }_{B}^{\dag }\right\rangle |^{2}>\left\langle \widehat{\Omega }%
_{A}^{\dag }\widehat{\Omega }_{A}\otimes \widehat{\Omega }_{B}^{\dag }%
\widehat{\Omega }_{B}\right\rangle $ then the state must be entangled, so
the correlation inequality violation is also a sufficiency test for
entanglement.\pagebreak

\section{Identical Particles and Entanglement}

\label{Section - Identical Particles and Entanglement}

We now take into account the situation where systems of \emph{identical
particles} are involved. This requires us to give special consideration to
the requirement that quantum states in such cases must conform to the \emph{%
symmetrisation principle} \cite{Dirac30a}. Further, entanglement is defined
as a property that involves systems with two (or more) \emph{sub-systems},
and the definition requires the specification of sub-systems that are \emph{%
distinguishable} from each other and on which\emph{\ measurements} can be
made. In addition, the sub-systems must be able to exist as separate systems
which can in principle be prepared in quantum states for that sub-system
alone. This feature is vital to the definition of separable (or
non-entangled) states on which the defintion of entangled states is based.
These key requirements that the sub-systems must be distinguishable,
susceptible to measurements and can exist in separate quantum states are
necessary for the concept of entanglement to make physical sense, and will
have important consequences for the choice of sub-systems when identical
particles are involved. These three key logical requirement for sub-systems
rule out considering labelled identical particles as sub-systems and lead to
the conclusion that sub-systems must be modes or sets of modes. \textbf{%
\bigskip }

\subsection{Symmetrisation Principle}

\label{SubSection - Symmetrization Principle}

Whether \emph{entangled} or \emph{not} the quantum states for systems of 
\emph{identical particles} must conform to the \emph{symmetrisation principle%
}, whereby for mixed states the overall density operator has to be invariant
under \emph{permutation operators,} or if pure states are involved, the
state vector is either unchanged (bosons) or changes sign (fermions) if the
permutation operator is odd. Either a first quantisation approach in which
the basis states are written as\textbf{\ }\emph{symmetrised products}\textbf{%
\ }of single particle states occupied by\textbf{\ }\emph{labeled} identical%
\textbf{\ }particles can be used, or a second quantisation approach where
the basis states are products of \emph{Fock states}\textbf{\ }for all single
particle states (modes), each Fock state specifying the\textbf{\ }\emph{%
number}\textbf{\ }of identical particles occupying the particular mode. In
first quantisation the symmetrisation process\textbf{\ }\emph{removes}%
\textbf{\ }any distinction between identical particles, whereas in second
quantisation only \emph{mode creation operators}\textbf{\ }are involved, and
these do not involve labeled particles. Symmetrization is built into the
definition of the Fock states. The two approaches are equivalent, but as we
will see the second quantisation approach is more suited to identifying
sub-systems and defining entanglement in systems of identical particles.

It is useful to clarify some of the issues involved by considering a simple
example. Since density operators can always be expressed in a diagonal form
involving their orthonormal eigenstates $\left\vert \Phi \right\rangle $
with real, positive eigenvalues $P(\Phi )$ as $\widehat{\rho }%
=\tsum\limits_{\Phi }$\ $P(\Phi )\,\left\vert \Phi \right\rangle
\left\langle \Phi \right\vert $ and each $\left\vert \Phi \right\rangle $
can always be written as a linear combination of basis vectors $\left\vert
\Psi \right\rangle $, we will focus on these basis vectors and their forms
in both first and second quantisation. We consider a system with $N=2$
particles, which may be \emph{identical} and are labeled $1$ and $2$, or
they may be \emph{distinguishable} and labeled $\alpha $ and $\beta $.%
\textbf{\ }In each case a particle has a choice of two modes which it may
occupy. Thus there are two distinct single particle states (modes)
designated as $\left\vert A\right\rangle $\ and $\left\vert B\right\rangle $%
\ in the\textbf{\ }\emph{identical}\textbf{\ }particle case, and four
distinct single particle states (modes) designated as $\left\vert A_{\alpha
}\right\rangle $\textbf{, }$\left\vert B_{\alpha }\right\rangle $\textbf{\ }%
and\textbf{\ }$\left\vert A_{\beta }\right\rangle $\textbf{, }$\left\vert
B_{\beta }\right\rangle $\ in the\textbf{\ }\emph{distinguishable}\textbf{\ }%
particle case for particles $\alpha $\ and $\beta $\ respectively. The
notation in first quantisation is that $\left\vert C(i)\right\rangle $
refers to a vector in which particle $i$ is in mode $\left\vert
C\right\rangle $. The notation in second quantisation is that $\left\vert
n\right\rangle _{C}$ refers to a vector where there are $n$ particles in
mode $\left\vert C\right\rangle $.

For the case of the \emph{identical} particles we consider \emph{basis states%
} for two \emph{bosons} or for two \emph{fermions}, which are written in
terms of \emph{first quantization} as 
\begin{eqnarray}
\left\vert \Psi \right\rangle _{boson} &=&\frac{1}{\sqrt{2}}(\left\vert
A(1)\right\rangle \otimes \left\vert B(2)\right\rangle +\left\vert
B(1)\right\rangle \otimes \left\vert A(2)\right\rangle )
\label{Eq.TwoBosonsTwoModes} \\
\left\vert \Psi \right\rangle _{fermion} &=&\frac{1}{\sqrt{2}}(\left\vert
A(1)\right\rangle \otimes \left\vert B(2)\right\rangle -\left\vert
B(1)\right\rangle \otimes \left\vert A(2)\right\rangle )
\label{Eq.TwoFermionsTwoModes}
\end{eqnarray}%
and clearly satisfy the symmetrization principle. In \emph{second
quantization} the basis state in both the fermion and boson cases is 
\begin{equation}
\left\vert \Psi \right\rangle _{boson,\,fermion}=\left\vert 1\right\rangle
_{A}\otimes \left\vert 1\right\rangle _{B}
\label{Eq.SecondQnTwoIdentTwoModes}
\end{equation}%
In both first and second quantisation this basis state involves one
identical particle in mode $\left\vert A\right\rangle $ and the other in
mode $\left\vert B\right\rangle $.

These examples highlight two possibilities for specifying \emph{sub-systems}
for systems of \emph{identical} particles. The two possibilities have \emph{%
differing} consequences in terms of whether specific pure states are
regarded as separable or entangled in terms of the general form in Eq.(\ref%
{Eq.NonEntangledPureState}) for separable pure states, depending on whether
the first or second quantisation approach is used. The \emph{first option}
is to regard the\emph{\ labeled identical particles} as sub-systems - in
which case using first quantisation the boson or fermion basis states in
Eqs.(\ref{Eq.TwoBosonsTwoModes}) and (\ref{Eq.TwoFermionsTwoModes}) would be
regarded as \emph{entangled} states of the two sub-systems consisting of
particle $1$ and particle $2$ \cite{Peres93a}, \cite{Sorensen01a}, \cite%
{Hyllus12a}. This is a more mathematical approach, and suffers from the
feature that the sub-systems are not distinguishable and measurements cannot
be made on specifically labelled identical particles. In the case of
identical particles the option of regarding labeled identical particles as
the sub-systems leads to the concept of \emph{entanglement due to
symmetrisation}. In the textbook by Peres (\cite{Peres93a}, see pp126-128)
it is stated that "two particles of the same type are \emph{always}
entangled". Peres obviously considers such entanglement is a result of
symmetrization. The\emph{\ second option} would be to regard the \emph{modes}
or single particle states as sub-systems \cite{TerraCunha07a} - in which
case using second quantisation the basis state for both fermions or bosons
in Eq.(\ref{Eq.SecondQnTwoIdentTwoModes}) would be regarded as a \emph{%
separable} state of two sub-systems consisting of modes $\left\vert
A\right\rangle $\textbf{\ }and\textbf{\ }$\left\vert B\right\rangle $. This
is a more physically based approach, and has the advantage that the
sub-systems are distinguishable and measurements can be made on specific
modes. Noting that in the example the \emph{same} quantum state is involved
with one identical particle in mode $\left\vert A\right\rangle $ and the
other in mode $\left\vert B\right\rangle $, the different categorisation is
disconcerting. It indicates that a choice must be made in regard to defining
sub-systems when identical particles are involved (see SubSection \ref%
{SubSection - Sub-Systems - Particles or Modes ?}).

Now consider the case where the particles are \emph{distinguishable}. Each
distinguishable particle\textbf{\ }$\alpha $\textbf{, }$\beta $\textbf{\ }%
has its own unique set of modes\textbf{\ }$A_{\alpha }$\textbf{, }$B_{\alpha
}$\textbf{, }$A_{\beta }$\textbf{, }$B_{\beta }$\textbf{\ }. There are two
cases in which one particle\textbf{\ }$\alpha $\textbf{\ }occupies mode%
\textbf{\ }$\left\vert A_{\alpha }\right\rangle $\textbf{\ }or\textbf{\ }$%
\left\vert B_{\alpha }\right\rangle $\textbf{\ }and the other particle%
\textbf{\ }$\beta $\textbf{\ }occupies mode\textbf{\ }$\left\vert A_{\beta
}\right\rangle $\textbf{\ }or\textbf{\ }$\left\vert B_{\beta }\right\rangle $%
\textbf{. }Basis states analogous to the previous ones are given in\emph{\
first quantization} as%
\begin{equation}
\left\vert \Psi \right\rangle _{dist}=\left\vert A_{\alpha }(\alpha
)\right\rangle \otimes \left\vert B_{\beta }(\beta )\right\rangle \qquad
or\qquad \left\vert \Psi \right\rangle _{dist}=\left\vert B_{\alpha }(\alpha
)\right\rangle \otimes \left\vert A_{\beta }(\beta )\right\rangle
\label{Eq.TwoDistPartTwoModes}
\end{equation}%
The somewhat surplus particle labels $(\alpha )$ and $(\beta )$ have been
added for comparison with (\ref{Eq.TwoBosonsTwoModes}) and (\ref%
{Eq.TwoFermionsTwoModes}). The states (\ref{Eq.TwoDistParticlesTwoModes})
are not required to satisfy the symmetrization principle since the particles
are not identical. Each may be either a boson or a fermion. In \emph{second
quantisation} the basis states are 
\begin{eqnarray}
\left\vert \Psi \right\rangle _{dist} &=&(\left\vert 1\right\rangle
_{A_{\alpha }}\otimes \left\vert 0\right\rangle _{B_{\alpha }})\otimes
(\left\vert 0\right\rangle _{A_{\beta }}\otimes \left\vert 1\right\rangle
_{B_{\beta }})  \nonumber \\
&&\qquad or\qquad  \nonumber \\
\left\vert \Psi \right\rangle _{dist} &=&(\left\vert 0\right\rangle
_{A_{\alpha }}\otimes \left\vert 1\right\rangle _{B_{\alpha }})\otimes
(\left\vert 1\right\rangle _{A_{\beta }}\otimes \left\vert 0\right\rangle
_{B_{\beta }})  \label{Eq.SecondQnTwoDistTwoModesEach}
\end{eqnarray}%
In both first and second quantisation, the first case corresponds to
particle $\alpha $ being in mode $\left\vert A_{\alpha }\right\rangle $ and
particle $\beta $ being in mode $\left\vert B_{\beta }\right\rangle $ with
the other two modes empty, and the second case corresponds to particle $%
\alpha $ being in mode $\left\vert B_{\alpha }\right\rangle $ and particle $%
\beta $ being in mode $\left\vert A_{\beta }\right\rangle $ with the other
two modes empty.

These examples also highlight two possibilities for specifying sub-systems
for systems of \emph{distinguishable} particles. In this case the two
possibilities have \emph{similar} consequences in terms of whether specific
pure states are regarded as separable or entangled, based on the general
form in Eq.(\ref{Eq.NonEntangledPureState}) for separable pure states,
irrespective of whether the first or second quantisation approach is used.
Here the \emph{first option} is to regard the\emph{\ labeled distinguishable
particles} as sub-systems - in which case using first quantisation the boson
or fermion basis states in Eqs.(\ref{Eq.TwoDistPartTwoModes}) would be
regarded as \emph{separable} states of the two sub-systems consisting of
particle $\alpha $ and particle $\beta $. The\emph{\ second option} would be
to regard the \emph{modes} or single particle states as sub-systems - in
which case using second quantisation the basis state for both fermions or
bosons in Eq.(\ref{Eq.SecondQnTwoDistTwoModesEach}) would be regarded as a 
\emph{separable} state of four sub-systems consisting of modes $\left\vert
A_{\alpha }\right\rangle $, $\left\vert B_{\alpha }\right\rangle $\textbf{\ }%
and\textbf{\ }$\left\vert A_{\beta }\right\rangle $, $\left\vert B_{\beta
}\right\rangle $. Both expressions refer to the same quantum state, and the
same result regarding separability\ is obtained in both first and second
quantisation, even though the number of sub-systems differ. It indicates
that either option may be chosen in regard to defining sub-systems when
distinguishable particles are involved. However, it is \emph{simpler} if the
same option - particles or modes as sub-systems - is made for treating
either identical or distinguishable particle systems and we will adopt this
approach.

To highlight the distinction between the identical and distinguishable
particles situation, we note that for the two distinguishable particle case
treated previously we can also form entangled states from the basis states (%
\ref{Eq.TwoDistPartTwoModes}) or (\ref{Eq.SecondQnTwoDistTwoModesEach})

\begin{equation}
\left\vert \Psi \right\rangle =\frac{1}{\sqrt{2}}(\left\vert A_{\alpha
}(\alpha )\right\rangle \otimes \left\vert B_{\beta }(\beta )\right\rangle
\pm \left\vert B_{\alpha }(\alpha )\right\rangle \otimes \left\vert A_{\beta
}(\beta )\right\rangle )  \label{Eq.TwoDistParticlesTwoModes}
\end{equation}%
which are similar in mathematical form to (\ref{Eq.TwoBosonsTwoModes}) and (%
\ref{Eq.TwoFermionsTwoModes}) when written in first quantisation, and which
are given by \ 
\begin{equation}
\left\vert \Psi \right\rangle =\frac{1}{\sqrt{2}}((\left\vert 1\right\rangle
_{A_{\alpha }}\otimes \left\vert 0\right\rangle _{B_{\alpha }})\otimes
(\left\vert 0\right\rangle _{A_{\beta }}\otimes \left\vert 1\right\rangle
_{B_{\beta }})\pm (\left\vert 0\right\rangle _{A_{\alpha }}\otimes
\left\vert 1\right\rangle _{B_{\alpha }})\otimes (\left\vert 1\right\rangle
_{A_{\beta }}\otimes \left\vert 0\right\rangle _{B_{\beta }}))
\label{Eq.SecondQnTwoDistTwoModes}
\end{equation}%
when written in second quantisation. However, in this case both the first
and second quantisation forms are clearly cases of \emph{entangled} states.
Whether they are regarded as entangled states of two sub-systems consisting
of particle $\alpha $ and particle $\beta $ (first option) or entangled
states of the four sub-systems consisting of modes $\left\vert A_{\alpha
}\right\rangle $\textbf{, }$\left\vert B_{\alpha }\right\rangle $\textbf{\ }%
and\textbf{\ }$\left\vert A_{\beta }\right\rangle $\textbf{, }$\left\vert
B_{\beta }\right\rangle $ (second option) depends on whether particle or
modes are chosen as sub-systems.

Note however that \emph{not} all basis states result in separable/entangled
distinctions even in the case of identical particles. For the same two mode,
two particle case as considered previously for bosons the basis vectors $%
\left\vert A(1)\right\rangle \otimes \left\vert A(2)\right\rangle $ or $%
\left\vert B(1)\right\rangle \otimes \left\vert B(2)\right\rangle $ (first
quantisation) or equivalently $\left\vert 2\right\rangle _{A}\otimes
\left\vert 0\right\rangle _{B}$ or $\left\vert 0\right\rangle _{A}\otimes
\left\vert 2\right\rangle _{B}$ (second quantisation) would be regarded as
separable states irrespective of whether particle or modes were chosen as
the sub-systems. Entangled states such as $(\left\vert A(1)\right\rangle
\otimes \left\vert A(2)\right\rangle \pm \left\vert B(1)\right\rangle
\otimes \left\vert B(2)\right\rangle )/\sqrt{2}$ (first quantisation) or
equivalently $(\left\vert 2\right\rangle _{A}\otimes \left\vert
0\right\rangle _{B}\pm \left\vert 0\right\rangle _{A}\otimes \left\vert
2\right\rangle _{B})/\sqrt{2}$ (second quantisation) can also be formed from
the two doubly occupied basis states. There are no analogous states for
fermions due to the Pauli principle.

It is worth noting that these examples illustrate the general point that
just the \emph{mathematical form} of the state vector or the density
operator alone is not enough to determine whether a separable or an
entangled state is involved. The meaning of the factors involved also has to
be taken into account. Failure to realise this may lead to states being
regarded as separable when they are not (see SubSection \ref{SubSection -
SSR Separate Modes} for further examples).

In the above discussion the symmetrisation principle was complied with both
in the first and second quantisation \ treatments. It should be noted
however that some authors disregard the symmetrisation principle. In
describing Bose-Einstein condensates (\cite{Sorensen01a}, \cite{Pezze09a})
consider states of the form 
\begin{equation}
\widehat{\rho }=\sum_{R}P_{R}\,\widehat{\rho }_{R}^{1}\otimes \widehat{\rho }%
_{R}^{2}\otimes \widehat{\rho }_{R}^{3}\otimes ...
\label{Eq.NonEntStateIdenticalAtoms}
\end{equation}%
as defining non-entangled states, where $\widehat{\rho }_{R}^{i}$ is a
density operator for particle $i$. However such a state would not in general
be allowed\textbf{, }since the symmetrisation principle would be violated
unless the $\widehat{\rho }_{R}^{i}$ were related. For example, consider the
state for two identical bosonic atoms given by 
\begin{equation}
\widehat{\rho }=P_{\sigma \xi }\,\widehat{\sigma }^{1}\otimes \widehat{\xi }%
^{2}+P_{\theta \eta }\,\widehat{\theta }^{1}\otimes \widehat{\eta }^{2}
\label{Eq.NonEntangStateTwoBosons}
\end{equation}%
and apply the permutation $\widehat{P}=\widehat{P}(1\leftrightarrow 2)$. The
invariance of $\widehat{\rho }$ in general requires $\,\widehat{\sigma }=%
\widehat{\xi }$ and $\widehat{\theta }=\widehat{\eta }$, giving $\widehat{%
\rho }=P_{\sigma }\,\widehat{\sigma }^{1}\otimes \widehat{\sigma }%
^{2}+P_{\theta }\,\widehat{\theta }^{1}\otimes \widehat{\theta }^{2}$. This
is a statistical mixture of two states, one with \emph{both} atoms in state $%
\widehat{\sigma }$, the other with\textbf{\ }both atoms in state $\widehat{%
\theta }$. Thus only special cases of (\ref{Eq.NonEntangStateTwoBosons}) are
compatible with the symmetrisation principle. Of course if the atoms were
all different (atom $1$ a Rb$^{87}$ atom, atom $2$ a Na$^{23\text{ }}$atom,
..) then the expression (\ref{Eq.NonEntangStateTwoBosons}) would be a valid
non-entangled state, but there the atomic sub-systems are distinguishable
and symmetrisation is not required. Such authors are really ignoring the
symmetrisation principle, and in addition are treating the individual
identical particles in the BEC as separate sub-systems - a viewpoint we have
described previously and will discuss further in the next SubSubSection. For
the present we just point out that valid quantum states must comply with the
symmetrisation principle. \bigskip

\subsubsection{Sub-Systems - Particles or Modes ?}

\label{SubSection - Sub-Systems - Particles or Modes ?}

As highlighted in the previous SubSection \ref{SubSection - Symmetrization
Principle}, when the quantum system involves identical particles the very
definition of entanglement itself requires special care in regard to
identifying legitimate sub-systems. There is a long-standing debate on the
issue, with at present two schools of thought - see reviews such as \cite%
{Horodecki09a} or \cite{Tichy11a}. As explained in the previous SubSection,
the \emph{first} approach is to identify mathematically labelled individual
identical \emph{particles} as the sub-systems \cite{Peres93a}, \cite%
{Sorensen01a}, \cite{Pezze09a}, \cite{Hyllus12a}. Sub-systems may of course
also be sets of such individually labeled particles. This approach leads to
the conclusion that \emph{symmetrisation} creates entanglement of identical
particles. The \emph{second} approach is to identify single particle states
or \emph{modes} that the identical particles may occupy as the sub-systems 
\cite{TerraCunha07a}. The sub-systems may of course also be sets of
distinguishable modes. This approach leads to the conclusion that it is 
\emph{interaction processes} between modes that creates entanglement of
distinguishable modes.

The approach based on \emph{particle entanglement} is still being used \cite%
{Hyllus12a}. As explained in SubSection \ref{SubSection - Symmetrization
Principle} this is not the same as mode entanglement so tests and measures
for particle entanglement will differ from those for mode entanglement. A
further discussion about the distinction is given in \cite{Amico08a}. In a
recent paper Killoran et al \cite{Killoran14a} considered original states
such as $(\left\vert a0(1)\right\rangle \otimes \left\vert
a1(2)\right\rangle \pm \left\vert a0(2)\right\rangle \otimes \left\vert
a1(1)\right\rangle )/\sqrt{2}$\ involving two modes $a0$\ and $a1$\ - which
were considered (based on first quantisation) as an\textbf{\ }\emph{entangled%
}\textbf{\ }state for two sub-systems consisting of particles $1$\ and $2$,
but would be considered (in second quantisation) as a\textbf{\ }\emph{%
separable}\textbf{\ }state\textbf{\ }$\left\vert 1\right\rangle _{a0}\otimes
\left\vert 1\right\rangle _{a1}$\textbf{\ }for two sub-systems consisting of
modes $a0$\ and $a1$. In addition there were two modes $b0$\ and $b1$ which
are intially unoccupied. The particles may be bosons or fermions. They
envisaged converting such an input state using\textbf{\ }\emph{interferometer%
}\textbf{\ }processes which couple $A$\ modes $a0$\ and $a1$\ to previously
unoccupied $B$\ modes $b0$\ and $b1$, into an output state - which is
different. Projective measurements would then be made on the output state,
based on having known numbers of particles in each of the $A$\ mode pairs $%
a0 $\ and $a1$\ and in the $B$\ mode pairs $b0$\ and $b1$.\textbf{\ }The
projected state with one particle in the $A$\ modes and one particle in the $%
B$\ modes would be of the form (in second quantisation) $(\left\vert
1\right\rangle _{a0}\otimes \left\vert 0\right\rangle _{a1}\otimes
\left\vert 0\right\rangle _{b0}\otimes \left\vert 1\right\rangle _{b1}\pm
\left\vert 0\right\rangle _{a0}\otimes \left\vert 1\right\rangle
_{a1}\otimes \left\vert 1\right\rangle _{b0}\otimes \left\vert
0\right\rangle _{b1})/\sqrt{2}$\ , which is a bipartite entangled state for
the two pairs of modes $A$ and $B$\ and is\textbf{\ }\emph{mathematically}%
\textbf{\ }of the same form as the first quantisation\emph{\ form} for the
original $A$\ modes state considered as an example of \emph{particle}
entanglement if the correspondences $\left\vert a0(1)\right\rangle
\rightarrow \left\vert 1\right\rangle _{a0}\otimes \left\vert 0\right\rangle
_{a1}$,\ $\left\vert a1(2)\right\rangle \rightarrow \left\vert
0\right\rangle _{b0}\otimes \left\vert 1\right\rangle _{b1}$, $\left\vert
a0(2)\right\rangle \rightarrow \left\vert 1\right\rangle _{b0}\left\vert
0\right\rangle _{b1}$\ and $\left\vert a1(1)\right\rangle \rightarrow $\ $%
\left\vert 0\right\rangle _{a0}\left\vert 1\right\rangle _{a1}$\ are made.
Even the minus sign is obtained in the the fermion case. Details are given
in Appendix \ref{Appendix - Extracting Entang due Symm}$.$Killoran et al
stated that this represented a way of \emph{extracting} the original
symmetrization generated entanglement. However, another point of view is
that the two mode interferometer process \emph{created} an entangled state
from a non-entangled state, and as the final measurements are still based on
entanglement of modes it is hard to justify the claim that entanglement due
to symmetrization exists as a directly observable basic feature in composite
quantum systems - though the mapping identified in \cite{Killoran14a}
is.mathematically correct. Furthermore, all quantum states for identical
particles are required to be symmetrized, so if symmetrization causes
entanglement it differs from the numerous other controllable processes that
produce entanglement by coupling the sub-systems. Since the idea of
extracting entanglement due to symmetrisation is of current interest, a
fuller discussion of the approach by Killoran et al \cite{Killoran14a} is
set out in Appendix \ref{Appendix - Extracting Entang due Symm}.

However, it is generally recognised that sub-systems consisting of \emph{%
individually labeled} identical particles are \emph{not }amenable to \emph{%
measurements}. What is distinguishable for systems of identical bosons or
fermions is \emph{not} the individual particles themselves - which do not
carry labels, boson $1$, boson $2$, etc. - but the \emph{single particle
states }or \emph{modes} that the bosons may occupy. For bosonic or fermionic
atoms with several hyperfine components, each component will have its own
set of modes. For photons the modes may be specified via wave vectors and
polarisations. Although the quantum pure states can be specified via
symmetrized products of single particle states occupied by specific
particles using a \emph{first quantization} approach, it is more suitable to
use \emph{second quantization}. Here, a basis set for the quantum states of
such sub-systems are the \emph{Fock states} $\left\vert n\right\rangle _{A}$
($n=0,1,2,\,...$) etc, which specify the number of identical particles
occupying the mode $A$, etc., so in this approach the mode is the sub-system
and the Fock states give different quantum states for this sub-system.
Symmetrization is built into the definition of the Fock states, so the
symmetrisation principle is\emph{\ automatically} adhered to. If the atoms
were fermions rather than bosons the Pauli exclusion principle would of
course restrict $n=0,1$ only. In this second quantization approach
situations with differing \emph{numbers} of identical particles are
recognised as being different \emph{states} of a system consisting of a set
of modes, not different \emph{systems} as would be the case in first
quantisation. The overall system will be associated with quantum states
represented in the theory by density operators and state vectors in \emph{%
Fock space}, which includes states with total numbers of identical particles
ranging from zero in the vacuum state right up to infinity. Finally, the
artificial concept of entanglement due to symmetrisation is replaced by the
physically realistic concept of entanglement due to mode coupling.

The point of view in which the possible \emph{sub-systems} \emph{\ }$A$, $B$%
, etc are \emph{modes} (or \emph{sets} of modes) rather than \emph{particles}
has been adopted by several authors (\cite{Simon02a}, \cite{Hines03a}, \cite%
{TerraCunha07a}), \cite{Vedral07a}, \cite{Benatti10a}, \cite{Benatti11a} and
will be the approach used here - as in \cite{Dalton14a}. To emphasise - what
are or are not entangled in the present treatment involving systems of
identical particles\textbf{\ }are \emph{distinguishable modes} not \emph{%
labelled-indistinguishable-particles}. Overall, the system is a collection
of modes, not particles. Particles are associated with mode\emph{\
occupancies}, and therefore related to specifying the quantum states of the
system, rather than the system itself. In terms of this approach, for
non-interacting identical particles at zero temperature, the ground states
for Bose-Einstein condensates and Fermi gases trapped in a harmonic
potential provide examples of non-entangled states for bosonic and fermionic
atoms respectively, when the sub-systems are chosen as the harmonic
oscillator (HO) modes. In the bosonic case all the bosons occupy the lowest
energy HO state, in the fermionic case one fermion occupies each HO state
from the lowest up to a high energy state (the Fermi energy) until all the
fermions are accommodated. On the other hand, if one particle position
states spatially localised in two different regions are chosen as two
sub-systems, then the same zero temperature state for the identical particle
system is spatially entangled, as pointed out by Goold et al \cite{Goold09a}%
. Note that in this approach states where there is only a \emph{single atom}
may still be entangled states - for example with two spatial modes $A,B$ the
states which are a quantum superposition of the atom in each of these modes,
such as the Bell state $(\left\vert 1\right\rangle _{A}\left\vert
0\right\rangle _{B}+\left\vert 0\right\rangle _{A}\left\vert 1\right\rangle
_{B})/\sqrt{2}$ are entangled states. For entangled states associated with
the EPR paradox or for quantum teleportation, the mode functions may be 
\emph{localised} in well-separated spatial regions - spooky action at a
distance - but spatially overlapping mode functions apply in other
situations. This distinction is important in discussions of quantum
non-locality. Atoms in states with overall spin zero only have one internal
state, but two mode systems can be created for their spatial motion using
double-well trap potentials. If the wells are separated then two spatially
separated modes can be created for studies of quantum non-locality. On the
other hand atoms with spin 1/2 have two internal states, which constitute a
two mode system. However these two modes may be associated with the same or
overlapping spatial wave functions, in which case studies of quantum
non-locality are precuded. These latter situation can however still lead to
what is referred to as intrasystem entanglement \cite{Aiello14a}.
Furthermore, as well as being distinguishable the modes can act as \emph{%
separate systems}, with other modes being ignored. For interacting bosonic
atoms this is much harder to accomplish experimentally than for the case of
photons, where the relatively slow processes in which photons are destroyed
in one EM field mode and created in another may require the presence of
atoms as intermediaries. Two bosonic atoms in one mode may collide and
rapidly disappear into other modes. However, atomic boson interactions can
be made very small via \emph{Feshbach resonance} methods. Near absolute zero
the basic physics of a BEC in a single trap potential is describable via a 
\emph{one mode theory}. Hence with $A$, $B$, .. signifying distinct modes,
the general non-entangled state is given in Eq. (\ref{Eq.NonEntangledState})
though the present paper mainly involves only two modes.

As pointed out in SubSection \ref{SubSection - Symmetrization Principle}, in
the case of systems consisting entirely of \emph{distinguishable}\textbf{\ }%
particles the sub-systems may still be regarded as sets of modes, namely
those single particle states associated with the particular distinguishable
particle. In this case the particle descriptor (He atom, Na atom, ..) is
synonomous with its collection of modes. Here all the sub-system states are
one particle states. \textbf{\bigskip }

\subsubsection{Multi-Mode Sub-Systems}

As well as the simple case where the sub-systems are all \emph{individual}
modes, the concept of entanglement may be \emph{extended} to situations
where the sub-systems are \emph{sets of modes}, rather than individual
modes, In this case entanglement or non-entanglement will be of these
distinct sets of modes. Such a case in considered in SubSection \textbf{4.3 }%
of paper II, where \emph{pairs} of modes associated with distinct lattice
sites are considered as the sub-systems. Another example is treated in He et
al \cite{He12a}, which involves a double well potential with each well
associated with two bosonic modes, these pairs of modes being the two
sub-systems. Entanglement criteria for the mode pairs based on local spin
operators associated with each potential well are considered (sse
SubSections \textbf{4.2 and 5.3 }of paper II). A further example is treated
by Heaney et al \cite{Heaney10a}, again involving \ four modes associated
with a double well potential. As in the previous example, each mode pair is
associated with the same well in the potential, but here a Bell entanglement
test was obtained for pairs of modes in the different wells. The concept of 
\emph{entanglement of sets of modes} is a straightforward extension of the
basic concept of entanglement of individual modes.

\subsection{Super-Selection Rule}

\label{SubSection - Super-Selection Rule}

As well as the symmetrisation principle there is a further requirement that
quantum states of systems of identical particles must satisfy - these are
known as \emph{super-selection rules}. These rules restrict the allowed
quantum states of such systems to those in which the \emph{coherences}
between states with differing numbers of particles are zero. This applies at
the global level for the overall quantum state, but also - as will be
discussed in a later sub-section - to the sub-system states involved in the
definition of separable or non-entangled states. The justfication of the SSR
at both the global and local level will be considered both in terms of
simple physics arguements and in terms of reference frames. Examples of SSR
and non-SSR compliant states will be given, both for overall states and for
separable states. The validity of the SSR for the case of massive bosons or
fermions is generally accepted, but in the case of photons there is doubt
regarding their applicability -as will be discussed below. As pointed\textbf{%
\ }out in the Introduction, in the case of systems consisting entirely of
single\emph{\ distinguishable}\textbf{\ }particles the sub-systems may still
be regarded as sets of modes, namely those single particle states associated
with the particular distinguishable particle. Here all the sub-system states
are one particle states and the overall system is an $N$ particle state, so
the local and global particle number super-selection rules, though true are
irrelevant.

\subsubsection{Global Particle Number SSR}

The question of what\textbf{\ }quantum states - entangled or not - are
possible in the \emph{non-relativistic quantum physics} of a system of
identical \emph{bosonic} particles - such as bosonic \emph{atoms} or \emph{%
photons} - has been the subject of much discussion. Whether \emph{entangled}
or \emph{not} it is generally accepted that there is a \emph{super-selection
rule} that prohibits \emph{quantum superposition states} of the form 
\begin{equation}
\left\vert \Phi \right\rangle =\tsum\limits_{N=0}^{\infty }C_{N}\,\left\vert
N\right\rangle \qquad \widehat{\rho }=\tsum\limits_{N=0}^{\infty }\left\vert
C_{N}\right\vert ^{2}\,\left\vert N\right\rangle \left\langle N\right\vert
+\tsum\limits_{N=0}^{\infty }\tsum\limits_{M=0}^{\infty }(1-\delta
_{N,M})C_{N}\,C_{M}^{\ast }\,\left\vert N\right\rangle \left\langle
M\right\vert  \label{Eq.ForbiddenStates}
\end{equation}%
being \emph{quantum} states when they involve Fock states $\left\vert
N\right\rangle $ with differing total numbers $N$ of particles. The density
operator for such a state would involve \emph{coherences} between states
with differing $N$. Although such superpositions - such as the \emph{Glauber
coherent state }$\left\vert \alpha \right\rangle $, where $C_{N}=\exp
(-|\alpha |^{2}/2)\,\alpha ^{N}/\sqrt{N!}$ - do have a useful \emph{%
mathematical} role, they do \emph{not} represent actual quantum states
according to the super-selection rule. The papers by Sanders et al \cite%
{Sanders03a} and Cable et al \cite{Cable05a} \ are examples of applying the
SSR for optical fields, but also using the mathematical features of coherent
states to treat phenomena such as interference between independent lasers.
The super-selection rule indicates that the most \emph{general quantum state}
for a system of identical bosonic particles can only be of the form%
\begin{eqnarray}
\widehat{\rho } &=&\tsum\limits_{N=0}^{\infty }\tsum\limits_{\Phi }P_{\Phi
N}\,\left( \left\vert \Phi _{N}\right\rangle \left\langle \Phi
_{N}\right\vert \right)  \nonumber \\
\left\vert \Phi _{N}\right\rangle &=&\tsum\limits_{i}C_{i}^{N}\,\left\vert
N\,i\right\rangle  \label{Eq.PhysicalState}
\end{eqnarray}%
where $\left\vert \Phi _{N}\right\rangle $ is a quantum superposition of
states $\left\vert N\,i\right\rangle $ each of which involves exactly $N$
particles, and where different states with the same $N$ are designated as $%
\left\vert N\,i\right\rangle $. This state $\widehat{\rho }$ is a
statistical mixture of states, each of which contains a specific number of
particles. Such a SSR is referred to as a \emph{global} SSR, as it applies
to the system as a whole. Mathematically, the global particle number SSR can
be expressed as 
\begin{equation}
\lbrack \widehat{N},\widehat{\rho }]=0  \label{Eq.GloabalSSR}
\end{equation}%
where $\widehat{N}$ is the \emph{total number} operator.

\subsubsection{Examples of Global Particle Number SSR Compliant States}

Examples of a state vector $\left\vert \Phi _{N}\right\rangle $ for an
entangled pure state \cite{Hines03a} and a density operator $\widehat{\rho }$
for a non-entangled mixed \cite{Goold09a} state for a \emph{two mode bosonic}
system, both of which are possible quantum states are 
\begin{eqnarray}
\left\vert \Phi _{N}\right\rangle
&=&\tsum\limits_{k=0}^{N}C(N,k)\,\left\vert k\right\rangle _{A}\otimes
\left\vert N-k\right\rangle _{B}  \label{Eq.EntangledTwoModePureState} \\
\widehat{\rho } &=&\tsum\limits_{k=0}^{N}P(k)\,\left\vert k\right\rangle
_{A}\left\langle k\right\vert _{A}\,\otimes \left\vert N-k\right\rangle
_{B}\left\langle N-k\right\vert _{B}
\label{Eq.NonEntangledTwoModeMixedState}
\end{eqnarray}%
The entangled pure state is a superposition of product states with $k$
bosons in mode $A$ and the remaining $N-k$ bosons in mode $B$. Every term in
the superposition is associated with the same total boson number $N$. The
non-entangled mixed state is a statistical mixture of product states also
with $k$ bosons in mode $A$ and the remaining $N-k$ bosons in mode $B$.
Every term in the statistical mixture is associated with the same total
boson number $N$. For the case of a \emph{two mode fermionic} system the
Pauli exclusion principle restricts the number of possible fermions to two,
with at most one fermion in each mode. Expressions for a state with exactly $%
N=2$ fermions are%
\begin{eqnarray}
\left\vert \Phi _{2}\right\rangle &=&\left\vert 1\right\rangle _{A}\otimes
\left\vert 1\right\rangle _{B}  \label{Eq.TwoModeFermionStateVector} \\
\widehat{\rho } &=&\left\vert 1\right\rangle _{A}\left\langle 1\right\vert
_{A}\,\otimes \left\vert 1\right\rangle _{B}\left\langle 1\right\vert _{B}
\label{Eq.TwoModeFermionDensOpr}
\end{eqnarray}%
Neither state is entangled and both are the same pure state since $\widehat{%
\rho }=\left\vert \Phi _{2}\right\rangle \left\langle \Phi _{2}\right\vert $%
. Although the super-selection rules and symmetrisation principle also
applies to fermions, as indicated in the Introduction this paper is focused
on bosonic systems, and it will be assumed that the modes are bosonic unless
indicated otherwise.

\emph{Bell states}\textbf{\ }\cite{Barnett09a}, \cite{Vedral07a} for $N=1$\
bosons provide important examples of entangled two mode pure quantum states
that are compliant with the global particle number SSR. The modes are
designated $A,B$\ and the Fock states are in general $\left\vert
n_{A},n_{B}\right\rangle $. These Bell states may be written\textbf{\ }%
\begin{eqnarray}
\left\vert \Psi _{AB}^{-}\right\rangle &=&\frac{1}{\sqrt{2}}(\left\vert
0_{A},1_{B}\right\rangle -\left\vert 1_{A},0_{B}\right\rangle )  \nonumber \\
\left\vert \Psi _{AB}^{+}\right\rangle &=&\frac{1}{\sqrt{2}}(\left\vert
0_{A},1_{B}\right\rangle +\left\vert 1_{A},0_{B}\right\rangle )
\label{Eq.BellStates}
\end{eqnarray}%
Neither of these states is separable. There are also two other two mode Bell
states given by\textbf{\ }%
\begin{eqnarray}
\left\vert \Phi _{AB}^{-}\right\rangle &=&\frac{1}{\sqrt{2}}(\left\vert
0_{A},0_{B}\right\rangle -\left\vert 1_{A},1_{B}\right\rangle )  \nonumber \\
\left\vert \Phi _{AB}^{+}\right\rangle &=&\frac{1}{\sqrt{2}}(\left\vert
0_{A},0_{B}\right\rangle +\left\vert 1_{A},1_{B}\right\rangle )
\label{Eq.OtherBellStates}
\end{eqnarray}%
These however are not compliant with the global particle number SSR. Linear
combinations $(\left\vert \Phi _{AB}^{-}\right\rangle +\left\vert \Phi
_{AB}^{+}\right\rangle )/\sqrt{2}=\left\vert 0_{A},0_{B}\right\rangle $\ and 
$(-\left\vert \Phi _{AB}^{-}\right\rangle +\left\vert \Phi
_{AB}^{+}\right\rangle )/\sqrt{2}=\left\vert 1_{A},1_{B}\right\rangle $\ are
global particle number SSR compliant and also separable, corresponding to
states with $N=0$\ and $N=2$\ bosons respectively.

\subsubsection{Super-Selection Rules and Conservation Laws}

\label{SubSubSystem - SSR and Conservation}

It is important to realise that such super-selection rules \cite{Wick52a}
are \emph{different constraints} to those imposed by \emph{conservation laws}%
, as emphasised by Bartlett et al \cite{Bartlett03a}. \textbf{\ }For
example, the conservation law on total particle number \emph{only} leads to
the requirement on the superposition state $\left\vert \Phi \right\rangle $
that the $|C_{N}|^{2}$ are time independent, it does \emph{not} require only
one $C_{N}$ being non-zero. They are however related, as is discussed in
Section \ref{SubSubSection - Link SSR Ref Frames} and Appendix \ref{Appendix
- Reference Frames and SSR} where the super-selection rules based on
particle number are related to invariances of the density operator under
changes of phase reference frames. This involves considering groups of phase
changing operators $\widehat{T}(\theta _{a})=\exp (i\widehat{N}_{a}\theta
_{a})$ when considering local particle number SSR for single modes in the
context of separable states, or $\widehat{T}(\theta )=\exp (i\widehat{N}%
\theta )$ when considering global particle number SSR in the context of
multimode entangled states. Super-selection rules are broad in their scope,
forbidding quantum superpositions of states of systems with differing
charge, differing baryon number and differing statistics. Thus a combined
system of a hydrogen atom and a helium ion does not exist in quantum states
that are linear combinations of hydrogen atom states and helium ion states -
the super-selection rules on both charge and baryon number preclude such
states. The basis quantum states for such a combined system would involve
symmetrised tensor products of hydrogen atom and helium ion states, not
linear combinations - symmetrisation being required because the system
contains two identical electrons. On the other hand, super-selection rules
do not prohibit quantum superpositions of states of systems with differing
energy, angular or linear momenta - other physical quantities that may also
be conserved. Thus in a hydrogen atom quantum superpositions of states with
differing energy and angular momentum quantum numbers are allowed quantum
states.

However, conservation laws on total particle number (such as apply in the
case of massive bosons) are relevant to showing that multi-mode states
generated via total particle number conserving processes from an initial
separable state will be global SSR compliant if the sub-systems in the
initial state are local particle number SSR compliant, and will not be if
the initial state involves a sub-system state that is not local particle
number SSR compliant. For simplicity we consider two sub-systems $A$ and $B$
with the initial state 
\begin{equation}
\widehat{\rho }(0)=\sum_{R}P_{R}\,\widehat{\rho }_{R}^{A}\otimes \widehat{%
\rho }_{R}^{B}
\end{equation}%
If $\widehat{U}(t)$ is the evolution operator where $\widehat{\rho }(t)=%
\widehat{U}(t)\widehat{\rho }(0)\widehat{U}(t)^{\dag }$ and the processes
are number conserving then $[\widehat{N},\widehat{U}(t)]=0$. We then have%
\begin{eqnarray}
\lbrack \widehat{N},\widehat{\rho }(t)] &=&\widehat{U}(t)\,[\widehat{N},%
\widehat{\rho }(0)]\,\widehat{U}(t)^{\dag }  \nonumber \\
&=&\widehat{U}(t)\sum_{R}P_{R}\,\left( [\widehat{N}_{A},\,\widehat{\rho }%
_{R}^{A}]\otimes \widehat{\rho }_{R}^{B}+\widehat{\rho }_{R}^{A}\otimes
\lbrack \widehat{N}_{B},\,\widehat{\rho }_{R}^{B}]\right) \widehat{U}%
(t)^{\dag }
\end{eqnarray}%
Hence if $\widehat{\rho }_{R}^{A}$\ and $\widehat{\rho }_{R}^{B}$\ are local
particle number SSR compliant, then $[\widehat{N}_{A},\,\widehat{\rho }%
_{R}^{A}]$ and $[\widehat{N}_{B},\,\widehat{\rho }_{R}^{B}]$ are zero,
showing that $[\widehat{N},\widehat{\rho }(t)]=0$ so the state is global
particle number SSR compliant. On the other hand if $[\widehat{N},\widehat{%
\rho }(t)]=0$ we see that $[\widehat{N},\widehat{\rho }(0)]=\sum_{R}P_{R}\,%
\left( [\widehat{N}_{A},\,\widehat{\rho }_{R}^{A}]\otimes \widehat{\rho }%
_{R}^{B}+\widehat{\rho }_{R}^{A}\otimes \lbrack \widehat{N}_{B},\,\widehat{%
\rho }_{R}^{B}]\right) =0$. By taking $Tr_{A}$\ and $Tr_{B}$ of this result
gives $\sum_{R}P_{R}\,[\widehat{N}_{A},\,\widehat{\rho }_{R}^{A}]=%
\sum_{R}P_{R}\,[\widehat{N}_{B},\,\widehat{\rho }_{R}^{B}]=0$. This shows
that both of the reduced density operators $\sum_{R}P_{R}\widehat{\rho }%
_{R}^{A}$\ and $\sum_{R}P_{R}\widehat{\rho }_{R}^{B}$\ must be local
particle number SSR compliant, which amounts to requiring the sub-system
density operators to be local particle number SSR compliant. This situation
applies even when there is coupling between modes, provided the interaction
is number conserving - such as a coupling given by $\widehat{V}=\lambda 
\widehat{a}\widehat{b}^{\dag }+HC$. Analogous results apply for systems of
massive bosons if there are more than two modes involved, where again global
SSR compliance involves the total particle number since even with
interactions there is total number conservation. For example with three
modes in coupled BECs, interactions of the form $\widehat{V}=\lambda (%
\widehat{c})^{2}\widehat{a}^{\dag }\widehat{b}^{\dag }+HC$ in which two
bosons are annihilated in mode $C$ and one boson is created in each of modes 
$A$ and $B$ are consistent with total particle number conservation and lead
to global SSR involving the total particle number.

Although outside the focus of this paper, it is worth pointing out that
somewhat different considerations apply to photons. Single non-interacting
modes, such as are discussed in the context of separable states do have a
conservation law for the photon number in that mode. The applicability (or
otherwise) of the local particle number SSR for the sub-system density
operators in a separable state is discussed in Section \ref{SubSubSection -
SSR and Photons}. In the case of interacting photonic modes there may be no
conservation law associated with total photon number and it may be thought
that no global SSR would apply. However, other global SSR involving
combinations of the mode photon numbers may still apply. As an example, we
consider a three mode situation in a non-degenerate parametric amplifier,
where the basic generation process involves one pump photon of frequency $%
\omega _{C}=\omega _{A}+\omega _{B}$ being destroyed and one photon created
in each of modes $A$ and $B$. The interaction term is $\widehat{V}=\lambda 
\widehat{c}\widehat{a}^{\dag }\widehat{b}^{\dag }+HC$. It is
straight-forward to show the a total \emph{quanta} number operator $\widehat{%
N}_{tot}=\widehat{N}_{A}+\widehat{N}_{B}+2\widehat{N}_{C}$ commutes with the
Hamiltonian. The situation is analogous to the atom-molecule system treated
in Appendix \ref{Appendix - Super-Selection Rule Violations ?}. Thus $%
\widehat{N}_{tot}$ is conserved and we can then consider a group of phase
changing operators $\widehat{T}(\theta )=\exp (i\widehat{N}_{tot}\theta )$
and show that there could be a global SSR for the three mode system, but now
involving the total quanta number $N_{A}+N_{B}+2N_{C}$. The pure state which
is often used in a quantum treatment of the non-degenerate parametric
amplifier $\left\vert \Psi \right\rangle =$\ $\tsum\limits_{n}C_{n}\left%
\vert n\right\rangle _{A}\otimes \left\vert n\right\rangle _{B}\otimes
\left\vert N-n\right\rangle _{C}$ is global SSR compliant in terms of the
modified $\widehat{N}_{tot}$, since in every term $N_{A}+N_{B}+2N_{C}=2N$
and there are no coherences between terms with different $N_{tot}$. For the
non-degenerate parametric amplifier case an analogous treatment to that for
number conserving processes shows that if\textbf{\ }%
\begin{equation}
\widehat{\rho }(0)=\sum_{R}P_{R}\,\widehat{\rho }_{R}^{A}\otimes \widehat{%
\rho }_{R}^{B}\otimes \widehat{\rho }_{R}^{C}
\end{equation}%
then using $[\widehat{N}_{tot},\widehat{U}(t)]=0$ we have 
\begin{eqnarray}
\lbrack \widehat{N}_{tot},\widehat{\rho }(t)] &=&\widehat{U}(t)\,[\widehat{N}%
_{tot},\widehat{\rho }(0)]\,\widehat{U}(t)^{\dag }  \nonumber \\
&=&\widehat{U}(t)\sum_{R}P_{R}\,\left( 
\begin{array}{c}
\lbrack \widehat{N}_{A},\,\widehat{\rho }_{R}^{A}]\otimes \widehat{\rho }%
_{R}^{B}\otimes \widehat{\rho }_{R}^{C}+\widehat{\rho }_{R}^{A}\otimes
\lbrack \widehat{N}_{B},\,\widehat{\rho }_{R}^{B}]\otimes \widehat{\rho }%
_{R}^{C} \\ 
+2\widehat{\rho }_{R}^{A}\otimes \widehat{\rho }_{R}^{B}\otimes \lbrack 
\widehat{N}_{C},\,\widehat{\rho }_{R}^{C}]%
\end{array}%
\right) \widehat{U}(t)^{\dag }  \nonumber \\
&&
\end{eqnarray}%
Hence if $\widehat{\rho }_{R}^{A}$, \ $\widehat{\rho }_{R}^{B}$ and $%
\widehat{\rho }_{R}^{C}$\ are local particle number SSR compliant, then $[%
\widehat{N}_{tot},\widehat{\rho }(t)]=0$ so the state is SSR compliant, but
with global total quanta number $\widehat{N}_{tot}$. On the other hand if $[%
\widehat{N}_{tot},\widehat{\rho }(t)]=0$ we find that $\sum_{R}P_{R}\,[%
\widehat{N}_{A},\,\widehat{\rho }_{R}^{A}]=\sum_{R}P_{R}\,[\widehat{N}_{B},\,%
\widehat{\rho }_{R}^{B}]=\sum_{R}P_{R}\,[\widehat{N}_{C},\,\widehat{\rho }%
_{R}^{C}]=0$. This shows that each of the reduced density operators $%
\sum_{R}P_{R}\widehat{\rho }_{R}^{A}$\ , $\sum_{R}P_{R}\widehat{\rho }%
_{R}^{B}$ and $\sum_{R}P_{R}\widehat{\rho }_{R}^{C}$\ must be local particle
number SSR compliant, which amounts to requiring the sub-system density
operators to be local particle number SSR compliant.

\subsubsection{Global SSR Compliant States and Quantum Correlation Functions}

We now prove a theorem concerning \emph{quantum correlation functions} for
bosonic systems with two modes $A$ and $B$.

\textit{Theorem. }If a state is global particle number SSR compliant then
all quantum correlation functions $\left\langle (\widehat{a}^{\dag })^{n}(%
\widehat{a})^{m}(\widehat{b}^{\dag })^{l}(\widehat{b})^{k}\right\rangle $
for which $n+l\neq m+k$ must be zero.

\textit{Proof: }If the state is global particle number SSR compliant then if
we choose a complete orthonormal set of Fock states $\left\vert N,\,\alpha
\right\rangle $ with $\alpha =1,2,\,...\,,d_{N}$ listing states which are
eigenstates of the total number operator $\widehat{N}$ with eigenvalue $N$
we can write the density operator in the form%
\begin{equation}
\widehat{\rho }=\dsum\limits_{N}\dsum\limits_{\alpha ,\beta }P_{\alpha
,\beta }^{N}\,\left\vert N,\,\alpha \right\rangle \left\langle N,\,\beta
\right\vert  \label{Eq.GlobalSSRCompliant}
\end{equation}%
where since $Tr\widehat{\rho }=1$ we must have 
\begin{equation}
1=\dsum\limits_{N}\dsum\limits_{\alpha }P_{\alpha ,\alpha }^{N}
\label{Eq.Trace}
\end{equation}%
Now $(\widehat{a}^{\dag })^{n}(\widehat{a})^{m}(\widehat{b}^{\dag })^{l}(%
\widehat{b})^{k}\,\left\vert N\,\alpha \right\rangle $ must be a linear
combination of Fock states with $N$ replaced by $N+n+l-m-k$ so we can write 
\begin{equation}
(\widehat{a}^{\dag })^{n}(\widehat{a})^{m}(\widehat{b}^{\dag })^{l}(\widehat{%
b})^{k}\,\left\vert N\,,\alpha \right\rangle =\dsum\limits_{\gamma
}C_{\alpha ,\gamma }^{N}(n,m,l,k)\,\left\vert (N+n+l-m-k),\gamma
\right\rangle
\end{equation}%
Hence 
\begin{equation}
\left\langle (\widehat{a}^{\dag })^{n}(\widehat{a})^{m}(\widehat{b}^{\dag
})^{l}(\widehat{b})^{k}\right\rangle =Tr\left(
\dsum\limits_{N}\dsum\limits_{\alpha ,\beta }P_{\alpha ,\beta
}^{N}\,\dsum\limits_{\gamma }C_{\alpha ,\gamma }^{N}(n,m,l,k)\,\left\vert
(N+n+l-m-k),\gamma \right\rangle \left\langle N,\,\beta \right\vert \right)
\end{equation}%
But $Tr(\left\vert (N+n+l-m-k),\gamma \right\rangle \left\langle N,\,\beta
\right\vert )=0$ unless $n+l-m-k=0$. Hence 
\begin{equation}
\left\langle (\widehat{a}^{\dag })^{n}(\widehat{a})^{m}(\widehat{b}^{\dag
})^{l}(\widehat{b})^{k}\right\rangle =0\quad if\;n+l\neq m+k
\label{Eq.GlobalSSRCompliantResult}
\end{equation}%
which is the required theorem.

\subsubsection{Testing the Super-Selection Rules}

The last result for the general two mode quantum correlation function $%
\left\langle (\widehat{a}^{\dag })^{n}(\widehat{a})^{m}(\widehat{b}^{\dag
})^{l}(\widehat{b})^{k}\right\rangle $ is relevant to the various
experimental measurements that are discussed in the accompanying paper II.
For example, as we will see $\left\langle \widehat{S}_{x}\right\rangle
_{\rho }$ is a combination of $\left\langle (\widehat{a}^{\dag })^{n}(%
\widehat{a})^{m}(\widehat{b}^{\dag })^{l}(\widehat{b})^{k}\right\rangle $
with $n=1,m=0,l=0,k=1$ and $n=0,m=1,l=1,k=0$, and $\left\langle \Delta 
\widehat{S}_{x}^{2}\right\rangle _{\rho }$ $=\left\langle \widehat{S}%
_{x}^{2}\right\rangle _{\rho }-\left\langle \widehat{S}_{x}\right\rangle
_{\rho }^{2}$ would involve terms such as $\left\langle (\widehat{a}^{\dag
})^{n}(\widehat{a})^{m}(\widehat{b}^{\dag })^{l}(\widehat{b}%
)^{k}\right\rangle $ with $n=2,m=0,l=0,k=2$ and $n=0,m=2,l=2,k=0$, and $%
n=1,m=1,l=1,k=$ $1$ from $\left\langle \widehat{S}_{x}^{2}\right\rangle
_{\rho }$. All of these have $n+l=m+k$, so they can be non-zero for globally
SSR compliant states. The question then arises - what sort of quantity of
the form $\left\langle (\widehat{a}^{\dag })^{n}(\widehat{a})^{m}(\widehat{b}%
^{\dag })^{l}(\widehat{b})^{k}\right\rangle $ could be used to see if the
quantum state was not globally SSR compliant? The answer is seen in terms of
two corollaries to the last theorem.

\textit{Corollary 1. }If we find that any of the quantum correlation
functions $\left\langle (\widehat{a}^{\dag })^{n}(\widehat{a})^{m}(\widehat{b%
}^{\dag })^{l}(\widehat{b})^{k}\right\rangle $ are non-zero when $n+l\neq
m+k $ then the state is not global particle number SSR compliant.

This result indicates what type of measurement is needed to see if SSR non
compliant states exist. Quantities of the type $\left\langle (\widehat{a}%
^{\dag })^{n}(\widehat{a})^{m}(\widehat{b}^{\dag })^{l}(\widehat{b}%
)^{k}\right\rangle $ are measured for which $n+l\neq m+k$. If we find \emph{%
any} that are non-zero we can then conclude that we have found a state which
is\emph{\ not} global particle number SSR compliant.

\textit{Corollary . }Measurements of the QCF $\left\langle (\widehat{a}%
^{\dag })^{n}(\widehat{a})^{m}(\widehat{b}^{\dag })^{l}(\widehat{b}%
)^{k}\right\rangle $ when $n+l=m+k$ cannot determine whether or not a state
is global particle number SSR non-compliant.

If the state is \emph{non-compliant} then its density operator must contain
a contribution which allows for \emph{non-zero coherences} between Fock
states with different $N$. We can therefore write the density operator as 
\begin{equation}
\widehat{\rho }=\dsum\limits_{N}\dsum\limits_{\alpha ,\beta }P_{\alpha
,\beta }^{N}\,\left\vert N,\,\alpha \right\rangle \left\langle N,\,\beta
\right\vert +\dsum\limits_{N\neq M}\dsum\limits_{\alpha ,\beta }P_{\alpha
,\beta }^{N,M}\,\left\vert N,\,\alpha \right\rangle \left\langle M,\,\beta
\right\vert  \label{Eq.NonSSRCompliant}
\end{equation}%
where the second term is the \emph{SSR\ non-compliant contribution}.

A similar calculation to before for the situation when $n+l=m+k$ gives%
\begin{eqnarray}
\left\langle (\widehat{a}^{\dag })^{n}(\widehat{a})^{m}(\widehat{b}^{\dag
})^{l}(\widehat{b})^{k}\right\rangle &=&Tr\left(
\dsum\limits_{N}\dsum\limits_{\alpha ,\beta }P_{\alpha ,\beta
}^{N}\,\dsum\limits_{\gamma }C_{\alpha ,\gamma }^{N}(n,m,l,k)\,\left\vert
N\,,\,\gamma \right\rangle \left\langle N\,,\beta \right\vert \right) 
\nonumber \\
&&+Tr\left( \dsum\limits_{N\neq M}\dsum\limits_{\alpha ,\beta }P_{\alpha
,\beta }^{N}\,\dsum\limits_{\gamma }C_{\alpha ,\gamma
}^{N}(n,m,l,k)\,\left\vert N\,,\gamma \right\rangle \left\langle M,\,\beta
\right\vert \right)  \nonumber \\
&&
\end{eqnarray}%
But $Tr(\left\vert N\,,\gamma \right\rangle \left\langle M\,,\beta
\right\vert )=0$ for $N\neq M$, so the non-compliant contribution gives zero
and as $Tr(\left\vert N\,,\gamma \right\rangle \left\langle N,\,\beta
\right\vert )=\delta _{\gamma ,\beta }$ we end up with 
\begin{equation}
\left\langle (\widehat{a}^{\dag })^{n}(\widehat{a})^{m}(\widehat{b}^{\dag
})^{l}(\widehat{b})^{k}\right\rangle =\left(
\dsum\limits_{N}\dsum\limits_{\alpha ,\beta }P_{\alpha ,\beta
}^{N}\,C_{\alpha ,\beta }^{N}(n,m,l,k)\right)  \label{Eq.NonCompResult}
\end{equation}%
which is entirely dependent on the contribution to the density operator that
is globally SSR compliant. Measurements of this type with $n+l=m+k$ would 
\emph{not} respond to the presence of contribution to the density operator
that is not globally SSR compliant. The BS measurements discussed in this
paper are all of this type, so will not test the super-selection rule.

Hence the \emph{conclusion} is that a quantum correlation function of the
form $\left\langle (\widehat{a}^{\dag })^{n}(\widehat{a})^{m}(\widehat{b}%
^{\dag })^{l}(\widehat{b})^{k}\right\rangle $ must be measured for cases
where $n+l\neq m+k$ \emph{and} a non-zero measurement result must be found.
If it is, then we would have demonstrated that the state is not globally SSR
compliant. The simplest case would be to find a non-zero result for $%
\left\langle \widehat{a}\right\rangle _{\rho }$ or $\left\langle \widehat{b}%
\right\rangle _{\rho }$.

Similar considerations apply to \emph{local SSR compliance} in the
sub-system states. For sub-system $a$ a QCF of the form $\left\langle (%
\widehat{a}^{\dag })^{n}(\widehat{a})^{m}\right\rangle $ must be measured
for cases where $n\neq m$ and a non-zero measurement result must be found.
If it is, then we would have demonstrated that the state is not locally SSR
compliant. The simplest case would be to find a non-zero result for $%
\left\langle \widehat{a}\right\rangle _{\rho }$. \pagebreak

\subsubsection{SSR Justification and No Suitable Phase Reference}

There are two types of justification for applying the super-selection rules
for systems of identical particles. The first approach is based on simple
considerations and will be outlined below in this subsection. The second
approach \cite{Aharonov67a}, \cite{Bartlett03a}, \cite{Sanders03a}, \cite%
{Kitaev04a}, \cite{van Enk05a}, \cite{Bartlett06a}, \cite{Bartlett07a}, \cite%
{Vaccaro08a}, \cite{White09a}, \cite{Tichy11a} is more sophisticated and
involves linking the absence or presence of SSR to whether or not there is a
suitable \emph{reference frame} in terms of which the quantum state is
described, and is outlined in the next subsection and Appendix \ref{Appendix
- Reference Frames and SSR}. The key idea is that SSR are a consequence of
considering the description of a quantum state by an external observer
(Charlie) whose phase reference frame has an unknown phase difference from
that of an observer ((Alice) more closely linked to the system being
studied. Thus, whilst Alice's description of the quantum state may violate
the SSR, the description of the \emph{same} quantum state by Charlie will
not. In the main part of this paper the density operator $\widehat{\rho }$
used to describe the various quantum states will be that of the external
observer (Charlie).

\subsubsection{SSR Justication and Physics Considerations}

\label{SubSubSection - SSR and Physics}

A number of \emph{straightforward reasons} have been given in the
Introduction for why it is appropriate to apply the superselection rule to
exclude quantum superposition states of the form (\ref{Eq.ForbiddenStates})
as quantum states for systems of identical particles, and these will now be
considered in more detail.

Firstly, no way is known for creating such states. The Hamiltonian for such
a system commutes with the total boson number operator, resulting in the $%
\left\vert C_{N}\right\vert ^{2}$ remaining constant, so the quantum
superposition state would need to have existed initially. In the simplest
case of non-interacting bosonic atoms, the Fock states are also energy
eigenstates, such Fock states involve total energies that differ by energies
of order the rest mass energy $mc^{2}$, so a coherent superposition of
states with such widely differing energies would at least seem unlikely in a 
\emph{non-relativistic theory, }though for massless photons this would not
be an issue as the energy differences are of order the photon energy $\hbar
\omega $. The more important question is: Is there a non-relativistic
quantum process could lead to the creation of such a state? Processes such
as the dissociation of $M$ diatomic molecules into up to $2M$ bosonic atoms
under Hamiltonian evolution involve entangled atom-molecule states of the
form 
\begin{equation}
\left\vert \Phi \right\rangle =\tsum\limits_{m=0}^{M}C_{m}\,\left\vert
M-m\right\rangle _{mol}\otimes \left\vert 2m\right\rangle _{atom}
\label{Eq.MolecDissn}
\end{equation}%
but the reduced density operator for the bosonic atoms is 
\begin{equation}
\widehat{\rho }_{atoms}=\tsum\limits_{m=0}^{M}\left\vert C_{m}\right\vert
^{2}\,\left( \left\vert 2m\right\rangle \left\langle 2m\right\vert \right)
_{atom}  \label{Eq.RDOAtoms}
\end{equation}%
which is a statistical mixture of states with differing atom numbers with no
coherence terms between such states. Such statistical mixtures are valid
quantum states, corresponding to a lack of a priori knowledge of how many
atoms have been produced. To obtain a quantum superposition state for the
atoms \emph{alone}, the atom-molecule state vector would need to evolve at
some time into the form 
\begin{equation}
\left\vert \Phi \right\rangle =\tsum\limits_{m=0}^{M}B_{m}\,\left\vert
M-m\right\rangle _{mol}\otimes \tsum\limits_{n=0}^{M}A_{2n}\left\vert
2n\right\rangle _{atom}  \label{Eq.ProposedEvolvedState}
\end{equation}%
where the separate atomic system is in the required quantum superposition
state. However if such a state existed there would be terms with at least
one non-zero product of\textbf{\ }coefficient $B_{m}A_{2n}$ involving
product states $\left\vert M-m\right\rangle _{mol}\otimes \left\vert
2n\right\rangle _{atom}$ with $n\neq m$ if the state $\left\vert \Phi
\right\rangle $ is not just in the entangled form (\ref{Eq.MolecDissn}).
However, the presence of such a term would mean that the conservation law
involving the number of molecules plus two times the number of atoms was
violated. This is impossible, so such an evolution is not allowed.

Secondly, no way is known for measuring all the properties of such states,
even if they existed. If a state such as (\ref{Eq.ForbiddenStates}) did
exist then the amplitudes $C_{N}$ would oscillate with frequencies that
differ by frequencies of order $mc^{2}/\hbar $ (the Compton frequency, which
is $\gtrsim $ $10^{25}$ Hz for massive bosons) even if boson-boson
interactions were included \ To distinguish the phases of the $C_{N}\,$\ in
order to verify the existence of the state, \emph{measurement operators}
would need to include terms that also oscillate at relativistic frequencies,
and no such measurement operators are known.

Thirdly, there is no need to invoke the existence of such states in order to
understand coherence and interference effects..It is sometimes thought that
states involving quantum superpositions of number states are needed for
discussing \emph{coherence} and \emph{interference properties} of BECs, and
some papers describe the state via the Glauber coherent states. However, as
Leggett \cite{Leggett01a} has pointed out (see also Bach et al \cite{Bach04a}%
, Dalton and Ghanbari \cite{Dalton12a}), a highly occupied number state for
a single mode with $N$ bosons has coherence properties of high order $n$, as
long as $n\ll N$. The introduction of a Glauber coherent state is \emph{not}
required to account for coherence effects. Even the well-known presence of
spatial interference patterns produced when two independent BECs are
overlapped can be accounted for via treating the BECs as Fock states. The
interference pattern is built up as a result of successive boson position
measurements \cite{Javainainen96a}, \cite{Sanders03a}, \cite{Cable05a}.

\subsubsection{SSR Justification and Galilean Frames ?}

Finally, in addition to the previous reasons there is an arguement that has
been proposed based on the requirement that the dynamical equations for such
non-relativistic quantum systems should be invariant under a \emph{Galilean
transformation} which has been proposed \cite{Stenholm02a} as a proof of the
super-selection rule for atom number. This approach is linked to the
reference frame based justification of SSR (see Appendix \ref{Appendix -
Reference Frames and SSR}). However, whilst the paper shows that under a
Galilean transformation - corresponding to describing the system from the
point of view of an observer moving with a constant velocity $\mathbf{v}$
with respect to the original observer, and where the two observers have
identical clocks - the terms in a superposition state with different numbers 
$N$ of massive bosons would oscillate like $\exp i\left( \frac{1}{2}Nm%
\mathbf{v}^{2}t\right) /\hbar $, and may be expected if the \emph{same}
quantum state is described by a moving observer. This feature alone does not
seem to require the super-selection rule, since here the moving observer's
reference frame has a well-defined velocity with respect to that attached to
the system. However, the moving observer's reference frame may actually have
an unknown relative velocity, in which case a twirling operation resulting
in the elimination of number state coherences could be involved (see
Appendix \ref{Appendix - Reference Frames and SSR}). This will be not be
considered further at this stage.

On the other hand, an approach of this kind involving \emph{rotation symmetry%
} would seem to rule out such states as quantum superpositions of a boson
(spin $0$) and a fermion.(spin $1/2$). Let such a state be prepared in the
form $(\left\vert F\right\rangle +\left\vert B\right\rangle )/\sqrt{2}$.
Consider an observer whose cartesian reference frame is $X,Y,Z$. This is a
classical system that can be rotated in space. If the observer rotates with
his frame through $2\pi $ about any axis they are then back in the same
position, but the observer now sees the state as $(-\left\vert
F\right\rangle +\left\vert B\right\rangle )/\sqrt{2}.$ This state is
apparently orthogonal to the one observed before the rotation, and this is
paradoxical since the observer would be in the same position. Thus there is
a super-selection rule excluding states such as $(\left\vert F\right\rangle
+\left\vert B\right\rangle )/\sqrt{2}.$ A similar argument based on the 
\emph{time reversal} anti-unitary operator was given by Wick et al \cite%
{Wick52a}.

\subsubsection{SSR and Photons}

\label{SubSubSection - SSR and Photons}

Though this paper is focused on massive bosonic atoms the question is
whether similar SSR also apply to the optical quantum EM field, which
involve \emph{massless} bosons - \emph{photons}. Here the situation is not
so clear\textbf{\ }and we therefore merely present the differing viewpoints
in the current literature. Some of the same general reasons for applying the
super-selection rule to systems of identical massive bosons also apply here,
though the details differ, but others do not. The situation depends also on
whether optical or microwave photons are involved. The issue is whether for
individual photon modes, states can be prepared that are not local particle
number SSR compliant, and if so can the effects of the non-SSR\ compliant
terms be observed and furthermore do we need to invoke the existence of non
SSR\ compliant states to understand interference and coherence effects. As
we will see, some SSR non compliant feature needs to be present in order to
prepared allegedly non SSR compliant states. Another way of looking at the
issue is to ask whether phase reference systems exist for photon modes. In
addition, there is the issue for multi-mode situations whether states can be
prepared, observed or are needed that are not global particle number SSR\
compliant. As we will see, SSR may now involve a modified total particle
number involving combinations of the mode numbers diferent to the total
particle number - because it is these combinations that are conserved in
mode interaction processes. In the approach adopted in this paper, the
states in question are those described by so-called external observers - not
hypothetical observers that are somehow attached to phase reference systems
internal to the experiment (see below, Section \ref{SubSection - Challenges
to SSR}). The SSR issue for systems involving massless photons is
particularly important in regard to describing entanglement. As explained
below in Section \ref{SubSection - SSR Separate Modes}, if separate
sub-system states in photonic systems can be prepared with density operators
that violate the local particle number SSR, then these so-called "separable
but non-local" states \cite{Verstraete03a} would be classified as separable
rather than as entangled states. Some of the tests for entanglement
described in Paper II for systems of massive boson (such as spin squeezing
in any spin component) would then no longer apply for photonic systems,
though others (such as the Hillery spin variance test) which do not depend
on SSR would still apply.

We first consider the requirement of showing how a non SSR compliant states
can be prepared for single modes. In the case of photons, Molmer \cite%
{Molmer97a} has argued that the quantum state for a single mode optical
laser field operating well above threshold is not a Glauber coherent state,
and the density operator would be a statistical mixture of the form (\ref%
{Eq.PhysicalState}), with $\left\vert \Phi _{N}\right\rangle =\left\vert
N\right\rangle $ and $P_{\Phi N}=\exp (-\overline{N})\,\overline{N}^{N}/N!$.
Here the density operator is a statistical mixture of photon number states
with a Poisson distribution, or equivalently a statistical mixture of equal
amplitude\ coherent states $\left\vert \alpha \right\rangle $ with $\alpha =%
\sqrt{\overline{N}}\exp (i\phi )$ and all phases $\phi $ having equal
probability. In either form, the quantum state is SSR compliant. In terms of
possible processes for preparing states for single mode optical laser
fields, this feature is confirmed in theories for single mode lasers
involving atomic gain media energised via incoherent pumping processes -
there is no well defined optical phase that is imposed on the process. The
Scully-Lamb theory (see Mandel and Wolf \cite{Mandel95a}, p935) gives the
above threshold steady state density operator for the laser mode can be
written in the form of a statistical mixture of number states (somewhat
broader than for a Poisson distribution), which again is an SSR compliant
state with no well-defined optical phase. Further detailed discussion of
laser light generation processes by Pegg and Jeffers \cite{Pegg05a} confirms
this. An alternative approach is presented by Wiseman et al \cite{Wiseman02a}%
, \cite{Wiseman02b}, in which the optical laser is treated via a master
equation, but where monitoring of the laser environment (difficult!) is
required to determine whether certain pure state ensembles - such as those
involving coherent states - are physically realisable. The conclusion
reached is that for finite self energy the coherent state ensemble is not
physically realisable, the closest ensemble being that involving squeezed
states, though for zero self energy coherent state ensembles are obtained.
On the other hand, microwave photons in single mode high Q cavities can be
generated by oscillating electric currents having a well-defined phase. In
this case, as shown in experiments on the Jaynes-Cummings model by Rempe et
al \cite{Rempe87a} and Brune et al \cite{Brune96a} demonstrating collapses
and revivals of Rydberg atom population differences, it is possible to
create Glauber coherent states in microwave cavity modes, and the presence
of these states are necessary to explain the collapse and revival effects.

We next consider cases where interacting photon modes are involved. The two
mode squeezed states generated for example in a non-degenerate parametric
amplifier are often written in the form $\tsum\limits_{n}C_{n}\left\vert
n\right\rangle _{A}\left\vert n\right\rangle _{B}$, corresponding to the
basic generation process in which a pump photon of frequency $\omega
_{C}=\omega _{A}+\omega _{B}$ is destroyed and one photon is created in each
of modes $A$ and $B$. Such a state is not even global SSR compliant, but is
used in describing various quantum information processes as well as
describing two mode squeezing. However, whilst mathematically convenient for
treating such applications this do not demonstrate that this two mode pure
state has actually been created. This state vector is in fact based on a
very simplified version of the process, in which the pump mode is treated
classically. If it is treated quantum mechanically and there were $N$\
photons initially in mode $C$, the interaction term $\widehat{V}=\lambda 
\widehat{c}\widehat{a}^{\dag }\widehat{b}^{\dag }+HC$ would result in a
global SSR compliant state vector like $\tsum\limits_{n}C_{n}\left\vert
N-n\right\rangle _{C}\left\vert n\right\rangle _{A}\left\vert n\right\rangle
_{B}$, but now involving a total \emph{quanta} number $\widehat{N}_{tot}=%
\widehat{N}_{A}+\widehat{N}_{B}+2\widehat{N}_{C}$\ (see Section \ref%
{SubSubSystem - SSR and Conservation} for details). The state describing
modes $A$ and $B$ alone would be $\widehat{\rho }_{AB}=\tsum%
\limits_{n}|C_{n}|^{2}\left\vert n\right\rangle _{A}\left\langle
n\right\vert _{A}\otimes \left\vert n\right\rangle _{B}\left\langle
n\right\vert _{B}$. This state is global SSR\ compliant in terms of $%
\widehat{N}_{A,B}=\widehat{N}_{A}+\widehat{N}_{B}$ but is not the same as
the pure state $\tsum\limits_{n}C_{n}\left\vert n\right\rangle
_{A}\left\vert n\right\rangle _{B}$\ - which is not SSR compliant. Even more
elaborate quantum treatments allowing for irreversible damping processes for
all three modes (see for example McNeil et al \cite{McNeil83a}) that result
in non SSR compliant steady state solutions, include assumptions such as the
pump mode being coupled to a laser mode that is treated classically - and
thus begging the question of whether non SSR compliant states were prepared,
since the classical treatment of the laser mode is itself SSR\ non
compliant. We are unaware of any situation where non SSR compliant states
are claimed to have been created for \emph{optical photons}, where the
theoretical treatment of the preparation process has not assumed the
presence of non SSR compliant states for some key sub-system involved -
usually in an input pump mode. In Section \ref{SubSubSystem - SSR and
Conservation} we considered a preparation process for the non-degenerate
parametric amplifier involving conservation of $\widehat{N}_{tot}$\ starting
from an initial separable state in which all the sub-system density
operators are local particle number SSR compliant and show that this results
in a quantum state that is global SSR\ compliant in terms of total quanta
number $\widehat{N}_{tot}$. There must therefore be some non SSR compliant
feature in the initial state (which could include pump modes) to produce
global non SSR compliant states, so then the issue shifts back to how these
non SSR compliant states are prepared in the first place.

Second, there is the requirement of being able to measure the non SSR
compliant terms. For the free quantum EM\ field there is a conservation law
for the photon number in each mode,\ so in a pure state such as in Eq. (\ref%
{Eq.ForbiddenStates}) the $|C_{N}|^{2}$ would be time independent. However,
for photons the $C_{N}$ would oscillate with frequencies that only differ by
non-relativistic frequencies of order $\omega $ rather than the Compton
frequency that applies for massive bosons, so the arguement against being
able to detect coherent states based on this frequency being so large that
the oscillations cannot be followed do not necessarily apply. Clocks that
can follow microwave oscillations are common-place, and even at optical
frequencies the development of atomic clocks based on optical atomic
transitions that may enable optical frequency oscilations to be observed. So
this consideration does not rule out non SSR compliant states, even for
optical photons.

Finally, we consider the requirement of non SSR\ compliant states being
needed to explain interference, coherence effects etc. We need to
distinguish the sitation where it is mathematically convenient to invoke SSR
non compliant states to explain these effects from the situation where it is
essential to do so. Thus it may be convenient to explain the presence of
interference patterns in position measurements for bosons from two
independent BECs by choosing Glauber coherent states to represent their
states, but as pointed out in Section \ref{SubSubSection - SSR and Physics}
such interference patterns are accounted in terms of Fock states, together
with quantum interference of probability amplitudes associated with bosons
being taken from the two different sources (see Refs. \cite{Javainainen96a}, 
\cite{Sanders03a}, \cite{Cable05a}). In fact, the more detailed feature that
although the separation of the peaks is well defined the actual position of
the peaks are random, is inconsistent with the Glauber coherent state
description. We also point out below (see Section \ref{SubSubSection - Coh
Super Atom Mol} and Appendix \ref{Appendix - Super-Selection Rule Violations
?}) that the interpretation of Ramsey fringes in a proposed experiment to
detect a coherent superposition of an atom and a molecule does not show that
such a state was created or that the BEC involved had to be described by a
Glauber coherent state. Many experiments in which coherence, interference
effects are observed do not depend on SSR non compliant states being
created. Optical interference and coherence effects can also be explained
without invoking Glauber coherent states, as shown by Molmer \cite{Molmer97a}
and in other papers such as \cite{Sanders03a}. In regard to two mode
squeezing in the non-degenerate parametric amplifier described above, the
observation of squeezing effects is often discussed in terms of SSR non
compliant states of the form $\tsum\limits_{n}C_{n}\left\vert n\right\rangle
_{A}\left\vert n\right\rangle _{B}$\ (see Ref. \cite{Reid09a} for example).
However, as may be seen from the experimental paper of Ou et al \cite{Ou92a}%
, the way in which two mode squeezing is observed in the non-degenerate
parametric amplifier involves generating the pump field by frequency
doubling from a lower frequency laser. That lower frequency laser is also
used to provide the local oscillator fields for the homodyne measurements on
modes $A$ and $B$\ used to detect squeezing - these modes are coupled to the
local oscillator fields using beam splitters. The original lower frequency
laser acts as an internal phase reference for the overall experiment, as may
be seen in Fig 2 of Ref \cite{Ou92a} which involves the relative phase
between the local oscillator and the squeezed input field. But as there is
no external phase reference system involved, only the relative phases of the 
$A$, $B$ and local oscillator modes are well defined, and not the overall
phase as would be required for preparing non SSR compliant states. Again the
convenient use of SSR non compliant states to understand experiments in
which only internal phase references are involved does not show that SSR non
compliant states are necessary to interpret the experiments. A similar
arrangement occurs for the degenerate parametric amplifier experiment of Wu
et al \cite{Wu86a}, where Fig 2 clearly shows how the local oscillator field
derives from the original Nd-YAG laser.

Perhaps the best way to approach the question of whether SSR compliance is
required for optical photons for example (see next sub-section and
SubSection \ref{AppendixSubSection - Situation B} in Appendix \ref{Appendix
- Reference Frames and SSR}) involves the consideration of phase reference
frames. The quantum state of a single mode laser may be described as a
Glauber coherent state by an observer (Alice) with one reference frame, but
would be described as a statistical mixture of photon number states by
another observer (Charlie) with a different reference frame whose phase
reference is completely unrelated to the previous one. This arguement
against the presence of coherent state in Charlie's viewpoint is only
overcome if inter-related phase reference frames at the relevant photon
frequencies actually exist.

\subsection{Reference Frames and Violations of Superselection Rules}

\label{SubSection - Challenges to SSR}

Challenges to the requirement for quantum states to be consistent with
super-selection rules have occured since the 1960's when Aharonov and
Susskind \cite{Aharonov67a} suggested that coherent superpositions of
different charge eigenstates could be created. It is argued that
super-selection rules are not a fundamental requirement of quantum theory,
but the restrictions involved could be lifted if there is a suitable system
that acts as a \emph{reference} for the coherences involved - \cite%
{Aharonov67a}, \cite{Bartlett03a}, \cite{Sanders03a}, \cite{Kitaev04a}, \cite%
{van Enk05a}, \cite{Bartlett06a}, \cite{Bartlett07a}, \cite{Vaccaro08a}, 
\cite{White09a}, \cite{Tichy11a} provide discussions regarding reference
systems and SSR.

\subsubsection{Linking SSR and Reference Frames}

\label{SubSubSection - Link SSR Ref Frames}

The discussion of the super-selection rule issue in terms of reference
systems is quite complex and too lengthy to be covered in the body of this
paper. However, in view of the wide use of the reference frame approach a
full outline is presented in Appendix \ref{Appendix - Reference Frames and
SSR}. The key idea is that there are two observers - Alice and Charlie - who
are describing the same prepared system in terms of their own reference
frames and hence their descriptions involve two different quantum states.
The reference systems are \emph{macroscopic systems} in states where the
behaviour is essentially \emph{classical}, such as large magnets that can be
used to define \emph{cartesian axes} or BEC in Glauber coherent states that
are introduced to define a \emph{phase reference}. The relationship between
the two reference systems is represented by a \emph{group} of \emph{unitary
transformation operators} listed as $\widehat{T}(g)$, where the particular
transformation (translation or rotation of cartesian axes, phase change of
phase references, ..) that changes Alice's reference system into Charlie's
is denoted by $g$. Alice describes the quantum state via her density
operator, whereas Charlie is the \emph{external} observer whose
specification of the \emph{same} quantum state via his density operator is
of most interest. There are two cases of importance, \emph{Situation A} -
where the relationship between Alice's and Charlie's reference frame is is 
\emph{known} and specified by a \emph{single} parameter $g$, and \emph{%
Situation B} - where on the other hand the relationship between frames is
completely \emph{unknown}, all possible transformations $g$ must be given
equal weight. Situation A is not associated with SSR, whereas Situation B
leads to SSR. The relationship between Alice's and Charlie's density
operators is given in terms of the transformation operators (see Eq. (\ref%
{Eq.AliceCharlieStatesSitnA}) for Situation A and Eq. (\ref%
{Eq.AliceCharlieStatesSitnB}) for Situation B). In Situation B there is
often a qualitative change between Alice's and Charlie's description of the
same quantum state, with pure states as described by Alice becoming mixed
states when described by Charlie. It is Situation B with the \emph{U(1)}
transformation group - for which \emph{number operators} are the \emph{%
generators} - that is of interest for the \emph{single} or \emph{multi-mode}
systems involving \emph{identical bosons} on which the present paper
focuses. An example of the qualitative change of behaviour for the single
mode case is that \emph{if} it is \emph{assumed} that Alice could prepare
the system in a Glauber coherent pure state - which involves SSR breaking
coherences between differing number states - then Charlie would describe the
same state as a Poisson statistical mixture of number states - which is
consistent with the operation of the SSR. Thus the SSR applies in terms of
external observer Charlie's description of the state. This is how the
dispute on whether the state for single mode laser is a coherent state or a
statistical mixture is resolved - the two descriptions apply to different
observers - Alice and Charlie. On the other hand there are quantum states
such as Fock states and Bell states which are described the same way by both
Alice and Charlie, even in Situation B. The general justification of the SSR
for Charlie's density operator description of the quantum state in Situation
B is derived in terms of the \emph{irreducible representations} of the
transformation group, there being no coherences between states associated
with differing irreducible representations (see Eq. (\ref%
{Eq.CharlieDensOprSSRForm})). For the particular case of the \emph{U(1)}
transformation group the irreducible representations are associated with the
total \emph{boson number} for the system or sub-system, hence the SSR that
prohibits coherences between states where this number differs. Finally, it
is seen that if Alice describes a general non-entangled state of sub-systems
- which being separable have their own reference frames - then Charlie will
also describe the state as a non-entangled state and with the same
probability for each product state (see Eqs. (\ref%
{Eq.AliceDensOprNonEntState}) and (\ref{Eq.CharlieDensOprNonEntState})). For
systems involving \emph{identical bosons} Charlie's description of the
sub-system density operators will only involve density operators that
conform to the SSR. This is in accord with the key idea of the present paper.

\subsubsection{Can Coherent Superpositions of Atom and Molecules Occur ?}

\label{SubSubSection - Coh Super Atom Mol}

Based around the reference frame approach Dowling et al \cite{Dowling06a}
and Terra Cunha et al \cite{TerraCunha07a} propose processes using a BEC as
a reference system that would create a coherent superposition of an atom and
a molecule, or a boson and a fermion \cite{Dowling06a}. Dunningham et al 
\cite{Dunningham11a} consider a scheme for observing a superposition of a
one boson state and the vacuum state. Obviously if super-selection rules can
be overcome in these instances, it might be possible to \emph{produce}
coherent superpositions of Fock states with differing particle numbers such
as Glauber coherent states, though states with $\overline{N}$ $\sim $ $%
10^{8} $ would presumably be difficult to produce. However, detailed
considerations of such papers indicate that the states actually produced in
terms of Charlie's description are statistical mixtures consistent with the
super-selection rules rather than coherent superpositions, which are only
present in Alice's description of the state\textbf{\ }(see Appendix \ref%
{Appendix - Reference Frames and SSR}). Also, although coherence and
interference effects are demonstrated, these can also be accounted for
without invoking the presence of coherent superpositions that violate the
super-selection rule. As the paper by Dowling et al \cite{Dowling06a}
entitled "Observing a coherent superposition of an atom and a molecule." is
a good example of where the super-selection rules are challenged, the key
points are described in Appendix \ref{Appendix - Super-Selection Rule
Violations ?}. Essentially the process involves one atom $A$ interacting
with a BEC of different atoms $B$ leading to the creation of one molecule $%
AB $, with the BEC being depleted by one $B$ atom. There are three stages in
the process, the first being with the interaction that turns separate atoms $%
A$ and $B$ into the molecule $AB$ turned on at Feshbach resonance for a time 
$t$ related to the interaction strength and the mean number of bosons in the
BEC reference system, the second being free evolution at large Feshbach
detuning $\Delta $ for a time $\tau $ leading to a phase factor $\phi
=\Delta \,\tau $, the third being again with the interaction turned on at
Feshbach resonance for a further time $t$. However, it is pointed out in
Appendix \ref{Appendix - Super-Selection Rule Violations ?} that Charlie's
description of the state produced for the atom plus molecule system is
merely a statistical mixture of a state with one atom and no molecules and a
state with no atom and one molecule, the mixture coefficients depending on
the phase $\phi $ imparted during the process. However a coherent
superposition is seen in Alice's description of the final state, though this
is not surprising since a SSR violating initial state was assumed. The
feature that in Charlie's description of the final state no coherent
superposition of an atom and a molecule is produced in the process is not
really surprising, because of the averaging over phase differences in going
from Alice's reference frame to Charlie's. It is the dependence on the phase 
$\phi $ imparted during the process that demonstates coherence (Ramsey
interferometry) effects, but it is shown in Appendix \ref{Appendix -
Super-Selection Rule Violations ?} that exactly the same results can be
obtained via a treatment in which states which are coherent superpositions
of an atom and a molecules are never present, the initial BEC state being
chosen as a Fock state. In terms of the description by an external observer
(Charlie) the claim of violating the super-selection rule has not been
demonstrated via this particular process.

\subsubsection{Detection of SSR Violating States}

Whether such super-selection rule violating states can be \emph{detected}
has also not been justified. For example, consider the state given by a
superposition of a one boson state and the vacuum state (as discussed in 
\cite{Dunningham11a}). We consider an interferometric process in which one
mode $A$ for a two mode BEC interferometer is initially in the state $\alpha
\left\vert 0\right\rangle +\beta \left\vert 1\right\rangle $, and the other
mode $B$ is initially in the state $\left\vert 0\right\rangle $ - thus $%
\left\vert \Psi (i)\right\rangle =(\alpha \left\vert 0\right\rangle +\beta
\left\vert 1\right\rangle )_{A}\otimes \left\vert 0\right\rangle _{B}$ in
the usual occupancy number notation, where $|\alpha |^{2}+|\beta |^{2}=1$.
The modes are first coupled by a beam splitter, then a free evolution stage
occurs for time $\tau $ associated with a phase difference $\phi =\Delta
\tau $ (where $\Delta =\omega _{B}-\omega _{A}$ is the mode frquency
difference), the modes are then coupled again by the beam splitter and the
probability of an atom being found in modes $A$, $B$ finally being measured.
The probabilities of finding one atom in modes $A$, $B$ respectively are
found to only depend on $|\beta |^{2}$ and $\phi $. Details are given in
Appendix \ref{Appendix - Super-Selection Rule Violations ?}. There is no
dependence on the relative phase between $\alpha $ and $\beta $, as would be
required if the superposition state $\alpha \left\vert 0\right\rangle +\beta
\left\vert 1\right\rangle $ is to be specified. Exactly the same detection
probabilities are obtained if the initial state is the mixed state $\widehat{%
\rho }(i)=|\alpha |^{2}(\left\vert 0\right\rangle _{A}\left\langle
0\right\vert _{A}\otimes \left\vert 0\right\rangle _{B}\left\langle
0\right\vert _{B})+|\beta |^{2}(\left\vert 1\right\rangle _{A}\left\langle
1\right\vert _{A}\otimes \left\vert 0\right\rangle _{B}\left\langle
0\right\vert _{B})$, in which the vacuum state for mode $A$ occurs with a
probability $|\alpha |^{2}$ and the one boson state for mode $A$ occurs with
a probability $|\beta |^{2}$. In this example the proposed coherent
superposition associated with the super-selection rule violating state would
not be detected in this interferometric process, nor in the more elaborate
scheme discussed in \cite{Dunningham11a}.

Of course, the claim that in isolated systems of\emph{\ massive} particles
it is not possible in \emph{non-relativistic quantum physics} to create
states that violate the particle number SSR - either for the sub-system
states in a separable state or for any quantum state of the overall system -
can be questioned. Ideally the claim should be \emph{tested} by experiment,
in particular when the number of particles is large in view of the interest
in \emph{macroscopic entanglement} since the Schrodinger cat was first
described. The simplest situation would be to test whether states that
violate the (local) particle number SSR could be created for a \emph{single
mode system}. Clearly, a specific \emph{proposal} for an experiment in which
the SSR \emph{could} be violated is required, but to our knowledge no such
proposal has been presented. Bose-Einstein condensates, in which all the
bosons can occupy a single mode would seem an ideal candidate as a suitable
bosonic system, and the \emph{Glauber coherent state} is an example of a non
SSR compliant state. For fermions, the Pauli exclusion principle would limit
the number of fermions in a one mode system to be zero or one, but coherent
superpositions of a zero and one fermion state are examples of non-SSR
compliant pure states. As pointed out above, some authors such as \cite%
{Hillery06a}, \cite{Hillery09a}, \cite{Verstraete03a}, \cite{Schuch04a},
base their definition of entanglement by allowing for the possible presence
of non-SSR compliant sub-system states when defining separable states. The
approach in Refs. \cite{Dalton14a} is based on the \emph{physical assumption}
that states that are non-compliant with particle number SSR - both local and
global - do\emph{\ not} come into the realm of non-relativistic quantum
physics, in which the concept of entanglement is useful. Until \emph{clear
evidence} is presented that non-SSR- compliant states\emph{\ can} be
prepared, and in view of the theoretical reasons why they \emph{cannot}, it
seems preferable to base the \emph{theory of entanglement} on their \emph{%
absence} when defining separable and entangled quantum states.

\subsection{Super-Selection Rule - Separate Sub-Systems}

\label{SubSection - SSR Separate Modes}

In this sub-section the important case of SSR in \emph{separable} states
will be dealt with, since this is key to understanding what entangled states
are allowed in systems involving identical particles. This forms the basis
for the treatment of entanglement tests presented in the second part of this
review (Paper II).

\subsubsection{Local Particle Number SSR}

We now consider the role of the super-selection rule for the case of \emph{%
non-entangled} states. The global super-selection rule on \emph{total
particle number} has restricted the physical quantum state for a system of
identical bosons to be of the form (\ref{Eq.PhysicalState}). Such states may
or may not be entangled states of the modes involved. The question is - do
similar restrictions involving the \emph{sub-system particle number} apply
to the modes, considered as \emph{separate} sub-systems in the definition of
non-entangled states ? The viewpoint in this paper is that this is so. Note
that applying the SSR on the separate sub-system density operators $\widehat{%
\rho }_{R}^{A}$, $\widehat{\rho }_{R}^{B}$, ... is \emph{only} in the
context of non-entangled states. Such a SSR is referred to as a \emph{local }%
SSR, as it applies to each of the separate sub-systems. Mathematically, the
local particle number SSR can be expressed as 
\begin{equation}
\lbrack \widehat{N}_{X},\widehat{\rho }_{R}^{X}]=0  \label{Eq.LocalSSR}
\end{equation}%
where $\widehat{N}_{X}$ is the \emph{number} operator for sub-system $%
X=A,B,\,...\,$.The SSR restriction is based on the proposition that the
density operators $\widehat{\rho }_{R}^{A}$, $\widehat{\rho }_{R}^{B}$, ...
for the separate sub-systems $A$, $B$, ... should themselves represent
possible \emph{quantum states} for each of the sub-systems, considered as a 
\emph{separate system} and thus be required to satisfy the super-selection
rule that forbids quantum superpositions of Fock states with differing boson
numbers. Note that if the local particle number SSR applies in each
sub-system the global particle number SSR applies to any separable state.
The proof is trivial and just requires showing that $[\widehat{N},\widehat{%
\rho }_{sep}]=0$\ 

The justification of applying the\textbf{\ }\emph{local particle number
super-selection rule}\textbf{\ }to the density operators $\widehat{\rho }%
_{R}^{a}$, $\widehat{\rho }_{R}^{b}$, ... for the sub-system quantum states
that occur in any separable state is simply that these are possible quantum
states of the sub-systems when the latter are considered as separate quantum
systems before being combined as in the Werner protocol \cite{Werner89a} to
form the separable state. Hence all the justifications based either on
simple physical considerations or phase reference systems that were
previously invoked for the density operator $\widehat{\rho }$\ of any
general quantum states of the combined sub-systems to establish the\ \emph{%
global particle number super-selection rule}\textbf{\ }apply equally well
here. No more need be said\textbf{. }It is contended that expressions for
the non-entangled quantum state $\widehat{\rho }$ in which $\widehat{\rho }%
_{R}^{A}$, $\widehat{\rho }_{R}^{B}$, $\widehat{\rho }_{R}^{C}$... were 
\emph{not} allowed quantum states for the sub-systems would only be of
mathematical interest.

Applying the local particle number SSR to the sub-system density operators
for non-entangled states is discussed in papers by Bartlett et al \cite%
{Bartlett06b}, \cite{Bartlett07a} as one of several \emph{operational
approaches} for defining entangled states. As pointed out above other
authors \cite{Verstraete03a}, \cite{Schuch04a} define separable (and hence
entangled) states differently by specifically allowing sub-system density
operators that are \emph{not} consistent with the local particle number SSR,
though the overall density operator is globally SSR consistent. The
corresponding overall states are termed \emph{separable but non-local}, and
states that they would regard as separable would here be regarded as
entangled. Examples of such states are given in Eqs. (\ref%
{Eq.VerstraeteState}) and (\ref{Eq.TwoModeCoherentStateMixture}). There are
also other authors \cite{Hillery06a}, \cite{Hillery09a} who define separable
(and hence entangled) states via (\ref{Eq.NonEntangledState}) but leave
unspecified whether the sub-system density operators are consistent or
inconsistent with the local particle number SSR. Note that any inequalities
involving measured quantities that are found for separable states in which
local SSR compliance is \emph{neglected} must also apply to separable states
where it is \emph{required}. The consequent implications for entanglement
tests where local particle number SSR compliance is required is discussed in
the accompanying paper II. Hence, in this paper we are advocating a \emph{%
revision} to a \emph{widely held} \emph{notion} of entanglement in identical
particle systems, the consequence being that the set of entangled states is
now much \emph{larger}. This is a \emph{key idea} in this paper - not only
should super-selection rules on particle numbers be applied to the the \emph{%
overall} quantum state, entangled or not, but it \emph{also} should be
applied to the density operators that describe states of the modal \emph{%
sub-systems} involved in the general definition of \emph{non-entangled}
states. The reasons for adopting this viewpoint have been discussed above -
basically it is because in separable states the sub-system density operators
must represent possible quantum states for the sub-systems considered as
isolate quantum systems, so the general reasons for applying the SSR will
apply to these density operators also. Apart from the papers by Bartlett et
al \cite{Bartlett06b}, \cite{Bartlett07a} we are not aware that this
definition of non-entangled states has been invoked previously, indeed the
opposite approach has been proposed \cite{Verstraete03a}, \cite{Schuch04a}.
However, the idea of considering whether sub-system states should satisfy
the local particle number SSR has been presented in several papers -\emph{\ }%
\cite{Verstraete03a}, \cite{Schuch04a}, \cite{Bartlett06b}, \cite%
{Bartlett07a}, \cite{Vaccaro08a}, \cite{White09a}, \cite{Paterek11a}, mainly
in the context of pure states for bosonic systems, though in these papers
the focus is on issues other than the definition of entanglement, such as
quantum communication protocols \cite{Verstraete03a}, multicopy distillation 
\cite{Bartlett06b}, mechanical work and accessible entanglement \cite%
{Vaccaro08a}, \cite{White09a} and Bell inequality violation \cite{Paterek11a}%
. However, there are a number of papers that do not apply the SSR to the
sub-system density operators, and those that do have not studied the
consequences for various entanglement tests. These tests are also discussed
in the accompanying paper II.

\subsubsection{Criterion for Local and Global SSR in Separable States}

\textit{Theorem. }A necessary and sufficient condition for all separable
states for a given set of sub-system density operators $\widehat{\rho }%
_{R}^{a},\widehat{\rho }_{R}^{b}$ to be global particle number SSR compliant
is that all such sub-system states are local particle number SSR compliant.

We first note that 
\begin{equation}
\lbrack \widehat{N},\widehat{\rho }]=\dsum\limits_{R}P_{R}([\widehat{n}_{a},%
\widehat{\rho }_{R}^{a}]\otimes \widehat{\rho }_{R}^{b}+\widehat{\rho }%
_{R}^{a}\otimes \lbrack \widehat{n}_{b},\widehat{\rho }_{R}^{b}])
\label{Eq.NumberCommutatorSepState}
\end{equation}

\textit{Necessity: \ }If the state $\widehat{\rho }$ is globally SSR
compliant then $[\widehat{N},\widehat{\rho }]=0$. Taking the trace of both
sides of (\ref{Eq.NumberCommutatorSepState}) over sub-system space $b$,
using $Tr_{b}(\widehat{\rho }_{R}^{b})=1$ and $Tr_{b}([\widehat{n}_{b},%
\widehat{\rho }_{R}^{b}])=0$ and then repeating the process for sub-system
space $a$ gives%
\begin{equation}
0=\dsum\limits_{R}P_{R}([\widehat{n}_{a},\widehat{\rho }_{R}^{a}])\qquad
0=\dsum\limits_{R}P_{R}([\widehat{n}_{b},\widehat{\rho }_{R}^{b}])
\end{equation}%
which are operator equation in sub-system spaces $a$ and $b$ respectively.

The $P_{R}$ are not independent, satisfying $\dsum\limits_{R}P_{R}=1$. By
choosing a particular $P_{S}$ we can write the last equation for sub-system $%
a$ as 
\begin{equation}
0=\dsum\limits_{R\neq S}P_{R}([\widehat{n}_{a},\widehat{\rho }%
_{R}^{a}])+(1-\dsum\limits_{R\neq S}P_{R})([\widehat{n}_{a},\widehat{\rho }%
_{S}^{a}])
\end{equation}%
where the remaining $P_{R}$ are now independent. Differentiating the last
equation with respect to $P_{R}$ then gives%
\begin{equation}
0=[\widehat{n}_{a},\widehat{\rho }_{R}^{a}]-[\widehat{n}_{a},\widehat{\rho }%
_{S}^{a}]
\end{equation}%
for any two different $R$ and $S$. Thus all the $[\widehat{n}_{a},\widehat{%
\rho }_{S}^{a}]$ must be the same. Using $0=\dsum\limits_{R}P_{R}([\widehat{n%
}_{a},\widehat{\rho }_{R}^{a}])$ again with equal $[\widehat{n}_{a},\widehat{%
\rho }_{R}^{a}]$ and $\dsum\limits_{R}P_{R}=1$ we then see that all $[%
\widehat{n}_{a},\widehat{\rho }_{R}^{a}]$ must be zero. Similar
considerations show that $[\widehat{n}_{b},\widehat{\rho }_{R}^{b}]=0$.

As these results apply for any choice of the $P_{R}$ and of the $\widehat{%
\rho }_{R}^{a}$, $\widehat{\rho }_{R}^{b}$ we can then conclude that 
\begin{equation}
\lbrack \widehat{n}_{a},\widehat{\rho }_{R}^{a}]=0\qquad \lbrack \widehat{n}%
_{b},\widehat{\rho }_{R}^{b}]  \label{Eq.SubSysLocalSSR}
\end{equation}%
which establishes that the sub-system states are local particle number SSR
compliant.

Note that the proof depended on the choice of the $P_{R}$ being arbitrary
apart from $\dsum\limits_{R}P_{R}=1$. If the $P_{R}$ are fixed then although
we can show that $0=\dsum\limits_{R}P_{R}([\widehat{n}_{a},\widehat{\rho }%
_{R}^{a}])=\dsum\limits_{R}P_{R}([\widehat{n}_{b},\widehat{\rho }_{R}^{b}])$%
, the steps leading to $[\widehat{n}_{a},\widehat{\rho }_{R}^{a}]=[\widehat{n%
}_{b},\widehat{\rho }_{R}^{b}]=0$ do not follow. The four sub-system states
in Section \ref{SubSection - Two Mode Coherent State Mixture} where all $%
P_{R}=1/4$ are not local particle number SSR compliant even though the
overall state is global particle number SSR compliant. This would not be the
case if any of the $P_{R}$ differed from $1/4$.

\textit{Sufficiency: }If the sub-system states are local particle number SSR
compliant then $[\widehat{n}_{a},\widehat{\rho }_{R}^{a}]=[\widehat{n}_{b},%
\widehat{\rho }_{R}^{b}]=0$. It then follows from (\ref%
{Eq.NumberCommutatorSepState}) that 
\begin{equation}
\lbrack \widehat{N},\widehat{\rho }]=0  \label{Eq.GlobalSepStateSSR}
\end{equation}%
which establishes that the separable state is global particle number SSR
compliant. This conclusion applies for arbitrary $P_{R}$.

\subsubsection{Global but not Local Particle Number SSR Compliant States}

\label{SubSection - Two Mode Coherent State Mixture}

However, it should be noted that some authors \cite{Verstraete03a}, \cite%
{Schuch04a} consider sub-system density operators in the context of two mode
systems which comply with the global particle number SSR but not the local
particle number SSR. Such a case involving four \emph{zero and one boson
superpositions} is presented by Verstraete et al \cite{Verstraete03a}, \cite%
{Schuch04a}. The overall density operator is a statistical mixture 
\begin{eqnarray}
\widehat{\rho } &=&\frac{1}{4}(\left\vert \psi _{1}\right\rangle
\left\langle \psi _{1}\right\vert )_{A}\otimes \left\vert \psi
_{1}\right\rangle \left\langle \psi _{1}\right\vert )_{B}+\frac{1}{4}%
(\left\vert \psi _{i}\right\rangle \left\langle \psi _{i}\right\vert
)_{A}\otimes \left\vert \psi _{i}\right\rangle \left\langle \psi
_{i}\right\vert )_{B}  \nonumber \\
&&+\frac{1}{4}(\left\vert \psi _{-1}\right\rangle \left\langle \psi
_{-1}\right\vert )_{A}\otimes \left\vert \psi _{-1}\right\rangle
\left\langle \psi _{-1}\right\vert )_{B}+\frac{1}{4}(\left\vert \psi
_{-i}\right\rangle \left\langle \psi _{-i}\right\vert )_{A}\otimes
\left\vert \psi _{-i}\right\rangle \left\langle \psi _{-i}\right\vert )_{B} 
\nonumber \\
&&  \label{Eq.VerstraeteState}
\end{eqnarray}%
where $\left\vert \psi _{\omega }\right\rangle =(\left\vert 0\right\rangle
+\omega \left\vert 1\right\rangle )/\sqrt{2}$, with $\omega =1,i,-1,-i$. The 
$\left\vert \psi _{\omega }\right\rangle $ are superpositions of zero and
one boson states and consequently the local particle number SSR is violated
by each of the sub-system density operators $\left\vert \psi _{\omega
}\right\rangle \left\langle \psi _{\omega }\right\vert )_{A}$ and $%
\left\vert \psi _{\omega }\right\rangle \left\langle \psi _{\omega
}\right\vert )_{B}$. Although the expression in Eq.(\ref{Eq.VerstraeteState}%
) is of the form in Eq.(\ref{Eq.NonEntangledState}), the subsystem density
operators $\left\vert \psi _{\omega }\right\rangle \left\langle \psi
_{\omega }\right\vert )_{A}$ and $\left\vert \psi _{\omega }\right\rangle
\left\langle \psi _{\omega }\right\vert )_{B}$ do not comply with the local
particle number SSR, so in the present paper and in \cite{Dalton14a} the
state would be regarded as \emph{entangled}. However, Verstraete et al \cite%
{Verstraete03a}, \cite{Schuch04a} regard it as separable. They refer to such
a state as \emph{separable but nonlocal}.

On the other hand, the global particle number SSR is obeyed since the
density operator can also be wriiten as 
\begin{eqnarray}
\widehat{\rho } &=&\frac{1}{4}(\left\vert 0\right\rangle \left\langle
0\right\vert )_{A}\otimes \left\vert 0\right\rangle \left\langle
0\right\vert )_{B}+\frac{1}{4}(\left\vert 1\right\rangle \left\langle
1\right\vert )_{A}\otimes \left\vert 1\right\rangle \left\langle
1\right\vert )_{B}  \nonumber \\
&&+\frac{1}{2}(\left\vert \Psi _{+}\right\rangle \left\langle \Psi
_{+}\right\vert )_{AB}  \label{Eq.VerstraeteState2}
\end{eqnarray}%
where $\left\vert \Psi _{+}\right\rangle _{AB}=(\left\vert 0\right\rangle
_{A}\left\vert 1\right\rangle _{B}+\left\vert 1\right\rangle _{A}\left\vert
0\right\rangle _{B})/\sqrt{2}$. This is a statistical mixture of $N=0,1,2$
boson states. Note that Eq.(\ref{Eq.VerstraeteState2}) indicates that the
state could be prepared as a mixed state containing two terms that comply
with the local particle number SSR in each of the sub-systems plus a term
which is an entangled state of the two sub-systems. The presence of an
entangled state in such an obvious preparation process challenges the
description of the state as being separable.

To further illustrate some of the points made about super-selection rules -
local and global - it is useful to consider a second specific case also
presented by Verstraete et al \cite{Verstraete03a}, \cite{Schuch04a}. This 
\emph{mixture} of \emph{two mode coherent states }is represented by the two
mode density operator%
\begin{eqnarray}
\widehat{\rho } &=&\tint \frac{d\theta }{2\pi }\,\left\vert \alpha ,\alpha
\right\rangle \left\langle \alpha ,\alpha \right\vert  \nonumber \\
&=&\tint \frac{d\theta }{2\pi }\,\left( \left\vert \alpha \right\rangle
\left\langle \alpha \right\vert \right) _{A}\otimes \left( \left\vert \alpha
\right\rangle \left\langle \alpha \right\vert \right) _{B}
\label{Eq.TwoModeCoherentStateMixture}
\end{eqnarray}%
where $\left\vert \alpha \right\rangle _{C}$ is a one mode coherent state
for mode $C=A,B$ with $\alpha =|\alpha |\,\exp (-i\theta )$, and modes $A,B$
are associated with bosonic annihilation operators $\widehat{a}$, $\widehat{b%
}$. The magnitude $|\alpha |$ is fixed.

This density operator \emph{appears} to be that for a non-entangled state of
modes $A,B$ in the form%
\begin{equation}
\widehat{\rho }=\tsum\limits_{R}P_{R}\,\widehat{\rho }_{R}^{A}\otimes 
\widehat{\rho }_{R}^{B}  \label{Eq.NonEntState}
\end{equation}%
with $\tsum\limits_{R}P_{R}\rightarrow \tint \frac{d\theta }{2\pi }$ and $%
\widehat{\rho }_{R}^{A}\rightarrow \left( \left\vert \alpha \right\rangle
\left\langle \alpha \right\vert \right) _{A}$ and $\widehat{\rho }%
_{R}^{B}\rightarrow \left( \left\vert \alpha \right\rangle \left\langle
\alpha \right\vert \right) _{B}$. However although this choice of $\widehat{%
\rho }_{R}^{A}$, $\widehat{\rho }_{R}^{B}$ satisfy the Hermitiancy, unit
trace, positivity features they do \emph{not} conform to the requirement of
satisfying the (\emph{local}) sub-system boson number \emph{super-selection
rule}. From Eq. (\ref{Eq.TwoModeCoherentStateMixture}) we have%
\begin{eqnarray}
\left\langle n\right\vert \left( \left\vert \alpha \right\rangle
\left\langle \alpha \right\vert \right) \left\vert m\right\rangle _{A}
&=&\exp (-|\alpha |^{2})\frac{\alpha ^{n}}{\sqrt{n!}}\frac{(\alpha )^{\ast
\,m}}{\sqrt{m!}}  \nonumber \\
\left\langle p\right\vert \left( \left\vert \alpha \right\rangle
\left\langle \alpha \right\vert \right) \left\vert q\right\rangle _{B}
&=&\exp (-|\alpha |^{2})\frac{\alpha ^{p}}{\sqrt{p!}}\frac{(\alpha )^{\ast
\,q}}{\sqrt{q!}}  \label{Eq.MatrixElements}
\end{eqnarray}%
so clearly for each of the separate modes there are \emph{coherences}
between Fock states with differing boson occupation numbers. In the approach
in the present paper the density operator in Eq. (\ref%
{Eq.TwoModeCoherentStateMixture}) does \emph{not} represent a non-entangled
state. However, in the papers of Verstraete et al \cite{Verstraete03a}, \cite%
{Schuch04a}, Hillery et al \cite{Hillery06a}, \cite{Hillery09a} and others
it would represent an allowable non-entangled (separable) state. Indeed,
Verstraete et al \cite{Verstraete03a} specifically state ".., this state is 
\emph{obviously} separable, though the states $\left\vert \alpha
\right\rangle $ are incompatible with the (local) super-selection rule.".
Verstraete et al \cite{Verstraete03a} introduce the state defined in Eq. (%
\ref{Eq.TwoModeCoherentStateMixture}) as an example of a state that is
separable (in their terms) but which cannot be prepared locally, because it
is incompatible with the local particle number super-selection rule.

The \emph{mixture} of \emph{two mode coherent states} does of course satisfy
the \emph{total} or \emph{global} boson number super-selection rule. The
matrix elements between two mode Fock states are%
\begin{eqnarray}
(\left\langle n\right\vert _{A}\otimes \left\langle p\right\vert _{B})\,%
\widehat{\rho }\,(\left\vert m\right\rangle _{A}\otimes \left\vert
q\right\rangle _{B}) &=&\exp (-2|\alpha |^{2})\frac{|\alpha |^{n+m}}{\sqrt{n!%
}\sqrt{m!}}\frac{|\alpha |^{p+q}}{\sqrt{p!}\sqrt{q!}}\tint \frac{d\theta }{%
2\pi }\,\exp (-i(n-m+p-q)\theta )  \nonumber \\
&=&\exp (-2|\alpha |^{2})\frac{|\alpha |^{n+m}}{\sqrt{n!}\sqrt{m!}}\frac{%
|\alpha |^{p+q}}{\sqrt{p!}\sqrt{q!}}\,\delta _{n+p,m+q}
\label{Eq.OverallMatrixElements}
\end{eqnarray}%
These overall matrix elements are zero unless $n+p=m+q$, showing that there
are \emph{no coherences} between two mode Fock states where the total boson
number differs. The mixture of two mode coherent states has the interesting
feature of providing an example of a two mode state which satisfies the
global but not the local super-selection rule.

The \emph{reduced density operators} for modes $A,B$ are 
\[
\widehat{\rho }_{A}=\tint \frac{d\theta }{2\pi }\,\left( \left\vert \alpha
\right\rangle \left\langle \alpha \right\vert \right) _{A}\qquad \widehat{%
\rho }_{B}=\tint \frac{d\theta }{2\pi }\,\left( \left\vert \alpha
\right\rangle \left\langle \alpha \right\vert \right) _{B} 
\]%
and a straightforward calculation gives 
\[
\widehat{\rho }_{A}=\exp (-|\alpha |^{2})\dsum\limits_{n}\frac{|\alpha |^{2n}%
}{n!}\left( \left\vert n\right\rangle \left\langle n\right\vert \right)
_{A}\qquad \widehat{\rho }_{B}=\exp (-|\alpha |^{2})\dsum\limits_{p}\frac{%
|\alpha |^{2p}}{p!}\left( \left\vert p\right\rangle \left\langle
p\right\vert \right) _{B} 
\]%
which are statistical mixtures of Fock states with the expected Poisson
distribution associated with coherent states. This shows that the reduced
density operators \emph{are} consistent with the separate mode local
super-selection rule, whereas the density operators $\widehat{\rho }%
_{R}^{A}=\left( \left\vert \alpha \right\rangle \left\langle \alpha
\right\vert \right) _{A}$ , $\widehat{\rho }_{R}^{B}$ $=\left( \left\vert
\alpha \right\rangle \left\langle \alpha \right\vert \right) _{B}$ are \emph{%
not} . Later we will revisit this example in the context of entanglement
tests.

Note that if a twirling operation (see Eq.(\ref{Eq.BECDensityOpr3})) were to
be applied to mode $A$, the result would be equivalent to applying two
independent twirling operations to each mode. In this case the density
operator for each mode is a Poisson statistical mixture of number states, so
each mode has a density operator that complies with the local particle
number SSR.

\subsubsection{Particle Entanglement Measure}

Wiseman et al have also treated entanglement for pure states \cite%
{Wiseman03a} and mixed states \cite{Dowling06a} in identical particle
systems, applying both the symmetrization principle and super-selection
rules, invoking the arguement that \emph{without} a phase reference the
quantum state must be comply with the local (and global) particle number
SSR. This is essentially the same approach as in \cite{Bartlett06a}, \cite%
{Bartlett07a}, \cite{Dalton14a} and in the present paper. For two mode
systems the \emph{observable} system density operator $\widetilde{\widehat{%
\rho }}$ is obtained from the density operator $\widehat{\rho }$ that would
apply \emph{if} such a phase reference existed via the expression 
\begin{equation}
\widetilde{\widehat{\rho }}=\tsum\limits_{n_{A}n_{B}}\,\widehat{\Pi }%
_{n_{A}n_{B}}\,\widehat{\rho }\,\widehat{\Pi }_{n_{A}n_{B}}=\tsum%
\limits_{n_{A}n_{B}}\,\widehat{\rho }^{(n_{A}n_{B})}
\label{Eq.ObservableDensityOpr}
\end{equation}%
where $\widehat{\Pi }_{n_{A}n_{B}}=$ $\widehat{\Pi }_{n_{A}n_{B}}^{2}$ is a
projector onto sub-system states with $n_{A}$, $n_{B}$ particles in modes $A$%
, $B$ respectively. Note that $\widehat{\rho }^{(n_{A}n_{B})}=\widehat{\Pi }%
_{n_{A}n_{B}}\,\widehat{\rho }\,\widehat{\Pi }_{n_{A}n_{B}}$ is not
normalised to unity. In fact the probability that there are $n_{A}$, $n_{B}$
particles in modes $A$, $B$ respectively is given by $P_{n_{A}n_{B}}=Tr(%
\widehat{\Pi }_{n_{A}n_{B}}\,\widehat{\rho }\,\widehat{\Pi }%
_{n_{A}n_{B}})=Tr(\widehat{\rho }^{(n_{A}n_{B})})$, so $Tr(\widetilde{%
\widehat{\rho }})=\tsum\limits_{n_{A}n_{B}}P_{n_{A}n_{B}}=1$. For \emph{%
separable} states defined here as in Eq.(\ref{Eq.NonEntangledState}), the
expression in (\ref{Eq.ObservableDensityOpr}) for the density operator is
the \emph{same} as that used here, since with $\widehat{\rho }$ given by Eq.(%
\ref{Eq.NonEntangledState}) and with $\tsum\limits_{n_{A}n_{B}}\,\,\widehat{%
\Pi }_{n_{A}n_{B}}\,(\widehat{\rho }_{R}^{A}\otimes \widehat{\rho }%
_{R}^{B})\,\,\widehat{\Pi }_{n_{A}n_{B}}=\widehat{\rho }_{R}^{A}\otimes 
\widehat{\rho }_{R}^{B}$ it is easy to show that $\widetilde{\widehat{\rho }}%
=\widehat{\rho }_{sep}$. For general mixed states Wiseman et al introduce in
Ref. \cite{Dowling06a} the idea of \emph{particle entanglement} by defining
its measure $E_{P}(\widehat{\rho }\,)$ by 
\begin{equation}
E_{P}(\widehat{\rho }\,)=\tsum\limits_{n_{A}n_{B}}\,P_{n_{A}n_{B}}E_{M}(%
\widehat{\rho }^{(n_{A}n_{B})})=E_{P}(\widetilde{\widehat{\rho }}\,)
\label{Eq.ParticleEntMeasure}
\end{equation}%
where $E_{M}(\widehat{\rho }^{(n_{A}n_{B})})$ is a measure of the\emph{\
mode entanglement} associated with the (unnormalised) state $\widehat{\rho }%
^{(n_{A}n_{B})}$. This might be taken as the \emph{entropy} of mode
entanglement $E_{M}(\widehat{\sigma })=-Tr(\widehat{\sigma }_{A}\,\ln 
\widehat{\sigma }_{A})$ for normalised density operators $\widehat{\sigma }$%
, where the reduced density operator for mode $A$ is $\widehat{\sigma }%
_{A}=Tr_{B}(\widehat{\sigma })$. Note that from $\widehat{\Pi }_{n_{A}n_{B}}%
\widehat{\Pi }_{m_{A}m_{B}}=\delta _{n_{A}m_{A}}\delta _{n_{B}m_{B}}$ $%
\widehat{\Pi }_{n_{A}n_{B}}$ the particle entanglement measure $E_{P}(%
\widehat{\rho }\,)$ is the same for $\widetilde{\widehat{\rho }}$, the
observable density operator for the system. In the case of the separable
state for modes $A$, $B$ given in (\ref{Eq.NonEntangledState}) it is
straightforward to show that 
\begin{eqnarray}
\widehat{\rho }_{sep}^{(n_{A}n_{B})} &=&\dsum\limits_{R}P_{R}\,(\widehat{\Pi 
}_{n_{A}}\widehat{\rho }_{R}^{A}\,\widehat{\Pi }_{n_{A}})\otimes (\widehat{%
\Pi }_{n_{B}}\widehat{\rho }_{R}^{B}\,\widehat{\Pi }_{n_{B}})
\label{Eq.SepStatesProjected} \\
P_{n_{A}n_{B}}^{sep} &=&\dsum\limits_{R}P_{R}\,P_{n_{A}}(\widehat{\rho }%
_{R}^{A})\,P_{n_{B}}(\widehat{\rho }_{R}^{B})  \label{Eq.SepStatesProbs}
\end{eqnarray}%
where $\widehat{\Pi }_{n_{A}}$ and $\widehat{\Pi }_{n_{B}}$ are projectors
onto sub-system states in modes $A$, $B$ respectively with $n_{A}$ and $%
n_{B} $ particles in the respective modes ($\widehat{\Pi }_{n_{A}n_{B}}=%
\widehat{\Pi }_{n_{A}}\otimes \widehat{\Pi }_{n_{B}}$), with $P_{n_{A}}(%
\widehat{\rho }_{R}^{A})=Tr_{A}(\widehat{\Pi }_{n_{A}}\widehat{\rho }%
_{R}^{A})$ and $P_{n_{B}}(\widehat{\rho }_{R}^{B})=Tr_{B}(\widehat{\Pi }%
_{n_{B}}\widehat{\rho }_{R}^{B})$ being the probabilities of finding $n_{A}$
and $n_{B}$ particles in the respective modes when the corresponding
sub-system states are $\widehat{\rho }_{R}^{A}$ and $\widehat{\rho }_{R}^{B}$%
. Since the state $\widehat{\rho }_{sep}^{(n_{A}n_{B})}$ is clearly a
separable state of the form (\ref{Eq.NonEntangledState}) for the modes $A$, $%
B$, the corresponding measure of mode entanglement must be zero. It then
follows from the general expression (\ref{Eq.ParticleEntMeasure}) that the
particle entanglement measure is also zero for the separable state. This is
as expected. 
\begin{equation}
E_{P}(\widehat{\rho }_{sep}\,)=0  \label{Eq.PartEntMeasureSepState}
\end{equation}%
For the pure states considered in \cite{Wiseman03a} we note that among them
is the two boson state $\left\vert 1\right\rangle _{A}\otimes \left\vert
1\right\rangle _{B}$ which has one boson in each of the two modes $A$, $B$.
The particle entanglement measure $E_{P}(\widehat{\rho }\,)$ is zero for
this state (where $\widehat{\rho }\,=(\left\vert 1\right\rangle \left\langle
1\right\vert )_{A}\otimes (\left\vert 1\right\rangle \left\langle
1\right\vert )_{B}$), consistent with it being a separable rather than an
entangled state. This indicates that Wiseman et al \cite{Wiseman03a} do 
\emph{not} consider that entanglement occurs due to symmetrization, as the
first quantization form for the state might indicate. However, finding $%
E_{P}(\widehat{\rho }\,)$ to be zero does not always shows that the state is
separable, as the case of the \emph{relative phase state} (defined in
Appendix \textbf{J }of paper II,\textbf{\ }see also \cite{Dalton12a}) shows.
As is shown there, $E_{P}(\widehat{\rho }\,)=0$ for the relative phase
state, yet the state is clearly an entangled one. Just as some entangled
states have zero spin squeezing, some entangled states may be associated
with a zero particle entanglement measure. Nevertheless a non-zero result
for the particle entanglement measure\textbf{\ }$E_{P}(\widehat{\rho }\,)$ 
\emph{shows}\textbf{\ }that the state must be\textbf{\ }\emph{entangled}%
\textbf{\ - }again we have a\textbf{\ }\emph{sufficiency}\textbf{\ }test.
However, as in the case of other entanglement measures the problem with
using the particle entanglement measure to\ detect entangled states is that
there is no obvious way to measure it experimentally.

\subsubsection{General Form of Non-Entangled States}

To summarise: basically the sub-systems are \emph{single modes} that the
identical bosons can occupy, the super-selection rule for identical bosons,
massive or otherwise, prohibits states which are coherent superpositions of
states with different numbers of bosons, and the only physically allowable $%
\widehat{\rho }_{R}^{A}$, $\widehat{\rho }_{R}^{B}$, ..for the separate mode
sub-systems that are themselves compatiible with the local particle number
SSR are allowed. For single mode sub-systems these can be written as
statistical mixtures of states with definite numbers of bosons in the form 
\begin{equation}
\widehat{\rho }_{R}^{A}=\sum_{n_{A}}P_{n_{A}}^{A}\left\vert
n_{A}\right\rangle \left\langle n_{A}\right\vert \qquad \widehat{\rho }%
_{R}^{B}=\sum_{n_{B}}P_{n_{B}}^{B}\left\vert n_{B}\right\rangle \left\langle
n_{B}\right\vert \qquad ..  \label{Eq.PhysicalStatesSubSys}
\end{equation}

However, in cases where the sub-systems are \emph{pairs of modes} the
density operators $\widehat{\rho }_{R}^{A}$, $\widehat{\rho }_{R}^{B}$,
..for the separate sub-systems are still required to conform to the
symmetrisation principle and the super-selection rule. The forms for $%
\widehat{\rho }_{R}^{A}$, $\widehat{\rho }_{R}^{B}$, .. are now of course
more complex, as entanglement \emph{within} the pairs of modes $A_{1}$, $%
A_{2}$ associated with sub-system $A$, the pairs of modes $B_{1}$, $B_{2}$
associated with sub-system $B$, etc is now possible within the definition
for the general non-entangled state Eq. (\ref{Eq.NonEntangledState}) for
these \emph{pairs} of modes. Within each pair of modes $A_{1}$, $A_{2}$
statistical mixtures of states with differing total numbers $N_{A}$ bosons
in the two modes are possible and the sub-system density operators are based
on states of the form given in Eq. (\ref{Eq.EntangledTwoModePureState}). We
have%
\begin{eqnarray}
\left\vert \Phi _{N_{A}}\right\rangle _{A}
&=&\tsum\limits_{k=0}^{N_{A}}C_{A\Phi }(N_{A},k)\,\left\vert k\right\rangle
_{A_{1}}\otimes \left\vert N_{A}-k\right\rangle _{A_{2}}  \nonumber \\
\widehat{\rho }_{R}^{A} &=&\dsum\limits_{N_{A}=0}^{\infty
}\dsum\limits_{\Phi }P_{\Phi N_{A}}\left\vert \Phi _{N_{A}}\right\rangle
_{A}\left\langle \Phi _{N_{A}}\right\vert _{A}
\label{Eq.PhysStateSubSystPairA}
\end{eqnarray}%
with analogous expressions for the density operators $\widehat{\rho }%
_{R}^{B} $ etc for the other pairs of modes. Note that $\left\vert \Phi
_{N_{A}}\right\rangle _{A}$ only involves quantum superpositions of states
with the same total number of bosons $N_{A\text{. }}$The expression (\textbf{%
208}) in Appendix \textbf{B} of paper II is of this form.

\subsection{Bipartite Systems}

We now consider the bipartite case where there are just two sub-systems
involved. The simplest case is where each sub-system involves only a single
mode, such as for two modes in a double well potential when only a single
hyperfine state is involve. Another important case is where each sub-system
contains two modes, such as in the double well case where modes with two
different hyperfine states are involved.

\subsubsection{Two Single Modes - Coherence Terms}

The general non-entangled state for modes $\widehat{a}$ and $\widehat{b}$ is
given by 
\begin{equation}
\widehat{\rho }=\sum_{R}P_{R}\,\widehat{\rho }_{R}^{A}\otimes \widehat{\rho }%
_{R}^{B}  \label{Eq.SeparableStateAB}
\end{equation}%
and as a consequence of the requirement that $\widehat{\rho }_{R}^{A}$ and $%
\widehat{\rho }_{R}^{B}$ are allowed quantum states for modes $\widehat{a}$
and $\widehat{b}$ satisying the super-selection rule, it follows that 
\begin{eqnarray}
\left\langle (\widehat{a})^{n}\right\rangle _{a} &=&Tr(\widehat{\rho }%
_{R}^{A}(\widehat{a})^{n})=0\qquad \left\langle (\widehat{a}^{\dag
})^{n}\right\rangle _{a}=Tr(\widehat{\rho }_{R}^{A}(\widehat{a}^{\dag
})^{n})=0  \nonumber \\
\left\langle (\widehat{b})^{m}\right\rangle _{b} &=&Tr(\widehat{\rho }%
_{R}^{B}(\widehat{b})^{m})=0\qquad \left\langle (\widehat{b}^{\dag
})^{m}\right\rangle _{b}=Tr(\widehat{\rho }_{R}^{B}(\widehat{b}^{\dag
})^{m})=0  \nonumber \\
&&  \label{Eq.CondNonEntStateAB}
\end{eqnarray}%
Thus coherence terms are zero. As we will see these results will limit spin
squeezing to entangled states of modes $\widehat{a}$ and $\widehat{b}$. Note
that similar results also apply when non-entangled states for the original
modes $\widehat{c}$ and $\widehat{d}$ are considered - $\left\langle (%
\widehat{c})^{n}\right\rangle _{c}=0$, etc..

\subsubsection{Two Pairs of Modes - Coherence Terms}

\label{SubSystem - Two SubSystems of Pairs}

In this case the general non-entangled state where $A$ and $B$ are pairs of
modes - $\widehat{a}_{1}$, $\widehat{a}_{2}$ associated with sub-system $A$,
and modes $\widehat{b}_{1}$, $\widehat{b}_{2}$ associated with sub-system $B$%
, the overall density operator is of the form (\ref{Eq.SeparableStateAB}).
Consistent with the requirement that the sub-system density operators $%
\widehat{\rho }_{R}^{A}$, $\widehat{\rho }_{R}^{B}$ conform to the
symmetrisation principle and the super-selection rule, these density
operators will \emph{not} in general represent separable states for their
single mode sub-systems $\widehat{a}_{1}$, $\widehat{a}_{2}$ or $\widehat{b}%
_{1}$, $\widehat{b}_{2}$ - and may even be entangled states. As a result
when considering \emph{non-entangled} states for the sub-systems $A$ and $B$
we now have 
\begin{eqnarray}
\left\langle (\widehat{a}_{i}^{\dag }\widehat{a}_{j})^{n}\right\rangle _{A}
&=&Tr(\widehat{\rho }_{R}^{A}(\widehat{a}_{i}^{\dag }\widehat{a}%
_{j})^{n})\neq 0\qquad i,j=1,2  \nonumber \\
\left\langle (\widehat{b}_{i}^{\dag }\widehat{b}_{j})^{n}\right\rangle _{A}
&=&Tr(\widehat{\rho }_{R}^{B}(\widehat{b}_{i}^{\dag }\widehat{b}%
_{j})^{n})\neq 0\qquad i,j=1,2  \label{Eq.MeanAnnihCreatOprsModePairs}
\end{eqnarray}%
in general. In this case where the sub-systems are \emph{pairs} of modes the
spin squeezing entanglement tests as in \textbf{Eqs.(50) - (52)} \textbf{in
paper II }for sub-systems consisting of \emph{single} modes cannot be
applied, as we will see. Nevertheless, there are still tests of bipartite
entanglement involving spin operators.\pagebreak

\section{Discussion and Summary of Key Results}

\label{Section - Discussion & Summary of Key Results}

This paper is mainly concerned with two mode entanglement for systems of
identical massive bosons, though multimode entanglement is also considered.
These bosons may be atoms or molecules as in cold quantum gases. In the
present paper we focus on the definition and general features of
entanglement, whilst in the accompanying paper we consider spin squeezing
and other tests for entanglement.

The present paper starts with the \emph{general definition} of \emph{%
entanglement} for a system consisting of several \emph{sub-systems}, and
highlights the distinctive \emph{features} of entangled states in regard to
measurement probabilities for joint measurements on the sub-systems in
contrast to the results for non-entangled or \emph{separable} states. The
relationship between entanglement and \emph{hidden variable theory} is then
explored followed by a discussion of key \emph{paradoxes} such as EPR and
violations of Bell inequalities. The notion of entanglement \emph{measures}
and entanglement \emph{tests} was briefly introduced, the latter being
covered more fully in the accompanying Paper II.

The paper then focuses on entanglement for systems of identical massive
particles in the regime of non-relativistic quantum physics. A careful
analysis is first given regarding the proper definition of a non-entangled
state for systems of identical particles, and hence by implication the
proper definition of an entangled state. Noting that entanglement is
meaningless until the subsystems being entangled are specified, it is
pointed out that whereas it is not possible to distinguish identical
particles and hence the individual particles are not legitimate sub-systems,
the same is not the case for the single particle states or modes, so the 
\emph{modes} are then the the rightful \emph{sub-systems} to be considered
as being entangled or not. In this approach where the sub-systems are modes,
situations where there are differing numbers of identical particles are
treated as different quantum states, not as differing physical systems, and
the \emph{symmetrisation principle} required of quantum states for identical
particle systems will be satisfied by using Fock states to describe the
states.

Furthermore, it is argued that the overall quantum states should conform to
the superselection rule that excludes quantum superposition states of the
form (\ref{Eq.ForbiddenStates}) as allowed quantum states for systems of
identical particles - massive or otherwise. Although the justication of the
SSR in terms of observers and their \emph{reference frames} formulated by
other authors has also been presented for completeness, a number of fairly 
\emph{straightforward reasons} were given for why it is appropriate to apply
this superselection rule for massive bosons, which may be summarised as: 1.
No way is known for creating such states; 2 \ No way is known for measuring
all the properties of such states, even if they existed; and 3. There is no
need to invoke the existence of such states in order to understand coherence
and interference effects. Invoking the existence of states that as far as we
know cannot be made or measured, and for which there are no known physical
effects that require their presence seems a rather unnecessary feature to
add to the non-relativistic quantum physics of many body systems, and
considerations based on the general principle of simplicity (Occam's razor)
would suggest not doing this until a clear physical justification for
including them is found. As two mode fermionic systems are restricted to
states with at most two fermions, the focus of the paper is then on bosonic
systems, where large numbers of bosons can occupy two mode systems.

However, although there is related work involving local particle number
super-selection rules, \emph{this paper differs} from a number of others by 
\emph{extending} the \emph{super-selection rule} to also apply to the
density operators $\widehat{\rho }_{R}^{A}$, $\widehat{\rho }_{R}^{B}$, ...
for the \emph{mode sub-systems} $A$, $B$, ... that occur in the definition (%
\ref{Eq.NonEntangledState}) of a \emph{general non-entangled} state for
systems of identical particles. Hence it follows that the definition of 
\emph{entangled states} will differ in this paper from that which would
apply if density operators $\widehat{\rho }_{R}^{A}$, $\widehat{\rho }%
_{R}^{B}$, ... allowed for coherent superpositions of number states within
each mode. In fact more states are regarded as entangled in terms of the
definition in the present paper. Indeed, if \emph{further restrictions} are
placed on the sub-system density operators - such as requiring them to
specify a fixed number of bosons - the set of entangled states is further
enlarged. The \emph{simple justification} for our viewpoint on applying the 
\emph{local particle number super-selection rule} has three aspects.
Firstly, since experimental arrangements in which only one bosonic mode is
involved can be created, the same reasons (see last paragraph) justify
applying the super-selection rule to this mode system as applied for the
system as a whole. Secondly, measurements can be carried out on the separate
modes, and the joint probability for the outcomes of these measurements
determined. For a non-entangled state the joint probability (\ref%
{Eq.JointProbNonEntState}) for these measurements depends on all the density
operators $\widehat{\rho }_{R}^{A}$, $\widehat{\rho }_{R}^{B}$, ... for the
mode sub-systems as well as the probability $P_{R}$ for the product state $%
\widehat{\rho }_{R}^{A}\otimes \widehat{\rho }_{R}^{B}\otimes \,...$
occuring when the general mixed non-entangled state is prepared, which can
be accomplished by local preparations and classical communication. For the
non-entangled state the form of the joint probability $P_{AB..}(i,j,\,...)$
for measurements on all the sub-systems is given by the products of the
individual sub-system probabilities $P_{A}^{R}(i)=Tr(\widehat{\Pi }_{i}^{A}\,%
\widehat{\rho }_{R}^{A})$, etc that measurements on the sub-systems $%
A,B,\,...\,$yield the outcomes $\lambda _{i}^{A}$ etc when the sub-systems
are in states $\widehat{\rho }_{R}^{A}$, $\widehat{\rho }_{R}^{B}$, ... ,
the overall products being weighted by the probability $P_{R}$ that a
particular product state is prepared. If $\widehat{\rho }_{R}^{A}$, $%
\widehat{\rho }_{R}^{B}$, ... did not represent allowed quantum states then
the interpretation of the joint probability as this statistical average
would be unphysical \ Thirdly, attempts to allow the density operators $%
\widehat{\rho }_{R}^{A}$, $\widehat{\rho }_{R}^{B}$, ... for the mode
sub-systems to violate the super-selection rule provided that the reduced
density operators $\widehat{\rho }_{A}$, $\widehat{\rho }_{B}$ for the
separate modes are consistent with it are shown not to be possible in
general.

As well as the above justifications for applying the super-selection rule to
both the overall multi-mode state for systems of identical particles and the
separate sub-system states in the definition of non-entangled states, a more
sophisticated justification based on considering SSR to be the consequence
of describing the quantum state by an observer (Charlie) whose phase
reference is unknown has also been presented in detail in Appendix \ref%
{Appendix - Reference Frames and SSR} for completeness. For the sub-systems 
\emph{local reference frames} are involved. The SSR is seen as a special
case of a general SSR which forbids quantum states from exhibiting
coherences between states associated with \emph{irreducible representations}
of the transformation group that relates reference frames, and which may be
the \emph{symmetry group} for the system.

In regard to entanglement measures, we discussed the particle entanglement
measure of Wiseman et al \cite{Wiseman03a}, \cite{Dowling06b} and found that
a non-zero result for the particle entanglement measure \emph{shows}\textbf{%
\ }that the state must be\textbf{\ }\emph{entangled.}\textbf{\ }However, as
for other entanglement measures the problem with using the particle
entanglement measure to\ detect entangled states is that there is no obvious
way to measure it experimentally. On the other hand, as will be seen in the
accompanying Paper II, the quantities involved in entanglement tests can be
measured experimentally.

\pagebreak

\bigskip

\section{Acknowledgements}

The authors thank S. M. Barnett, F. Binder, Th. Busch, J. F. Corney, P. D.
Drummond, M. Hall, L. Heaney, J. Jeffers, U. Marzolini,\textbf{\ }K. Molmer,
D. Oi, M. Plenio, K. Rzazewski, T. Rudolph, J. A. Vaccaro, V. Vedral and H.
W. Wiseman for helpful discussions. BJD thanks the Science Foundation of
Ireland for funding this research via an E\ T S Walton Visiting Fellowship
and E Hinds for the hospitality of the Centre for Cold Matter, Imperial
College, London during this work. MDR acknowledges support from the
Australian Research Council via a Discovery Project Grant. \pagebreak

\section{Appendix 1 - Projective Measurements and Conditional Probabilities}

\label{Appendix - Projective Measurements}

\subsubsection{Projective Measurements}

For simplicity, we will only consider \emph{projective} (or von Neumann)
measurements rather than more general measurements involving \emph{positive
operator measurements} (POM). If $\widehat{\Omega }$ is a physical quantity
associated with the system, with eigenvalues $\lambda _{i}$ and with $%
\widehat{\Pi }_{i}$ the projector onto the subspace with eigenvalue $\lambda
_{i}$ then the probability $P(i)$ that measurement of $\widehat{\Omega }$
leads to the value $\lambda _{i}$ is given by \cite{Isham95a} 
\begin{equation}
P(i)=Tr(\widehat{\Pi }_{i}\widehat{\rho })  \label{Eq.ProbMeast}
\end{equation}%
For projective measurements $\widehat{\Pi }_{i}=\widehat{\Pi }_{i}^{2}=%
\widehat{\Pi }_{i}^{\dag }$ and $\dsum\limits_{i}\widehat{\Pi }_{i}=1$,
together with $\widehat{\Omega }\widehat{\Pi }_{i}=\widehat{\Pi }_{i}%
\widehat{\Omega }=\lambda _{i}\widehat{\Pi }_{i}$.

Following the measurement which leads to the value $\lambda _{i}$ the
density operator is different and given by%
\begin{equation}
\widehat{\rho }_{cond}(\widehat{\Omega },i)=(\widehat{\Pi }_{i}\widehat{\rho 
}\widehat{\Pi }_{i})/P(i)  \label{Eq.DensOprAfterMeast}
\end{equation}%
This is known as the \emph{reduction of the wave function}, and can be
viewed in two ways. From an ontological point of view a quantum projective
measurement \emph{changes} the quantum state significantly because the
interaction with the measurement system is not just a small perturbation, as
it can be in classical physics. From the epistomological point of view we
know what value the physical quantity $\widehat{\Omega }$ now has, so if
measurement of $\widehat{\Omega }$ were to be \emph{repeated} immediately it
would be expected -- with a probability of unity - that the value would be $%
\lambda _{i}$. The new density operator $\widehat{\rho }_{cond}(\widehat{%
\Omega },i)$ satisfies this requirement. It also satisfies the standard
requirements of Hermitiancy, unit trace, positivity - as is easily shown.

To show this formally we have for the \emph{mean value} for $\widehat{\Omega 
}$ following the measurement%
\begin{eqnarray}
\left\langle \widehat{\Omega }\right\rangle _{i} &=&Tr(\widehat{\Omega }\,%
\widehat{\rho }_{cond}(\widehat{\Omega },i))  \nonumber \\
&=&Tr(\widehat{\Omega }\,(\widehat{\Pi }_{i}\widehat{\rho }\widehat{\Pi }%
_{i}))/P(i)  \nonumber \\
&=&\lambda _{i}Tr(\widehat{\Pi }_{i}\widehat{\rho })/P(i)  \nonumber \\
&=&\lambda _{i}  \label{Eq.MeanAfterMeast}
\end{eqnarray}%
whilst for the \emph{variance}%
\begin{eqnarray}
\left\langle \Delta \widehat{\Omega }^{2}\right\rangle _{i} &=&Tr((\widehat{%
\Omega }-\left\langle \widehat{\Omega }\right\rangle _{i})^{2}\,\widehat{%
\rho }_{cond}(\widehat{\Omega },i))  \nonumber \\
&=&Tr(\widehat{\Omega }^{2}\,\widehat{\rho }_{red}(i))-\left\langle \widehat{%
\Omega }\right\rangle _{i}^{2}  \nonumber \\
&=&\lambda _{i}^{2}-\lambda _{i}^{2}  \nonumber \\
&=&0  \label{Eq.VarianceAfterMeast}
\end{eqnarray}%
which is zero as expected.

If following the measurement of $\widehat{\Omega }$ the results of the
measurement were discarded then the density operator after the measurement
is 
\begin{equation}
\widehat{\rho }_{cond}(\widehat{\Omega })=\dsum\limits_{i}P(i)\,\widehat{%
\rho }_{cond}(\widehat{\Omega },i)=\dsum\limits_{i}\widehat{\Pi }_{i}%
\widehat{\rho }\widehat{\Pi }_{i}  \label{Eq.DensOprAfterMeastDiscarded}
\end{equation}%
which is the sum of the $\widehat{\rho }_{cond}(\widehat{\Omega },i)$ each
weighted by the probability $P(i)$ of the result $\lambda _{i}$ occuring.
Note that the expression for $\widehat{\rho }_{cond}(\widehat{\Omega })$ is
not the same as the original density operator $\widehat{\rho }$. This is to
be expected from both the epistimological and ontological points of view,
since although we do \emph{not} know \emph{what} value $\lambda _{i}$ has
occurred, it is known that \emph{a} definite value for $\widehat{\Omega }$
has been found, or that measurement process has destroyed any \emph{%
coherences} that previously existed between different eigenstates of $%
\widehat{\Omega }$. We note that $\widehat{\rho }_{cond}(\widehat{\Omega })$
also satisfies the standard requirements of Hermitiancy, unit trace,
positivity - as is easily shown.

\subsubsection{Conditional Probabilities}

Suppose we follow the measurement of $\widehat{\Omega }$ resulting in
eigenvalue $\lambda _{i}$ with a measurement of $\widehat{\Lambda }$
resulting in eigenvalue $\mu _{j}$ where the projector associated with the
latter measurement is $\widehat{\Xi }_{j}$. Then the \emph{conditional
probabiltity} of measuring $\widehat{\Lambda }$ resulting in eigenvalue $\mu
_{j}$ following the measurement of $\widehat{\Omega }$ that resulted in
eigenvalue $\lambda _{i}$ would be 
\begin{eqnarray}
P(j|i) &=&Tr(\widehat{\Xi }_{j}\widehat{\rho }_{cond}(\widehat{\Omega },i)) 
\nonumber \\
&=&Tr(\widehat{\Xi }_{j}(\widehat{\Pi }_{i}\widehat{\rho }\widehat{\Pi }%
_{i}))/P(i)  \nonumber \\
&=&Tr((\widehat{\Xi }_{j}\widehat{\Pi }_{i})\,\widehat{\rho }\,(\widehat{\Pi 
}_{i}\widehat{\Xi }_{j}))/P(i)  \label{Eq.CondProbMeastJAfterI}
\end{eqnarray}%
where the cyclic properties of the trace and the idempotent property of the
projector have been used. If the measurements had taken place in the reverse
order the conditional probabiltity of measuring $\widehat{\Omega }$
resulting in eigenvalue $\lambda _{i}$ following the measurement of $%
\widehat{\Lambda }$ that resulted in eigenvalue $\mu _{j}$ would be%
\begin{equation}
P(i|j)=Tr((\widehat{\Pi }_{i}\widehat{\Xi }_{j})\,\widehat{\rho }\,(\widehat{%
\Xi }_{j}\widehat{\Pi }_{i}))/P(j)  \label{Eq.CondProbMeastIAfterJ}
\end{equation}

We note that the actual probability of measuring $\lambda _{i}$ then $\mu
_{j}$ would be the \emph{joint probability} 
\begin{equation}
P(j\;after\;i)=P(j|i)\,P(i)=Tr((\widehat{\Xi }_{j}\widehat{\Pi }_{i})\,%
\widehat{\rho }\,(\widehat{\Pi }_{i}\widehat{\Xi }_{j}))
\label{Eq.JointProbJAfterI}
\end{equation}%
whilst the actual probability of measuring $\mu _{j}$ then $\lambda _{i}$
would be the joint probability%
\begin{equation}
P(i\;after\;j)=P(i|j)\,P(j)=Tr((\widehat{\Pi }_{i}\widehat{\Xi }_{j})\,%
\widehat{\rho }\,(\widehat{\Xi }_{j}\widehat{\Pi }_{i}))
\label{Eq. JointProbIAfterJ}
\end{equation}%
and we note that in general these two joint probabilities are different.

If however, the two physical quantities \emph{commute}, then there are a
complete set of simultaneous eigenvectors $\left\vert \lambda _{i},\mu
_{j}\right\rangle $ for $\widehat{\Omega }$ and $\widehat{\Lambda }$. It is
then straightforward to show that $\widehat{\Pi }_{i}\widehat{\Xi }_{j}=%
\widehat{\Xi }_{j}\widehat{\Pi }_{i}$, in which case $P(j\;after\;i)=P(i%
\;after\;j)=P(i,j)$, so it does not matter which order the measurements are
carried out. The overall result 
\begin{eqnarray}
P(i,j) &=&P(j|i)\,P(i)=P(i|j)\,P(j)  \nonumber \\
&=&Tr(\widehat{\Pi }_{i}\widehat{\Xi }_{j}\,\widehat{\rho }\,\widehat{\Xi }%
_{j}\widehat{\Pi }_{i})  \nonumber \\
&=&Tr(\widehat{\Pi }_{i}\widehat{\Xi }_{j}\,\widehat{\rho })
\label{Eq.BayesThm}
\end{eqnarray}%
is an expression of \emph{Bayes theorem}.

A case of particular importance where this occurs is in situations involving
two or more distinct sub-systems, in which the operators $\widehat{\Omega }$
and $\widehat{\Lambda }$ are associated with different sub-systems. For two
sub-systems $A$ and $B$ the operators $\widehat{\Omega }$ and $\widehat{%
\Lambda }$ are of the form $\widehat{\Omega }_{A}$ and $\widehat{\Omega }%
_{B} $, or more strictly $\widehat{\Omega }_{A}\otimes \widehat{1}_{B}$ and $%
\widehat{1}_{A}\otimes \widehat{\Omega }_{B}$. It is easy to see that $(%
\widehat{\Omega }_{A}\otimes \widehat{1}_{B})(\widehat{1}_{A}\otimes 
\widehat{\Omega }_{B})=\widehat{\Omega }_{A}\otimes \widehat{\Omega }_{B}=(%
\widehat{1}_{A}\otimes \widehat{\Omega }_{B})(\widehat{\Omega }_{A}\otimes 
\widehat{1}_{B})$, so the operators commute and results such as in Bayes
theorem (\ref{Eq.BayesThm}) apply.

\subsubsection{Conditional Mean and Variance}

To determine the \emph{conditioned mean value} of $\widehat{\Lambda }$ after
measurement of $\widehat{\Omega }$ has led to the eigenvalue $\lambda _{i}$
we use $\widehat{\rho }_{cond}(\widehat{\Omega },i)$ rather than $\,\widehat{%
\rho }$ in the mean formula $\left\langle \widehat{\Lambda }\right\rangle
=Tr(\widehat{\Lambda }\widehat{\rho })$. Hence%
\begin{eqnarray}
\left\langle \widehat{\Lambda }\right\rangle _{i} &=&Tr(\widehat{\Lambda }%
\widehat{\rho }_{cond}(\widehat{\Omega },i))  \nonumber \\
&=&Tr(\widehat{\Lambda }\,(\widehat{\Pi }_{i}\widehat{\rho }\widehat{\Pi }%
_{i}))/P(i)  \label{Eq.CondMean}
\end{eqnarray}%
Now 
\begin{equation}
\widehat{\Lambda }=\dsum\limits_{j}\mu _{j}\widehat{\Xi }_{j}
\label{Eq.OprProjectors}
\end{equation}%
so that 
\begin{eqnarray}
\left\langle \widehat{\Lambda }\right\rangle _{i} &=&\dsum\limits_{j}\mu
_{j}\,Tr(\widehat{\Xi }_{j}\,\widehat{\Pi }_{i}\widehat{\rho }\widehat{\Pi }%
_{i})/P(i)  \nonumber \\
&=&\dsum\limits_{j}\mu _{j}\,Tr(\widehat{\Xi }_{j}\widehat{\Pi }_{i}\widehat{%
\rho }\widehat{\Pi }_{i}\widehat{\Xi }_{j})/P(i)  \nonumber \\
&=&\dsum\limits_{j}\mu _{j}\,P(j|i)  \label{Eq.CondMeanResult}
\end{eqnarray}%
using $\widehat{\Xi }_{j}=\widehat{\Xi }_{j}^{2}$, the cyclic trace
properties and Eq.(\ref{Eq.CondProbMeastJAfterI}). Hence the conditional
mean value is as expected, with the conditional probability $P(j|i)$
replacing $P(j)$ in the averaging process.

For the \emph{conditioned variance} of $\widehat{\Lambda }$ after
measurement of $\widehat{\Omega }$ has led to the eigenvalue $\lambda _{i}$
we use $\widehat{\rho }_{cond}(\widehat{\Omega },i)$ rather than $\,\widehat{%
\rho }$ and the conditioned mean $\left\langle \widehat{\Lambda }%
\right\rangle _{i}$ rather than $\left\langle \widehat{\Lambda }%
\right\rangle $ in the variance formula $\left\langle \Delta \widehat{%
\Lambda }^{2}\right\rangle =Tr((\widehat{\Lambda }-\left\langle \widehat{%
\Lambda }\right\rangle )^{2}\widehat{\rho })$. Hence%
\begin{eqnarray}
\left\langle \Delta \widehat{\Lambda }^{2}\right\rangle _{i} &=&Tr((\widehat{%
\Lambda }-\left\langle \widehat{\Lambda }\right\rangle _{i})^{2}\widehat{%
\rho }_{cond}(\widehat{\Omega },i))  \nonumber \\
&=&Tr((\widehat{\Lambda }-\left\langle \widehat{\Lambda }\right\rangle
_{i})^{2}(\widehat{\Pi }_{i}\widehat{\rho }\widehat{\Pi }_{i}))/P(i)
\label{Eq.CondVariance}
\end{eqnarray}%
Now 
\begin{equation}
(\widehat{\Lambda }-\left\langle \widehat{\Lambda }\right\rangle
_{i})^{2}=\dsum\limits_{j}(\mu _{j}-\left\langle \widehat{\Lambda }%
\right\rangle _{i})^{2}\widehat{\Xi }_{j}  \label{Eq.FluctnProjectors}
\end{equation}%
so that 
\begin{eqnarray}
\left\langle \Delta \widehat{\Lambda }^{2}\right\rangle _{i}
&=&\dsum\limits_{j}(\mu _{j}-\left\langle \widehat{\Lambda }\right\rangle
_{i})^{2}\,Tr(\widehat{\Xi }_{j}\,\widehat{\Pi }_{i}\widehat{\rho }\widehat{%
\Pi }_{i})/P(i)  \nonumber \\
&=&\dsum\limits_{j}(\mu _{j}-\left\langle \widehat{\Lambda }\right\rangle
_{i})^{2}\,P(j|i)  \label{Eq.CondVarianceResult}
\end{eqnarray}%
using the same steps as for the conditioned mean. Hence the conditional
variance is as expected, with the conditional probability $P(j|i)$ replacing 
$P(j)$ in the averaging process.

\subsection{Detailed Inequalities for EPR Situation}

For\emph{\ separable states} the \emph{conditional probability} that
measurement of $\widehat{p}_{A}$ on sub-system $A$ leads to eigenvalue $%
p_{A} $ given that measurement of $\widehat{p}_{B}$ on sub-system $B$ leads
to eigenvalue $p_{B}$ is obtained from Eq.(\ref{Eq.CondProbNonEntangledState}%
) as 
\begin{equation}
P(\widehat{p}_{A},p_{A}|\widehat{p}_{B},p_{B})=\sum_{R}P_{R}\,P_{A}^{R}(%
\widehat{p}_{A},p_{A})P_{B}^{R}(\widehat{p}_{B},p_{B})/\sum_{R}P_{R}%
\,P_{B}^{R}(\widehat{p}_{B},p_{B})
\end{equation}%
where 
\begin{equation}
P_{A}^{R}(\widehat{p}_{A},p_{A})=Tr_{A}(\widehat{\Pi }_{p_{A}}^{A}\widehat{%
\rho }_{R}^{A})\qquad P_{B}^{R}(\widehat{p}_{B},p_{B})=Tr_{B}(\widehat{\Pi }%
_{p_{B}}^{B}\widehat{\rho }_{R}^{B})
\end{equation}%
are the probabilities for position measurements in the separate sub-systems.
The probability that measurement of $\widehat{p}_{B}$ on sub-system $B$
leads to eigenvalue $p_{B}$ is 
\begin{equation}
P(\widehat{p}_{B},p_{B})=\sum_{R}P_{R}\,P_{B}^{R}(\widehat{p}_{B},p_{B})
\end{equation}

The \emph{mean} result for measurement of $\widehat{p}_{A}$ for this \emph{%
conditional} measurement is from Eq.(\ref{Eq.CondMean0}) 
\begin{eqnarray}
\left\langle \widehat{p}_{A}\right\rangle _{\widehat{p}_{B},p_{B}}
&=&\dsum\limits_{p_{A}}p_{A}\,P(\widehat{p}_{A},p_{A}|\widehat{p}_{B},p_{B})
\nonumber \\
&=&\sum_{R}P_{R}\,\left\langle \widehat{p}_{A}\right\rangle _{R}P_{B}^{R}(%
\widehat{p}_{B},p_{B})/P(\widehat{p}_{B},p_{B})
\end{eqnarray}%
where 
\begin{equation}
\left\langle \widehat{p}_{A}\right\rangle
_{R}=\dsum\limits_{p_{A}}p_{A}P_{A}^{R}(\widehat{p}_{A},p_{A})
\end{equation}%
is the \emph{mean} result for measurement of $\widehat{p}_{A}$ when the
sub-system is in state $\widehat{\rho }_{R}^{A}$.

The \emph{conditional variance} for measurement of $\widehat{p}_{A}$ for the
conditional measurement of $\widehat{p}_{B}$ on sub-system $B$ which led to
eigenvalue $p_{B}$ is from Eq.(\ref{Eq.CondVariance0}) 
\begin{eqnarray}
\left\langle \Delta \widehat{p}_{A}^{2}\right\rangle _{\widehat{p}%
_{B},p_{B}} &=&\dsum\limits_{p_{A}}(p_{A}-\left\langle \widehat{p}%
_{A}\right\rangle _{\widehat{p}_{B},p_{B}})^{2}\,P(\widehat{p}_{A},p_{A}|%
\widehat{p}_{B},p_{B})  \nonumber \\
&=&\sum_{R}P_{R}\,\left\langle \Delta \widehat{p}_{A}^{2}\right\rangle _{%
\widehat{p}_{B},p_{B}}^{R}P_{B}^{R}(\widehat{p}_{B},p_{B})/P(\widehat{p}%
_{B},p_{B})
\end{eqnarray}%
where 
\[
\left\langle \Delta \widehat{p}_{A}^{2}\right\rangle _{\widehat{p}%
_{B},p_{B}}^{R}=\dsum\limits_{p_{A}}(p_{A}-\left\langle \widehat{p}%
_{A}\right\rangle _{\widehat{p}_{B},p_{B}})^{2}\,P_{A}^{R}(\widehat{p}%
_{A},p_{A}) 
\]%
is a variance for measurement of $\widehat{p}_{A}$ for when the sub-system
is in state $\widehat{\rho }_{R}^{A}$ but now with the fluctuation about the
mean $\left\langle \widehat{p}_{A}\right\rangle _{\widehat{p}_{B},p_{B}}$
for measurements conditional on measuring $\widehat{p}_{B}$.

However, for each sub-system state $R$ the quantity $\left\langle \Delta 
\widehat{p}_{A}^{2}\right\rangle _{\widehat{p}_{B},p_{B}}^{R}$ is \emph{%
minimised} if $\left\langle \widehat{p}_{A}\right\rangle _{\widehat{p}%
_{B},p_{B}}$ is replaced by the unconditioned mean $\left\langle \widehat{p}%
_{A}\right\rangle _{R}$ just determined from $\widehat{\rho }_{R}^{A}$. Thus
we have an inequality%
\begin{equation}
\left\langle \Delta \widehat{p}_{A}^{2}\right\rangle _{\widehat{p}%
_{B},p_{B}}^{R}\geq \left\langle \Delta \widehat{p}_{A}^{2}\right\rangle ^{R}
\end{equation}%
where 
\begin{equation}
\left\langle \Delta \widehat{p}_{A}^{2}\right\rangle
^{R}=\dsum\limits_{p_{A}}(p_{A}-\left\langle \widehat{p}_{A}\right\rangle
)^{2}\,P_{A}^{R}(\widehat{p}_{A},p_{A})
\end{equation}%
is the \emph{normal variance} for measurement of $\widehat{p}_{A}$ for when
the sub-system is in state $\widehat{\rho }_{R}^{A}$.

Now if the measurements of $\widehat{p}_{B}$ are \emph{unrecorded} then the 
\emph{conditioned variance} is 
\begin{eqnarray}
\left\langle \Delta \widehat{p}_{A}^{2}\right\rangle _{\widehat{p}_{B}}
&=&\tsum\limits_{x_{B}}\left\langle \Delta \widehat{p}_{A}^{2}\right\rangle
_{\widehat{x}_{B},x_{B}}P(\widehat{x}_{B},x_{B})  \nonumber \\
&=&\tsum\limits_{x_{B}}\sum_{R}P_{R}\,\left\langle \Delta \widehat{p}%
_{A}^{2}\right\rangle _{\widehat{x}_{B},x_{B}}^{R}P_{B}^{R}(\widehat{x}%
_{B},x_{B})
\end{eqnarray}%
which in view of inequality (\ref{Eq.VarianceIneqEPR}) satisfies 
\begin{eqnarray}
\left\langle \Delta \widehat{p}_{A}^{2}\right\rangle _{\widehat{p}_{B}}
&\geq &\tsum\limits_{p_{B}}\sum_{R}P_{R}\,\left\langle \Delta \widehat{p}%
_{A}^{2}\right\rangle ^{R}\,P_{B}^{R}(\widehat{p}_{B},p_{B})  \nonumber \\
&=&\sum_{R}P_{R}\,\left\langle \Delta \widehat{p}_{A}^{2}\right\rangle ^{R}
\end{eqnarray}%
using $\tsum\limits_{p_{B}}P_{B}^{R}(\widehat{p}_{B},p_{B})=1$. Thus the
variance for measurement of momentum $\widehat{p}_{A}$ conditioned on
unrecorded measurements for momentum $\widehat{p}_{B}$ satisfies an
inequality that only depends on the variances for measurements of $\widehat{p%
}_{A}$ in the possible sub-system $A$ states $\widehat{\rho }_{R}^{A}$.

\pagebreak

\section{Appendix 2 - Inequalities}

\label{Appendix - Inequalities}

These inequalities are examples of Schwarz inequalities.

\subsection{Integral Inequality}

If $C(\lambda ),D(\lambda )$ are real, positive functions of $\lambda $ and $%
P(\lambda )$ is another real, positive function then we can show that%
\begin{equation}
\tint d\lambda \,P(\lambda )C(\lambda ).\tint d\lambda \,P(\lambda
)D(\lambda )\geq \left( \tint d\lambda \,P(\lambda )\sqrt{C(\lambda
)D(\lambda )}\right) ^{2}  \label{Eq.IntegralInequality0}
\end{equation}

To show this write $x=\tint d\lambda \,P(\lambda )C(\lambda )$ and $y=\tint
d\lambda \,P(\lambda )D(\lambda )$. Then 
\begin{eqnarray}
xy &=&\tint d\lambda \,P(\lambda )C(\lambda )\tint d\mu \,P(\mu )D(\mu ) 
\nonumber \\
&=&\tint \tint d\lambda \,d\mu \,P(\lambda )P(\mu )C(\lambda )D(\mu ) 
\nonumber \\
&=&\tint d\lambda \,P(\lambda )^{2}C(\lambda )D(\lambda )+\tint \tint
d\lambda \,d\mu \,(1-\delta (\lambda -\mu ))\,P(\lambda )P(\mu )C(\lambda
)D(\mu )  \nonumber \\
&&  \label{Eq.R1}
\end{eqnarray}%
Also, write $z=\left( \tint d\lambda \,P(\lambda )\sqrt{C(\lambda )D(\lambda
)}\right) ^{2}$. Then%
\begin{eqnarray}
z &=&\tint d\lambda \,P(\lambda )\sqrt{C(\lambda )D(\lambda )}\,\tint d\mu
\,P(\mu )\sqrt{C(\mu )D(\mu )}\,  \nonumber \\
&=&\tint \tint d\lambda \,d\mu \,P(\lambda )P(\mu )\,\sqrt{C(\lambda
)D(\lambda )}\sqrt{C(\mu )D(\mu )}  \nonumber \\
&=&\tint d\lambda \,P(\lambda )^{2}C(\lambda )D(\lambda )+\tint \tint
d\lambda \,d\mu \,(1-\delta (\lambda -\mu ))\,P(\lambda )P(\mu )\,\sqrt{%
C(\lambda )D(\lambda )}\sqrt{C(\mu )D(\mu )}  \nonumber \\
&&  \label{Eq.R2}
\end{eqnarray}%
so that 
\begin{eqnarray}
xy-z &=&\tint \tint d\lambda \,d\mu \,(1-\delta (\lambda -\mu ))\,P(\lambda
)P(\mu )\,\left( C(\lambda )D(\mu )-\sqrt{C(\lambda )D(\lambda )}\sqrt{C(\mu
)D(\mu )}\right)  \nonumber \\
&=&\frac{1}{2}\tint \tint d\lambda \,d\mu \,(1-\delta (\lambda -\mu
))\,P(\lambda )P(\mu )\,\left( C(\lambda )D(\mu )+C(\mu )D(\lambda )-2\sqrt{%
C(\lambda )D(\mu )}\sqrt{C(\mu )D(\lambda )}\right)  \nonumber \\
&=&\frac{1}{2}\tint \tint d\lambda \,d\mu \,(1-\delta (\lambda -\mu
))\,P(\lambda )P(\mu )\,\left( \sqrt{C(\lambda )D(\mu )}-\sqrt{C(\mu
)D(\lambda )}\right) ^{2}  \nonumber \\
&\geq &0  \label{Eq.R3}
\end{eqnarray}%
which proves the result.

For the special case where $D(\lambda )=1$ and where $\tint d\lambda
\,P(\lambda )=1$ we get the simpler result%
\begin{equation}
\tint d\lambda \,P(\lambda )C(\lambda )\geq \left( \tint d\lambda
\,P(\lambda )\sqrt{C(\lambda )}\right) ^{2}  \label{Eq.IntegralInequality}
\end{equation}

\subsection{Sum Inequality}

If $C_{R}$ and $D_{R}$ are real, positive quantities for various $R$ and $%
P_{R}$ is another real, positive quantity then we can show that%
\begin{equation}
\tsum\limits_{R}\,P_{R}\,C_{R}\,\tsum\limits_{R}\,P_{R}\,D_{R}\geq \left(
\tsum\limits_{R}\,P_{R}\,\sqrt{C_{R}D_{R}}\right) ^{2}
\label{Eq.SumInequality0}
\end{equation}%
To prove this write $x=\tsum\limits_{R}\,P_{R}\,C_{R}$ and $%
y=\tsum\limits_{R}\,P_{R}\,D_{R}$ Then 
\begin{eqnarray}
xy &=&\tsum\limits_{R}\,P_{R}\,C_{R}\,\tsum\limits_{S}\,P_{S}\,D_{S} 
\nonumber \\
&=&\tsum\limits_{R}\tsum\limits_{S}\,P_{R}\,P_{S}\,C_{R}D_{S}  \nonumber \\
&=&\tsum\limits_{R}\,P_{R}^{2}\,C_{R}D_{R}+\tsum\limits_{R}\tsum\limits_{S}%
\,(1-\delta _{RS})\,P_{R}\,P_{S}\,C_{R}D_{S}  \label{Eq.R4}
\end{eqnarray}%
Also, write $z=\left( \tsum\limits_{R}\,P_{R}\,\sqrt{C_{R}D_{R}}\right) ^{2}$%
. Then%
\begin{eqnarray}
z &=&\left( \tsum\limits_{R}\,P_{R}\,\sqrt{C_{R}D_{R}}\right) \left(
\tsum\limits_{S}\,P_{S}\,\sqrt{C_{S}D_{S}}\right)  \nonumber \\
&=&\tsum\limits_{R}\tsum\limits_{S}\,P_{R}\,P_{S}\,\sqrt{C_{R}D_{R}}\sqrt{%
C_{S}D_{S}}  \nonumber \\
&=&\tsum\limits_{R}\,P_{R}^{2}\,C_{R}D_{R}+\tsum\limits_{R}\tsum\limits_{S}%
\,(1-\delta _{RS})\,P_{R}\,P_{S}\,\sqrt{C_{R}D_{R}}\sqrt{C_{S}D_{S}}
\label{Eq.R5}
\end{eqnarray}%
so that 
\begin{eqnarray}
xy-z &=&\tsum\limits_{R}\tsum\limits_{S}\,\,P_{R}\,P_{S}\,(1-\delta
_{RS})\,\left( C_{R}D_{S}-\sqrt{C_{R}D_{R}}\sqrt{C_{S}D_{S}}\right) 
\nonumber \\
&=&\frac{1}{2}\tsum\limits_{R}\tsum\limits_{S}\,\,P_{R}\,P_{S}\,(1-\delta
_{RS})\,\left( C_{R}D_{S}+C_{S}D_{R}-2\sqrt{C_{R}D_{S}}\sqrt{C_{S}D_{R}}%
\right)  \nonumber \\
&=&\frac{1}{2}\tsum\limits_{R}\tsum\limits_{S}\,\,P_{S}\,P_{R}\,(1-\delta
_{RS})\,\,\left( \sqrt{C_{R}D_{S}}-\sqrt{C_{S}D_{R}}\right) ^{2}  \nonumber
\\
&\geq &0  \label{Eq.R6}
\end{eqnarray}%
which proves the result.

For the special case where $D_{R}=1$ and where $\tsum\limits_{R}\,P_{R}=1$
we get the simpler result%
\begin{equation}
\tsum\limits_{R}\,P_{R}\,C_{R}\,\geq \left( \tsum\limits_{R}\,P_{R}\,\sqrt{%
C_{R}}\right) ^{2}  \label{Eq.SumInequality}
\end{equation}%
This inequality is used in \cite{Hillery06a}.

\pagebreak

\section{Appendix 3 - Spin EPR Paradox}

\label{Appendix - Spin EPR Paradox}

\subsection{Local Spin Operators}

For two sub-systems $1$ and $2$ there are numerous possibilities for
defining separate commuting \emph{spin operators }for the two systems. One
situation of interest is where each sub-system is associated with two \emph{%
bosonic modes}, the standard annihilation operators being $\widehat{a}_{1}$
and $\widehat{b}_{1}$ for system $1$ and $\widehat{a}_{2}$ and $\widehat{b}%
_{2}$ for system $2$. The \emph{local spin operators} for each sub-system
can be defined as 
\begin{eqnarray}
\widehat{S}_{x}^{1} &=&(\widehat{b}_{1}^{\dag }\widehat{a}_{1}+\widehat{a}%
_{1}^{\dag }\widehat{b}_{1})/2\qquad \widehat{S}_{y}^{1}=(\widehat{b}%
_{1}^{\dag }\widehat{a}_{1}-\widehat{a}_{1}^{\dag }\widehat{b}_{1})/2i\qquad 
\widehat{S}_{z}^{1}=(\widehat{b}_{1}^{\dag }\widehat{b}_{1}-\widehat{a}%
_{1}^{\dag }\widehat{a}_{1})/2  \nonumber \\
\widehat{S}_{x}^{2} &=&(\widehat{b}_{2}^{\dag }\widehat{a}_{2}+\widehat{a}%
_{2}^{\dag }\widehat{b}_{2})/2\qquad \widehat{S}_{y}^{2}=(\widehat{b}%
_{2}^{\dag }\widehat{a}_{2}-\widehat{a}_{2}^{\dag }\widehat{b}_{2})/2i\qquad 
\widehat{S}_{z}^{2}=(\widehat{b}_{2}^{\dag }\widehat{b}_{2}-\widehat{a}%
_{2}^{\dag }\widehat{a}_{2})/2  \nonumber \\
&&  \label{Eq.LocalSpinOprs2}
\end{eqnarray}%
These satisfy the usual angular momentum commutation rules and those or the
different sub-systems commute. The squares of the local vector spin
operators are related to the total number operators $\widehat{N}_{1}=%
\widehat{b}_{1}^{\dag }\widehat{b}_{1}+\widehat{a}_{1}^{\dag }\widehat{a}%
_{1} $ and $\widehat{N}_{2}=\widehat{b}_{2}^{\dag }\widehat{b}_{2}+\widehat{a%
}_{2}^{\dag }\widehat{a}_{2}$ as $\tsum\limits_{\alpha }(\widehat{S}_{\alpha
}^{1})^{2}=($ $\widehat{N}_{1}/2)(\widehat{N}_{1}/2+1)$ and $%
\tsum\limits_{\alpha }(\widehat{S}_{\alpha }^{2})^{2}=($ $\widehat{N}_{2}/2)(%
\widehat{N}_{2}/2+1)$. The\emph{\ total spin operators} are 
\begin{equation}
\widehat{S}_{\alpha }=\widehat{S}_{\alpha }^{1}+\widehat{S}_{\alpha
}^{2}\qquad \alpha =x,y,z  \label{Eq.TotalSpinOprs2}
\end{equation}%
and these satisfy the usual angular momentum commutation rules.

\subsection{Conditional Variances}

The question is whether the conditional variances $\left\langle \Delta 
\widehat{S}_{x1}^{2}\right\rangle _{\widehat{S}_{x2}}$\ for measuring $%
\widehat{S}_{x1}$\ for sub-system $1$\ having measured $\widehat{S}_{x2}$\
for sub-system $2$, and $\left\langle \Delta \widehat{S}_{y1}^{2}\right%
\rangle _{\widehat{p}_{B}}$\ for measuring $\widehat{S}_{y1}$\ for
sub-system $1$\ having measured $\widehat{S}_{y2}$\ for sub-system $2$\
violate the Heisenberg Uncertainty Principle. 
\begin{equation}
\left\langle \Delta \widehat{S}_{x1}^{2}\right\rangle _{\widehat{S}%
_{x2}}\left\langle \Delta \widehat{S}_{y1}^{2}\right\rangle _{\widehat{S}%
_{y2}}<\frac{1}{4}|\left\langle \widehat{S}_{z1}\right\rangle |^{2}
\label{Eq.EPRViolationSpin}
\end{equation}%
where the measurements on sub-system $2$\ are left unrecorded. If this
inequality holds we have an\emph{\ EPR\ violation}.

For\emph{\ separable states} the \emph{conditional probability} that
measurement of $\widehat{S}_{x1}$ on sub-system $1$ leads to eigenvalue $%
s_{x1}$ given that measurement of $\widehat{S}_{x2}$ on sub-system $2$ leads
to eigenvalue $s_{x2}$ is obtained from Eq.(\ref%
{Eq.CondProbNonEntangledState}) as 
\begin{equation}
P(\widehat{S}_{x1},s_{x1}|\widehat{S}_{x2},s_{x2})=\sum_{R}P_{R}\,P_{1}^{R}(%
\widehat{S}_{x1},s_{x1})P_{2}^{R}(\widehat{S}_{x2},s_{x2})/\sum_{R}P_{R}%
\,P_{2}^{R}(\widehat{S}_{x2},s_{x2})
\end{equation}%
where 
\begin{equation}
P_{1}^{R}(\widehat{S}_{x1},s_{x1})=Tr_{1}(\widehat{\Pi }_{s_{x1}}^{1}%
\widehat{\rho }_{R}^{1})\qquad P_{2}^{R}(\widehat{S}_{x2},s_{x2})=Tr_{2}(%
\widehat{\Pi }_{s_{x2}}^{2}\widehat{\rho }_{R}^{2})
\end{equation}%
are the probabilities for position measurements in the separate sub-systems.
The probability that measurement of $\widehat{S}_{x2}$ on sub-system $2$
leads to eigenvalue $s_{x2}$ is 
\begin{equation}
P(\widehat{S}_{x2},s_{x2})=\sum_{R}P_{R}\,P_{2}^{R}(\widehat{S}_{x2},s_{x2})
\end{equation}

The \emph{mean} result for measurement of $\widehat{S}_{x1}$ for this \emph{%
conditional} measurement is from Eq.(\ref{Eq.CondMean0}) 
\begin{eqnarray}
\left\langle \widehat{S}_{x1}\right\rangle _{\widehat{S}_{x2},s_{x2}}
&=&\dsum\limits_{s_{x1}}s_{x1}\,P(\widehat{S}_{x1},s_{x1}|\widehat{S}%
_{x2},s_{x2})  \nonumber \\
&=&\sum_{R}P_{R}\,\left\langle \widehat{S}_{x1}\right\rangle _{R}P_{2}^{R}(%
\widehat{S}_{x2},s_{x2})/P(\widehat{S}_{x2},s_{x2})
\end{eqnarray}%
where 
\begin{equation}
\left\langle \widehat{S}_{x1}\right\rangle
_{R}=\dsum\limits_{s_{x1}}s_{x1}P_{1}^{R}(\widehat{S}_{x1},s_{x1})
\end{equation}%
is the \emph{mean} result for measurement of $\widehat{S}_{x1}$ when the
sub-system is in state $\widehat{\rho }_{R}^{1}$.

The \emph{conditional variance} for measurement of $\widehat{S}_{x1}$ for
the conditional measurement of $\widehat{S}_{x2}$ on sub-system $2$ which
led to eigenvalue $s_{x2}$ is from Eq.(\ref{Eq.CondVariance0}) 
\begin{eqnarray}
\left\langle \Delta \widehat{S}_{x1}^{2}\right\rangle _{\widehat{S}%
_{x2},s_{x2}} &=&\dsum\limits_{s_{x1}}(s_{x1}-\left\langle \widehat{S}%
_{x1}\right\rangle _{\widehat{S}_{x2},s_{x2}})^{2}\,P(\widehat{S}%
_{x1},s_{x1}|\widehat{S}_{x2},s_{x2})  \nonumber \\
&=&\sum_{R}P_{R}\,\left\langle \Delta \widehat{S}_{x1}^{2}\right\rangle _{%
\widehat{S}_{x2},s_{x2}}^{R}P_{2}^{R}(\widehat{S}_{x2},s_{x2})/P(\widehat{S}%
_{x2},s_{x2})
\end{eqnarray}%
where 
\[
\left\langle \Delta \widehat{S}_{x1}^{2}\right\rangle _{\widehat{S}%
_{x2},s_{x2}}^{R}=\dsum\limits_{s_{x1}}(s_{x1}-\left\langle \widehat{S}%
_{x1}\right\rangle _{\widehat{S}_{x2},s_{x2}})^{2}\,P_{1}^{R}(\widehat{S}%
_{x1},s_{x1}) 
\]%
is a variance for measurement of $\widehat{S}_{x1}$ for when the sub-system
is in state $\widehat{\rho }_{R}^{1}$ but now with the fluctuation about the
mean $\left\langle \widehat{S}_{x1}\right\rangle _{\widehat{S}_{x2},s_{x2}}$
for measurements conditional on measuring $\widehat{S}_{x2}$.

However, for each sub-system state $R$ the quantity $\left\langle \Delta 
\widehat{S}_{x1}^{2}\right\rangle _{\widehat{S}_{x2},s_{x2}}^{R}$ is \emph{%
minimised} if $\left\langle \widehat{S}_{x1}\right\rangle _{\widehat{S}%
_{x2},s_{x2}}$ is replaced by the unconditioned mean $\left\langle \widehat{S%
}_{x1}\right\rangle _{R}$ just determined from $\widehat{\rho }_{R}^{1}$.
Thus we have an inequality%
\begin{equation}
\left\langle \Delta \widehat{S}_{x1}^{2}\right\rangle _{\widehat{S}%
_{x2},s_{x2}}^{R}\geq \left\langle \Delta \widehat{S}_{x1}^{2}\right\rangle
^{R}  \label{Eq.VarianceIneqEPRSpin}
\end{equation}%
where 
\begin{equation}
\left\langle \Delta \widehat{S}_{x1}^{2}\right\rangle
^{R}=\dsum\limits_{s_{x1}}(s_{x1}-\left\langle \widehat{S}_{x1}\right\rangle
)^{2}\,P_{1}^{R}(\widehat{S}_{x1},s_{x1})
\end{equation}%
is the \emph{normal variance} for measurement of $\widehat{S}_{x1}$ for when
the sub-system is in state $\widehat{\rho }_{R}^{1}$.

Now if the measurements of $\widehat{S}_{x2}$ are \emph{unrecorded} \textbf{-%
} as would be the case from the point of view of the experimenter on
spatially well-separated sub-system $1$ when measurements on this sub-system
take place at the same time - then the \emph{conditioned variance} is 
\begin{eqnarray}
\left\langle \Delta \widehat{S}_{x1}^{2}\right\rangle _{\widehat{S}_{x2}}
&=&\tsum\limits_{s_{x2}}\left\langle \Delta \widehat{S}_{x1}^{2}\right%
\rangle _{\widehat{S}_{x2},s_{x2}}P(\widehat{S}_{x2},s_{x2})  \nonumber \\
&=&\tsum\limits_{s_{x2}}\sum_{R}P_{R}\,\left\langle \Delta \widehat{S}%
_{x1}^{2}\right\rangle _{\widehat{S}_{x2},s_{x2}}^{R}P_{2}^{R}(\widehat{S}%
_{x2},s_{x2})
\end{eqnarray}%
which in view of inequality (\ref{Eq.VarianceIneqEPR}) satisfies 
\begin{eqnarray}
\left\langle \Delta \widehat{S}_{x1}^{2}\right\rangle _{\widehat{S}_{x2}}
&\geq &\tsum\limits_{s_{x2}}\sum_{R}P_{R}\,\left\langle \Delta \widehat{S}%
_{x1}^{2}\right\rangle ^{R}\,P_{2}^{R}(\widehat{S}_{x2},s_{x2})  \nonumber \\
&=&\sum_{R}P_{R}\,\left\langle \Delta \widehat{S}_{x1}^{2}\right\rangle ^{R}
\end{eqnarray}%
using $\tsum\limits_{s_{x2}}P_{2}^{R}(\widehat{S}_{x2},s_{x2})=1$. Thus the
variance for measurement of spin $\widehat{S}_{x1}$ conditioned on
unrecorded measurements for spin $\widehat{S}_{x2}$ satisfies an inequality
that only depends on the variances for measurements of $\widehat{S}_{x1}$ in
the possible sub-system $1$ states $\widehat{\rho }_{R}^{1}$.

Now exactly the same treatment can be carried out for the variance of \emph{%
spin} $\widehat{S}_{y1}$ \textbf{also }conditioned on unrecorded
measurements of measurements for momentum $\widehat{S}_{y2}$. Details are
given in Appendix \ref{Appendix - Projective Measurements}. We have with%
\begin{eqnarray*}
\left\langle \Delta \widehat{S}_{y1}^{2}\right\rangle _{\widehat{S}_{y2}}
&=&\tsum\limits_{s_{y2}}\left\langle \Delta \widehat{S}_{y1}^{2}\right%
\rangle _{\widehat{S}_{y2},p_{sy2}}P(\widehat{S}_{y2},s_{y2}) \\
\left\langle \Delta \widehat{S}_{y1}^{2}\right\rangle _{\widehat{S}%
_{y2},s_{y2}} &=&\dsum\limits_{s_{y1}}(s_{y1}-\left\langle \widehat{S}%
_{y1}\right\rangle _{\widehat{S}_{y2},s_{y2}})^{2}\,P(\widehat{S}%
_{y2},s_{y2}|\widehat{S}_{y2},s_{y2}) \\
\left\langle \widehat{S}_{y1}\right\rangle _{\widehat{S}_{y2},s_{y2}}
&=&\dsum\limits_{s_{y1}}s_{y1}\,P(\widehat{S}_{y1},s_{y1}|\widehat{S}%
_{y2},s_{y2})
\end{eqnarray*}%
the inequality 
\begin{equation}
\left\langle \Delta \widehat{S}_{y1}^{2}\right\rangle _{\widehat{S}%
_{y2}}\geq \sum_{R}P_{R}\,\left\langle \Delta \widehat{S}_{y1}^{2}\right%
\rangle ^{R}  \label{Eq.InequaltiyEPR2Spin}
\end{equation}%
with 
\begin{equation}
\left\langle \Delta \widehat{S}_{y1}^{2}\right\rangle
^{R}=\dsum\limits_{s_{y1}}(s_{y1}-\left\langle \widehat{S}_{y1}\right\rangle
)^{2}\,P_{1}^{R}(\widehat{S}_{y1},s_{y1})
\end{equation}%
is the normal variance for measurement of $\widehat{S}_{y1}$ for when the
sub-system is in state $\widehat{\rho }_{R}^{1}$.

We now multiply the two conditional variances, which it is important to note
were associated with two \emph{different conditioned states} based on two 
\emph{different} measurements - spin $\widehat{S}_{x2}$ and spin $\widehat{S}%
_{y2}$ - carried out on sub-system $2$. 
\begin{equation}
\left\langle \Delta \widehat{S}_{x1}^{2}\right\rangle _{\widehat{S}%
_{x2}}\left\langle \Delta \widehat{S}_{y1}^{2}\right\rangle _{\widehat{S}%
_{y2}}\geq \sum_{R}P_{R}\,\left\langle \Delta \widehat{S}_{x1}^{2}\right%
\rangle ^{R}\sum_{S}P_{S}\,\left\langle \Delta \widehat{S}%
_{y1}^{2}\right\rangle ^{S}
\end{equation}%
However, from the general inequality in Eq.(\ref{Eq.SumInequality0})%
\begin{equation}
\tsum\limits_{R}\,P_{R}\,C_{R}\,\tsum\limits_{R}\,P_{R}\,D_{R}\geq \left(
\tsum\limits_{R}\,P_{R}\,\sqrt{C_{R}D_{R}}\right) ^{2}
\end{equation}%
we then have%
\begin{equation}
\left\langle \Delta \widehat{S}_{x1}^{2}\right\rangle _{\widehat{S}%
_{x2}}\left\langle \Delta \widehat{S}_{y1}^{2}\right\rangle _{\widehat{S}%
_{y2}}\geq \left( \tsum\limits_{R}\,P_{R}\,\sqrt{\left\langle \Delta 
\widehat{S}_{x1}^{2}\right\rangle ^{R}\left\langle \Delta \widehat{S}%
_{y1}^{2}\right\rangle ^{R}}\right) ^{2}
\end{equation}%
But we know from the HUP that for \emph{any} given state $\widehat{\rho }%
_{R}^{1}$ that $\left\langle \Delta \widehat{S}_{x1}^{2}\right\rangle
^{R}\left\langle \Delta \widehat{S}_{y1}^{2}\right\rangle ^{R}\geq \frac{1}{4%
}|\left\langle \widehat{S}_{z1}\right\rangle ^{R}|^{2}$, so for the
conditioned variances associated with a \emph{separable} state 
\begin{eqnarray}
\left\langle \Delta \widehat{S}_{x1}^{2}\right\rangle _{\widehat{S}%
_{x2}}\left\langle \Delta \widehat{S}_{y1}^{2}\right\rangle _{\widehat{S}%
_{y2}} &\geq &\frac{1}{4}\left( \tsum\limits_{R}\,P_{R}\;|\left\langle 
\widehat{S}_{z1}\right\rangle ^{R}|\right) ^{2}  \nonumber \\
&>&\frac{1}{4}\left( \tsum\limits_{R}\,P_{R}\;\left\langle \widehat{S}%
_{z1}\right\rangle ^{R}\right) ^{2}  \nonumber \\
&=&\frac{1}{4}|\left\langle \widehat{S}_{z1}\right\rangle |^{2}
\label{Eq.HUPresultSepStatesSpin}
\end{eqnarray}%
showing that for a separable state the conditioned variances $\left\langle
\Delta \widehat{S}_{x1}^{2}\right\rangle _{\widehat{S}_{x2}}$ and $%
\left\langle \Delta \widehat{S}_{y1}^{2}\right\rangle _{\widehat{S}_{y2}}$%
\emph{still satisfy} the HUP. It is important to note that these variances
were associated with two different conditioned states based on two different
measurements - spin $\widehat{S}_{x2}$ and spin $\widehat{S}_{y2}$ - carried
out on sub-system $2$, the results of which the observer for sub-system $1$
would be unaware of. Thus if the EPR violations \textbf{as }defined in Eq.(%
\ref{Eq.EPRViolationSpin}) are to occur then the state must be entangled. 
\textbf{\pagebreak }

\section{Appendix 4 - Extracting Entanglement due to Symmetrisation}

\label{Appendix - Extracting Entang due Symm}

\subsection{Two Particle Case - Bosons}

The approach of Killoran et al \cite{Killoran14a} can be first applied to
the simple case of $N=2$ bosons initially in the $A$ modes $a0$ and $a1$ and
were discussed in SubSection \ref{SubSection - Sub-Systems - Particles or
Modes ?}. Here we present the detailed derivation of the results. The $B$
modes $b0$ and $b1$ are initially unoccupied.

The occupied state is 
\begin{equation}
\left\vert \Phi _{A}\right\rangle =\frac{1}{\sqrt{2}}\{\left\vert
a0(1)\right\rangle \left\vert a1(2)\right\rangle +\left\vert
a0(2)\right\rangle \left\vert a1(1)\right\rangle \}
\end{equation}%
in first quantisation. This is regarded by Killoran et al \cite{Killoran14a}
as an entangled state for sub-systems consisting of particle $1$ and
particle $2$. In second quantisation the occupied state $\left\vert \Phi
_{A}\right\rangle $ and the unoccupied state $\left\vert \Phi
_{B}\right\rangle $ are given by%
\begin{eqnarray}
\left\vert \Phi _{A}\right\rangle &=&\left\vert 1\right\rangle
_{a0}\left\vert 1\right\rangle _{a1}\qquad \left\vert \Phi _{B}\right\rangle
=\left\vert 0\right\rangle _{b0}\left\vert 0\right\rangle _{b1}  \nonumber \\
\left\vert \Phi _{A}\right\rangle &=&\frac{(\widehat{a}_{0}^{\dag })}{\sqrt{1%
}}\frac{(\widehat{a}_{1}^{\dag })}{\sqrt{1}}\left\vert 0\right\rangle
_{a0}\left\vert 0\right\rangle _{a1}\qquad \left\vert \Phi _{B}\right\rangle
=\left\vert 0\right\rangle _{b0}\left\vert 0\right\rangle _{b1}
\end{eqnarray}%
These are regarded as separable states for the $A$ modes $a0$ and $a1$ and
separable states for the $B$ modes $b0$ and $b1$.

In second quantisation we consider the effect of the beam splitter on an 
\emph{input} state 
\begin{equation}
\left\vert \Phi _{in}\right\rangle =\left\vert \Phi _{A}\right\rangle
\otimes \left\vert \Phi _{B}\right\rangle
\end{equation}%
The effect is to produce an \emph{output} state given by 
\begin{eqnarray}
\left\vert \Phi _{out}\right\rangle &=&\widehat{U}\left\vert \Phi
_{in}\right\rangle  \nonumber \\
&=&\frac{(r\widehat{b}_{0}^{\dag }+t\widehat{a}_{0}^{\dag })}{\sqrt{1}}\frac{%
(r\widehat{b}_{1}^{\dag }+t\widehat{a}_{1}^{\dag })}{\sqrt{1}}\left\vert
0\right\rangle _{a0}\left\vert 0\right\rangle _{a1}\otimes \left\vert
0\right\rangle _{b0}\left\vert 0\right\rangle _{b1}  \nonumber \\
&=&r^{2}(\left\vert 0\right\rangle _{a0}\left\vert 0\right\rangle
_{a1}\otimes \left\vert 1\right\rangle _{b0}\left\vert 1\right\rangle
_{b1})+rt(\left\vert 0\right\rangle _{a0}\left\vert 1\right\rangle
_{a1}\otimes \left\vert 1\right\rangle _{b0}\left\vert 0\right\rangle
_{b1}+\left\vert 1\right\rangle _{a0}\left\vert 0\right\rangle _{a1}\otimes
\left\vert 0\right\rangle _{b0}\left\vert 1\right\rangle _{b1})  \nonumber \\
&&+t^{2}(\left\vert 1\right\rangle _{a0}\left\vert 1\right\rangle
_{a1}\otimes \left\vert 0\right\rangle _{b0}\left\vert 0\right\rangle _{b1})
\end{eqnarray}

Measurements can then be done on the output state based on projecting the
state onto eigenstates for the number operators for the $A$ and $B$
mode-based sub-systems. The projectors $\widehat{\Pi }^{A}(N_{A})$ for
sub-system $A$ onto eigenstates with $N_{A}=0,1,2$ bosons are given by%
\begin{eqnarray}
\widehat{\Pi }^{A}(0) &=&\left\vert 0\right\rangle _{a0}\left\vert
0\right\rangle _{a1}\left\langle 0\right\vert _{a0}\left\langle 0\right\vert
_{a1}  \nonumber \\
\widehat{\Pi }^{A}(1) &=&(\left\vert 1\right\rangle _{a0}\left\vert
0\right\rangle _{a1}\left\langle 1\right\vert _{a0}\left\langle 0\right\vert
_{a1}+\left\vert 0\right\rangle _{a0}\left\vert 1\right\rangle
_{a1}\left\langle 0\right\vert _{a0}\left\langle 1\right\vert _{a1}) 
\nonumber \\
\widehat{\Pi }^{A}(2) &=&(\left\vert 2\right\rangle _{a0}\left\vert
0\right\rangle _{a1}\left\langle 2\right\vert _{a0}\left\langle 0\right\vert
_{a1}+\left\vert 1\right\rangle _{a0}\left\vert 1\right\rangle
_{a1}\left\langle 1\right\vert _{a0}\left\langle 1\right\vert
_{a1}+\left\vert 0\right\rangle _{a0}\left\vert 2\right\rangle
_{a1}\left\langle 0\right\vert _{a0}\left\langle 2\right\vert _{a1}) 
\nonumber \\
&&
\end{eqnarray}%
with corresponding expressions for projectors $\widehat{\Pi }^{B}(N_{B})$
for sub-system $B$.

To demonstrate entanglement extraction for particle based sub-systems with
particle $1$ in one sub-system, and particle $2$ in the other sub-system we
choose projectors corresponding to there being one particle in the $A$ modal
sub-system and one particle being in the $B$ modal sub-system. Thus the
output state is projected onto the states with $N_{A}=1$ and $N_{B}=1$ and
we get after normalising%
\begin{eqnarray}
\left\vert \Phi _{out}(1,1)\right\rangle &=&\mathcal{N}\left( \widehat{\Pi }%
^{A}(1)\otimes \widehat{\Pi }^{B}(1)\right) \left\vert \Phi
_{out}\right\rangle  \nonumber \\
&=&\frac{1}{\sqrt{2}}(\left\vert 1\right\rangle _{a0}\left\vert
0\right\rangle _{a1}\otimes \left\vert 0\right\rangle _{b0}\left\vert
1\right\rangle _{b1}+\left\vert 0\right\rangle _{a0}\left\vert
1\right\rangle _{a1}\otimes \left\vert 1\right\rangle _{b0}\left\vert
0\right\rangle _{b1})  \nonumber \\
&&  \label{Eq.ProjectedOutput2}
\end{eqnarray}%
This is still a bipartite entangled state of the of two modal sub-systems, $%
A $ and $B$.

If we construct a mathematical correspondence of the form%
\begin{eqnarray}
\left\vert a0(1)\right\rangle &\rightarrow &\left\vert 1\right\rangle
_{a0}\left\vert 0\right\rangle _{a1}\qquad \left\vert a1(2)\right\rangle
\rightarrow \left\vert 0\right\rangle _{b0}\left\vert 1\right\rangle _{b1} 
\nonumber \\
\left\vert a1(1)\right\rangle &\rightarrow &\left\vert 0\right\rangle
_{a0}\left\vert 1\right\rangle _{a1}\qquad \left\vert a0(2)\right\rangle
\rightarrow \left\vert 1\right\rangle _{b0}\left\vert 0\right\rangle _{b1}
\end{eqnarray}%
we see that the projected output state given in (\ref{Eq.ProjectedOutput2})
as a bipartite entangled state of the of two modal sub-systems, $A$ and $B$,
has the same \emph{mathematical} form as the bipartite entangled state of
the of two particle sub-systems containing particle $1$ and particle $2$.
respectively.

\subsection{Two Particle Case - Fermions}

Here the details for the simple case of $N=2$ fermions initially in the $C$
modes $c0$ and $c1$ are presented, following the same approach as in the
previous SubSection. The $D$ modes $d0$ and $d1$ are initially unoccupied.
Fermion modes are denoted $c$ and $d$ to distinguish them from bosonic modes 
$a$ and $b$.

The occupied state is 
\begin{equation}
\left\vert \Phi _{C}\right\rangle =\frac{1}{\sqrt{2}}\{\left\vert
c0(1)\right\rangle \left\vert c1(2)\right\rangle -\left\vert
c0(2)\right\rangle \left\vert c1(1)\right\rangle \}
\end{equation}%
in first quantisation. This is regarded by Killoran et al \cite{Killoran14a}
as an entangled state for sub-systems consisting of particle $1$ and
particle $2$. In second quantisation the occupied state $\left\vert \Phi
_{C}\right\rangle $ and the unoccupied state $\left\vert \Phi
_{D}\right\rangle $ are given by%
\begin{eqnarray}
\left\vert \Phi _{C}\right\rangle &=&\left\vert 1\right\rangle
_{c0}\left\vert 1\right\rangle _{c1}\qquad \left\vert \Phi _{D}\right\rangle
=\left\vert 0\right\rangle _{d0}\left\vert 0\right\rangle _{d1}  \nonumber \\
\left\vert \Phi _{C}\right\rangle &=&\frac{(\widehat{c}_{0}^{\dag })}{\sqrt{1%
}}\frac{(\widehat{c}_{1}^{\dag })}{\sqrt{1}}\left\vert 0\right\rangle
_{c0}\left\vert 0\right\rangle _{c1}\qquad \left\vert \Phi _{D}\right\rangle
=\left\vert 0\right\rangle _{d0}\left\vert 0\right\rangle _{d1}
\end{eqnarray}%
These are regarded as separable states for the $C$ modes $c0$ and $c1$ and
separable states for the $D$ modes $d0$ and $d1$.

In second quantisation we consider the effect of the beam splitter on an 
\emph{input} state 
\begin{equation}
\left\vert \Phi _{in}\right\rangle =\left\vert \Phi _{C}\right\rangle
\otimes \left\vert \Phi _{D}\right\rangle
\end{equation}%
The effect is to produce an \emph{output} state given by 
\begin{eqnarray}
\left\vert \Phi _{out}\right\rangle &=&\widehat{U}\left\vert \Phi
_{in}\right\rangle  \nonumber \\
&=&\frac{(r\widehat{d}_{0}^{\dag }+t\widehat{c}_{0}^{\dag })}{\sqrt{1}}\frac{%
(r\widehat{d}_{1}^{\dag }+t\widehat{c}_{1}^{\dag })}{\sqrt{1}}\left\vert
0\right\rangle _{c0}\left\vert 0\right\rangle _{c1}\otimes \left\vert
0\right\rangle _{d0}\left\vert 0\right\rangle _{d1}  \nonumber \\
&=&r^{2}(\left\vert 0\right\rangle _{c0}\left\vert 0\right\rangle
_{c1}\otimes \left\vert 1\right\rangle _{d0}\left\vert 1\right\rangle
_{d1})+rt(-\left\vert 0\right\rangle _{c0}\left\vert 1\right\rangle
_{c1}\otimes \left\vert 1\right\rangle _{d0}\left\vert 0\right\rangle
_{d1}+\left\vert 1\right\rangle _{c0}\left\vert 0\right\rangle _{c1}\otimes
\left\vert 0\right\rangle _{d0}\left\vert 1\right\rangle _{d1})  \nonumber \\
&&+t^{2}(\left\vert 1\right\rangle _{c0}\left\vert 1\right\rangle
_{c1}\otimes \left\vert 0\right\rangle _{d0}\left\vert 0\right\rangle _{d1})
\end{eqnarray}%
Note the minus sign in the second term - this is due to the fermion creation
operators anti-commuting $\widehat{d}_{0}^{\dag }\widehat{c}_{1}^{\dag }=-%
\widehat{c}_{1}^{\dag }\widehat{d}_{0}^{\dag }$.

Measurements can then be done on the output state based on projecting the
state onto eigenstates for the number operators for the $C$ and $D$
mode-based sub-systems. The projectors $\widehat{\Pi }^{C}(N_{C})$ for
sub-system $C$ onto eigenstates with $N_{C}=0,1,2$ bosons are given by%
\begin{eqnarray}
\widehat{\Pi }^{C}(0) &=&\left\vert 0\right\rangle _{c0}\left\vert
0\right\rangle _{c1}\left\langle 0\right\vert _{c0}\left\langle 0\right\vert
_{c1}  \nonumber \\
\widehat{\Pi }^{C}(1) &=&(\left\vert 1\right\rangle _{c0}\left\vert
0\right\rangle _{c1}\left\langle 1\right\vert _{c0}\left\langle 0\right\vert
_{c1}+\left\vert 0\right\rangle _{c0}\left\vert 1\right\rangle
_{c1}\left\langle 0\right\vert _{c0}\left\langle 1\right\vert _{c1}) 
\nonumber \\
\widehat{\Pi }^{C}(2) &=&(\left\vert 2\right\rangle _{c0}\left\vert
0\right\rangle _{c1}\left\langle 2\right\vert _{c0}\left\langle 0\right\vert
_{c1}+\left\vert 1\right\rangle _{c0}\left\vert 1\right\rangle
_{c1}\left\langle 1\right\vert _{c0}\left\langle 1\right\vert
_{c1}+\left\vert 0\right\rangle _{c0}\left\vert 2\right\rangle
_{c1}\left\langle 0\right\vert _{c0}\left\langle 2\right\vert _{c1}) 
\nonumber \\
&&
\end{eqnarray}%
with corresponding expressions for projectors $\widehat{\Pi }^{D}(N_{D})$
for sub-system $D$.

To demonstrate entanglement extraction for particle based sub-systems with
particle $1$ in one sub-system, and particle $2$ in the other sub-system we
choose projectors corresponding to there being one particle in the $C$ modal
sub-system and one particle being in the $D$ modal sub-system. Thus the
output state is projected onto the states with $N_{C}=1$ and $N_{D}=1$ and
we get after normalising%
\begin{eqnarray}
\left\vert \Phi _{out}(1,1)\right\rangle &=&\mathcal{N}\left( \widehat{\Pi }%
^{C}(1)\otimes \widehat{\Pi }^{D}(1)\right) \left\vert \Phi
_{out}\right\rangle  \nonumber \\
&=&\frac{1}{\sqrt{2}}(\left\vert 1\right\rangle _{c0}\left\vert
0\right\rangle _{c1}\otimes \left\vert 0\right\rangle _{d0}\left\vert
1\right\rangle _{d1}-\left\vert 0\right\rangle _{c0}\left\vert
1\right\rangle _{c1}\otimes \left\vert 1\right\rangle _{d0}\left\vert
0\right\rangle _{d1})  \nonumber \\
&&  \label{Eq.ProjectedOutput3}
\end{eqnarray}%
This is still a bipartite entangled state of the of two modal sub-systems, $%
C $ and $D$.

If we construct a mathematical correspondence of the form%
\begin{eqnarray}
\left\vert c0(1)\right\rangle &\rightarrow &\left\vert 1\right\rangle
_{c0}\left\vert 0\right\rangle _{c1}\qquad \left\vert c1(2)\right\rangle
\rightarrow \left\vert 0\right\rangle _{d0}\left\vert 1\right\rangle _{d1} 
\nonumber \\
\left\vert c1(1)\right\rangle &\rightarrow &\left\vert 0\right\rangle
_{c0}\left\vert 1\right\rangle _{c1}\qquad \left\vert c0(2)\right\rangle
\rightarrow \left\vert 1\right\rangle _{d0}\left\vert 0\right\rangle _{d1}
\end{eqnarray}%
we see that the projected output state given in (\ref{Eq.ProjectedOutput3})
as a bipartite entangled state of the of two modal sub-systems, $C$ and $D$,
has the same \emph{mathematical} form as the bipartite entangled state of
the of two particle sub-systems containing particle $1$ and particle $2$.
respectively - even down to the correct minus sign in the second term.

\subsection{Three Particle Case - Bosons}

The key ideas in the approach by Killoran et al \cite{Killoran14a} are more
fully illustrated by considering one of their specific cases, namely a
quantum state with $N=3$ identical bosons for a system with four modes - two 
$A$ modes $a0$ and $a1$ and two $B$ modes $b0$ and $b1$ - in which there are
two bosons in mode $a0$ and one boson in mode $a1$. The other modes $b0$ and 
$b1$ are initially un-occupied. The modes $a0$ and $b0$ could be two
different spatial modes for a bosonic atom in one hyperfine state, and $a1$
and $b1$ could be two different spatial modes for a bosonic atom in another
hyperfine state. With particles labelled $1,2$ and $3$ the quantum state in
terms of first quantisation is given by 
\begin{eqnarray}
\left\vert \Phi _{A}\right\rangle &=&\frac{1}{\sqrt{12}}\{\left\vert
a0(1)\right\rangle \left\vert a0(2)\right\rangle \left\vert
a1(3)\right\rangle +\left\vert a0(1)\right\rangle \left\vert
a0(3)\right\rangle \left\vert a1(2)\right\rangle  \nonumber \\
&&+\left\vert a0(2)\right\rangle \left\vert a0(1)\right\rangle \left\vert
a1(3)\right\rangle +\left\vert a0(2)\right\rangle \left\vert
a0(3)\right\rangle \left\vert a1(1)\right\rangle  \nonumber \\
&&+\left\vert a0(3)\right\rangle \left\vert a0(1)\right\rangle \left\vert
a1(2)\right\rangle +\left\vert a0(3)\right\rangle \left\vert
a0(2)\right\rangle \left\vert a1(1)\right\rangle \}  \label{Eq.StateFirstQn}
\end{eqnarray}%
In first quantisation there is no state for the $B$ modes, since the vacuum
state is not recognised as a quantum state of anything. Following the
approach of regarding labelled identical particles as sub-systems we
consider a\emph{\ bipartite} division of the three particle system with the
first sub-system as consisting of particle $1$ and $2$ and the second
sub-system consisting of particle $3$. The same state $\left\vert \Phi
_{A}\right\rangle $ can be written as 
\begin{eqnarray}
\left\vert \Phi _{A}\right\rangle &=&\frac{1}{\sqrt{3}}\{\left\vert
a0(1)\right\rangle \left\vert a0(2)\right\rangle \}\left\vert
a1(3)\right\rangle  \nonumber \\
&&+\sqrt{\frac{2}{3}}\frac{1}{\sqrt{2}}\{\left\vert a0(1)\right\rangle
\left\vert a1(2)\right\rangle +\left\vert a0(2)\right\rangle \left\vert
a1(1)\right\rangle \}\left\vert a0(3)\right\rangle
\label{Eq.StateFirstQnForm2}
\end{eqnarray}%
In this form the state appears to be an \emph{entangled} state for the two
sub-systems. The first term (which has amplitude $1/\sqrt{3}$) represents a
state for the sub-system of particle $1$ and $2$ with both particles in
single particle state $\left\vert a0\right\rangle $ and a state for the
sub-system of particle $3$ with this particles in single particle state $%
\left\vert a1\right\rangle $. The second term (which has amplitude $\sqrt{2/3%
}$) represents a state for the sub-system of particles $1$ and $2$ with one
particles in single particle state $\left\vert a0\right\rangle $ and the
other in single particle state $\left\vert a1\right\rangle ,$ and a state
for the sub-system of particle $3$ with this particle in single particle
state $\left\vert a0\right\rangle $. It is this entanglement which Killoran
et al \cite{Killoran14a} wish to extract by applying a beam splitter to the
state $\left\vert \Phi _{A}\right\rangle $, the \emph{beam splitter} being
associated with a unitary operator $\widehat{U}$ whose effect is to
transform the single particle states $\left\vert ak\right\rangle $ and $%
\left\vert bk\right\rangle $ into linear combinations of each other involving%
\emph{\ reflection} and \emph{transmission} coefficients $r$, $t$ as follows 
\begin{eqnarray}
\widehat{U}\left\vert ak\right\rangle &=&r\left\vert bk\right\rangle
+t\left\vert ak\right\rangle \qquad k=0,1  \nonumber \\
\widehat{U}\left\vert bk\right\rangle &=&t\left\vert bk\right\rangle
-r\left\vert ak\right\rangle \qquad k=0,1  \label{Eq.UnitaryBSOperator}
\end{eqnarray}%
For simplicity $r$, $t$ are assumed to be real with $r^{2}+t^{2}=1$. The
beam splitter is just assumed to couple spatial modes of the same hyperfine
state. This unitary operator applies irrespective of the particular particle
occupying the one particle states.

In second quantisation the occupied state $\left\vert \Phi _{A}\right\rangle 
$ and the unoccupied state $\left\vert \Phi _{B}\right\rangle $ are given by%
\begin{eqnarray}
\left\vert \Phi _{A}\right\rangle &=&\left\vert 2\right\rangle
_{a0}\left\vert 1\right\rangle _{a1}\qquad \left\vert \Phi _{B}\right\rangle
=\left\vert 0\right\rangle _{b0}\left\vert 0\right\rangle _{b1}  \nonumber \\
\left\vert \Phi _{A}\right\rangle &=&\frac{(\widehat{a}_{0}^{\dag })^{2}}{%
\sqrt{2}}\frac{(\widehat{a}_{1}^{\dag })}{\sqrt{1}}\left\vert 0\right\rangle
_{a0}\left\vert 0\right\rangle _{a1}\qquad \left\vert \Phi _{B}\right\rangle
=\left\vert 0\right\rangle _{b0}\left\vert 0\right\rangle _{b1}
\label{Eq.StateSEcondQnForm}
\end{eqnarray}%
where the Fock states are also written in terms of mode creation operators
and vacuum states for the modes. In second quantisation it is clear that $%
\left\vert \Phi _{A}\right\rangle $ and $\left\vert \Phi _{B}\right\rangle $
themselves are respectively \emph{separable} states for the $A$ and $B$
modes. In second quantisation the effect of the unitary operator associated
with the beam splitter follows from (\ref{Eq.UnitaryBSOperator}) noting that 
$\left\vert ak\right\rangle \equiv \widehat{a}_{k}^{\dag }\left\vert
0\right\rangle $ and $\left\vert bk\right\rangle \equiv \widehat{b}%
_{k}^{\dag }\left\vert 0\right\rangle $ and is given by%
\begin{equation}
\widehat{U}\,\widehat{a}_{k}^{\dag }\,\widehat{U}^{-1}=r\widehat{b}%
_{k}^{\dag }+t\widehat{a}_{k}^{\dag }\qquad \widehat{U}\,\widehat{b}%
_{k}^{\dag }\,\widehat{U}^{-1}=t\widehat{b}_{k}^{\dag }-r\widehat{a}%
_{k}^{\dag }\qquad k=0,1  \label{Eq.UnitaryBSForm2}
\end{equation}%
In paper II we show that two mode beam splitters can indeed be described by
equations analogous to (\ref{Eq.UnitaryBSForm2}). In second quantisation we
consider the effect of the beam splitter on an \emph{input} state 
\begin{equation}
\left\vert \Phi _{in}\right\rangle =\left\vert \Phi _{A}\right\rangle
\otimes \left\vert \Phi _{B}\right\rangle  \label{Eq.InputState}
\end{equation}%
The effect is to produce an \emph{output} state given by 
\begin{eqnarray}
\left\vert \Phi _{out}\right\rangle &=&\widehat{U}\left\vert \Phi
_{in}\right\rangle  \nonumber \\
&=&\frac{(r\widehat{b}_{0}^{\dag }+t\widehat{a}_{0}^{\dag })^{2}}{\sqrt{2}}%
\frac{(r\widehat{b}_{1}^{\dag }+t\widehat{a}_{1}^{\dag })}{\sqrt{1}}%
\left\vert 0\right\rangle _{a0}\left\vert 0\right\rangle _{a1}\otimes
\left\vert 0\right\rangle _{b0}\left\vert 0\right\rangle _{b1}  \nonumber \\
&=&r^{3}(\left\vert 0\right\rangle _{a0}\left\vert 0\right\rangle
_{a1}\otimes \left\vert 2\right\rangle _{b0}\left\vert 1\right\rangle
_{b1})+r^{2}t(\left\vert 0\right\rangle _{a0}\left\vert 1\right\rangle
_{a1}\otimes \left\vert 2\right\rangle _{b0}\left\vert 0\right\rangle _{b1}+%
\sqrt{2}(\left\vert 1\right\rangle _{a0}\left\vert 0\right\rangle
_{a1}\otimes \left\vert 1\right\rangle _{b0}\left\vert 1\right\rangle _{b1})
\nonumber \\
&&+rt^{2}((\left\vert 2\right\rangle _{a0}\left\vert 0\right\rangle
_{a1}\otimes \left\vert 0\right\rangle _{b0}\left\vert 1\right\rangle _{b1}+%
\sqrt{2}(\left\vert 1\right\rangle _{a0}\left\vert 1\right\rangle
_{a1}\otimes \left\vert 1\right\rangle _{b0}\left\vert 0\right\rangle
_{b1})+t^{3}(\left\vert 2\right\rangle _{a0}\left\vert 1\right\rangle
_{a1}\otimes \left\vert 0\right\rangle _{b0}\left\vert 0\right\rangle _{b1})
\nonumber \\
&&  \label{Eq.OutputState}
\end{eqnarray}%
Note this state is normalised to unity as $\left\langle \Phi _{out}|\Phi
_{out}\right\rangle =(r^{2}+t^{2})^{3}=1$. The input state is a bipartite 
\emph{separable} state of two modal sub-systems, one containing the two $A$
modes $a0$ and $a1$ and the other the two $B$ modes $b0$ and $b1$. The
output state terms each are eigenstates of number operators $\widehat{N}_{A}=%
\widehat{a}_{0}^{\dag }\widehat{a}_{0}+\widehat{a}_{1}^{\dag }\widehat{a}%
_{1} $ and $\widehat{N}_{B}=\widehat{b}_{0}^{\dag }\widehat{b}_{0}+\widehat{b%
}_{1}^{\dag }\widehat{b}_{1}$ with eigenvalues $N_{A}=0$, $N_{B}=3$ for the $%
r^{3}$ term, $N_{A}=1$, $N_{B}=2$ for the $r^{2}t$ term, $N_{A}=2$, $N_{B}=1$
for the $rt^{2}$ term, $N_{A}=3$, $N_{B}=0$ for the $t^{3}$ term. The same
result as in (\ref{Eq.OutputState}) can also be obtained using (\ref%
{Eq.UnitaryBSOperator}) in conjunction with the first quantisation form of
the input state given by (\ref{Eq.StateFirstQn}) though the algebra is more
complex. In contrast to the input state, the output state is a bipartite 
\emph{entangled} state of two modal sub-systems, one containing the two $A$
modes $a0$ and $a1$ and the other the two $B$ modes $b0$ and $b1$. Both
input and output states are states with the same total of $N=3$ bosons.

Measurements can then be done on the output state based on projecting the
state onto eigenstates for the number operators for the $A$ and $B$
mode-based sub-systems. The projectors $\widehat{\Pi }^{A}(N_{A})$ for
sub-system $A$ onto eigenstates with $N_{A}=0,1,2,3$ bosons are given by%
\begin{eqnarray}
\widehat{\Pi }^{A}(0) &=&\left\vert 0\right\rangle _{a0}\left\vert
0\right\rangle _{a1}\left\langle 0\right\vert _{a0}\left\langle 0\right\vert
_{a1}  \nonumber \\
\widehat{\Pi }^{A}(1) &=&(\left\vert 1\right\rangle _{a0}\left\vert
0\right\rangle _{a1}\left\langle 1\right\vert _{a0}\left\langle 0\right\vert
_{a1}+\left\vert 0\right\rangle _{a0}\left\vert 1\right\rangle
_{a1}\left\langle 0\right\vert _{a0}\left\langle 1\right\vert _{a1}) 
\nonumber \\
\widehat{\Pi }^{A}(2) &=&(\left\vert 2\right\rangle _{a0}\left\vert
0\right\rangle _{a1}\left\langle 2\right\vert _{a0}\left\langle 0\right\vert
_{a1}+\left\vert 1\right\rangle _{a0}\left\vert 1\right\rangle
_{a1}\left\langle 1\right\vert _{a0}\left\langle 1\right\vert
_{a1}+\left\vert 0\right\rangle _{a0}\left\vert 2\right\rangle
_{a1}\left\langle 0\right\vert _{a0}\left\langle 2\right\vert _{a1}) 
\nonumber \\
\widehat{\Pi }^{A}(3) &=&(\left\vert 3\right\rangle _{a0}\left\vert
0\right\rangle _{a1}\left\langle 3\right\vert _{a0}\left\langle 0\right\vert
_{a1}+\left\vert 2\right\rangle _{a0}\left\vert 1\right\rangle
_{a1}\left\langle 2\right\vert _{a0}\left\langle 1\right\vert
_{a1}+\left\vert 1\right\rangle _{a0}\left\vert 2\right\rangle
_{a1}\left\langle 1\right\vert _{a0}\left\langle 2\right\vert
_{a1}+\left\vert 0\right\rangle _{a0}\left\vert 3\right\rangle
_{a1}\left\langle 0\right\vert _{a0}\left\langle 3\right\vert _{a1}) 
\nonumber \\
&&  \label{Eq.Projectors}
\end{eqnarray}%
with corresponding expressions for projectors $\widehat{\Pi }^{B}(N_{B})$
for sub-system $B$.

To demonstrate entanglement extraction for particle based sub-systems with
particles $1$ and $2$ in one sub-system, and particle $3$ in the other
sub-system we choose projectors corresponding to there being two particles
in the $A$ modal sub-system and one particle being on the $B$ modal
sub-system. Thus the output state is projected onto the states with $N_{A}=2$
and $N_{B}=1$ and we get after normalising%
\begin{eqnarray}
\left\vert \Phi _{out}(2,1)\right\rangle &=&\mathcal{N}\left( \widehat{\Pi }%
^{A}(2)\otimes \widehat{\Pi }^{B}(1)\right) \left\vert \Phi
_{out}\right\rangle  \nonumber \\
&=&\frac{1}{\sqrt{3}}\left\vert 2\right\rangle _{a0}\left\vert
0\right\rangle _{a1}\otimes \left\vert 0\right\rangle _{b0}\left\vert
1\right\rangle _{b1}+\sqrt{\frac{2}{3}}(\left\vert 1\right\rangle
_{a0}\left\vert 1\right\rangle _{a1}\otimes \left\vert 1\right\rangle
_{b0}\left\vert 0\right\rangle _{b1}  \nonumber \\
&&  \label{Eq.ProjectedOutput}
\end{eqnarray}%
This is still a bipartite entangled state of the of two modal sub-systems, $%
A $ and $B$.

If we construct a mathematical correspondence of the form%
\begin{eqnarray}
\left\vert a0(1)\right\rangle \left\vert a0(2)\right\rangle &\rightarrow
&\left\vert 2\right\rangle _{a0}\left\vert 0\right\rangle _{a1}\qquad
\left\vert a1(3)\right\rangle \rightarrow \left\vert 0\right\rangle
_{b0}\left\vert 1\right\rangle _{b1}  \nonumber \\
\frac{1}{\sqrt{2}}\{\left\vert a0(1)\right\rangle \left\vert
a1(2)\right\rangle +\left\vert a0(2)\right\rangle \left\vert
a1(1)\right\rangle \} &\rightarrow &\left\vert 1\right\rangle
_{a0}\left\vert 1\right\rangle _{a1}\qquad \left\vert a0(3)\right\rangle
\rightarrow \left\vert 1\right\rangle _{b0}\left\vert 0\right\rangle _{b1} 
\nonumber \\
&&  \label{Eq.Corresp}
\end{eqnarray}%
we see that the projected output state given in (\ref{Eq.ProjectedOutput})
as a bipartite entangled state of the of two modal sub-systems, $A$ and $B$,
has the same \emph{mathematical} form as the bipartite entangled state of
the of two particle sub-systems containing particles $1$ and $2$ and
particle $3$. respectively. This type of result is proved in more general
cases in \cite{Killoran14a} - here we have exhibited the key features of
their approach in a particular case.

It is on this basis that Killoran et al \cite{Killoran14a} assert that the
action of the beam splitter is to extract the entanglement due to
symmetrisation that was present in the quantum state $\left\vert \Phi
_{A}\right\rangle $ for the particle sub-systems containing particles $1$
and $2$ and particle $3$. respectively. It is of course not their ingeneous
mathematical derivation that is in dispute - it is the interpretation. From
the point of view of sub-systems being modes, not particles the action of
the beam splitter is to create an entangled state of the two modal
sub-systems, $A$ and $B$ from a state that was separable. That this
entangled output state can be projected onto eigenstates of the number
operators for the two modal sub-systems, $A$ and $B$ which have the same
mathematical form as the presumed entangled initial state for the particle
sub-systems containing particles $1$ and $2$ and particle $3$. respectively
is of course interesting, but it does not show that labeled identical
particles can be regarded as physically accessible sub-systems. Apart from
the logical issue that sub-systems must be both distinguishable from each
other via physical measurements, it is noteworthy that the approach of
Killoran et al \cite{Killoran14a} rested on physical processes that involved
coupling modes, not identified indistinguishable particles.\pagebreak

\section{Appendix 5 - Reference Frames and Super-Selection Rules}

\label{Appendix - Reference Frames and SSR}

Several papers such as \cite{Kitaev04a}, \cite{Bartlett06a}, \cite%
{Bartlett07a}, \cite{Vaccaro08a}, \cite{Tichy11a}, \cite{White09a}, \cite%
{Paterek11a} explain the link between \emph{reference frames} and \emph{%
super-selection rules} (SSR). In this Appendix we present the key ideas
involved.

\subsection{Two Observers with Different Reference Frames}

The first point to appreciate is that there are \emph{two observers} - Alice
and Charlie - who are involved in describing the \emph{same} \emph{quantum
system}, which has been prepared via some physical process \ We will refer
to Charlie as the \emph{external} observer, Alice the \emph{internal}
observer. The system could be a \emph{multi-mode} system involving identical
particles, it could just be a \emph{single mode} system or it could even be
a \emph{single particle} with or without spin. Alice and Charlie each
describe quantum states in terms of their own \emph{reference frames}, which
might be a set of \emph{coordinate axes} for the case of the spin or
position states for the single particle system, or it could be a \emph{large
quantum system} with a well-defined reference \emph{phase} in the case of
multi-mode or single mode systems involving identical particles. Alice and
Charlie may each choose from a set of possible reference frames - for the
single particle case there are an infinite number of difference choices of
coordinate axes for example, related to each other via \emph{rotations}
and/or \emph{translations}. In \emph{Situation A} - which \emph{is} \emph{not%
} associated with \emph{SSR} - Alice and Charlie \emph{do know} the
relationship between their two reference frames (and can communicate this
relationship via \emph{classical communications}) - such as in the case of
the single particle system when the relative orientation of their two
different coordinate axes are known. In \emph{Situation B} - which \emph{is}
associated with \emph{SSR} - Alice and Charlie \emph{do not know} the
relationship between their two reference frames - such as in the multi-mode
or single mode system involving identical particles when the relative phase
between their two large quantum phase reference systems is not known. Alice
and Charlie describe the same \textbf{system} via density operators $%
\widehat{\sigma }$ and $\widehat{\rho }$, and the key question is the \emph{%
relationship} between these two operators in situations A and B and for
various types of reference frames. In terms of the notation in \cite%
{Bartlett07a} $\rho \rightarrow $ $\widehat{\sigma }$ and $\widetilde{\rho }%
\rightarrow \widehat{\rho }$. In some situations the assumption that Alice
even possesses a well-defined reference frame may be invalid, in which case
it is important to realise that it is \emph{Charlie's} \emph{quantum state}
which is of \emph{most interest} for describing the system. This description
may differ from what hypothetical observer Alice would regard as the quantum
state.

\subsection{Symmetry Groups}

A particular relationship going from Alice's to Charlie's reference frame is
specified by the \emph{parameter} $g$, which in turn defines a \emph{unitary
transformation operator} $\widehat{T}(g)$ that acts in the system space.
Particular examples will be listed below. If there was a third observer -
Donald - and the relationship going from Charlie's to Donald's reference
frame is specified by the parameter $h$, which in turn defines a unitary
operator $\widehat{T}(h)$, then if we symbolise the relationship going from
Alice's to Donald's reference frame by the parameter $hg$, it follows that $%
\widehat{T}(hg)=\widehat{T}(h)\widehat{T}(g)$. This shows that the unitary
operators satisfy one of the requirements to constitute a \emph{group},
referred to generally as the \emph{transformation} group. The other
requirements are easily confirmed. The unitary operator $\widehat{T}(0)=%
\widehat{1}$ corresponding to the case where no change of reference frame
occurs (specified by the parameter $0$) exists, and satisfies the
requirement that $\widehat{T}(0g)=\widehat{T}(0)\widehat{T}(g)=\widehat{T}%
(g0)=\widehat{T}(g)\widehat{T}(0)$. The unitary operator $\widehat{T}%
(g^{-1})=\widehat{T}(g)^{\dag }$ corresponding to the relationship specified
as $g^{-1}$ that converts Charlie's reference frame back to that of Alice
exists, and satisfies the requirement that $\widehat{T}(0)=\widehat{T}%
(g^{-1})\widehat{T}(g)=\widehat{T}(g)\widehat{T}(g^{-1})$. Hence all the
group properties are satisfied.

A few examples are as follows:

1. \emph{Translation group} - single spinless particle system, with $%
\widehat{p}$, $\widehat{x}$.the momentum, position vector operators. Here $%
\underrightarrow{a}$ is a vector giving the translation of Charlie's
cartesian axes reference frame from that of Alice, thus $g\equiv $.$%
\underrightarrow{a}$. The unitary translation operator is $\widehat{T}(%
\underrightarrow{a})=\exp (i\widehat{p}\cdot \underrightarrow{a}/\hbar )$.

2. \emph{Rotation group} - single particle system, with $\widehat{J}$ the
angular momentum vector operators. Here $\underrightarrow{u}$ is a unit
vector giving the axis and rotation angle $\phi $ for rotating Alice's
cartesian axes reference frame into that of Charlie, thus $g\equiv $.$%
\underrightarrow{u}$.$,\phi $. The unitary rotation operator is $\widehat{T}(%
\underrightarrow{u},\phi )=\exp (i\phi \widehat{J}\cdot \underrightarrow{u}%
/\hbar )$.

3. \emph{Particle number U(1) group} - single mode bosonic system, with $%
\widehat{a}$ the mode annihilation operator and $\widehat{N}_{a}=\widehat{a}%
^{\dag }\widehat{a}$ the mode number operator. Here $\theta _{a}$ is the
phase change Alice's to Charlie's reference frame. The unitary operator is $%
\widehat{T}(\theta _{a})=\exp (i\widehat{N}_{a}\theta _{a})$.

4. \emph{Particle number U(1) group} - multi-mode bosonic system, with $%
\widehat{a}$ as a typical mode annihilation operator and $\widehat{N}%
=\dsum\limits_{a}\widehat{a}^{\dag }\widehat{a}$ the total number operator.
Here $\theta $ is the phase change from Alice's to Charlie's reference
frame. The unitary operator is $\widehat{T}(\theta )=\exp (i\widehat{N}%
\theta )$.

In these examples the system operators $\widehat{p}$, $\widehat{J}$, $%
\widehat{N}_{a}$, $\widehat{N}$ etc are the \emph{generators} of the
respective groups. In many situations the generators commute with the
Hamiltonian for the system (or more generally with the evolution operator
that describes time evolution of the quantum state), in which case the group
of unitary operators $\widehat{T}(g)$ is the \emph{symmetry group}, and the
generators are \emph{conserved} physical quantities.

\subsection{Relationships - Situation A}

In \emph{Situation A}, where the relationship between the reference frames
for Alice and Charlie is \emph{known} and specified by a \emph{single}
parameter $g$, Alice's description of the state $\widehat{\sigma }$ is
related to Charlie's description $\widehat{\rho }$ for the same state via
the unitary transformation 
\begin{equation}
\widehat{\rho }=\widehat{T}(g)\,\widehat{\sigma }\,\widehat{T}(g)^{-1}
\label{Eq.AliceCharlieStatesSitnA}
\end{equation}%
Note that this is a \emph{passive} transformation - no change of state is
involved, just the same state being described by two different observers.

As an example, consider the \emph{spinless particle} and the \emph{%
translation} group. If $\left\vert \underrightarrow{x}\right\rangle $ is a
position eigenstate then $\widehat{T}(\underrightarrow{a})\left\vert 
\underrightarrow{x}\right\rangle =\left\vert \underrightarrow{x}-%
\underrightarrow{a}\right\rangle $. A pure quantum position eigenstate
described by Alice as $\,\widehat{\sigma }=\left\vert \Phi \right\rangle
\left\langle \Phi \right\vert $ with state vector $\left\vert \Phi
\right\rangle =\left\vert \underrightarrow{x}\right\rangle $ would be
described by Charlie as $\widehat{\rho }=\left\vert \Psi \right\rangle
\left\langle \Psi \right\vert $ but now with $\left\vert \Psi \right\rangle
=\left\vert \underrightarrow{x}-\underrightarrow{a}\right\rangle $, which is
also a pure quantum position eigenstate but with eigenvalue $%
\underrightarrow{x}-\underrightarrow{a}$. This is as expected since Alices's
cartesian axes have been translated by $\underrightarrow{a}$ to the origin
of Charlie's axes without change of orientation. In the case of momentum
eigenstates $\left\vert \underrightarrow{p}\right\rangle $ we have $\widehat{%
T}(\underrightarrow{a})\left\vert \underrightarrow{p}\right\rangle =\exp (i%
\underrightarrow{p}\cdot \underrightarrow{a}/\hbar )\left\vert 
\underrightarrow{p}\right\rangle $, so a pure quantum momentum eigenstate
described by Alice with $\left\vert \Phi \right\rangle =\left\vert 
\underrightarrow{p}\right\rangle $ would be described by Charlie with $%
\left\vert \Psi \right\rangle =\exp (i\underrightarrow{p}\cdot 
\underrightarrow{a}/\hbar \,\left\vert \underrightarrow{p}\right\rangle $,
which is also a pure momentum eigenstate with the same eigenvalue $%
\underrightarrow{p}$. Alice and Charlie describe the pure momentum
eigernstate with the same density operator $\widehat{\rho }=\widehat{\sigma }
$, the phase factor cancels.

For more general pure states, consider a quantum state described by Alice as 
$\,\widehat{\sigma }=\left\vert \Phi \right\rangle \left\langle \Phi
\right\vert $ with state vector $\left\vert \Phi \right\rangle =\dint d%
\underrightarrow{x}\,\phi (\underrightarrow{x})\,\left\vert \underrightarrow{%
x}\right\rangle $. States of this form can represent \emph{localised} states
when $\phi (\underrightarrow{x})$ is only significant in confined spatial
regions, or they can represent \emph{delocalised} states, such as momentum
eigenstates $\left\vert \underrightarrow{p}\right\rangle $ when $\phi (%
\underrightarrow{x})=(2\pi \hbar )^{-3/2}\exp (i\underrightarrow{p}\cdot 
\underrightarrow{x}/\hbar )$. We see that Charlie also describes a pure
quantum state $\widehat{\rho }=\left\vert \Psi \right\rangle \left\langle
\Psi \right\vert $ but now with $\left\vert \Psi \right\rangle =\widehat{T}(%
\underrightarrow{a})\,\left\vert \Phi \right\rangle =\dint d\underrightarrow{%
x}\,\phi (\underrightarrow{x}+\underrightarrow{a})\,\left\vert 
\underrightarrow{x}\right\rangle =\dint d\underrightarrow{x}\,\psi (%
\underrightarrow{x})\,\left\vert \underrightarrow{x}\right\rangle $, so the
wavefunction is now $\psi (\underrightarrow{x})=\phi (\underrightarrow{x}+%
\underrightarrow{a})$.

Note that if Alice's state vector was written in terms of momentum
eigenstates $\left\vert \Phi \right\rangle =\dint d\underrightarrow{p}\,%
\widetilde{\phi }(\underrightarrow{p})\,\left\vert \underrightarrow{p}%
\right\rangle $, then Charlie's state vector $\left\vert \Psi \right\rangle
=\dint d\underrightarrow{p}\,\widetilde{\psi }(\underrightarrow{p}%
)\,\left\vert \underrightarrow{p}\right\rangle $ has a momentum wave
function $\widetilde{\psi }(\underrightarrow{p})=\exp (i\underrightarrow{p}%
\cdot \underrightarrow{a}/\hbar )\,\widetilde{\phi }(\underrightarrow{p})$
related to that of Alice by a phase factor. Note that a state which is a
quantum superposition of momentum eigenstates as described by Alice is also
described as a quantum superposition of momentum eigenstates by Charlie. A
similar feature applies in all situation A cases, and is related to SSR 
\emph{not} applying in situation A.

The case of the \emph{particle} with \emph{spin} and the \emph{rotation}
group is outlined in Ref. \cite{Bartlett06a}.

\subsection{Relationships - Situation B}

\label{AppendixSubSection - Situation B}

In \emph{Situation B}, where on the other hand the relationship between
frames is completely \emph{unknown}, all possible transformations $g$ must
be given \emph{equal weight}, and hence the relationship between Alice's and
Charlie's description of the same state becomes 
\begin{eqnarray}
\widehat{\rho } &=&\dint w(g)dg\,\widehat{T}(g)\,\widehat{\sigma }\,\widehat{%
T}(g)^{-1}  \nonumber \\
&=&\mathcal{G\,[}\widehat{\sigma }]  \label{Eq.AliceCharlieStatesSitnB}
\end{eqnarray}%
where $\dint w(g)dg$ is a symbolic integral over the parameter $g$, which
includes a weight factor $w(g)$ so that $\dint w(g)dg=1$. This linear
process connecting $\widehat{\sigma }$ to $\widehat{\rho }$ is the "$%
\mathcal{G}$- \emph{twirling}" operation. Again, this is a paassive
transformation.

It is straightforward to show that for any fixed parameter $h$ that 
\begin{equation}
\widehat{T}(h)\,\widehat{\rho }\,\widehat{T}(h)^{-1}=\widehat{\rho }
\label{Eq.CharlieStateInvarSitnB}
\end{equation}%
showing that Charlie's density operator is $\mathcal{G}$ invariant under the
transformation group - unlike the case for Situation A.

As an example, consider the \emph{single mode} bosonic system and the \emph{%
U(1)} group. If $\left\vert n_{a}\right\rangle $ is a Fock state then $%
\widehat{T}(\theta _{a})\,\left\vert n_{a}\right\rangle =\exp (in_{a}\theta
_{a})\,\left\vert n_{a}\right\rangle $. Consider a pure quantum state
described by Alice as the \emph{Glauber coherent state} $\widehat{\sigma }%
=\left\vert \Phi \right\rangle \left\langle \Phi \right\vert $ with state
vector $\left\vert \Phi (\beta )\right\rangle
=\dsum\limits_{n_{a}}\,C(n_{a},\beta )\,\left\vert n_{a}\right\rangle $,
where $C(n_{a},\beta )=\exp (-|\beta |^{2}/2)\,\beta ^{n_{a}}\,/\sqrt{%
(n_{a})!}$. It is straightforward to show that 
\begin{equation}
\widehat{T}(\theta _{a})\,\left\vert \Phi (\beta )\right\rangle =\left\vert
\Phi (\beta \exp (i\theta _{a}))\right\rangle  \label{Eq.GlauberStateTransfn}
\end{equation}%
so that the Glauber coherent state is transformed into another Glauber
coherent state, but with $\beta $ changed via a phase factor to $\beta \exp
(i\theta _{a})$. The quantum state described by Charlie is given by%
\begin{eqnarray}
\widehat{\rho } &=&\dint \frac{d\theta _{a}}{2\pi }\left\vert \Phi (\beta
\exp (i\theta _{a}))\right\rangle \left\langle \Phi (\beta \exp (i\theta
_{a}))\right\vert  \label{Eq.CharlieStateGlauberForm1} \\
&=&\dint \frac{d\theta _{a}}{2\pi }\dsum\limits_{n_{a}}\dsum\limits_{m_{a}}%
\,C(n_{a},\beta )\,C(m_{a},\beta )^{\ast }\,\widehat{T}(\theta
_{a})\,\left\vert n_{a}\right\rangle \left\langle m_{a}\right\vert \,%
\widehat{T}(\theta _{a})^{\dag }  \nonumber \\
&=&\dsum\limits_{n_{a}}\dsum\limits_{m_{a}}\,C(n_{a},\beta )\,C(m_{a},\beta
)^{\ast }\,\left\vert n_{a}\right\rangle \left\langle m_{a}\right\vert
\,\dint \frac{d\theta _{a}}{2\pi }\exp (i[n_{a}-m_{a}]\theta _{a})  \nonumber
\\
&=&\dsum\limits_{n_{a}}\,|C(n_{a},\beta )|^{2}\,\left\vert
n_{a}\right\rangle \left\langle n_{a}\right\vert  \nonumber \\
&=&\dsum\limits_{n_{a}}\,\exp (-|\beta |^{2})\,\frac{(|\beta |^{2})^{n_{a}}}{%
(n_{a})!}\,\left\vert n_{a}\right\rangle \left\langle n_{a}\right\vert
\label{Eq.CharlieStateGlauberForm2}
\end{eqnarray}%
which is a \emph{mixed state} consisting of a \emph{Poisson distribution} of 
\emph{Fock states} with mean occupation number $\overline{n}_{a}=|\beta
|^{2} $. In view of the first expression for $\widehat{\rho }$ it can also
be thought of as a mixed state consisting of Glauber coherent states each
with the same amplitude $|\beta |\,=\,\sqrt{\overline{n}_{a}}$, but with all
phases $(\arg \beta +\theta _{a})$ equally probable. Thus, whereas Alice
describes the state as a pure state that is a quantum superposition of Fock
states with differing occupancy numbers, Charlie describes the same state as
a mixed state involving a statistical mixture of number states. The former
violates the SSR whereas the latter does not. A similar feature applies in
all situation B cases, and is related to SSR applying in Situation B.
Whether Alice could ever prepare such a state in the first place is
controversial - see the discussion presented above in SubSections \ref%
{SubSection - Super-Selection Rule} and \ref{SubSection - SSR Separate Modes}%
. However, \emph{assuming} she could, the quantum state as described by
Charlie is a mixed state.

The situation just studied relates of course to the debate \cite{Molmer97a}
regarding whether the quantum state for a \emph{single mode laser} operating
well above threshold should be described by a Glauber coherent state or as a
Poisson statistical mixture of photon number states. The first viewpoint
(Alice) describes the state from the point of view of an internal observer
with a reference frame, the second (Charlie) describes the same state from
the point of view of an external observer for whose reference frame
relationship to that of the internal observer is unknown. The debate is
regarded by \cite{Bartlett06a} as settled on the basis that both viewpoints
are valid, they are just at cross purposes because they refer to
descriptions of the same quantum state by two different observers.

It should not be thought however that the quantum state would always be
described in such a fundamentally different manner for all Situation B
cases. As an example, consider the \emph{multi-mode bosonic} system and the 
\emph{U(1)} group. Consider the pure quantum state described by Alice as the
multi-mode $N$ boson \emph{Fock state} $\widehat{\sigma }=\left\vert \Phi
\right\rangle \left\langle \Phi \right\vert $ with state vector $\left\vert
\Phi (N)\right\rangle =\left\vert n_{1}n_{2}...n_{a}...;N\right\rangle
=\dprod\limits_{a}\left\vert n_{1}\right\rangle \left\vert
n_{2}\right\rangle ..\left\vert n_{a}\right\rangle ...$, where $%
N=\dsum\limits_{a}n_{a}$. We have $\widehat{T}(\theta )\left\vert
n_{1}n_{2}...n_{a}...;N\right\rangle =\exp (iN\theta )\,\left\vert
n_{1}n_{2}...n_{a}...;N\right\rangle $, so that the same state would be
described by Charlie as $\widehat{\rho }=\left\vert \Psi \right\rangle
\left\langle \Psi \right\vert $ and with $\left\vert \Psi \right\rangle
=\left\vert n_{1}n_{2}...n_{a}...;N\right\rangle $. This is also a
multi-mode $N$ boson Fock state with exactly the same occupancies. The
product $\exp (iN\theta )\,\exp (-iN\theta )$ of phase factors averages out
to unity and here $\widehat{\rho }=\widehat{\sigma }$, so Alice and Charlie
both describe the multi-mode Fock states in the same way. Another example
for \emph{two mode bosonic} systems and the \emph{U(1)} group\ is provided
by the one boson \emph{Bell states} (the BS\ notation used here is
non-conventional). These are entangled two mode states that Alice would
describe via the state vectors $\left\vert \Phi ^{\pm }\right\rangle
=(\left\vert 10\right\rangle \pm \left\vert 01\right\rangle )/\sqrt{2}$. We
have $\widehat{T}(\theta )\,\left\vert \Phi ^{\pm }\right\rangle =\exp
(i\theta )\,\left\vert \Phi ^{\pm }\right\rangle $, so that the same state
would be described by Charlie with $\left\vert \Psi ^{\pm }\right\rangle
=(\left\vert 10\right\rangle \pm \left\vert 01\right\rangle )/\sqrt{2}$.
Again the product of phase factors averages to unity and $\widehat{\rho }=%
\widehat{\sigma }$, so Alice and Charlie both describe the quantum states as
Bell states, and in the same form.

\subsection{Dynamical and Measurement Considerations}

Discussions of the relationship between equations governing the dynamical
behaviour of Alice's and Charlie's density operators depend on whether the
evolution is just governed by a Hamiltonian or whether master equations
describing evolution affected by interactions with an external environment
are involved. Such matters will not be treated in detail here, nor will the
issue of relating Alice's and Charlie's measurements. The latter issue is
dealt with in \cite{Bartlett07a}.

However, in the case where Alice describes the \emph{Hamiltonian evolution}
of her density operator via the Liouville - von-Neumann equation 
\begin{equation}
i\hbar \frac{\partial }{\partial t}\widehat{\sigma }=[\widehat{H},\widehat{%
\sigma }]  \label{Eq.LVNAlice}
\end{equation}%
where in Alice's frame the Hamiltonian is $\widehat{H}$, and where in
addition the transformation group is also the \emph{symmetry group} so that $%
\widehat{T}(g)\widehat{H}\widehat{T}(g)^{-1}=\widehat{H}$ for all $g$, it is
easy to see that for both Situations A and B, Charlie's density operator
will evolve via the same LVN equation%
\begin{equation}
i\hbar \frac{\partial }{\partial t}\widehat{\rho }=[\widehat{H},\widehat{%
\rho }]  \label{Eq.LVNCharlie}
\end{equation}%
Thus both Alice and Charlie will describe the same dynamical evolution,
though of course the initial (and hence evolved) states may differ in the
two cases.

\subsection{Nature of Reference Frames}

Reference frames of differing types are involved for the various
transformation groups. The common feature is that they are thought of as 
\emph{actual physical systems} themselves which are either macroscopic \emph{%
classical} systems or macroscopic \emph{quantum} systems in states
associated with the \emph{classical limit}. They are intended to be \emph{%
essentially unaffected} by the presence of the systems for which they are
acting as reference frames. In some cases relatively uncontroversial
examples exist, such as for the \emph{cartesian axes} associated with the 
\emph{translation} and \emph{rotation} groups associated with the single
particle system. The physical reference system may be a large \emph{magnet}
whose magnetic field points in a well defined direction and defines a $z$
axis, combined with an \emph{electrostatic generator} whose electric field
is in another well defined direction at right angles that defines an $x$
axis. In other cases the existence of suitable reference frames is less
clear.

In this SubSection we will describe possible phase reference frames as if
they are entirely separated (or uncorrelated) with the system of interest.
In terms of the treatment by Bartlett et al \cite{Bartlett06a}, \cite%
{Bartlett07a} these are \emph{non-implicated} reference frames. In the next
SubSection and in the next Appendix phase reference frames that are
correlated with the system of interest will be described - these are the
so-called \emph{implicated} reference frames of Bartlett et al.

For the large quantum system with a well-defined reference \emph{phase}
associated with the \emph{U(1)} group in the case of multi-mode or single
mode systems involving identical particles, the usual choice is a single
mode bosonic system such as a single mode \emph{BEC} or a \emph{laser} with
a large mean occupancy, and which is thought of as being prepared in a
Glauber coherent state $\left\vert \Phi (\alpha )\right\rangle $ in order to
provide the \emph{phase reference frame}, the reference phase being $\arg
\alpha $. Whether such a reference frame really exists is controversial. The
discussion presented above in SubSections \ref{SubSection - Super-Selection
Rule} and \ref{SubSection - SSR Separate Modes} raises the question of
whether such a phase reference state could ever be prepared, so this choice
of a physical phase reference is rather unsatisfactory. However, from the
point of view of this presentation we \emph{assume} it does, so that - as in
the previous example - Alice can describe the reference state as another
coherent state. Again, whether Alice could ever prepare such a state is
questionable.

Another possibility for a physical phase reference is a \emph{macroscopic }%
low frequency\emph{\ harmonic oscillator}, whose quantum energy eigenstates $%
\left\vert n\right\rangle $ - with $n=0,1,\,...\,,n_{\max }$ and energies $%
n\,\hbar \omega $ can be used to construct phase eigenstates $\left\vert
\theta _{p}\right\rangle $ with $p=0,1,\,...\,,n_{\max }$ and $\theta
_{p}=p\times 2\pi /(n_{\max }+1)$, and which are defined by \cite{Barnett89a}%
\begin{equation}
\left\vert \theta _{p}\right\rangle =\frac{1}{\sqrt{n_{\max }+1}}%
\dsum\limits_{n=0}^{n_{\max }}\,\exp (in\theta _{p})\,\left\vert
n\right\rangle  \label{Eq.PhaseState}
\end{equation}%
These states are orthonormal. The separation between the equally spaced
phase angles $\Delta \theta =2\pi /(n_{\max }+1)$ can be made very small if $%
n_{\max }$ is large enough. Under the effect of the harmonic oscillator
Hamiltonian $\widehat{H}=\hbar \omega \widehat{N}$, where $\widehat{N}$ is
the number operator, the phase state $\left\vert \theta _{p}\right\rangle $
evolves into $\left\vert \theta _{p}-\omega \Delta t\right\rangle $ during a
time interval $\Delta t$, so if the time intervals are chosen so that $%
\omega \Delta t=2\pi /(n_{\max }+1)$, the phase angle $\theta _{p}$ changes
into $\theta _{p-1}$. Thus the system behaves like a backwards running \emph{%
clock }\cite{Pegg91a}, the phase angles $\theta _{p}$ defining the positions
of the hands. If the clock initially has phase $\theta _{p}$ the probability
of finding the clock to have phase $\theta _{q}$ after a time interval $%
\Delta t$ is given by 
\begin{equation}
P(\theta _{q},\theta _{p},\Delta t)=\frac{1}{(n_{\max }+1)^{2}}\frac{\sin
^{2}((n_{\max }+1)\Delta /2)}{\sin ^{2}(\Delta /2)}  \label{Eq.PhaseProb}
\end{equation}%
where $\Delta =\theta _{p}-\theta _{q}-\omega \Delta t$. For times $\Delta t$
such that $\omega \Delta t\ll 2\pi /(n_{\max }+1)$ the probability of the
phase remaining as $\theta _{p}$ is close to unity. Thus if the phase state $%
\left\vert \theta _{p}\right\rangle $ is used as a phase reference, it will
remain stable for a time $\Delta t$ satisfying the last inequality. For $%
\Delta t\sim 100\mu s$ and $n_{\max }\sim 10^{4\text{ }}$so that phase is
defined to $\sim 10^{-3}$ radians, an oscillator frequency $\omega \sim
10^{0}$ s$^{-1}$ would suffice for this phase reference standard. Such
macroscopic oscillators do exist, though the process to prepare them in the
phase reference \emph{quantum} state $\left\vert \theta _{p}\right\rangle $
would be technically difficult. Whether such a system would be useful as a
phase reference for optical fields or a BEC is another issue

\subsection{Relational Description of Phase References}

In this SubSection phase reference frames that are correlated with the
system of interest will be described - these are the so-called \emph{%
implicated} reference frames of Bartlett et al \cite{Bartlett06a}, \cite%
{Bartlett07a}.

One such approach to describing phase references in the \emph{U(1)} group
case is via the concept of \emph{maps}. For simplicity consider a one mode
system \emph{S}, the basis vectors for which are Fock states $\left\vert
m\right\rangle _{S}$, where it is sufficient to restrict $%
m=0,1,\,...\,,m_{\max }$. The reference system $R$, will also be a one mode
system with Fock states $\left\vert n\right\rangle _{R}$, where $n$ is
large. Product states $\left\vert m\right\rangle _{S}\otimes \left\vert
n\right\rangle _{R}$ for the combined modes exist in the Hilbert space $%
H_{S}\otimes H_{R}$ and are eigenstates of the various number operators,
including the total number operator $\widehat{N}_{T}=\widehat{N}_{S}+%
\widehat{N}_{R}$ - where the eigenvalue is $l=m+n$. The product states may
be listed via $m=0,1,\,...\,,m_{\max }$ and $n=0,1,\,...\,$or $%
m=0,1,\,...\,,m_{\max }$ and $l=m,m+1,\,...\,$. Here we will describe how a 
\emph{coherent superpostion} of \emph{number states}, such as a Glauber
coherent state can be represented.

In the so-called \emph{internalisation} or \emph{quantisation} of the
reference frame the product state $\left\vert m\right\rangle _{S}\otimes
\left\vert n\right\rangle _{R}$ is mapped onto the product state $\left\vert
m\right\rangle _{S}\otimes \left\vert n-m\right\rangle _{R}$ where $n\geq
m_{\max }$. Thus 
\begin{equation}
\left\vert m\right\rangle _{S}\otimes \left\vert n\right\rangle
_{R}\rightarrow \left\vert m\right\rangle _{S}\otimes \left\vert
n-m\right\rangle _{R}  \label{Eq.IntMap}
\end{equation}%
Hence for a linear combination of system states given by 
\begin{equation}
\left\vert \Phi \right\rangle _{S}=\dsum\limits_{m=0}^{m_{\max
}}\,C_{m}\left\vert m\right\rangle _{S}  \label{Eq.SysState}
\end{equation}%
we have for the state $\left\vert \Phi \right\rangle _{S}\otimes \left\vert
n\right\rangle _{R}$ in $H_{S}\otimes H_{R}$ 
\begin{equation}
\left\vert \Phi \right\rangle _{S}\otimes \left\vert n\right\rangle
_{R}=\dsum\limits_{m=0}^{m_{\max }}\,C_{m}\left\vert m\right\rangle
_{S}\otimes \left\vert n\right\rangle _{R}\rightarrow
\dsum\limits_{m=0}^{m_{\max }}\,C_{m}\left\vert m\right\rangle _{S}\otimes
\left\vert n-m\right\rangle _{R}=\left\vert \Psi _{n}\right\rangle _{RS}
\label{Eq.MapCoherSuper}
\end{equation}%
The mapping results in an entangled state where there are $n$ bosons
distributed betweeen the two modes. This state $\left\vert \Psi
_{n}\right\rangle _{RS}$ is a pure state which is compatible with the SSR
and is in one-one correspondence with the original system state $\left\vert
\Phi \right\rangle _{S}$. Note that to create this state the reference state 
$\left\vert n\right\rangle _{R}$ must have more bosons in it than $m_{\max }$%
. The density operator for the original pure system $S$ state would be $%
\widehat{\sigma }_{S}=\left\vert \Phi \right\rangle _{S}\left\langle \Phi
\right\vert _{S}$, and we note that this state violates the SSR. The state $%
\left\vert \Phi \right\rangle _{S}$ would be essentially a Glauber coherent
state if $C_{m}=\exp (-|\alpha |^{2}/2)\alpha ^{m}/(\sqrt{m!})$, with $%
m_{\max }\gg |\alpha |^{2}$. However, for the mapped state $\left\vert \Psi
_{n}\right\rangle _{RS}$ the reduced density operator $\widehat{\rho }_{S}$
is given by 
\begin{eqnarray}
\widehat{\rho }_{S} &=&Tr_{R}(\left\vert \Psi _{n}\right\rangle
_{RS}\left\langle \Psi \right\vert _{RS})  \nonumber \\
&=&\dsum\limits_{m=0}^{m_{\max }}\,|C_{m}|^{2}\,\left\vert m\right\rangle
_{S}\left\langle m\right\vert _{S}  \label{Eq.RDOSystem}
\end{eqnarray}%
This is a mixed state and is compatible with the SSR. For the Glauber
coherent state $\left\vert \Phi \right\rangle _{S}$ this is the Poisson
distribution of number states. Hence the original SSR violating
superposition of number states for system $S$ is mapped onto a state in the
combined system for which the reduced density operator is a statistical
mixture and is consistent with the SSR. $\widehat{\sigma }_{S}$ would
correspond to Alice's description of the state, $\widehat{\rho }_{S}$ to
Charlie's.

In the alternative so-called \emph{externalisation} of the reference frame
the mapping is between product states, and is the reverse of the previous
mapping. The product state $\left\vert m\right\rangle _{S}\otimes \left\vert
n\right\rangle _{R}$ is mapped onto the product state $\left\vert
m\right\rangle _{S}\otimes \left\vert m+n\right\rangle _{R}$ in the Hilbert
space $H_{S}\otimes H_{R}$ where the former is spanned by vectors $%
\left\vert m\right\rangle _{S}$ and the latter by vectors $\left\vert
m+n\right\rangle _{R}$, and where $n\geq m_{\max }$. Thus 
\begin{equation}
\left\vert m\right\rangle _{S}\otimes \left\vert n\right\rangle
_{R}\rightarrow \left\vert m\right\rangle _{S}\otimes \left\vert
m+n\right\rangle _{R}  \label{Eq.ExtMap}
\end{equation}%
The mapping of the $H_{S}\otimes H_{R}$ state $\left\vert \Psi
_{n}\right\rangle _{RS}$ then is 
\begin{eqnarray}
\left\vert \Psi _{n}\right\rangle _{RS} &=&\dsum\limits_{m=0}^{m_{\max
}}\,C_{m}\left\vert m\right\rangle _{S}\otimes \left\vert n-m\right\rangle
_{R}  \nonumber \\
&\rightarrow &\dsum\limits_{m=0}^{m_{\max }}\,C_{m}\left\vert m\right\rangle
_{S}\otimes \left\vert n\right\rangle _{R}=\left(
\dsum\limits_{m=0}^{m_{\max }}\,C_{m}\left\vert m\right\rangle _{S}\right)
\otimes \left\vert n\right\rangle _{R}=\left\vert \Xi _{n}\right\rangle _{RS}
\nonumber \\
&&  \label{Eq.MapMapCoherSuper}
\end{eqnarray}%
The mapping results in a non-entangled state which is incompatible with the
SSR. The state in the subspace $H_{S}$ is a coherent superposition of number
states, whilst that in $H_{R}$ is a Fock state. The reduced density operator
in $H_{S}$ is $\widehat{\sigma }_{S}^{\#}$ given by 
\begin{eqnarray}
\widehat{\sigma }_{S}^{\#} &=&Tr_{R}(\left\vert \Xi _{n}\right\rangle
_{RS}\left\langle \Xi _{n}\right\vert _{RS})  \nonumber \\
&=&\dsum\limits_{m=0}^{m_{\max }}\dsum\limits_{k=0}^{m_{\max
}}\,C_{m}C_{k}^{\ast }\,\left\vert m\right\rangle _{S}\left\langle
k\right\vert _{S}  \label{Eq.RDORel}
\end{eqnarray}%
which is the same as $\widehat{\sigma }_{S}=\left\vert \Phi \right\rangle
_{S}\left\langle \Phi \right\vert _{S}$ and involves coherences between
different number states in contradiction to the SSR. Clearly this second
mapping just reverses the first one.

Of these two treatments of phase reference frames, the internalisation
version has a closer link to physics in that the pure state $\left\vert \Psi
_{n}\right\rangle _{RS}$ can in principle be created and does lead to a way
of creating a state that is in one-one correspondence with any SSR violating
pure state $\left\vert \Phi \right\rangle _{S}$, though it is in the form of
an entangled state of the $S$, $R$ sub-systems rather than just $S$ alone.
This is an important point to note - the original SSR violating state does
not exist as a state of a separate system, all that exists is an SSR
compatible entangled state that is in one-one correspondence with it.
However, the general process for creating a state such as $\left\vert \Psi
_{n}\right\rangle _{RS}$ is not explained. For simple cases such as $%
\left\vert \Phi \right\rangle _{S}=(\left\vert 0\right\rangle
_{S}+\left\vert 1\right\rangle _{S})/\sqrt{2}$ the creation of the required
state $\left\vert \Psi _{n}\right\rangle _{RS}=(\left\vert 0\right\rangle
_{S}\otimes \left\vert n\right\rangle _{R}+\left\vert 1\right\rangle
_{S}\otimes \left\vert n-1\right\rangle _{R})/\sqrt{2}$, where $n\geq 1$
would seem feasible via the ejection of one boson from a BEC in a Fock state 
$\left\vert n\right\rangle _{R}$ into a previously unoccupied mode. .

\subsection{Irreducible Matrix Representations and Super-selection Rules}

If $\left\vert i\right\rangle $ $(i=1,2,\,...)$ are a set of orthonormal
basis vectors in the system state space, then the group of unitary operators 
$\widehat{T}(g)$ is represented by a group of \emph{unitary matrices} $D(g)$%
\begin{equation}
\widehat{T}(g)\,\left\vert i\right\rangle
=\dsum\limits_{j}D_{ji}(g)\,\left\vert j\right\rangle
\label{Eq.MatrixRepnTransfnGrp}
\end{equation}%
with elements $D_{ji}(g)$, and such that $D(hg)=D(h)D(g)$ etc corresponding
to the group properties of the operators. This is a \emph{matrix
representation} of the transformation group.

The theory of such group representations and their application to quantum
systems is well established, following the pioneering work of Wigner in the
1930s. We can just use the results here. A key concept is that of \emph{%
irreducible} representations. Within the system state space we can in
general choose so-called irreducible sub-spaces, denoted as $\Gamma _{\alpha
}$ of dimension $d_{\alpha }$ and spanned by new orthonormal basis vectors $%
\left\vert \Gamma _{\alpha }\lambda \right\rangle $ $(\lambda
=1,2,\,...\,,d_{\alpha })$ such that 
\begin{equation}
\widehat{T}(g)\,\left\vert \Gamma _{\alpha }\lambda \right\rangle
=\dsum\limits_{\mu =1}^{d_{\alpha }}D_{\mu \lambda }^{\alpha
}(g)\,\left\vert \Gamma _{\alpha }\mu \right\rangle
\label{Eq.IrreducMatrixRepnTransfnGrp}
\end{equation}%
For each irreducible sub-space $\Gamma _{\alpha }$ there is \emph{no}
smaller sub-space for which the operation of all $\widehat{T}(g)$ just leads
to linear combinations of vectors within that sub-space. The $d_{\alpha
}\times d_{\alpha }$ matrices $D^{\alpha }(g)$ then form an irreducible
matrix representation for the transformation group. For different $\alpha $
the representations are said to be \emph{inequivalent}.

The irreducible matrices satisfy the so-called \emph{great orthogonality
theorem} \cite{Tinkham64a}%
\begin{equation}
\dint w(g)dg\,D_{\mu \lambda }^{\alpha }(g)D_{\xi \tau }^{\beta }(g)^{\ast }=%
\frac{1}{d_{\alpha }}\delta _{\alpha \beta }\delta _{\mu \xi }\delta
_{\lambda \tau }  \label{Eq.GreatOrthogThm}
\end{equation}%
The proof of this result is based on Schur's lemma.

The importance of the irreducible representations and the consequent
orthogonality theorem lies in its application to Situation B cases, where we
have seen that Charlie's density operator $\widehat{\rho }$ is invariant
under any of the transformations $\widehat{T}(h)\,\widehat{\rho }\,\widehat{T%
}(h)^{-1}=\widehat{\rho }$. Suppose we represent $\widehat{\rho }$ in terms
of the basis vectors $\left\vert \Gamma _{\alpha }\lambda \right\rangle $
associated with the irreducible representations%
\begin{equation}
\widehat{\rho }=\dsum\limits_{\alpha \lambda }\dsum\limits_{\beta \tau
}\,R_{\lambda \tau }^{\alpha \beta }\,\left\vert \Gamma _{\alpha }\lambda
\right\rangle \left\langle \Gamma _{\beta }\tau \right\vert
\label{Eq.CharlieDensOprIrredBasis}
\end{equation}%
where $R$ will be a Hermitian, positive definite matrix with unit trace
since it represents a density operator. Applying the transformation gives%
\begin{eqnarray}
\widehat{T}(h)\,\widehat{\rho }\,\widehat{T}(h)^{-1} &=&\dsum\limits_{\alpha
\lambda \mu }\dsum\limits_{\beta \tau \xi }\,R_{\lambda \tau }^{\alpha \beta
}\,D_{\mu \lambda }^{\alpha }(h)\,\left\vert \Gamma _{\alpha }\mu
\right\rangle \left\langle \Gamma _{\beta }\xi \right\vert \,D_{\xi \tau
}^{\beta }(h)^{\ast }  \nonumber \\
&=&\widehat{\rho }
\end{eqnarray}%
Averaging over $h$ and using the great orthogonality theorem gives%
\begin{equation}
\widehat{\rho }=\dsum\limits_{\alpha }\dsum\limits_{\mu }\,\left(
\dsum\limits_{\lambda }\frac{1}{d_{\alpha }}R_{\lambda \lambda }^{\alpha
\alpha }\,\right) \,\left\vert \Gamma _{\alpha }\mu \right\rangle
\left\langle \Gamma _{\alpha }\mu \right\vert
\end{equation}%
This is in the form of a mixed state involving irreducible state vectors $%
\left\vert \Gamma _{\alpha }\mu \right\rangle $ each occuring with a
probability $P_{\mu }^{\alpha }$ given by 
\begin{equation}
P_{\mu }^{\alpha }=\dsum\limits_{\lambda }\frac{1}{d_{\alpha }}R_{\lambda
\lambda }^{\alpha \alpha }\,=P^{\alpha }  \label{Eq.ProbCharlieDensOpr}
\end{equation}%
which is the same for all $\mu $ associated with a given irreducible
representation $\Gamma _{\alpha }$. This is clearly a positive real quantity
and since 
\begin{eqnarray}
\dsum\limits_{\alpha }\dsum\limits_{\mu }P_{\mu }^{\alpha }
&=&\dsum\limits_{\alpha }\dsum\limits_{\mu }\dsum\limits_{\lambda }\frac{1}{%
d_{\alpha }}R_{\lambda \lambda }^{\alpha \alpha }\,=\dsum\limits_{\alpha
}\dsum\limits_{\lambda }R_{\lambda \lambda }^{\alpha \alpha }  \nonumber \\
&=&Tr\,\widehat{\rho }=1  \label{Eq.ProbSumUnity}
\end{eqnarray}%
the probabilities sum to unity as required.

The final result for Charlie's density operator

\begin{equation}
\widehat{\rho }=\dsum\limits_{\alpha }\dsum\limits_{\mu }\,P^{\alpha
}\,\left\vert \Gamma _{\alpha }\mu \right\rangle \left\langle \Gamma
_{\alpha }\mu \right\vert  \label{Eq.CharlieDensOprSSRForm}
\end{equation}%
demonstrates the presence of a \emph{super-selection rule}. In Charlie's
description of the quantum state there are \emph{no coherences} between
states $\left\vert \Gamma _{\alpha }\mu \right\rangle $ associated with
differing irreducible representations of the transformation group. This
represents the general form of the SSR for all transformation groups in
Situation B cases.

As an example, consider the \emph{U(1) group} and the \emph{single mode}
bosonic system. Since the Fock states satisfy $\widehat{T}(\theta
_{a})\,\left\vert n_{a}\right\rangle =\exp (in_{a}\theta _{a})\,\left\vert
n_{a}\right\rangle $ they form the basis for the irreducible representations
of the U(1) group, the occupation number $n_{a}$ specifying the irreducible
representation and the $1\times 1$ matrices $\exp (in_{a}\theta _{a})$ being
the unitary matrices. Hence Charlie will describe the quantum state as 
\begin{equation}
\widehat{\rho }=\dsum\limits_{n_{a}}P(n_{a})\,\left\vert n_{a}\right\rangle
\left\langle n_{a}\right\vert  \label{Eq.CharlieSingleModeSSRForm}
\end{equation}%
which is a statistical mixture of Fock states with no coherences between
different Fock states. This result is of the same form as in Eq.(\ref%
{Eq.PhysicalStatesSubSys}) and is in accord with the SSR on boson number.

As another example, consider the \emph{U(1) group} and the \emph{multi-mode}
bosonic system. Here sums of products of Fock states 
\begin{equation}
\left\vert n_{1}n_{2}...n_{a}...;N\right\rangle =\dprod\limits_{a}\left\vert
n_{1}\right\rangle \left\vert n_{2}\right\rangle ..\left\vert
n_{a}\right\rangle ...\qquad N=\dsum\limits_{a}n_{a}
\label{Eq.FockStatesMultiMode}
\end{equation}%
such that the total occupancy is $N=\dsum\limits_{a}n_{a}$ can be used to
form irreducible representations for the transformation group in terms of
linear combinations of the products with the same $N$. Writing these linear
combinations as 
\begin{equation}
\left\vert \Psi _{N}^{\mu }\right\rangle
=\dsum\limits_{\{n_{1}n_{2}...n_{a}...\}}C_{\{n_{1}n_{2}...n_{a}...\}}^{N\mu
}\left\vert n_{1}n_{2}...n_{a}...;N\right\rangle
\label{Eq.MultiModeStatesSameN}
\end{equation}%
we have since $\widehat{T}(\theta )\left\vert
n_{1}n_{2}...n_{a}...;N\right\rangle =\exp (iN\theta )\,\left\vert
n_{1}n_{2}...n_{a}...;N\right\rangle $ we see that $\widehat{T}(\theta
)\,\left\vert \Psi _{N}^{\mu }\right\rangle =\exp (iN\theta )\,\left\vert
\Psi _{N}^{\mu }\right\rangle $ also, so the $\left\vert \Psi _{N}^{\mu
}\right\rangle $ define the irreducible basis states. The total occupancy $N$
specifies the irreducible representation, but here there are many
irreducible representations with the same $N$ depending on the various $\mu $%
. In this case Charlie will describe the state as 
\begin{equation}
\widehat{\rho }=\dsum\limits_{N}\dsum\limits_{\mu }P_{\mu }^{N}\,\left\vert
\Psi _{N}^{\mu }\right\rangle \left\langle \Psi _{N}^{\mu }\right\vert
\label{Eq.CharlieMultModeSSRForm}
\end{equation}%
which is a statistical mixture of multi-mode states $\left\vert \Psi
_{N}^{\mu }\right\rangle $ all with the same total occupancy $N$. Although
there are coherence terms between individual modal Fock states, there are no
coherences between states with different total occupancy. This result is of
the same form as in Eq.(\ref{Eq.PhysicalState}) and again is an example of a
super-selection rule operating in terms of Charlie's description of the
quantum state.

Finally, we note that in situation A where the relationship between the
frames is known and there is no invariance for Charlie's density operator,
we do not have SSR applying. For the \emph{single particle }case and the 
\emph{translation group} the momentum states $\left\vert \underrightarrow{p}%
\right\rangle $ define the irreducible representations, each specified by $%
\underrightarrow{p}$, and as we saw Charlie's description of the quantum
state involved linear combinations of these irreducible basis vectors, in
contradiction to the SSR.

\subsection{Non-Entangled States}

\label{Appendix SubSection - Non Ent States}

The essential feature of an \emph{non-entangled} or \emph{separable} state
is that the sub-systems are considered to be \emph{unrelated} to each other.
Hence, both for Alice and Charlie there will be \emph{separate reference
frames} for each sub-system, with transformation groups - $\widehat{T}%
_{A}(g_{a})$ for sub-system $A$, $\widehat{T}_{B}(g_{b})$ for sub-system $B$%
, etc which relate the reference systems of Alice to those of Charlie. The
transformations $g_{a}$, $g_{b}$, $..$ are different. The \emph{overall}
transformation operator would be of the form $\widehat{T}(g_{a},g_{b},\,...)=%
\widehat{T}_{A}(g_{a})\otimes \widehat{T}_{B}(g_{b})\otimes ...\,$. Alice
would describe a general non-entangled state as having a density operator 
\begin{equation}
\widehat{\sigma }=\sum_{R}P_{R}\,\widehat{\sigma }_{R}^{A}\otimes \widehat{%
\sigma }_{R}^{B}\otimes \widehat{\sigma }_{R}^{C}\otimes ...
\label{Eq.AliceDensOprNonEntState}
\end{equation}%
It then follows for Situation B where the reference frames for Alice and
Charlie are unrelated, that Charlie would describe the same state via the
density operator 
\begin{equation}
\widehat{\rho }=\sum_{R}P_{R}\,\widehat{\rho }_{R}^{A}\otimes \widehat{\rho }%
_{R}^{B}\otimes \widehat{\rho }_{R}^{C}\otimes ...
\label{Eq.CharlieDensOprNonEntState}
\end{equation}%
where

\begin{equation}
\widehat{\rho }_{R}^{C}=\dint w(g_{c})dg_{c}\,\widehat{T}_{C}(g_{c})\,%
\widehat{\sigma }_{R}^{C}\,\widehat{T}_{C}(g_{c})^{-1}\qquad C=A,B,\,...
\label{Eq,SubSysDensityOprSitnB}
\end{equation}%
Note that \emph{separate} twirl operations are applied to the different
sub-systems, as explicitly shown in the papers by Vaccaro et al \cite%
{Vaccaro08a} (see Section IIIA, Eqn. 3.3 therein) and Paterek et al \cite%
{Paterek11a} (see Section 6). This leads for general transformation groups
to the \emph{local group super-selection rule}, where the $\widehat{\rho }%
_{R}^{C}$ involve \emph{no coherences} between states associated with
differing irreducible representations of the transformation group. We see
that Charlie also describes a non-entangled state and with the same mixture
probability $P_{R}$ as for Alice. Thus non-entanglement or separability is a
feature that is the $\emph{same}$ for both Alice and Charlie, as ought to be
the case.

In the context of sub-systems consisting of \emph{modes} (or sets of modes)
occupied by \emph{identical bosons}, the case of interest is Situation B,
with each transformation group being U(1). Here the relationship between
Charlie's and Alice's \emph{phase reference} frames are unknown. Hence
irrespective of Alice's description of the sub-system states $\widehat{%
\sigma }_{R}^{A}$, $\widehat{\sigma }_{R}^{B}$, ... we see from the previous
section that Charlie will describe the separate sub-system states $\widehat{%
\rho }_{R}^{A}$, $\widehat{\rho }_{R}^{B}$, as statistical mixtures of
number states for the separate modes (or total number states for the sets of
modes in each sub-system). Thus from Charlie's point of view the separate
mode density operators will satisfy the SSR. Thus we see that the
introduction of reference frames and two observers - Charlie being the
external one whose description of the quantum states is of primary interest
- leads to the \emph{same SSR outcome} as the simpler considerations set out
in SubSections \ref{SubSection - Super-Selection Rule} and \ref{SubSection -
SSR Separate Modes}. Essentially the same considerations have been used in 
\cite{Bartlett06b}, \cite{Vaccaro08a} and the other papers to justify the 
\emph{local photon number superselection rule}. \pagebreak

\section{Appendix 6 - Super-Selection Rule Violations ?}

\label{Appendix - Super-Selection Rule Violations ?}

\subsection{Preparation of Coherent Superposition of an Atom and a Molecule ?%
}

A key paper dealing with the coherent superposition of an atom and a
molecule is that by Dowling et al \cite{Dowling06a}, entitled
\textquotedblleft Observing a coherent superposition of an atom and a
molecule\textquotedblright . Essentially the process involves one atom A
interacting with a BEC of different atoms B leading to the creation of one
molecule AB, with the BEC being depleted by one B atom.

\subsubsection{Hamiltonian}

The Hamiltonian is given by 
\begin{equation}
\widehat{H}=\hbar \omega _{A}\widehat{b}_{A}^{\dag }\widehat{b}_{A}+\hbar
\omega _{M}\widehat{b}_{M}^{\dag }\widehat{b}_{M}+\hbar \omega _{2}\widehat{b%
}_{2}^{\dag }\widehat{b}_{2}+\frac{\hbar \kappa }{2}(\widehat{b}_{M}^{\dag }%
\widehat{b}_{A}\widehat{b}_{2}+\widehat{b}_{M}\widehat{b}_{A}^{\dag }%
\widehat{b}_{2}^{\dag })  \label{Eq.HamiltonianDowling}
\end{equation}%
where $\widehat{b}_{A},\widehat{b}_{M}$ and $\widehat{b}_{2}$ are standard
bosonic annihilation operators for the atom, molecule and BEC modes
respectively, $\omega _{A},\omega _{M}$ and $\omega _{2}$ are the
corresponding mode frequencies and $\kappa $ defines the interaction
strength for the process where a molecule is created or destroyed from/to an
atom A and a BEC\ atom B. $\Delta $ is the frequency difference between the
molecular state AB and the two separate states for atoms A and B -- this is
zero on Feshbach resonance - and is given by 
\begin{equation}
\Delta =\omega _{M}-\omega _{A}-\omega _{2}  \label{Eq.FeshbachDetuning}
\end{equation}%
The Hamiltonian commutes with the total number operator $\widehat{N}_{tot}$,
where 
\begin{equation}
\widehat{N}_{tot}=2\,\widehat{b}_{M}^{\dag }\widehat{b}_{M}+\widehat{b}%
_{A}^{\dag }\widehat{b}_{A}+\widehat{b}_{2}^{\dag }\widehat{b}_{2}
\label{Eq.TotalNumberOpr}
\end{equation}%
where the molecule number operator is multipled by two.

\subsubsection{Initial State}

Initially the state of the system is given by the density operator Eqs (10)
and (11) in the paper 
\begin{eqnarray}
\widehat{W}_{0L} &=&\dint \frac{d\theta }{2\pi }\exp (-i\widehat{N}%
_{tot}\theta )\left\vert \Psi \right\rangle _{0L}\left\langle \Psi
\right\vert _{0L}\exp (+i\widehat{N}_{tot}\theta )
\label{Eq.InitialdensityOpr} \\
\left\vert \Psi \right\rangle _{0L} &=&\left\vert A\right\rangle \left\vert
\beta \right\rangle  \label{Eq.BasicInitialStateVector}
\end{eqnarray}%
where $\left\vert A\right\rangle $ is a state with one atom A and $%
\left\vert \beta \right\rangle $ is a Glauber coherent state for the BEC of
atoms B.The super-operator acting on the pure state $\left\vert \Psi
\right\rangle _{0L}\left\langle \Psi \right\vert _{0L}$ is called the \emph{%
twirling operator}, the group of unitary operators $\exp (-i\widehat{N}%
_{tot}\theta )$ depend on a \emph{phase} variable $\theta $ and are a
unitary representation of $U(1)$, the \emph{generator} being $\widehat{N}%
_{tot}$. These operators act as a \emph{symmetry group} for the system and
leave the Hamiltonian invariant. The \emph{initial state} is also given by 
\begin{eqnarray}
\widehat{W}_{0L} &=&\widehat{\rho }_{A-M}(0)\otimes \widehat{\rho }_{2}(0)
\label{Eq.InitialDensityOpr2} \\
\widehat{\rho }_{A-M}(0) &=&\left\vert A\right\rangle \left\langle
A\right\vert  \label{Eq.AtomMolDensityOpr} \\
\widehat{\rho }_{2}(0) &=&\dint \frac{d\theta }{2\pi }\exp (-i\widehat{n}%
_{2}\theta )\left\vert \beta \right\rangle \left\langle \beta \right\vert
\exp (+i\widehat{n}_{2}\theta )  \label{Eq.BECDensityOpr1} \\
&=&\dsum\limits_{n}p_{n}(<n>)\left\vert n\right\rangle \left\langle
n\right\vert  \label{Eq.BECDensityOpr2} \\
&=&\dint \frac{d\theta }{2\pi }\left\vert \beta \exp (-i\theta
)\right\rangle \left\langle \beta \exp (-i\theta )\right\vert
\label{Eq.BECDensityOpr3}
\end{eqnarray}%
where $\widehat{n}_{2}=\widehat{b}_{2}^{\dag }\widehat{b}_{2}$ is the number
uperator for the BEC mode and $p_{n}(<n>)=\{\exp (-<n>)\,<n>^{n}/n!\}$ is a
Poisson distribution, whose mean is $<n>=|\beta |^{2}$. Initially then there
is one atom A and the BEC is in a statistical mixture of number states with
a Poisson distribution, which is mathematically equivalent to a statistical
mixture of Glauber coherent states $\left\vert \beta \exp (-i\theta
)\right\rangle $ with the same amplitude $\sqrt{<n>}$ but with all phases $%
(\arg \beta +\theta )$ being equally weighted.

\subsubsection{Implicated Reference Frame}

In the paper by Dowling et al \cite{Dowling06a} the BEC is acting as an 
\emph{implicated phase reference frame} (see \cite{Bartlett06a}, \cite%
{Bartlett07a}). The state of the reference frame as described by Charlie is
given by 
\begin{equation}
\widehat{\rho }_{REF}=\,\widehat{\rho }_{2}(0)=\dint \frac{d\theta }{2\pi }%
\exp (-i\widehat{n}_{2}\theta )\left\vert \beta \right\rangle \left\langle
\beta \right\vert \exp (+i\widehat{n}_{2}\theta )  \label{Eq.RefFrame}
\end{equation}%
and from Eq. (\ref{Eq.HamiltonianDowling}), there is an interaction between
the reference BEC and the separate atom A and molecule M systems. However,
because $<n>=|\beta |^{2}$ is very large, the BEC is essentially unchanged
during the process, as reflected in the use of approximations in eqs (27),
(28) of the paper. Another implicated phase reference frame situation, but
involving a two mode reference frame is discussed in the paper by Paterek et
al \cite{Paterek11a}

Overall, in terms of the discussion in Appendix \ref{Appendix - Reference
Frames and SSR} $\widehat{W}_{0L}$ would be \emph{Charlie}'s description of
the initial state, whereas \emph{Alice} would describe it as $\left\vert
\Psi \right\rangle _{0L}\left\langle \Psi \right\vert _{0L}$. Presumably in
the paper by Dowling et al \cite{Dowling06a} what is referred to as the
"state of the laboratory" be Charlie's reference frame, and what they refer
to as the "internal reference frame" would refer to that of Alice. However,
whether Alice could actually prepare such a state as $\left\vert \Psi
\right\rangle _{0L}\left\langle \Psi \right\vert _{0L}$ is controversial -
see SubSections \ref{SubSection - Super-Selection Rule} and \ref{SubSection
- SSR Separate Modes}, though here this is assumed to be possible.

\subsubsection{Process - Alice and Charlie Descriptions}

There are three stages in the process, the first being with the interaction
that turns separate atoms A and B into the molecule AB turned on at Feshbach
resonance for a time $t=\pi /(2\kappa <n>)$, the second being free evolution
at large Feshbach detuning $\Delta $ for a time $\tau $ leading to a phase
factor $\phi =\Delta \tau $, the third being again with the interaction
turned on at Feshbach resonance for a further time $t=\pi /(2\kappa <n>)$.
The typical initial state $\left\vert \Psi \right\rangle _{0L}$ given by $%
\left\vert A\right\rangle \left\vert \beta \right\rangle $ (eq (11)) evolves
into $\left\vert \Psi \right\rangle _{3L}$ given by (see eq. (32) of paper) 
\begin{equation}
\left\vert \Psi \right\rangle _{3L}=\left( \sin (\frac{\phi }{2}%
)\,\left\vert A\right\rangle -\exp (i\arg \beta )\,\cos (\frac{\phi }{2}%
)\,\left\vert M\right\rangle \right) \left\vert \beta \right\rangle
\label{Eq.BasicFinalStateVector}
\end{equation}%
using approximations set out in eqs (27), (28) of the paper that depend on $%
<n>$ being large. Here $\left\vert M\right\rangle $ is a state with one
molecule AB. Thus it looks like a coherent superposition of an atom state $%
\left\vert A\right\rangle $ and a molecule state $\left\vert M\right\rangle $
has been prepared, the atom plus molecule system being disentangled from the
BEC. \emph{Alice} would describe the final state of the system as $%
\left\vert \Psi \right\rangle _{3L}\left\langle \Psi \right\vert _{3L}$, so
from her point of view a coherent superposition of an atom and a molecule
has been prepared.

However, for \emph{Charlie} the \emph{final state} of the system is
described by a density operator $\widehat{W}_{3L}$ which is reconstructed by
applying the twirling operator to $\left\vert \Psi \right\rangle
_{3L}\left\langle \Psi \right\vert _{3L}$ . Noting that%
\begin{equation}
\exp (-i\widehat{N}_{tot}\theta )\left\vert \Psi \right\rangle _{3L}=\left(
\exp (-i\theta )\,\sin (\frac{\phi }{2})\,\left\vert A\right\rangle -\exp
(-2i\theta )\,\exp (i\arg \beta )\,\cos (\frac{\phi }{2})\,\left\vert
M\right\rangle \right) \left\vert \beta \exp (-i\theta )\right\rangle
\label{Eq.TwirlingEffect}
\end{equation}%
and using 
\begin{equation}
Tr_{2}(\left\vert \beta \exp (-i\theta )\right\rangle \left\langle \beta
\exp (-i\theta )\right\vert )=\left\langle \beta \exp (-i\theta )|\beta \exp
(-i\theta )\right\rangle =1  \label{Eq.BECTrace}
\end{equation}%
we see that Charlie's final reduced density operator for the \emph{%
atom-molecule system} is 
\begin{eqnarray}
\widehat{\rho }_{A-M}(3) &=&Tr_{2}\widehat{W}_{3L}  \nonumber \\
&=&Tr_{2}\dint \frac{d\theta }{2\pi }\exp (-i\widehat{N}_{tot}\theta
)\left\vert \Psi \right\rangle _{3L}\left\langle \Psi \right\vert _{3L}\exp
(+i\widehat{N}_{tot}\theta )  \nonumber \\
&=&\dint \frac{d\theta }{2\pi }\left( \exp (-i\theta )\,\sin (\frac{\phi }{2}%
)\,\left\vert A\right\rangle -\exp (-2i\theta )\,\exp (i\arg \beta )\,\cos (%
\frac{\phi }{2})\,\left\vert M\right\rangle \right)  \nonumber \\
&&\times \left( \exp (+i\theta )\,\sin (\frac{\phi }{2})\,\left\langle
A\right\vert -\exp (+2i\theta )\,\exp (-i\arg \beta )\,\cos (\frac{\phi }{2}%
)\,\left\langle M\right\vert \right)  \nonumber \\
&=&\sin ^{2}(\frac{\phi }{2})\,\left\vert A\right\rangle \left\langle
A\right\vert +\cos ^{2}(\frac{\phi }{2})\,\left\vert M\right\rangle
\left\langle M\right\vert  \label{Eq.FinalRDOAtomMol}
\end{eqnarray}%
Thus the coherence terms like $\left\vert A\right\rangle \left\langle
M\right\vert $ and $\left\vert M\right\rangle \left\langle A\right\vert $ do
not appear in the final density operator when the average over $\theta $
(not $\beta $) is carried out.

For Charlie the density operator for the atom and molecule is of course a
statistical mixture of a state with one atom and no molecule and a state
with no atom and one molecule. The authors of \cite{Dowling06a} actually
point this out in the paragraph after eq (35) where (presumably for the case 
$\phi =\pi /4$) it is stated \textquotedblleft the state is found to be
\ldots\ an incoherent mixture of an atom and a molecule.\textquotedblright .
The probabilities for detecting an atom A or a molecule AB are as in eq (33)
of the paper. In terms of Charlie's description, the density operator at the
end of the preparation process does \emph{not} signify the existence of a
coherent superposition of an atom and a molecule, as the title to the paper
might be taken to imply. The existence of such a coherent superposition
would of course be present in Alice's description, but it is Charlie's
(laboratory) description that is more relevant.

\subsubsection{Interference Effects Without SSR Violation}

Note that \emph{interference effects} are still present since the atom or
molecule detection probabilities depend on the phase $\phi $ associated with
the free evolution stage of the process. However, as in many other
instances, the presence of coherence effects does not require the existence
of \emph{coherent superposition states} that violate the super-selection
rule. The authors actually point this out in the paragraph after eq (35),
where it is stated \textquotedblleft we have clearly predicted the standard
operational signature of coherence, namely Ramsey type fringes, but the
coherence is not present in our mathematical description of the
system.\textquotedblright\ What they are referring to is Charlie's
description of the final state - which indeed shows no such coherence, but
the belief that coherent superposition states are needed to predict
coherence effects is mistaken.

To drive this point home, the process can be treated with the initial state
for the BEC being given as a Fock state $\left\vert N\right\rangle $. With
the interaction being given as in Eq.(\ref{Eq.HamiltonianDowling}) (eq (14)
in the paper) the state vector is a simple linear combination of two terms%
\begin{equation}
\left\vert \Psi (t)\right\rangle =A(t)\left\vert A\right\rangle \,\left\vert
N\right\rangle +B(t)\left\vert M\right\rangle \,\left\vert N-1\right\rangle
\label{Eq.StateVectorFockMethod}
\end{equation}%
This is of course an entangled state. Coupled equations for the two
amplitudes $A(t)$ and $B(t)$ can easily be obtained and simple solutions
obtained for stages where the Feshbach detuning is either zero or large. The
state vector is continuous from one stage to the next , and the reduced
density operator at the end of the three stage process for the atom plus
molecule sub-system can be obtained. It is of the form%
\begin{eqnarray}
\widehat{\rho }_{A-M}(3) &=&Tr_{2}(\left\vert \Psi (3)\right\rangle
\left\langle \Psi (3)\right\vert )  \nonumber \\
&=&\sin ^{2}(\frac{\phi }{2})\,\left\vert A\right\rangle \left\langle
A\right\vert +\cos ^{2}(\frac{\phi }{2})\,\left\vert M\right\rangle
\left\langle M\right\vert  \label{Eq.FinalRDOAtomMolFockMethod}
\end{eqnarray}%
which is of course a statistical mixture of a state with one atom and no
molecule and a state with no atom and one molecule - and is exactly the same
result as obtained in the paper by Dowling et al.\cite{Dowling06a}. Note
that coherence effects in regard to the interferometric dependence on $\phi $
for measurements on the final state has been found without invoking either
the description of the BEC via Glauber coherent states or the presence of a
coherent superposition of an atomic and a molecular state. The result can
easily be extended for the case where the BEC is initially in a statistical
mixture of Fock states with differing $N$ occuring with a probability $P_{N}$%
. Each initial state $\left\vert A\right\rangle \,\left\vert N\right\rangle $
evolves as in Eq. (\ref{Eq.StateVectorFockMethod}). We then would have 
\begin{eqnarray}
\widehat{\rho }_{A-M}(3) &=&Tr_{2}(\tsum\limits_{N}P_{N}\,\left\vert \Psi
_{N}(3)\right\rangle \left\langle \Psi _{N}(3)\right\vert )  \nonumber \\
&=&\tsum\limits_{N}P_{N}\left( \sin ^{2}(\frac{\phi }{2})\,\left\vert
A\right\rangle \left\langle A\right\vert +\cos ^{2}(\frac{\phi }{2}%
)\,\left\vert M\right\rangle \left\langle M\right\vert \right)  \nonumber \\
&=&\sin ^{2}(\frac{\phi }{2})\,\left\vert A\right\rangle \left\langle
A\right\vert +\cos ^{2}(\frac{\phi }{2})\,\left\vert M\right\rangle
\left\langle M\right\vert  \label{Eq.FinalRDOStatMixtFocks}
\end{eqnarray}%
which is the same as before. Allowing for a statistical mixture of Fock
states makes no difference to the interferometric result.

\subsubsection{Conclusion}

Dowling et al \cite{Dowling06a} state in their abstract that
\textquotedblleft we demonstrate that it is possible to perform a
Ramsey-type interference experiment to exhibit a coherent superposition of a
single atom and a diatomic molecule\textquotedblright\ . However the
interferometric effects (involving the dependence on $\phi $) cannot be said
to exhibit the \emph{existence} of such a coherent superposition, since the
same interferometric results can be obtained \emph{without} ever introducing
such a quantum state. There is \emph{not} a convincing case that quantum
states that violate the super-selection rule forbidding the creation of
coherent superpositions of Fock states with differing particle numbers can
be \emph{created}, even in Alice's reference system. The fact that an SSR
violating state $\left\vert \Psi \right\rangle _{3L}\left\langle \Psi
\right\vert _{3L}$ is created in Alice's reference system is not surprising,
because in the process considered the initial state $\left\vert \beta
\right\rangle $ for the BEC was assumed as a factor in Alice's initial
state, and this was itself inconsistent with the SSR. Furthermore, such SSR
violating states are not \emph{needed} to describe coherence and
interference effects, so that justification for their physical existence
also fails.

\subsection{Detection of Coherent Superposition of a Vacuum and a One-Boson
State ?}

Whether such super-selection rule violating states can be detected has also
not been justified. For example, consider the state given by a superposition
of a one boson state and the vacuum state (as discussed in \cite%
{Dunningham11a}). Consider an interferometric process in which one mode $A$
for a two mode BEC interferometer is initially in the state $\alpha
\left\vert 0\right\rangle +\beta \left\vert 1\right\rangle $, and the other
mode $B$ is initially in the state $\left\vert 0\right\rangle $ - thus $%
\left\vert \Psi (i)\right\rangle =(\alpha \left\vert 0\right\rangle +\beta
\left\vert 1\right\rangle )_{A}\otimes \left\vert 0\right\rangle _{B}$ in
the usual occupancy number notation, where $|\alpha |^{2}+|\beta |^{2}=1$.
Modes $A$, $B$ could refer to two different hyperfine states of a bosonic
atom with non-relativistic energies $\hbar \omega _{A}$.and $\hbar \omega
_{B}$, mode annihilation operators $\widehat{a}$, $\widehat{b}$. The modes
are first coupled by a \emph{beam splitter}, which could be a resonant
microwave pulse that causes transitions between the two hyperfine states and
which can be described via a unitary operator $\widehat{U}_{BS}$ such that%
\begin{eqnarray}
\widehat{U}_{BS}(\left\vert 1\right\rangle _{A}\otimes \left\vert
0\right\rangle _{B}) &=&(\left\vert 1\right\rangle _{A}\otimes \left\vert
0\right\rangle _{B}-i\left\vert 0\right\rangle _{A}\otimes \left\vert
1\right\rangle _{B})/\sqrt{2}  \nonumber \\
\widehat{U}_{BS}(\left\vert 0\right\rangle _{A}\otimes \left\vert
1\right\rangle _{B}) &=&(-i\left\vert 1\right\rangle _{A}\otimes \left\vert
0\right\rangle _{B}+\left\vert 0\right\rangle _{A}\otimes \left\vert
1\right\rangle _{B})/\sqrt{2}  \nonumber \\
\widehat{U}_{BS}(\left\vert 0\right\rangle _{A}\otimes \left\vert
0\right\rangle _{B}) &=&(\left\vert 0\right\rangle _{A}\otimes \left\vert
0\right\rangle _{B}).  \label{Eq.BeamSplitter}
\end{eqnarray}%
After passing through the beam splitter the system is allowed to evolve
freely for a time $\tau $, the Hamiltonian being $\widehat{H}%
_{free}=(mc^{2}+\hbar \omega _{A})\widehat{a}^{\dag }\widehat{a}%
+(mc^{2}+\hbar \omega _{B})\widehat{b}^{\dag }\widehat{b}$ - where
collisional effects have been ignored and the rest mass energy included for
completeness. Following the free evolution stage, the modes are then coupled
again via a beam splitter, and the probability of an atom being found in
modes $A$, $B$ then being measured. A straightforward treatment of the
evolution shows that the final state is given by 
\begin{eqnarray}
\left\vert \Psi (f)\right\rangle &=&\alpha (\left\vert 0\right\rangle
_{A}\otimes \left\vert 0\right\rangle _{B})  \nonumber \\
&&+\beta \exp (-i\{mc^{2}/\hbar +\omega _{A}\}\tau )  \nonumber \\
&&\times \left( \frac{1-\exp (-i\Delta \tau )}{2}(\left\vert 1\right\rangle
_{A}\otimes \left\vert 0\right\rangle _{B})-i\frac{1+\exp (-i\Delta \tau )}{2%
}(\left\vert 0\right\rangle _{A}\otimes \left\vert 1\right\rangle
_{B})\right)  \nonumber \\
&&  \label{Eq.FinalState}
\end{eqnarray}%
where $\Delta =\omega _{B}-\omega _{A}$ is the detuning. The probabilities
of finding one atom in modes $A$, $B$ respectively are 
\begin{equation}
P_{10}=|\beta |^{2}\sin ^{2}(\Delta \tau /2)\qquad P_{01}=|\beta |^{2}\cos
^{2}(\Delta \tau /2)  \label{Eq.AtomDetnProb}
\end{equation}%
Thus whilst coherence effects occur depending on the phase difference $\phi
=\Delta \tau $ associated with the interferometric process, the overall
detection probabilities only depend on the initial state via $|\beta |^{2}$.
There is no dependence on the \emph{relative phase} between $\alpha $ and $%
\beta $, as would be required if the superposition state $\alpha \left\vert
0\right\rangle +\beta \left\vert 1\right\rangle $ is to be specified from
the measurement results. Exactly the same detection probabilities are
obtained if the initial state is the mixed state $\widehat{\rho }(i)=|\alpha
|^{2}(\left\vert 0\right\rangle _{A}\left\langle 0\right\vert _{A}\otimes
\left\vert 0\right\rangle _{B}\left\langle 0\right\vert _{B})+|\beta
|^{2}(\left\vert 1\right\rangle _{A}\left\langle 1\right\vert _{A}\otimes
\left\vert 0\right\rangle _{B}\left\langle 0\right\vert _{B})$, in which the
vacuum state for mode $A$ occurs with a probability $|\alpha |^{2}$ and the
one boson state for mode $A$ occurs with a probability $|\beta |^{2}$. In
this example the coherent superposition associated with the super-selection
rule violating state would not be detected in the interferometric process.
The paper by Dunningham et al \cite{Dunningham11a} considers first a
detection process that involves using a Glauber coherent state as one of the
input states. Similar interference effects as in Eq. (\ref{Eq.AtomDetnProb})
are obtained. A second detection process in which the single term Glauber
coherent state is replaced by a statistical mixture with all phases equally
weighted in considered next, leading to the same interference effects. This
again confirms that it is not necessary to invoke the existence of coherent
superpositions of number states in order to demonstate interference effects.
\pagebreak

\end{document}